\documentclass[dvips, journal]{IEEEtran}
\usepackage{amsmath,amsfonts}
\usepackage{algorithmic}
\usepackage{algorithm}
\usepackage{array}
\usepackage[caption=false,font=normalsize,labelfont=sf,textfont=sf]{subfig}
\usepackage{textcomp}
\usepackage{stfloats}
\usepackage{url}
\usepackage{verbatim}
\usepackage{graphicx}
\usepackage{cite}
\usepackage{balance}
\usepackage{enumitem}

\newcommand{\In}{\mathrm{in}}
\newcommand{\Out}{\mathrm{out}}
\newcommand{\ato}{\overset{\mathrm{a.s.}}{\to}}
\newcommand{\pto}{\overset{\mathrm{p}}{\to}}
\newcommand{\peq}{\overset{\mathrm{p}}{=}}
\newcommand{\pljto}{\overset{\mathrm{PL}(j)}{\to}}
\newcommand{\plto}{\overset{\mathrm{PL}(2)}{\to}}

\newcommand{\argmax}{\mathop{\mathrm{argmax}}\limits}
\newcommand{\argmin}{\mathop{\mathrm{argmin}}\limits}

\newtheorem{definition}{Definition}
\newtheorem{assumption}{Assumption}
\newtheorem{theorem}{Theorem}
\newtheorem{lemma}{Lemma}
\newtheorem{proposition}{Proposition}
\newtheorem{corollary}{Corollary}

\newtheorem{remark}{Remark}

\begin{document}

\title{Generalized Approximate Message-Passing for Compressed Sensing with Sublinear Sparsity}

\author{Keigo~Takeuchi,~\IEEEmembership{Member,~IEEE}
\thanks{
%Manuscript received April 19, 2021; revised August 16, 2021.
This work was supported in part by the Grant-in-Aid for Scientific 
Research~(B) (Japan Society for the Promotion of Science (JSPS)
KAKENHI) under Grant 23K20932. An earlier version of this 
paper will be presented in part at the 2025 IEEE International Conference on 
Acoustics, Speech and Signal Processing.
}
\thanks{K.~Takeuchi is with the Department of Electrical and Electronic Information Engineering, Toyohashi University of Technology, Toyohashi 441-8580, Japan (e-mail: takeuchi@ee.tut.ac.jp).}
}

\markboth{IEEE transactions on information theory}%
{Takeuchi: Generalized Approximate Message-Passing for Compressed Sensing with Sublinear Sparsity}

\IEEEpubid{0000--0000/00\$00.00~\copyright~2021 IEEE}

\maketitle

\begin{abstract}
This paper addresses the reconstruction of an unknown signal vector with sublinear sparsity from generalized linear measurements. Generalized approximate message-passing (GAMP) is proposed via state evolution in the sublinear sparsity limit, where the signal dimension $N$, measurement dimension $M$, and signal sparsity $k$ satisfy $\log k/\log N\to \gamma\in[0, 1)$ and $M/\{k\log (N/k)\}\to\delta$ as $N$ and $k$ tend to infinity. While the overall flow in state evolution is the same as that for linear sparsity, each proof step for inner denoising requires stronger assumptions than those for linear sparsity. The required new assumptions are proved for Bayesian inner denoising. When Bayesian outer and inner denoisers are used in GAMP, the obtained state evolution recursion is utilized to evaluate the prefactor $\delta$ in the sample complexity, called reconstruction threshold. If and only if $\delta$ is larger than the reconstruction threshold, Bayesian GAMP can achieve asymptotically exact signal reconstruction. In particular, the reconstruction threshold is finite for noisy linear measurements when the support of non-zero signal elements does not include a neighborhood of zero. As numerical examples, this paper considers linear measurements and 1-bit compressed sensing. Numerical simulations for both cases show that Bayesian GAMP outperforms existing algorithms for sublinear sparsity in terms of the sample complexity.
\end{abstract}


\section{Introduction}
\subsection{Compressed Sensing}
\IEEEPARstart{C}{onsider} the reconstruction of an $N$-dimensional 
$k$-sparse signal vector $\boldsymbol{x}$ from generalized linear 
measurements~\cite{Donoho06,Candes061,Barbier19,Ma23}  
\begin{equation} \label{measurement}
\boldsymbol{y} = g(\boldsymbol{z}, \boldsymbol{w}), \quad 
\boldsymbol{z} = \boldsymbol{A}\boldsymbol{x}. 
\end{equation}
In (\ref{measurement}), $\boldsymbol{y}\in\mathbb{R}^{M}$, 
$\boldsymbol{A}\in\mathbb{R}^{M\times N}$, and $\boldsymbol{w}\in\mathbb{R}^{M}$ 
denote the $M$-dimensional measurement vector, sensing matrix, and 
noise vector, respectively. The random variables $\{\boldsymbol{A}, 
\boldsymbol{x}, \boldsymbol{w}\}$ are independent. Throughout this paper, 
the signal vector is assumed to have the average power 
$\mathbb{E}[\|\boldsymbol{x}\|_{2}^{2}]=P$. Furthermore, 
$\boldsymbol{A}$ has independent and identically 
distributed (i.i.d.) standard Gaussian elements while the noise vector 
has the average power per element 
$M^{-1}\mathbb{E}[\|\boldsymbol{w}\|_{2}^{2}]=\sigma^{2}$. 
The nonlinearity $g:\mathbb{R}^{2}\to\mathbb{R}$ in (\ref{measurement}) 
allows us to treat practical issues, such as 1-bit compressed 
sensing~\cite{Boufounos08} and phase retrieval~\cite{Gerchberg72,Fienup82}. 
A goal of compressed sensing is to estimate the signal 
vector $\boldsymbol{x}$ from the knowledge about the measurement vector 
$\boldsymbol{y}$ and the sensing matrix $\boldsymbol{A}$.   

%linear sparsity
Theoretical research on compressed sensing has considered three asymptotic 
regions: linear sparsity $k/N\to\rho>0$, sublinear sparsity 
$\log k/\log N\to\gamma$ for some $\gamma\in[0, 1)$, 
and critical sublinear sparsity 
$k/N\to0$, $\log k/\log N\to1$ as all $k$, $M$, and $N$ tend to infinity. 
The two quantities $\log(N-k)$ and $\log(N/k)$ are different from each other 
in terms of the order for critical sublinear sparsity, which is outside  
the scope of this paper. See Remark~\ref{remark2} for the details. 
While the main focus of this paper is sublinear sparsity, for which 
$\log(N-k)={\cal O}(\log N)$ and $\log(N/k)={\cal O}(\log N)$ hold, 
prior research on linear sparsity is first reviewed to clarify the aim of 
this paper. 

\IEEEpubidadjcol

\subsection{Linear Sparsity}
\subsubsection{Linear Measurement}
%linear measurements 
For the linear measurement $g(z, w)=z+w$, several reconstruction algorithms 
were proposed, such as basis pursuit~\cite{Chen98}, iterative 
shrinkage-thresholding algorithm (ISTA)~\cite{Daubechies04}, 
fast ISTA (FISTA)~\cite{Beck09}, and iterative hard thresholding 
(IHT)~\cite{Blumensath09}. In particular, 
ISTA or FISTA---accelerated ISTA---is an iterative algorithm to realize  
the least absolute shrinkage and selection operator 
(Lasso)~\cite{Tibshirani96} while IHT solves a similar problem in which 
$\ell_1$ regularization in Lasso is replaced with $\ell_0$ regularization. 

The sample complexity of basis pursuit and Lasso was analyzed for linear 
sparsity. 
For basis pursuit, the replica method~\cite{Kabashima09}---non-rigorous tool 
in statistical mechanics---was used to show that, for the noiseless case 
$\boldsymbol{y}=\boldsymbol{A}\boldsymbol{x}$, there is a threshold 
$\alpha^{*}\in(0, 1)$ such that the mean-square error 
(MSE)\footnote{
For an estimator $\hat{\boldsymbol{x}}$ of $\boldsymbol{x}$,  
the MSE in this paper means the square error per the signal 
dimension $\|\boldsymbol{x} - \hat{\boldsymbol{x}}\|_{2}^{2}/N$ or its 
expectation. When the square error is not normalized, 
$\|\boldsymbol{x} - \hat{\boldsymbol{x}}\|_{2}^{2}$ is referred to as 
unnormalized square error. 
} converges to zero as all $k$, $M$, and $N$ tend to infinity while 
$k/N\to\rho>0$ and $M/N\to\alpha>\alpha^{*}$ hold. The threshold $\alpha^{*}$ 
is possible to evaluate numerically by solving saddle point 
equations~\cite{Kabashima09}. For Lasso, a similar but rigorous 
result was proved in \cite{Bayati12}. 

Approximate message-passing (AMP)~\cite{Donoho092} is a low-complexity and 
powerful algorithm for linear sparsity. AMP with the Bayes-optimal 
denoiser---called Bayes-optimal AMP---is regarded as an approximation of 
loopy belief propagation~\cite{Kabashima03}. 
Since AMP with suboptimal soft thresholding computes the Lasso 
solution~\cite{Bayati12}, Bayes-optimal AMP outperforms Lasso. 
Bayati and Montanari~\cite{Bayati11} 
proved the optimality~\cite{Reeves191,Barbier201} of Bayes-optimal AMP 
via state evolution, inspired by Bolthausen's conditioning 
technique~\cite{Bolthausen14}, when the state evolution recursion has a 
unique fixed point. In the conditioning technique, statistical 
properties of the current messages are evaluated via the conditional 
distribution of the sensing matrix given all previous messages. 

\subsubsection{Nonlinear Measurements}
Reconstruction algorithms for the linear measurement can be generalized 
to those for 1-bit compressed sensing: IHT, Lasso, and AMP were 
extended to binary IHT (BIHT)~\cite{Jacques13,Matsumoto22},  
generalized Lasso (GLasso)~\cite{Plan13,Plan16}, and generalized AMP 
(GAMP)~\cite{Rangan11,Kamilov12}, respectively. In particular, 
GLasso for nonlinear measurements was proved to be essentially equivalent 
to that for the linear measurement~\cite{Thrampoulidis15}. In this sense, 
1-bit compressed sensing does not change essential difficulties in 
compressed sensing for the linear measurement.  

The situation in phase retrieval is different from that in 1-bit 
compressed sensing. The main reason is that signal reconstruction 
reduces to non-convex optimization with multiple solutions. To circumvent 
this issue, qualitatively different algorithms were proposed, 
such as rank-1 matrix 
reconstruction~\cite{Chai11,Shechtman11,Candes13,Ohlsson12,Li13}, 
local search~\cite{Shechtman14,Candes15}, and alternating 
minimization~\cite{Netrapalli15}. Interestingly, AMP can be 
generalized to GAMP for phase retrieval~\cite{Schniter14}, which is regarded 
as a local search algorithm with multiple restarts to circumvent local minima. 
See \cite{Mondelli22} for spectral initialization in GAMP. 

GAMP~\cite{Rangan11} is a low-complexity and powerful algorithm 
for the generalized linear measurement~(\ref{measurement}). 
Like AMP, Bayes-optimal GAMP was 
proved to be optimal via state evolution~\cite{Rangan11,Javanmard13}. 
More precisely, the state evolution recursion for Bayes-optimal GAMP was 
proved in \cite{Cobo24} to converge toward a fixed point. 
See \cite{Takeuchi241} for an alternative proof via 
long-memory message-passing strategy~\cite{Takeuchi22}. Furthermore, 
fixed points of the state evolution 
recursion coincide with those of a potential function to characterize the 
Bayes-optimal performance~\cite{Barbier19}. Thus, Bayes-optimal GAMP can 
achieve the optimal performance when the state evolution recursion has a 
unique fixed point. 

\subsection{Sublinear Sparsity} 
\subsubsection{Linear Measurement}
Research on sublinear sparsity is next reviewed for the linear measurement. 
In a simple estimation problem of sparse signals for 
$\boldsymbol{A}=\boldsymbol{I}_{N}$, Donoho {\em et al.}~\cite{Donoho92} 
proved that soft-thresholding is optimal in terms of MSE 
for signals with sublinear sparsity---called nearly black signals. 
This result, as well as the equivalence between Lasso and AMP with soft 
thresholding~\cite{Bayati12}, implies effectiveness of Lasso for sublinear 
sparsity. See \cite{Johnstone04,Pas14,Rockova18} for Bayesian estimation of 
nearly black signals from Gaussian measurements with 
$\boldsymbol{A}=\boldsymbol{I}_{N}$. 

Information-theoretic analysis for compressed sensing was performed in 
\cite{Wainwright092,Fletcher09,Aeron10,Scarlett17,Gamarnik17,Reeves192}. 
These results imply that the optimum sample complexity for exact support 
recovery is $M={\cal O}(k\log(N/k))$ for sublinear sparsity. In particular, 
the optimum sample complexity was proved in \cite{Reeves192} to 
be $M/\{k\log (N/k)\}\to2/\log(1 + P/\sigma^{2})$ for constant non-zero signal 
elements in the high signal-to-noise ratio (SNR) regime 
$P/\sigma^{2}\to\infty$. 
Interestingly, basis pursuit~\cite{Donoho091} for the noiseless 
case and Lasso~\cite{Wainwright091} were proved to be optimal\footnote{
MSE is not appropriate for sublinear sparsity. For instance, consider the 
estimator $\hat{\boldsymbol{x}}=\boldsymbol{0}$, which achieves zero MSE 
$N^{-1}\|\boldsymbol{x}-\hat{\boldsymbol{x}}\|_{2}^{2}<Ck/N\to0$ for bounded 
non-zero elements $|x_{n}|<C$. Performance for sublinear 
sparsity should be measured with unnormalized quantities, such as 
unnormalized support recovery probability and unnormalized square error 
$\|\boldsymbol{x}-\hat{\boldsymbol{x}}\|_{2}^{2}$. Exact signal reconstruction 
means that unnormalized quantities tend to zero. In this sense, 
it makes no sense to compare \cite{Donoho091,Wainwright091} with 
\cite{Kabashima09,Bayati12}.
} in terms of the sample complexity scaling. See \cite{Okajima23} for 
another non-rigorous attempt to evaluate the prefactor in $k\log (N/k)$. 

Research on AMP for sublinear sparsity is limited. AMP was applied to 
rank-1 matrix reconstruction for critical sublinear sparsity in mean, i.e.\ 
$k={\cal O}(N(\log N)^{-\alpha})$ for $\alpha>0$~\cite{Barbier202}. 
AMP was proved to exhibit the so-called all-or-nothing phenomenon in terms of 
MSE. Truong~\cite{Truong23} proposed AMP for sublinear sparsity 
$k={\cal O}(N^{\alpha})$ and the sample complexity scaling 
$M={\cal O}(N^{\alpha})$ with some $\alpha\in(0, 1]$. State evolution was used 
to evaluate the asymptotic performance of AMP in terms of the square 
error divided by $N^{\alpha}$. However, exact signal reconstruction is not 
guaranteed for these AMP algorithms because the normalized performance 
measures were considered. 

\subsubsection{Nonlinear Measurements}
For an estimator $\hat{\boldsymbol{x}}$ of $\boldsymbol{x}$,  
the squared norm $\|\boldsymbol{x}/\|\boldsymbol{x}\|_{2} 
- \hat{\boldsymbol{x}}/\|\hat{\boldsymbol{x}}\|_{2}\|_{2}^{2}$ 
for the normalized error is an appropriate performance measure in 
1-bit compressed sensing when a priori information on the signal power
is not available. GLasso was proved to achieve the squared norm for the 
normalized error smaller than any $\epsilon>0$ 
if $M=C\epsilon^{-4}k\log (N/k)$ is satisfied for some 
$C>0$~\cite{Plan13,Plan16}. 
Similarly, BIHT can achieve the squared norm for the normalized error 
smaller than any $\epsilon>0$ 
for $M=C\epsilon^{-1}k\log(N/k)$~\cite{Matsumoto22}. Information theoretic 
analysis~\cite{Scarlett17} implies that there is some $C>0$ such that exact 
support recovery is impossible for all $M<Ck\log(N/k)$. Thus, GLasso and BIHT 
are optimal in terms of the sample complexity scaling in $k$ and  $N$.  

For phase retrieval, the generalized measurement~(\ref{measurement}) in the 
noiseless case was proved to be injective for $M\geq 4k-1$, i.e.\ 
$g(\boldsymbol{A}\boldsymbol{x},\boldsymbol{0})
\neq g(\boldsymbol{A}\boldsymbol{x}',\boldsymbol{0})$ holds 
for all $k$-sparse vectors $\boldsymbol{x}\neq\pm\boldsymbol{x}'$~\cite{Li13}.
This result does not contradict the optimum sample complexity 
$M={\cal O}(k\log(N/k))$ in the linear measurement~\cite{Wainwright092} 
since the injectivity does not imply exact phase recovery as 
$N\to\infty$. In fact, Eldar and Mendelson~\cite{Eldar14} proved that
$M>C(\log k + \log\log N)^{2}k\log (N/k)$ with some $C>0$ is sufficient for 
exact phase recovery. There is a slight gap between this sample complexity and 
that in practical algorithms, e.g.\ an algorithm in \cite{Li13} achieves 
exact phase recovery for $M={\cal O}(k^{2}\log (N/k))$. Since it is outside 
the scope of this paper, phase retrieval is not discussed anymore. 

In summary, existing non-AMP reconstruction algorithms, such as Lasso, are 
available for both linear and sublinear sparsity. However, AMP-type algorithms 
have been applied mainly to the case of linear sparsity. In existing state 
evolution~\cite{Rush18,Berthier19,Rangan192,Takeuchi20,Takeuchi21,Fan22}, 
as well as \cite{Barbier202,Truong23} for sublinear sparsity, 
the law of large numbers or its generalization~\cite{Truong23} in scaling 
was used to evaluate normalized performance measures, such as MSE. Thus, 
existing state evolution cannot produce meaningful results for unnormalized 
performance measures to guarantee exact signal reconstruction. To the best of 
author's knowledge, this paper presents the first rigorous result on exact 
signal reconstruction via GAMP for sublinear sparsity.  

\subsection{Contributions}
The main contributions of this paper are threefold: A first contribution is 
the formulation of GAMP for sublinear sparsity and 
its state evolution. As revealed in a second contribution, the denoiser 
in GAMP needs to handle outliers of Gaussian measurements 
based on extreme value theory while only normal observations around the mean 
are taken into account for conventional scaling. Nonetheless, the proposed 
GAMP has an equivalent formulation to that for linear sparsity from an 
algorithmic point of view. This implies the robustness of GAMP against 
the sparsity level of signals.

State evolution is used to analyze the asymptotic dynamics of GAMP in terms 
of the unnormalized square error. The overall flow in state evolution 
is essentially the same as that for linear sparsity. To justify each proof 
step for the denoiser, however, this paper utilizes extreme-value-theoretic 
results proved in a second contribution. When the support of non-zero 
signals does not include a neighborhood of zero, the all-or-nothing 
phenomenon occurs for noisy linear measurements: GAMP realizes asymptotically 
exact signal reconstruction in noisy compressed sensing for signals with 
sublinear sparsity if and only if $M/\{k\log (N/k)\}$ 
is larger than a reconstruction threshold $\delta^{*}>0$. 
The state evolution recursion can be utilized to evaluate the prefactor 
$\delta^{*}$ in the sample complexity scaling $M={\cal O}(k\log (N/k))$.  

A second contribution is Bayesian estimation of an unknown signal vector with 
sublinear sparsity from Gaussian measurements.  
The so-called spike and slab prior~\cite{Johnstone04} is utilized to formulate 
a Bayesian estimator, which is used as the denoiser in GAMP for sublinear 
sparsity. This paper proves extreme-value-theoretic results for the Bayesian 
estimator that are required in state evolution for GAMP. The results are 
qualitatively different from those in conventional scaling based on the law of 
large numbers. From a statistical-physics point of view, the difference can be 
understood as follows: Intensive quantities are evaluated in this paper 
while extensive quantities---proportional to the signal dimension and 
therefore require the normalization---were considered in conventional 
state evolution.

The last contributions are numerical results on GAMP for sublinear sparsity 
in the linear measurement and 1-bit compressed sensing. 
To the best of author's knowledge, normalized performance measures have been 
considered in existing numerical simulations for GAMP. This paper presents 
numerical results on the unnormalized square error of GAMP using Bayesian 
denoisers---called Bayesian GAMP. Numerical simulations show the superiority 
of Bayesian GAMP to existing algorithms for sublinear sparsity in terms of 
the sample complexity. 

Part of these contributions were presented in \cite{Takeuchi25}. 

\subsection{Organization}
The remainder of this paper is organized as follows: After summarizing the 
notation used in this paper, GAMP for sublinear sparsity is formulated 
in Section~\ref{sec3}. 
The main results of this paper are presented in Section~\ref{sec4}. 
State evolution is used for deriving dynamical systems---called state 
evolution recursion---to evaluate the asymptotic dynamics of GAMP 
for sublinear sparsity. 

Section~\ref{sec2} presents Bayesian estimation 
of an unknown signal vector with sublinear sparsity from Gaussian measurements. 
Technical results on the Bayesian estimation are proved to justify 
technical assumptions postulated in state evolution. 

Section~\ref{sec5} presents numerical results on Bayesian GAMP for sublinear 
sparsity. Numerical simulations show the superiority of Bayesian GAMP to 
existing algorithms for sublinear sparsity. 
Section~\ref{sec6} concludes this paper. Theoretical results in 
Sections~\ref{sec4} and \ref{sec2} are in part proved in the appendices. 

\subsection{Notation}
For a scalar function $f: \mathbb{R}\to\mathbb{R}$ and a vector 
$\boldsymbol{x}$, the notation $f(\boldsymbol{x})$ represents the 
element-wise application of $f$ to $\boldsymbol{x}$, i.e.\ 
$[f(\boldsymbol{x})]_{i}=f([\boldsymbol{x}]_{i})$. The indicator function 
is denoted by $1(\cdot)$ while $\boldsymbol{1}$ represents a vector of which 
the elements are all one. 
The Dirac delta function is written as $\delta(\cdot)$. 

The notation $\boldsymbol{X}\sim
\mathcal{N}(\boldsymbol{\mu}, \boldsymbol{\Sigma})$ means that a random 
vector $\boldsymbol{X}$ follows the Gaussian distribution with mean 
$\boldsymbol{\mu}$ and covariance $\boldsymbol{\Sigma}$. The probability 
density function (pdf) of a random variable $X$ is denoted by $p(X)$. 
In particular, $p_{\mathrm{G}}(X; v)$ represents the zero-mean Gaussian pdf 
with variance $v$. For a sequence of random variables $\{X_{n}\}$, 
the convergence in probability of $X_{n}$ to some $X$ is denoted by 
$X_{n}\pto X$ while $\peq$ represents the equivalence in probability, i.e.\ 
$X_{n}\peq X + o(1)$ means that for any $\epsilon>0$ the probability 
$\mathrm{Pr}(|X_{n} - X|>\epsilon)$ tends to zero as $n\to\infty$. Thus, 
$X_{n}\pto X$ is equivalent to $X_{n}\peq X + o(1)$. 

The transpose of a matrix $\boldsymbol{M}$ is denoted 
by $\boldsymbol{M}^{\mathrm{T}}$. The notation $M_{i,j}$ represents the 
$(i, j)$ element of $\boldsymbol{M}$. For a vector $\boldsymbol{v}_{t}$ with 
an index~$t$, the $n$th element of $\boldsymbol{v}_{t}$ is denoted by 
$v_{n,t}$. The norm $\|\cdot\|_{p}$ denotes the $\ell_{p}$ norm. The 
notation $\boldsymbol{o}(1)$ represents a vector with vanishing 
$\ell_{2}$ norm, i.e.\ $\boldsymbol{v}=\boldsymbol{o}(1)$ indicates  
$\|\boldsymbol{v}\|_{2}\to0$.

For a symmetric matrix $\boldsymbol{S}$, the minimum eigenvalue of 
$\boldsymbol{S}$ is denoted by $\lambda_{\mathrm{min}}(\boldsymbol{S})$.
For a matrix $\boldsymbol{M}$ with linearly independent columns, 
the pseudo-inverse of $\boldsymbol{M}$ is represented as 
$\boldsymbol{M}^{\dagger}=(\boldsymbol{M}^{\mathrm{T}}\boldsymbol{M})^{-1}
\boldsymbol{M}^{\mathrm{T}}$. The notation $\boldsymbol{P}_{\boldsymbol{M}}^{\parallel}
=\boldsymbol{M}\boldsymbol{M}^{\dagger}$ denotes the 
projection matrix onto the space spanned by the columns of $\boldsymbol{M}$ 
while $\boldsymbol{P}_{\boldsymbol{M}}^{\perp}=\boldsymbol{I} 
- \boldsymbol{P}_{\boldsymbol{M}}$ represents the projection onto its orthogonal 
complement.

\section{Generalized Approximate Message-Passing} \label{sec3}
GAMP is an iterative algorithm that exchanges messages between two 
modules---called inner and outer modules in this paper. In 
iteration~$t\in\{0,1,\ldots\}$ the outer module computes an estimator 
$\hat{\boldsymbol{z}}_{t}\in\mathbb{R}^{M}$ of 
$\boldsymbol{z}=\boldsymbol{A}\boldsymbol{x}$ in the generalized linear 
measurement~(\ref{measurement}) while the inner module computes an estimator 
$\hat{\boldsymbol{x}}_{t+1}\in\mathbb{R}^{N}$ of the signal 
vector $\boldsymbol{x}$. 

The main feature of GAMP is in the so-called Onsager correction. 
The linear mapping $\boldsymbol{A}\hat{\boldsymbol{x}}_{t}$ would be a naive 
estimator of $\boldsymbol{z}$ that can be computed with the message 
$\hat{\boldsymbol{x}}_{t}$ sent from the inner module. However, this naive 
approach results in an intractable distribution of the estimation error 
$\boldsymbol{A}\hat{\boldsymbol{x}}_{t} - \boldsymbol{z}$. Thus, the outer 
module in GAMP computes an Onsager-corrected message 
$\boldsymbol{z}_{t}\in\mathbb{R}^{M}$ to make the corresponding 
estimation error tractable. Similarly, the inner module computes Onsager 
correction $\boldsymbol{x}_{t}\in\mathbb{R}^{N}$ for the matched-filter 
output $\boldsymbol{A}^{\mathrm{T}}\hat{\boldsymbol{z}}_{t}$, which is a 
low-complexity estimator of $\boldsymbol{x}$. 

Let $f_{\Out, t}(z_{t}, y; v_{\In,t}): \mathbb{R}^{2}\to\mathbb{R}$ denote 
a denoiser used in the outer module with a parameter $v_{\In,t}>0$. 
The outer module in GAMP with the initial messages 
$\boldsymbol{z}_{0}=\boldsymbol{0}$ and 
$v_{\In,0}=\mathbb{E}[\|\boldsymbol{x}\|_{2}^{2}]$ iterates 
the following messages:  
\begin{equation} \label{z_t}
\boldsymbol{z}_{t} = \boldsymbol{A}\hat{\boldsymbol{x}}_{t}
+ \frac{\bar{\xi}_{\In, t-1}}{\xi_{\Out,t-1}}\hat{\boldsymbol{z}}_{t-1}, 
\end{equation}
\begin{equation} \label{v_in}
v_{\In,t} = v_{\Out,t-1}\bar{\xi}_{\In,t-1}, 
\end{equation}
\begin{equation} \label{z_hat_t}
\hat{\boldsymbol{z}}_{t} = f_{\Out,t}(\boldsymbol{z}_{t}, \boldsymbol{y}; 
v_{\In,t}), 
\end{equation}
\begin{equation} \label{xi_out}
\xi_{\Out,t} = \frac{1}{M}\boldsymbol{1}^{\mathrm{T}}
\frac{\partial f_{\Out,t}}{\partial z_{t}}(\boldsymbol{z}_{t}, \boldsymbol{y}; 
v_{\In,t}),
\end{equation}
where the partial derivative in (\ref{xi_out}) is taken with respect to 
the first variable. 
The message $v_{\In,t}$ corresponds to an estimator of the unnormalized 
square error $\|\hat{\boldsymbol{x}}_{t} - \boldsymbol{x}\|^{2}$ for 
GAMP, which is proved to coincide with the MSE 
$M^{-1}\|\boldsymbol{z}_{t} - \boldsymbol{z}\|^{2}$ via state evolution. 
The outer module sends the messages $\hat{\boldsymbol{z}}_{t}$ and 
$\xi_{\Out,t}$ to the inner module. 

Similarly, let $f_{\In,t}(x; v_{\Out,t}): \mathbb{R}\to\mathbb{R}$ denote a 
denoiser used in the inner module with a parameter $v_{\Out,t}$. The 
inner module in GAMP iterates the following 
messages with the initial message $\hat{\boldsymbol{x}}_{0}=\boldsymbol{0}$:
\begin{equation} \label{x_t}
\boldsymbol{x}_{t} = \hat{\boldsymbol{x}}_{t} 
- \frac{1}{M\xi_{\Out,t}}
\boldsymbol{A}^{\mathrm{T}}\hat{\boldsymbol{z}}_{t},
\end{equation}
\begin{equation} \label{v_out}
v_{\Out,t} = \frac{1}{M\xi_{\Out,t}^{2}}\left\|
 \hat{\boldsymbol{z}}_{t}
\right\|_{2}^{2}, 
\end{equation}
\begin{equation} \label{x_hat_t}
\hat{\boldsymbol{x}}_{t+1} = f_{\In,t}(\boldsymbol{x}_{t}; v_{\Out,t}). 
\end{equation}
The message $M^{-1}v_{\Out,t}$ corresponds to an estimator of the MSE 
$N^{-1}\|\boldsymbol{x}_{t} - \boldsymbol{x}\|^{2}$ for GAMP. In particular, 
$v_{\Out,t}$ in (\ref{v_out}) reduces to the well-known update rule 
$v_{\Out,t} = M^{-1}\|\boldsymbol{z}_{t} - \boldsymbol{y}\|_{2}^{2}$ 
for the linear measurement~\cite{Montanari12}. 
The inner module feeds the message $\hat{\boldsymbol{x}}_{t+1}$ back to the 
outer module to refine the estimation of $\boldsymbol{x}$. 

The deterministic parameter $\bar{\xi}_{\In,t-1}\in\mathbb{R}$ in (\ref{z_t}) 
and (\ref{v_in}) is designed shortly via state evolution. 
As a practical implementation of GAMP, this paper recommends 
the replacement of $\bar{\xi}_{\In,t}$ with 
\begin{equation} \label{xi_in}
\xi_{\In,t} = \frac{1}{M}\boldsymbol{1}^{\mathrm{T}}
f_{\In,t}'(\boldsymbol{x}_{t}, v_{\Out,t}), 
\end{equation}
where the derivative of $f_{\In,t}(x; v_{\Out,t})$ is taken with respect to $x$.  
The message $\xi_{\In,t}$ is expected to be a consistent estimator of 
$\bar{\xi}_{\In,t}$. 
However, this paper cannot prove the weak law of large numbers for 
$\xi_{\In,t}$ while the asymptotic unbiasedness of $\xi_{\In,t}$ can be 
justified, i.e.\ $\mathbb{E}[\xi_{\In,t}]\to\bar{\xi}_{\In,t}$. To analyze the 
dynamics of GAMP rigorously, the deterministic quantity $\bar{\xi}_{\In,t}$ 
is used in the formulation of GAMP, rather than $\xi_{\In,t}$ in (\ref{xi_in}). 

Interestingly, the Onsager correction in (\ref{z_t}) and (\ref{x_t}) is 
equivalent to that in GAMP for linear sparsity~\cite{Rangan11}, in which the 
sparsity~$k$, measurement dimension~$M$, and signal dimension~$N$ tend to 
infinity while $k/N\to\rho$ and $M/N\to\alpha$ hold for some $\rho\in(0, 1]$ 
and $\alpha>0$. On the other hand, 
GAMP in this paper postulates the sublinear sparsity 
limit---$k$, $M$, and $N$ tend to infinity with $M=\delta k\log (N/k)$ and 
$\log k/\log N\to \gamma$ for some $\delta>0$ and $\gamma\in[0, 1)$. 
When $\gamma>0$ holds, $k$ tends to infinity sublinearly in $N$ 
without assuming $k\to\infty$, i.e.\ $k={\cal O}(N^{\gamma})$. 
For $\gamma=0$, the condition $\log k/\log N\to0$ implies finite $k$ or 
$k=o(N)$. 

This paper excludes finite $k$ by assuming $k\to\infty$, as well as 
the critical sublinear sparsity $\gamma=1$. 
Finite sparsity $k<\infty$ never justifies the weak law of large 
numbers used for designing the Onsager correction in (\ref{z_t}). Therefore, 
GAMP is not applicable to finite sparsity~\cite{Okajima23}.

%\begin{remark} \label{remark1}
%The condition $\log k/\log N\to\gamma$ implies the sublinearity $k/N\to0$ for $\gamma\in[0, 1)$. In fact, for any $\epsilon\in(0, 1-\gamma)$ there is some $N_{0}\in\mathbb{N}$ such that $\log k/\log N<\gamma + \epsilon$ holds for all $N>N_{0}$. Selecting another $N_{0}'\in\mathbb{N}$ for any $\epsilon'>0$ such that $e^{-(1-\gamma - \epsilon)\log N_{0}'}<\epsilon'$ holds, for all $N>\max\{N_{0}, N_{0}'\}$ we have 
%\begin{equation}
%\frac{k}{N} 
%=e^{\log k - \log N} < e^{-(1-\gamma-\epsilon)\log N}
%< e^{-(1-\gamma-\epsilon)\log N_{0}'} < \epsilon'.
%\end{equation}
%Thus, $k/N\to0$ holds. 
%\end{remark}

\begin{remark} \label{remark2}
For the critical sublinear sparsity $\gamma=1$, 
$\log(N-k)$ and $\log(N/k)$ are different 
from each other in terms of the order. For instance, consider 
$\log k=\{1-(\log\log N)^{-1}\}\log N$, which implies $\log k/\log N\to1$. 
Furthermore, we have $\log(k/N)=-\log N/\log\log N\to-\infty$, which is 
equivalent to $k/N\to0$. Thus, this scaling is the critical sublinear sparsity. 
For this $k$, we find $\log(N/k) = \log N/\log\log N$ and 
$\log(N-k)=\log N + o(1)$, which are different from each other. It simplifies 
analysis of the inner denoiser not to distinguish the two quantities from 
each other. 
\end{remark}

All Lipschitz-continuous denoisers $f_{\Out,t}$ and $f_{\In,t}$ could 
be used for the linear sparsity~\cite{Rangan11}. However, the situation 
for the sublinear sparsity is different from that for the linear sparsity: 
It depends on $f_{\In,t}$ whether the unnormalized square error 
$\|f_{\In,t}(\boldsymbol{x}_{t}) - \boldsymbol{x}\|_{2}^{2}$ is bounded, 
while all Lipschitz-continuous outer denoisers $f_{\Out,t}$ can be 
used. For instance, $f_{\In,t}(x; v_{\Out,t})=N^{-1/4}$ results in diverging 
$\|f_{\In,t}(\boldsymbol{x}_{t}; v_{\Out,t}) - \boldsymbol{x}\|_{2}^{2}\to\infty$
in the sublinear sparsity limit while $f_{\In,t}(x; v_{\Out,t})
=0$ implies the boundedness of the unnormalized square error
under appropriate assumptions on $\boldsymbol{x}$. This paper postulates 
proper inner denoisers $f_{\In,t}$ such that the unnormalized square error 
is bounded in the sublinear sparsity limit. 

\section{Main Results} \label{sec4}
\subsection{Assumptions and Definitions}
The main result of this paper is state evolution for GAMP in the sublinear 
sparsity limit. Before presenting the main result, technical assumptions 
and definitions are summarized. 

\begin{assumption} \label{assumption_x}
The support $\mathcal{S}=\{n\in\{1,\ldots,N\} : x_{n}\neq0\}$ of the 
$k$-sparse signal vector $\boldsymbol{x}$ is a uniform and random sample of 
size~$k$ from $\{1,\ldots, N\}$ without replacement. 
Let $u_{n}=\sqrt{k}x_{n}$. The scaled non-zero elements 
$\{u_{n}: n\in\mathcal{S}\}$ are i.i.d.\ random variables. 
The common distribution of $u_{n}$---represented by a random variable $U$, 
i.e.\ $u_{n}\sim U$---does not depend on $N$ or $k$. Furthermore, 
$U$ has no probability mass at the origin and 
a second moment $\mathbb{E}[U^{2}]=P>0$. The cumulative distribution of $U$ 
is everywhere left-continuously or right-continuously differentiable and 
almost everywhere continuously differentiable.   
\end{assumption}

Existing AMP~\cite{Barbier202,Truong23} for sublinear sparsity postulated 
i.i.d.\ signal elements with vanishing occurrence probability of 
non-zero elements. As a result, the signal sparsity is a random variable. 
In this paper, on the other hand, the number of non-zero signal elements 
is exactly equal to $k$, as assumed in \cite{Reeves192}. 

No probability mass assumption $\mathrm{Pr}(U=0)=0$ at the origin is required 
not to change the signal sparsity $k$. 
The last assumption in Assumption~\ref{assumption_x} implies that 
$U$ can represent many practically important signals: $U$ has no singular 
components while it can have continuous or countable discrete 
components. The left-continuously or right-continuously differentiability 
implies the boundedness for the pdf of the continuous component. 

\begin{assumption} \label{assumption_A}
The sensing matrix $\boldsymbol{A}$ has independent standard Gaussian 
elements.  
\end{assumption}

The zero-mean and independence in Assumption~\ref{assumption_A} are important 
in the state evolution analysis of GAMP. As long as GAMP is used, 
the sensing matrix $\boldsymbol{A}$ cannot be weakened to non-zero 
mean~\cite{Caltagirone14} or ill-conditioned matrices~\cite{Rangan191}. 

Assumptions~\ref{assumption_x} and \ref{assumption_A} imply that each 
element of $\boldsymbol{z}=\boldsymbol{A}\boldsymbol{x}$ is ${\cal O}(1)$. 
Conventional normalization in the linear sparsity~\cite{Bayati11} is 
$\mathbb{E}[x_{n}^{2}]=P$ and $\mathbb{E}[A_{m,n}^{2}]=1/M$, which result in 
$z_{n}={\cal O}(1)$. For the sublinear sparsity, however, the power 
normalization is considered in the signal vector $\boldsymbol{x}$, i.e.\ 
the non-zero signal power $\mathbb{E}[x_{n}^{2}]=P/k$ for $n\in\mathcal{S}$ and 
$\mathbb{E}[A_{m,n}^{2}]=1$. This normalization 
is essentially equivalent to that in \cite{Reeves192} while \cite{Reeves192} 
evaluated the square error divided by $k$ for $\mathbb{E}[x_{n}^{2}]=P$ with 
$n\in\mathcal{S}$ and $\mathbb{E}[A_{m,n}^{2}]=1$. 

The normalization in this paper is different from that in AMP~\cite{Truong23}  
for sublinear sparsity with $k={\cal O}(N^{\alpha})$ and 
$M={\cal O}(N^{\alpha})$ for $\alpha\in(0, 1]$. In \cite{Truong23} 
the square error divided by $N^{\alpha}$ was evaluated for the non-zero 
signal power $P$ and $\mathbb{E}[A_{m,n}^{2}]=1/M$, which imply 
$\mathbb{E}[z_{n}^{2}]=kP/M={\cal O}(1)$. Owing to the normalization by 
$N^{\alpha}$ for strictly positive $\alpha\in(0, 1]$, a generalized law of 
large numbers~\cite{Truong23} could be used to extend conventional state 
evolution for the linear sparsity~\cite{Bayati11}. For the normalization in 
this paper, on the other hand, extreme-value-theoretic results in 
Section~\ref{sec2} are required to generalize state evolution for the 
linear sparsity. 

To assume properties of the noise vector, we need the notion of empirical 
convergence for normalized quantities~\cite{Bayati11}. Note that this 
empirical convergence is not used in analysis of the inner denoiser, which 
requires convergence for unnormalized quantities. 

\begin{definition}[Pseudo-Lipschitz Function]
A function $f:\mathbb{R}^{t}\to\mathbb{R}$ is said to be a pseudo-Lipschitz 
function of order~$j$ if there is some constant $L>0$ such that 
the following inequality holds for all 
$\boldsymbol{x}, \boldsymbol{y}\in\mathbb{R}^{t}$: 
\begin{equation}
|f(\boldsymbol{x}) - f(\boldsymbol{y})| \leq 
L(1 + \|\boldsymbol{x}\|_{2}^{j-1} + \|\boldsymbol{y}\|_{2}^{j-1})
\|\boldsymbol{x} - \boldsymbol{y}\|_{2}.
\end{equation}
\end{definition}

\begin{definition}[Empirical Convergence]
We say that the array of random vectors 
$\{\boldsymbol{z}_{1},\ldots,\boldsymbol{z}_{t}\in\mathbb{R}^{M}\}$ 
converges jointly to scalar random variables $\{Z_{1},\ldots,Z_{t}\}$ 
in the sense of $j$th-order pseudo-Lipschitz (PL) if 
the following convergence in probability holds:   
\begin{equation} \label{normalized_convergence}
\lim_{M\to\infty}\frac{1}{M}\sum_{m=1}^{M}f([\boldsymbol{z}_{1}]_{m},\ldots,
[\boldsymbol{z}_{t}]_{m}) 
\peq \mathbb{E}[f(Z_{1},\ldots,Z_{t})]
\end{equation}
for all piecewise $j$th-order pseudo-Lipschitz functions 
$f: \mathbb{R}^{t}\to\mathbb{R}$. 
The convergence~(\ref{normalized_convergence}) in the sense of 
$j$th-order PL is denoted by 
$(\boldsymbol{z}_{1},\ldots,\boldsymbol{z}_{t})
\pljto(Z_{1},\ldots,Z_{t})$. 
\end{definition}

\begin{assumption} \label{assumption_w}
The $\mathrm{PL}(2)$ convergence holds for the noise vector $\boldsymbol{w}$: 
$\boldsymbol{w}\plto W$ for some absolutely continuous random variable $W$ 
with variance $\sigma^{2}$. 
\end{assumption}

Assumption~\ref{assumption_w} is satisfied for i.i.d.\ absolutely continuous 
elements $w_{n}\sim W$ with variance $\sigma^{2}$. The absolute continuity 
is required to use piecewise pseudo-Lipschitz functions in the definition of 
the $\mathrm{PL}(2)$ convergence. If everywhere pseudo-Lipschitz functions 
were used, the absolute continuity could be weakened. 

\begin{assumption} \label{assumption_outer}
The composition $f_{\Out,t}(z_{t}, g(z, w); v_{\In,t})$ of the outer denoiser 
$f_{\Out,t}$ and the measurement function $g$ is a piecewise 
Lipschitz-continuous function of $(z_{t}, z, w)$. 
\end{assumption}

The Lipschitz-continuity is the standard assumption in state 
evolution~\cite{Bayati11}. The piecewiseness is required to treat practical 
measurement functions, such as clipping~\cite{Takeuchi241}. 
This relaxation does not affect conventional state evolution analysis since 
the composition $f_{\Out,t}(Z_{t}, g(Z, W); v_{\In,t})$ is considered for 
absolutely continuous random variables $Z_{t}$, $Z$, and $W$. In other words, 
$(Z_{t}, Z, W)$ occurs at singular points of the composition with zero 
probability. 

To present technical assumptions for the inner denoiser $f_{\In,t}$, we define 
state evolution recursion for GAMP. It is given via scalar zero-mean Gaussian 
random variables $Z$, $\{Z_{\tau}\}_{\tau=0}^{t}$ and zero-mean Gaussian random 
vectors $\{\boldsymbol{\omega}_{\tau}\in\mathbb{R}^{N}\}_{\tau=0}^{t}$, associated 
with $\boldsymbol{z}$, $\{\boldsymbol{z}_{\tau}\}_{\tau=0}^{t}$, and 
$\{\boldsymbol{x}_{\tau}\}_{\tau=0}^{t}$ 
in (\ref{measurement}), (\ref{z_t}), and (\ref{x_t}), respectively. 
The zero-mean Gaussian random variable $Z\sim\mathcal{N}(0, P)$ is 
independent of $W$ in Assumption~\ref{assumption_w}. 
The initial condition $Z_{0}=0$ is used. 

To define statistical properties of the other random variables, 
with the initial condition $\tilde{v}_{\In,0}=P$, 
we first define three variables $\bar{\xi}_{\Out,t}$, 
$\bar{v}_{\Out,t}$, and $\bar{\eta}_{t}$ as 
\begin{equation} \label{xi_out_bar}
\bar{\xi}_{\Out,t} = \mathbb{E}\left[
 \frac{\partial f_{\Out,t}}{\partial z_{t}}(Z_{t}, g(Z, W); \tilde{v}_{\In,t}) 
\right], 
\end{equation}
\begin{equation} \label{v_out_bar}
\bar{v}_{\Out,t} 
= \frac{1}{\bar{\xi}_{\Out,t}^{2}}\mathbb{E}\left[
 f_{\Out,t}^{2}\left(
  Z_{t}, g(Z, W); \tilde{v}_{\In,t}
 \right)
\right], 
\end{equation}
\begin{equation} \label{eta_bar}
\bar{\eta}_{t} = \frac{\bar{\zeta}_{t}}{\bar{\xi}_{\Out,t}},  
\end{equation}
with 
\begin{equation} \label{zeta_bar}
\bar{\zeta}_{t} 
= - \mathbb{E}\left[
 \left.
  \frac{\partial}{\partial u}f_{\Out,t}\left(
   Z_{t}, g(u, W); \tilde{v}_{\In,t}
  \right)
 \right|_{u=Z} 
\right],
\end{equation}
where $Z_{t}$ and $\tilde{v}_{\In,t}$ for $t>0$ are defined shortly. 
The variable $\bar{\xi}_{\Out,t}$ is 
the asymptotic alternative of $\xi_{\Out,t}$ in (\ref{xi_out}). The variable 
$\bar{v}_{\Out,t}$ is the asymptotic alternative of $v_{\Out,t}$ in (\ref{v_out}) 
and corresponds to the asymptotic MSE for the outer module. 
The variable $\bar{\eta}_{t}$ is used to represent effective signal amplitude 
in the inner module. 

We next define zero-mean Gaussian random vectors 
$\{\boldsymbol{\omega}_{\tau}\}_{\tau=0}^{t}$. They are independent 
of the signal vector $\boldsymbol{x}$ and have covariance 
\begin{align}
\mathbb{E}[\boldsymbol{\omega}_{\tau}\boldsymbol{\omega}_{t}^{\mathrm{T}}]
= \frac{1}{M\bar{\xi}_{\Out,\tau}\bar{\xi}_{\Out,t}}
&\mathbb{E}\left[
 f_{\Out,\tau}\left(
  Z_{\tau}, g(Z, W); \tilde{v}_{\In,\tau}
 \right)
\right. \nonumber \\
\cdot f_{\Out,t}&\left.
 \left(
  Z_{t}, g(Z, W); \tilde{v}_{\In,t}
 \right)
\right]\boldsymbol{I}_{N}. \label{covariance}
\end{align}
In particular, we have 
$\mathbb{E}[\boldsymbol{\omega}_{t}\boldsymbol{\omega}_{t}^{\mathrm{T}}]
=M^{-1}\bar{v}_{\Out,t}\boldsymbol{I}_{N}$ 
with $\bar{v}_{\Out,t}$ given in (\ref{v_out_bar}). 
Using $\boldsymbol{\omega}_{t}$, we define the variable 
$\bar{v}_{\In,t+1}$ in the inner module as  
\begin{equation} \label{v_in_bar}
\bar{v}_{\In, t+1} 
= \mu_{t+1, t+1} - 2\mu_{0,t+1} + P, 
\end{equation}
with 
\begin{align} 
\mu_{\tau+1, t+1} 
= \mathbb{E}\left[
 f_{\In,\tau}^{\mathrm{T}}\left(
  \bar{\eta}_{\tau}\boldsymbol{x} + \boldsymbol{\omega}_{\tau}; \bar{v}_{\Out,\tau}
 \right)
\right. \nonumber \\
\cdot\left. 
 f_{\In,t}\left(
  \bar{\eta}_{t}\boldsymbol{x} + \boldsymbol{\omega}_{t}; \bar{v}_{\Out,t}
 \right)
\right],\label{mu_tt}
\end{align}
\begin{equation} \label{mu_0t}
\mu_{0, t+1} 
= \mathbb{E}\left[
 \boldsymbol{x}^{\mathrm{T}}f_{\In,t}\left(
  \bar{\eta}_{t}\boldsymbol{x} + \boldsymbol{\omega}_{t}; \bar{v}_{\Out,t}
 \right)
\right],
\end{equation}
where $\bar{\eta}_{t}$ is defined in (\ref{eta_bar}). On the other hand, 
the variable $\tilde{v}_{\In,t+1}$ is given by
\begin{equation} \label{v_in_tilde}
\tilde{v}_{\In,t+1} = \bar{v}_{\Out,t}\bar{\xi}_{\In,t}, 
\end{equation}
with 
\begin{equation} \label{xi_in_bar}
\bar{\xi}_{\In,t}
= \lim_{N\to\infty}\frac{1}{M}\mathbb{E}\left[
 \boldsymbol{1}^{\mathrm{T}}
 f_{\In,t}'(\bar{\eta}_{t}\boldsymbol{x} + \boldsymbol{\omega}_{t}; 
 \bar{v}_{\Out,t})
\right],
\end{equation}
which is also used in the Onsager correction~(\ref{z_t}) of GAMP. 

The variable $\bar{v}_{\In, t+1}$ corresponds to the asymptotic unnormalized 
square error for the inner module while $\tilde{v}_{\In,t+1}$ is the 
asymptotic alternative of $v_{\In,t+1}$ in (\ref{v_in}). 

The state evolution result in this paper is qualitatively different from 
that in conventional state evolution~\cite{Bayati11,Truong23}. 
In conventional state evolution, the so-called decoupling 
principle holds: A scalar Gaussian measurement is used to represent the 
asymptotic MSE in the inner module. On the other hand, $\bar{v}_{\In,t+1}$ 
in (\ref{v_in_bar}) depends on the $N$-dimensional Gaussian measurement 
vector $\bar{\eta}_{t}\boldsymbol{x} + \boldsymbol{\omega}_{t}$ to obtain an 
extreme-value-theoretic result. 

Finally, we define the random variable $Z_{t+1}$,  
which is independent of $W$ and correlated with $Z$. More 
precisely, $Z_{t+1}$ is a zero-mean Gaussian random variable with 
covariance 
\begin{equation} 
\mathbb{E}[Z_{\tau+1}Z_{t+1}] 
= \mu_{\tau+1,t+1}, 
\label{Z_tt}
\end{equation}
\begin{equation}
\mathbb{E}[ZZ_{t+1}] 
= \mu_{0,t+1}, \label{Z_0t}
\end{equation}
with $\mu_{\tau+1,t+1}$ and $\mu_{0,t+1}$ given in (\ref{mu_tt}) and 
(\ref{mu_0t}), respectively. 

The definitions~(\ref{xi_out_bar})--(\ref{Z_0t}) provide state evolution 
recursion for GAMP. To solve $\bar{v}_{\In,t+1}$ in (\ref{v_in_bar}), we 
do not need the covariance~(\ref{covariance}) for $\tau\neq t$ or 
$\mu_{\tau+1,t+1}$ for $\tau\neq t$. They are needed to present assumptions 
for the inner denoiser.  

\begin{assumption} \label{assumption_SE}
There is some $T\in\mathbb{N}$ such that $\bar{v}_{\Out,t}>0$ and 
$\bar{v}_{\In,t+1}>0$ hold for all $t\in\{0,\ldots,T\}$.  
\end{assumption}

Assumption~\ref{assumption_SE} is a technical assumption to terminate GAMP 
iterations before exact signal reconstruction 
$\bar{v}_{\In,t_{0}+1}=0$ is achieved in some iteration $t_{0}\in\mathbb{N}$. 
Once exact signal reconstruction is achieved, Bolthausen's conditioning 
technique~\cite{Bolthausen14} cannot be applied to state evolution analysis 
for excessive iterations. In other words, state evolution analysis should be 
terminated just after exact signal reconstruction is achieved. This paper 
postulates Assumption~\ref{assumption_SE} explicitly while it was implicitly 
assumed in conventional state evolution. 

\begin{assumption} \label{assumption_inner}
Let $\nu_{N}\in\mathbb{R}$ denote a sequence that converges to zero 
as $N\to\infty$. Suppose that $\boldsymbol{a}\in\mathbb{R}^{N}$ is a random 
vector that is independent of 
$\{\boldsymbol{\omega}_{t'}, \boldsymbol{\omega}_{t}\}$ 
and satisfies $\lim_{N\to\infty}\mathrm{Pr}(\|\boldsymbol{a}\|_{2}<C)=1$ 
for some $C>0$. 
The inner denoiser $f_{\In,t}$ satisfies the following assumptions:
\begin{itemize}
\item The function $f_{\In,t}(\cdot; \bar{v}_{\Out,t})$ is almost everywhere 
differentiable with respect to the first variable. Furthermore, 
the sublinear sparsity limit in (\ref{xi_in_bar}) exists. 
\item The unnormalized error covariance 
$\{f_{\In,\tau}(\bar{\eta}_{\tau}\boldsymbol{x} + \boldsymbol{\omega}_{\tau}; 
\bar{v}_{\Out,\tau}) 
- \bar{\eta}_{\tau}\boldsymbol{x}\}^{\mathrm{T}}
\{f_{\In,t}(\bar{\eta}_{t}\boldsymbol{x} + \boldsymbol{\omega}_{t}; 
\bar{v}_{\Out,t}) 
- \bar{\eta}_{t}\boldsymbol{x}\}$ and $\boldsymbol{x}^{\mathrm{T}}
\{f_{\In,t}(\bar{\eta}_{t}\boldsymbol{x} + \boldsymbol{\omega}_{t}; 
\bar{v}_{\Out,t}) - \bar{\eta}_{t}\boldsymbol{x}\}$ converge in probability 
to their expectation in the sublinear sparsity limit for $\tau\in\{t, t'\}$, 
respectively. 
\item The difference $\|f_{\In,t}(\nu_{N}\boldsymbol{a} 
+ \bar{\eta}_{t}\boldsymbol{x} + \boldsymbol{\omega}_{t}; \bar{v}_{\Out,t})
- f_{\In,t}(\bar{\eta}_{t}\boldsymbol{x} + \boldsymbol{\omega}_{t}; 
\bar{v}_{\Out,t})\|_{2}$ converges in probability to zero 
in the sublinear sparsity limit. 
\item The difference 
$(\boldsymbol{\omega}_{\tau})^{\mathrm{T}}f_{\In,t}(\nu_{N}\boldsymbol{a} 
+ \bar{\eta}_{t}\boldsymbol{x} + \boldsymbol{\omega}_{t}; \bar{v}_{\Out,t})
- \mathbb{E}[\boldsymbol{\omega}_{\tau}^{\mathrm{T}}f_{\In,t}(
\bar{\eta}_{t}\boldsymbol{x} + \boldsymbol{\omega}_{t}; \bar{v}_{\Out,t})]$ 
converges in probability to zero 
in the sublinear sparsity limit for $\tau\in\{t, t'\}$. 
\end{itemize} 
\end{assumption}

Assumption~\ref{assumption_inner} is the main assumption of this paper. 
When a Bayesian estimator introduced in Section~\ref{sec2} is used as 
the inner denoiser $f_{\In,t}$, Assumption~\ref{assumption_inner} can be 
justified under mild assumptions. 
However, it is a challenging open issue to specify sufficient conditions 
for the inner denoiser $f_{\In,t}$ to prove  
Assumption~\ref{assumption_inner}. This paper has postulated 
Assumption~\ref{assumption_inner} to separate analysis of the inner denoiser 
from state evolution. Assumption~\ref{assumption_inner} should 
be justified for individual inner denoisers.  

\subsection{State Evolution} 
The following theorem provides state evolution results for GAMP with 
$\bar{\xi}_{\In,t}$ in (\ref{xi_in_bar}): 

\begin{theorem} \label{theorem_SE}
Define $\bar{\xi}_{\In,t}$ in (\ref{z_t}) and (\ref{v_in}) as (\ref{xi_in_bar}) 
and suppose that Assumptions~\ref{assumption_x}--\ref{assumption_inner} hold. 
Then, the unnormalized square error 
$\|\hat{\boldsymbol{x}}_{t+1} - \boldsymbol{x}\|_{2}^{2}$ for GAMP 
converges in probability to $\bar{v}_{\In,t+1}$---given in 
(\ref{v_in_bar})---in the sublinear sparsity limit for all 
$t\in\{0,\ldots, T\}$. 
\end{theorem}
\begin{IEEEproof}
See Appendix~\ref{proof_theorem_SE}. 
\end{IEEEproof}

Theorem~\ref{theorem_SE} implies asymptotic Gaussianity for the 
estimation errors of $f_{\In,t}$ that satisfies 
Assumption~\ref{assumption_inner}. More precisely, the unnormalized 
square error $\|\hat{\boldsymbol{x}}_{t+1} - \boldsymbol{x}\|_{2}^{2}$ can 
be described as $\bar{v}_{\In,t+1}$ in (\ref{v_in_bar}), which is defined with 
the effective Gaussian noise vector $\boldsymbol{\omega}_{t}$ as shown in 
(\ref{mu_tt}) and (\ref{mu_0t}). 

For the linear sparsity~\cite{Rangan11,Javanmard13,Takeuchi241}, the asymptotic 
Gaussianity was proved for all Lipschitz-continuous denoisers in 
terms of the $\mathrm{PL}(2)$ convergence. On the other hand, the asymptotic 
Gaussianity for the sublinear sparsity is used in a limited sense: The inner 
denoiser $f_{\In,t}$ needs to satisfy Assumption~\ref{assumption_inner}. 
Furthermore, only the unnormalized error covariance is considered as a 
second-order pseudo-Lipschitz function. It is a challenging open issue to 
specify the class of inner denoisers and a subset of second-order 
pseudo-Lipschitz functions that realize the asymptotic Gaussianity 
for the sublinear sparsity. Such research elucidates the precise meaning 
of asymptotic Gaussianity for the sublinear sparsity in probability theory. 

\section{Bayesian GAMP} \label{sec2}
\subsection{Bayesian Estimation of Nearly Black Signals} 
\subsubsection{Gaussian Measurements}
To design the inner denoiser that satisfies 
Assumption~\ref{assumption_inner}, this section addresses 
the Bayesian estimation of an unknown $N$-dimensional signal vector with 
sublinear sparsity---called nearly black signals~\cite{Donoho92}. 

Consider the estimation of a $k$-sparse signal vector 
$\boldsymbol{X}=[X_{1},\ldots,X_{N}]^{\mathrm{T}}$ with $X_{n}=k^{-1/2}A_{n}U_{n}$ 
from two kinds of correlated Gaussian measurements, 
\begin{align} 
\boldsymbol{Y} &= \boldsymbol{X} + \boldsymbol{\Omega}, 
\quad 
\boldsymbol{\Omega}\sim\mathcal{N}(\boldsymbol{0}, v_{k,N}\boldsymbol{I}_{N}), 
\nonumber \\
\boldsymbol{Y}' &= \boldsymbol{X} + \boldsymbol{\Omega}', 
\quad 
\boldsymbol{\Omega}'\sim\mathcal{N}(\boldsymbol{0}, v_{k,N}'\boldsymbol{I}_{N}), 
\label{AWGN_measurement}
\end{align}
with $v_{k,N}=v/\{k\log(N/k)\}$ and $v_{k,N}'=v'/\{k\log(N/k)\}$ 
for $v>0$ and $v'>0$. To circumvent division by zero, $N\geq2$ and 
$1\leq k<N$ are assumed. The jointly Gaussian-distributed noise vectors 
$\boldsymbol{\Omega}$ and $\boldsymbol{\Omega}'$ have the covariance matrix 
$\mathbb{E}[\boldsymbol{\Omega}'\boldsymbol{\Omega}^{\mathrm{T}}]
=\mathrm{cov}_{k,N}\boldsymbol{I}_{N}$ with 
$\mathrm{cov}_{k,N}=\mathrm{cov}/\{k\log(N/k)\}$ for 
$\mathrm{cov}\in\mathbb{R}$. Since the covariance matrix has to be 
non-negative definite, we have the inequality $vv' \geq \mathrm{cov}^{2}$. 

The measurement vectors $\boldsymbol{Y}$ and $\boldsymbol{Y}'$ correspond 
to $\boldsymbol{x} + \bar{\eta}_{t}^{-1}\boldsymbol{\omega}_{t}$ and 
$\boldsymbol{x} + \bar{\eta}_{\tau}^{-1}\boldsymbol{\omega}_{\tau}$ 
in Assumption~\ref{assumption_inner}, respectively, while 
$\boldsymbol{X}$ is associated with the signal vector $\boldsymbol{x}$. 
Furthermore, $\boldsymbol{\Omega}$ and $\boldsymbol{\Omega}'$ correspond 
to $\bar{\eta}_{t}^{-1}\boldsymbol{\omega}_{t}$ and 
$\bar{\eta}_{\tau}^{-1}\boldsymbol{\omega}_{\tau}$, so that we have 
$v=\bar{v}_{\Out,t}/(\delta\bar{\eta}_{t}^{2})$ and 
$v'=\bar{v}_{\Out,\tau}/(\delta\bar{\eta}_{\tau}^{2})$, 
because of $M=\delta k\log(N/k)$ and the covariance 
in (\ref{covariance}). 

As postulated in Assumption~\ref{assumption_x}, 
the binary variables $\mathcal{A}=\{A_{n}\in\{0, 1\}\}_{n=1}^{N}$ 
represent the support of the signal vector: $\mathcal{A}$ is assumed to be 
uniformly distributed on the space $\{\boldsymbol{a}\in\{0, 1\}^{N} : 
\boldsymbol{1}^{\mathrm{T}}\boldsymbol{a}=k\}$ with exactly $k$ ones. 
The other variables $\mathcal{U}=\{U_{n}\}_{n=1}^{N}$ are 
i.i.d.\ random variables that represent non-zero signals. 
The common distribution of $U_{n}$ is represented with a random variable 
$U$, i.e.\ $U_{n}\sim U$. The goal is to formulate a 
Bayesian estimator of $\boldsymbol{X}$ given $\boldsymbol{Y}$ or 
$\boldsymbol{Y}'$ and prove Assumption~\ref{assumption_inner} for the 
Bayesian estimator. 

\subsubsection{Bayesian Estimator}
To reduce the complexity of estimation, the prior distribution of $\mathcal{A}$ 
is approximated with i.i.d.\ distributions 
$\mathrm{Pr}(\boldsymbol{A}=\boldsymbol{a})\approx\prod_{n=1}^{N}
\mathrm{Pr}(A_{n}=a_{n})$, 
\begin{equation} \label{probability_A}
\mathrm{Pr}(A_{n}=a) 
= \begin{cases}
\rho & \hbox{for $a=1$,} \\
1-\rho & \hbox{for $a=0$,} 
\end{cases}
\end{equation}
with $\rho=k/N$. This approximation results in the so-called spike and 
slab prior~\cite{Johnstone04}. 
Note that the approximation is used only in the definition 
of estimation. In evaluating its properties, we use the true 
distribution of $\mathcal{A}$. 

Under this approximation, the estimation problem of 
$\boldsymbol{X}$ based on the measurement $\boldsymbol{Y}$ 
is separated into individual problems for all $n\in\{1,\ldots,N\}$, 
\begin{equation} \label{AWGN_measurement_each}
Y_{n} = X_{n} + \Omega_{n}, \quad \Omega_{n}\sim\mathcal{N}(0, v_{k,N}). 
\end{equation}
A Bayesian estimator is defined as the posterior mean estimator 
$\mathbb{E}[X_{n} | Y_{n}]$ of $X_{n}$ given $Y_{n}$. The Bayesian estimator 
is not called Bayes-optimal estimator since it is different from the true 
posterior mean estimator $\mathbb{E}[X_{n} | \boldsymbol{Y}]$. 
To formulate the posterior mean estimator, as well as the posterior variance, 
we evaluate the posterior $i$th moment of $X_{n}$ given $Y_{n}$, 
\begin{align}
\mathbb{E}[X_{n}^{i} | Y_{n}] 
&= \mathrm{Pr}(A_{n}=1 | Y_{n})\mathbb{E}[X_{n}^{i} | Y_{n}, A_{n}=1] 
\nonumber \\
&+ \mathrm{Pr}(A_{n}=0 | Y_{n})\mathbb{E}[X_{n}^{i} | Y_{n}, A_{n}=0]
\nonumber \\
&= k^{-i/2}\mathrm{Pr}(A_{n}=1 | Y_{n})\mathbb{E}[U_{n}^{i} | Y_{n}, A_{n}=1],
\label{posterior_mean_estimator_X_tmp}
\end{align}
where the last equality follows from $X_{n}=k^{-1/2}A_{n}U_{n}$. 

The posterior $i$th moment $\mathbb{E}[U_{n}^{i} | Y_{n}, A_{n}=1]$ 
reduces to that of $U$ given the scalar Gaussian measurement 
$Y=k^{-1/2}U+\Omega$ with 
$\Omega\sim\mathcal{N}(0, v_{k,N})$. Let $\tilde{Y}=\sqrt{k}Y$ and 
$\tilde{\Omega}=\sqrt{k}\Omega$, which result in $\tilde{Y}=U+\tilde{\Omega}$. 
Since $\tilde{\Omega}\sim\mathcal{N}(0,v_{\tilde{N}})$ holds, with 
$\tilde{N}=N/k$ and $v_{\tilde{N}}=v/\log \tilde{N}$, we have 
\begin{equation} \label{posterior_mean_estimator_U}
\mathbb{E}[U_{n}^{i} | Y_{n}, A_{n}=1]
= \mathbb{E}[U^{i} | \tilde{Y}=\sqrt{k}Y_{n}]
\equiv f_{U}^{(i)}(\sqrt{k}Y_{n}; v_{\tilde{N}}). 
\end{equation} 
In particular, we use the simplified notation $f_{U}^{(1)}=f_{U}$. 

For the posterior probability of $A_{n}$, 
under the approximation for $\mathcal{A}$ we have 
$\mathrm{Pr}(A_{n}=1|Y_{n}=y)=f_{A}(\sqrt{k}y; v_{\tilde{N}})$, given by
\begin{align} 
&f_{A}(\sqrt{k}y; v_{\tilde{N}}) 
\nonumber \\
&= \frac{k \mathbb{E}_{U}[p_{\mathrm{G}}(\sqrt{k}y - U; v_{\tilde{N}})]}
{k \mathbb{E}_{U}[p_{\mathrm{G}}(\sqrt{k}y - U; v_{\tilde{N}})] 
+ (N - k)p_{\mathrm{G}}(\sqrt{k}y; v_{\tilde{N}})}. 
\label{posterior_mean_estimator_A}
\end{align}
Before proving (\ref{posterior_mean_estimator_A}), we 
substitute (\ref{posterior_mean_estimator_U}) and 
(\ref{posterior_mean_estimator_A}) into (\ref{posterior_mean_estimator_X_tmp}) 
to find that the posterior $i$th moment of $X_{n}$ conditioned on $Y_{n}$ 
is given by $\mathbb{E}[X_{n}^{i} | Y_{n}=y]=f_{X}^{(i)}(y; v_{\tilde{N}})$, 
\begin{equation} \label{posterior_mean_estimator_X}
f_{X}^{(i)}(y; v_{\tilde{N}}) =  k^{-i/2}f_{A}(\sqrt{k}y; v_{\tilde{N}})
f_{U}^{(i)}(\sqrt{k}y; v_{\tilde{N}}).
\end{equation} 
In particular, we write the posterior mean $f_{X}^{(1)}$ simply as $f_{X}$. 

We prove (\ref{posterior_mean_estimator_A}) under the 
approximation for $\mathcal{A}$. Using Bayes' rule yields  
\begin{equation}
\mathrm{Pr}(A_{n}=1|Y_{n}) = \frac{\rho p(Y_{n}|A_{n}=1)}
{\rho p(Y_{n}|A_{n}=1) + (1 - \rho)p(Y_{n}|A_{n}=0)}, 
\end{equation}
with $\rho=k/N$. Since $Y_{n}=\Omega_{n}$ holds for $A_{n}=0$, we have  
\begin{equation}
p(Y_{n}=y|A_{n}=0) = p_{\mathrm{G}}(y; v_{k,N})
= \sqrt{k}p_{\mathrm{G}}(\sqrt{k}y; v_{\tilde{N}}),
\end{equation}
where the last equality follows from $v_{k,N}=v_{\tilde{N}}/k$. 
Similarly, using $Y_{n}=k^{-1/2}U_{n} + \Omega_{n}$ for $A_{n}=1$ yields 
\begin{align}
p(Y_{n}=y|A_{n}=1) 
&= \mathbb{E}_{U}\left[
 p_{\mathrm{G}}(y - k^{-1/2}U; v_{k,N}) 
\right]
\nonumber \\
&= \sqrt{k}\mathbb{E}_{U}\left[
 p_{\mathrm{G}}(\sqrt{k}y - U; v_{\tilde{N}}) 
\right].
\end{align}
Combining these results, we arrive at (\ref{posterior_mean_estimator_A}). 

\subsubsection{Derivative of the Bayesian Estimator} 
GAMP utilizes the derivative $f_{X}'$ of the Bayesian 
estimator~(\ref{posterior_mean_estimator_X}) with respect to $y$ 
for the Onsager correction. More precisely, the following quantity is used: 
\begin{equation} \label{xi_def}
\xi(\boldsymbol{Y}; v_{\tilde{N}}) = \frac{1}{\delta' k\log(N/k)}
\sum_{n=1}^{N}f_{X}'(Y_{n}; v_{\tilde{N}})
\end{equation}
for some $\delta'>0$, as shown in $\bar{\xi}_{\In,t}$ in (\ref{xi_in_bar}).  

We evaluate the quantity~(\ref{xi_def}). 
Since (\ref{posterior_mean_estimator_X}) is the posterior mean estimator of 
$X_{n}$ given the Gaussian measurement $Y_{n}=y$ in 
(\ref{AWGN_measurement_each}), we know that $f_{X}'$ is equal to the posterior 
variance divided by $v_{k,N}$~\cite{Rangan11}, i.e.\ 
$f_{X}'(y; v_{\tilde{N}})=v_{\mathrm{post}}(y; v_{\tilde{N}})/v_{k,N}$, 
\begin{equation} \label{posterior_variance_X}
v_{\mathrm{post}}(y; v_{\tilde{N}}) = f_{X}^{(2)}(y; v_{\tilde{N}}) 
- f_{X}^{2}(y; v_{\tilde{N}}),
\end{equation} 
with $f_{X}^{(i)}$ in (\ref{posterior_mean_estimator_X}). 
Substituting this expression into the definition of 
$\xi(\boldsymbol{Y}; v_{\tilde{N}})$ 
in (\ref{xi_def}) and using $v_{k,N}=v/\{k\log(N/k)\}$, we have 
\begin{equation} \label{xi}
\xi(\boldsymbol{Y}; v_{\tilde{N}}) = \frac{1}{\delta' v}\boldsymbol{1}^{\mathrm{T}}
v_{\mathrm{post}}(\boldsymbol{Y}; v_{\tilde{N}}). 
\end{equation}
In particular, the expectation $\mathbb{E}[\xi(\boldsymbol{Y}; v_{\tilde{N}})]$ 
is equal to the average unnormalized square error $\mathbb{E}[\|\boldsymbol{X} 
- f_{X}(\boldsymbol{Y}; v_{\tilde{N}})\|_{2}^{2}]/(\delta' v)$ divided by 
$\delta v$, which is proved to exist shortly. Thus, the first assumption 
in Assumption~\ref{assumption_inner} is justified when the Bayesian 
estimator $f_{X}$ in (\ref{posterior_mean_estimator_X}) is used 
as the inner denoiser.

\subsubsection{Properties} 
To justify the last three assumptions in Assumption~\ref{assumption_inner} 
for the inner denoiser, let $\nu_{N}\in\mathbb{R}$ denote a sequence that 
converges to zero as $N\to\infty$. Furthermore, we assume that a random vector 
$\boldsymbol{a}\in\mathbb{R}^{N}$ is independent of 
$\{\boldsymbol{\Omega}, \boldsymbol{\Omega}'\}$ and satisfies 
$\lim_{N\to\infty}\mathrm{Pr}(\|\boldsymbol{a}\|_{2}<C)=1$ for some $C>0$. 
We investigate properties of the Bayesian estimator for the mismatched 
measurement $\nu_{N}\boldsymbol{a} + \boldsymbol{Y}$ in the sublinear sparsity 
limit: $N$ and $k$ tend to infinity while 
$\log k/\log N$ converges to some $\gamma\in[0, 1)$. 

Before presenting the main results in this section, we summarize 
technical assumptions. 

\begin{assumption} \label{assumption_lemma}
\begin{itemize}
\item The random variable $U$ has no probability mass at the origin and a 
bounded second moment. The cumulative distribution of $U$ 
is everywhere left-continuously or right-continuously differentiable and 
almost everywhere continuously differentiable.
\item The function $f_{U}(\cdot; v)$ in 
(\ref{posterior_mean_estimator_U}) is uniformly Lipschitz-continuous 
for all $v>0$: There is some constant $L>0$ such that 
for all $y, y'\in\mathbb{R}$, and $v>0$ the following inequality holds: 
\begin{equation}
|f_{U}(y; v) - f_{U}(y'; v)| < L|y - y'|. 
\end{equation}
\item The function $f_{U}$ is consistent in the following sense: 
\begin{equation}
|y - f_{U}(y ;v)| \leq o(|y|)
\quad \hbox{as $v\to0$}
\end{equation} 
for all $y\in\mathbb{R}$. 
\item The function $f_{U}^{(2)}(\cdot; v)$ in 
(\ref{posterior_mean_estimator_U}) is uniformly bounded on any bounded 
interval $\mathcal{D}$ with respect to $v>0$: There is some constant 
$\bar{f}_{U}^{(2)}>0$ such that $|f_{U}^{(2)}(y; v)|<\bar{f}_{U}^{(2)}$ holds 
for all $y\in\mathcal{D}$ and $v>0$. 
\end{itemize}
\end{assumption}

The first assumption in Assumption~\ref{assumption_lemma} is equivalent to 
the corresponding assumption in Assumption~\ref{assumption_x}. 
The second assumption implies that $f_{U}(\cdot; v_{\tilde{N}})$ with 
$v_{\tilde{N}}=v/\log(N/k)$ is uniformly Lipschitz-continuous with respect to 
$N$ and $k$. This uniform Lipschitz-continuity is used to evaluate properties 
of the Bayesian estimator. The third assumption is slightly stronger than 
another definition of consistency $\lim_{v\to0}f_{U}(y; v) = y$. 
The non-trivial part in the last assumption is the uniform boundedness with 
respect to $v$ while the continuity of $f_{U}^{(2)}$ implies the boundedness 
on any bounded interval for each $v$. 

The last three assumptions are satisfied 
for Gaussian $U\sim\mathcal{N}(0,\sigma_{U}^{2})$. In this case, 
we have $f_{U}(y; v)= \sigma_{U}^{2}y/(\sigma_{U}^{2} + v)$ and 
$f_{U}^{(2)}(y; v) - f_{U}^{2}(y; v) = \sigma_{U}^{2}v/(\sigma_{U}^{2} + v)$. 
The function $f_{U}(y; v)$ satisfies the uniform Lipschitz-continuity with 
$L=1$ for all $v>0$ and the consistency 
$|y - f_{U}(y; v)|=v|y|/(\sigma_{U}^{2} + v)=o(|y|)$ as $v\to0$. Furthermore, 
we find the uniform boundedness $f_{U}^{(2)}(y; v)< \sigma_{U}^{2} 
+ y^{2}\leq \sigma_{U}^{2} + \max\{(\inf\mathcal{D})^{2}, 
(\sup\mathcal{D})^{2}\}$ for all $y\in\mathcal{D}$.  

To prove the second assumption in Assumption~\ref{assumption_inner}, 
we first evaluate the unnormalized square error for the mismatched 
measurement while a minimax normalized square error was evaluated for 
spike and slab priors~\cite{Johnstone04}. 

\begin{lemma} \label{lemma_MSE}
Suppose that Assumption~\ref{assumption_lemma} holds.   
If $\nu_{N}=0$ or $\mathrm{Pr}(|U|=\sqrt{2v})=0$ holds, then 
the unnormalized square error converges in probability to 
\begin{equation} \label{unnormalized_square_error}
\lim_{N\to\infty}\left\|
 \boldsymbol{X} - f_{X}(\nu_{N}\boldsymbol{a} + \boldsymbol{Y}; v_{\tilde{N}})
\right\|_{2}^{2}
\peq \mathbb{E}\left[
 U^{2}1(U^{2}<2v)
\right]
\end{equation}
in the sublinear sparsity limit. 
\end{lemma}
\begin{IEEEproof}
Since $\mathcal{A}$ is a uniform and random sample of size~$k$ from 
$\{1,\ldots, N\}$ without replacement, we use the Gaussian  
measurement~(\ref{AWGN_measurement}) to represent 
the unnormalized square error~(\ref{unnormalized_square_error}) as 
\begin{align}
\left\|
 \boldsymbol{X} - f_{X}(\nu_{N}\boldsymbol{a} + \boldsymbol{Y}; v_{\tilde{N}})
\right\|_{2}^{2}
&= \sum_{n\notin\mathcal{S}}f_{X}^{2}(\nu_{N}a_{n} + \Omega_{n}; v_{\tilde{N}})
\nonumber \\
&+ \frac{1}{k}\sum_{n\in\mathcal{S}}\Delta U_{n}^{2}(v_{\tilde{N}}), 
\label{unnormalized_square_error_tmp}
\end{align}
with 
\begin{equation} \label{Delta_U_n}
\Delta U_{n}(v_{\tilde{N}}) 
=  U_{n} - \sqrt{k}f_{X}(\nu_{N}a_{n} + k^{-1/2}U_{n} + \Omega_{n}; v_{\tilde{N}}), 
\end{equation}
where $\mathcal{S}\subset\{1,\ldots,N\}$ denotes the support of 
$\boldsymbol{X}$ with cardinality $|\mathcal{S}|=k$. 
Lemma~\ref{lemma_MSE} is obtained by proving the following results 
in the sublinear sparsity limit: 
\begin{equation} \label{former_limit_general}
\sum_{n\notin\mathcal{S}}\mathbb{E}\left[
 \left.
  |\Omega_{n}|^{j}f_{X}^{2}(\nu_{N}a_{n} + \Omega_{n}; v_{\tilde{N}})
 \right| \boldsymbol{a}
\right] 
\pto 0
\end{equation}
for all non-negative integers $j\in\{0, 1,\ldots\}$ and 
\begin{equation}
\frac{1}{k}\sum_{n\in\mathcal{S}}\Delta U_{n}^{2}(v_{\tilde{N}})
\pto \mathbb{E}\left[
 U^{2}1(U^{2}<2v)
\right] 
\label{latter_limit_p}
\end{equation}
in the sublinear sparsity limit.  

We prove the more general result~(\ref{former_limit_general}) than 
that for $j=0$, which is utilized in proving another lemma. 
The convergence~(\ref{former_limit_general}) in mean implies 
the convergence in probability for $j=0$, 
\begin{equation} \label{former_limit_p}
\sum_{n\notin\mathcal{S}}f_{X}^{2}(\nu_{N}a_{n} + \Omega_{n}; v_{\tilde{N}})
\pto 0. 
\end{equation}
See Appendix~\ref{proof_former_limit} for the proof of 
(\ref{former_limit_general}) without assuming $\nu_{N}=0$ or 
$\mathrm{Pr}(|U|=\sqrt{2v})=0$.  

The former convergence~(\ref{former_limit_general}) implies that false positive 
events---the estimator of $A_{n}$ predicts the presence of a non-zero element 
incorrectly---do not contribute to the unnormalized square error.  
The latter convergence~(\ref{latter_limit_p}) implies that the unnormalized 
square error is dominated by false negative events---the estimator of $A_{n}$ 
predicts the absence of a non-zero element incorrectly. 

For the latter convergence~(\ref{latter_limit_p}), let 
\begin{equation} \label{Delta_U_n_tilde}
\Delta \tilde{U}_{n}(v_{\tilde{N}}) 
=  \tilde{U}_{n}
- \sqrt{k}f_{X}(k^{-1/2}\tilde{U}_{n} + \Omega_{n}; v_{\tilde{N}}),  
\end{equation}
with $\tilde{U}_{n}=\nu_{N}\tilde{a}_{n} + U_{n}$ and 
$\tilde{a}_{n}=\sqrt{k}a_{n}$. It is sufficient to prove the following  
result: If $\nu_{N}=0$ or $\mathrm{Pr}(|U|=\sqrt{2v})=0$ holds, 
\begin{equation}
\frac{1}{k}\sum_{n\in\mathcal{S}}\mathbb{E}\left[
 \left.
  \Delta \tilde{U}_{n}^{2}(v_{\tilde{N}})
 \right| 
 \boldsymbol{a}
\right]
\pto \mathbb{E}\left[
 U^{2}1(U^{2}<2v)
\right] \label{latter_limit} 
\end{equation}
in the sublinear sparsity limit. 
See Appendix~\ref{proof_latter_limit} for the proof of 
(\ref{latter_limit}). 

We confirm that (\ref{latter_limit}) implies (\ref{latter_limit_p}). 
Using the definition of $\Delta U_{n}(v_{\tilde{N}})$ in (\ref{Delta_U_n}) 
yields 
\begin{align}
\mathbb{E}\left[
 \left.
  \Delta U_{n}^{2}(v_{\tilde{N}})
 \right| 
 \boldsymbol{a}
\right]
&= \mathbb{E}\left[
 \left.
  \Delta \tilde{U}_{n}^{2}(v_{\tilde{N}})
 \right| 
 \boldsymbol{a}
\right]
\nonumber \\
&- 2\nu_{N}\tilde{a}_{n}\mathbb{E}\left[
 \left.
  \Delta \tilde{U}_{n}(v_{\tilde{N}})
 \right| 
 \boldsymbol{a}
\right]
+ \nu_{N}^{2}\tilde{a}_{n}^{2}. 
\end{align}
For the last term, we use the boundedness in probability of 
$\|\boldsymbol{a}\|_{2}$ and $\nu_{N}\to0$ to have 
\begin{equation} \label{latter_limit_last_term}
\frac{1}{k}\sum_{n\in\mathcal{S}}\nu_{N}^{2}\tilde{a}_{n}^{2} 
= \nu_{N}^{2}\sum_{n\in\mathcal{S}}a_{n}^{2}\pto 0. 
\end{equation}
For the second term, we use the Cauchy-Schwarz inequality and 
$(\mathbb{E}[\Delta \tilde{U}_{n}(v_{\tilde{N}}) | \boldsymbol{a}])^{2}
\leq \mathbb{E}[\Delta \tilde{U}_{n}^{2}(v_{\tilde{N}}) | \boldsymbol{a}]$ 
to obtain 
\begin{align}
&\left|
 \frac{1}{k}\sum_{n\in\mathcal{S}}\nu_{N}\tilde{a}_{n}\mathbb{E}\left[
  \left.
   \Delta \tilde{U}_{n}(v_{\tilde{N}})
  \right| 
  \boldsymbol{a}
 \right]
\right|
\nonumber \\
&\leq \frac{1}{k}\left(
 \sum_{n\in\mathcal{S}}\nu_{N}^{2}\tilde{a}_{n}^{2}
 \sum_{n\in\mathcal{S}}\mathbb{E}\left[
  \left.
   \Delta \tilde{U}_{n}^{2}(v_{\tilde{N}}) 
  \right| 
  \boldsymbol{a}
 \right]
\right)^{1/2}\pto 0, 
\end{align}
where the last convergence follows from (\ref{latter_limit}) and 
(\ref{latter_limit_last_term}). 

We have already proved that (\ref{latter_limit}) implies 
\begin{equation} \label{latter_limit_tmp}
\frac{1}{k}\sum_{n\in\mathcal{S}}\mathbb{E}\left[
 \left.
  \Delta U_{n}^{2}(v_{\tilde{N}})
 \right| 
 \boldsymbol{a}
\right]
\pto \mathbb{E}\left[
 U^{2}1(U^{2}<2v)
\right].  
\end{equation}
Since (\ref{latter_limit_p}) is the arithmetic average of independent 
non-negative random variables $\{\Delta U_{n}^{2}\}$ in (\ref{Delta_U_n}), 
as well as (\ref{latter_limit_tmp}), 
we can use the weak law of large numbers to obtain 
\begin{equation}
\frac{1}{k}\sum_{n\in\mathcal{S}}\left\{
 \Delta U_{n}^{2}(v_{\tilde{N}})
 - \mathbb{E}\left[
  \left.
   \Delta U_{n}^{2}(v_{\tilde{N}})
  \right|
  \boldsymbol{a} 
 \right] 
\right\}
\pto 0. 
\end{equation}
Using the convergence~(\ref{latter_limit_tmp}), 
we arrive at (\ref{latter_limit_p}). 
\end{IEEEproof}

Lemma~\ref{lemma_MSE} implies $\bar{v}_{\In,t+1}={\cal O}(1)$ in the state 
evolution recursion~(\ref{v_in_bar}) when 
the Bayesian estimator is used as the inner denoiser. In evaluating 
$\bar{v}_{\In,t+1}$, we consider the case $\nu_{N}=0$, so that the other 
assumption $\mathrm{Pr}(|U|=\sqrt{2v})=0$ is not required. The case 
$\nu_{N}\neq 0$ has been considered as an intermediate step for proving the 
third assumption in Assumption~\ref{assumption_inner}. 

The unnormalized square error~(\ref{unnormalized_square_error}) is 
dominated by false negative events---the estimator of $A_{n}$ indicates the 
absence of a non-zero element incorrectly. The critical point $|U|=\sqrt{2v}$ 
is consistent with extreme value theory~\cite[p.~302]{David03} for 
$\log k/\log N\to\gamma=0$: 
We know that the expected maximum of $N$ independent 
Gaussian noise samples $\{k^{1/2}\Omega_{n}\sim\mathcal{N}(0, v_{\tilde{N}})\}$ is 
$\sqrt{2v_{\tilde{N}}\log N}\{1 + o(1)\}$. 
Thus, the signal component $U$ can be buried in noise when 
$|U|$ is smaller than $\sqrt{2v_{\tilde{N}}\log N}\to\sqrt{2v}$ for $\gamma=0$. 
When the mismatched measurement $\nu_{N}\boldsymbol{a} + \boldsymbol{Y}$ is 
considered, the unnormalized square error can be affected by the discrete 
component of $U$ at the critical point $|U|=\sqrt{2v}$. 

More generally, for $\log k/\log N\to\gamma\in(0, 1)$ we should focus on 
the expected $k$th maximum of the $N$ independent Gaussian noise samples in 
the sublinear sparsity limit. It may be equal to 
$\sqrt{2v_{\tilde{N}}\log(N/k)} + o(1)=\sqrt{2v}+o(1)$. In this sense, 
Lemma~\ref{lemma_MSE} is an extreme-value-theoretic result for the 
unnormalized square error while the normalization of the square error allows 
us to ignore outliers of noise samples far from zero. 

We next evaluate the unnormalized error covariance to justify the second 
assumption in Assumption~\ref{assumption_inner}. 

\begin{lemma} \label{lemma_covariance}
Suppose that Assumption~\ref{assumption_lemma} holds. 
Then, the unnormalized error 
covariance $\{\boldsymbol{X} - f_{X}(\boldsymbol{Y}'; v_{\tilde{N}}')\}^{\mathrm{T}}
\{\boldsymbol{X} - f_{X}(\boldsymbol{Y}; v_{\tilde{N}})\}$ and 
$\boldsymbol{X}^{\mathrm{T}}\{\boldsymbol{X} 
- f_{X}(\boldsymbol{Y}; v_{\tilde{N}})\}$ 
converge in probability to their expectation in the sublinear sparsity limit. 
\end{lemma}
\begin{IEEEproof}
We only prove the former convergence in probability since the latter 
convergence can be proved in the same manner. 
Repeating the derivation of (\ref{unnormalized_square_error_tmp}) yields  
\begin{align}
&\{\boldsymbol{X} - f_{X}(\boldsymbol{Y}'; v_{\tilde{N}}')\}^{\mathrm{T}}
\{\boldsymbol{X} - f_{X}(\boldsymbol{Y}; v_{\tilde{N}})\}
\nonumber \\
&= \sum_{n\notin\mathcal{S}}f_{X}(\Omega_{n}'; v_{\tilde{N}}')
f_{X}(\Omega_{n}, v_{\tilde{N}}) 
+ \frac{1}{k}\sum_{n\in\mathcal{S}}\Delta U_{n}(v_{\tilde{N}}')
\Delta U_{n}(v_{\tilde{N}}), \label{unnormalized_error_covariance}
\end{align}
with $\Delta U_{n}(v_{\tilde{N}})$ in (\ref{Delta_U_n}), where 
$\Omega_{n}$ in $\Delta U_{n}(v_{\tilde{N}}')$ is replaced with $\Omega_{n}'$.  
Using the Cauchy-Schwarz inequality for the first term in 
(\ref{unnormalized_error_covariance}), we have 
\begin{align}
&\left\{
 \sum_{n\notin\mathcal{S}}f_{X}(\Omega_{n}'; v_{\tilde{N}}')
f_{X}(\Omega_{n}, v_{\tilde{N}}) 
\right\}^{2}
\nonumber \\
&\leq \sum_{n\notin\mathcal{S}}f_{X}^{2}(\Omega_{n}'; v_{\tilde{N}}')
\sum_{n\notin\mathcal{S}}f_{X}^{2}(\Omega_{n}, v_{\tilde{N}}) 
\pto 0,
\end{align}
where the last convergence follows from (\ref{former_limit_p}). 

The second term in (\ref{unnormalized_error_covariance}) is the sum of 
i.i.d.\ random variables. Furthermore, we can prove that the first absolute 
moment of each term is bounded by using the Cauchy-Schwarz inequality and 
(\ref{latter_limit_tmp}) for $\nu_{N}=0$. 
Thus, we use the weak law of large numbers to find 
that the second term in (\ref{unnormalized_error_covariance}) converges in 
probability to its expectation in the sublinear sparsity limit. 
Combining these results, we arrive at Lemma~\ref{lemma_covariance}. 
\end{IEEEproof}

Lemma~\ref{lemma_MSE} for $\nu_{N}=0$ and Lemma~\ref{lemma_covariance} justify 
the second assumption in Assumption~\ref{assumption_inner} when the Bayesian 
estimator is used as the inner denoiser. 
This paper does not need any closed-form expression of the asymptotic 
error covariance while it may be possible to derive. 
The existence of the limits is sufficient for state evolution. 

We next prove the third assumption in Assumption~\ref{assumption_inner} 
for the Bayesian estimator. 

\begin{lemma} \label{lemma_difference}
Suppose that Assumption~\ref{assumption_lemma} holds. 
If $\mathrm{Pr}(|U|=\sqrt{2v})=0$ holds, then we have  
\begin{equation}
\| f_{X}(\nu_{N}\boldsymbol{a} + \boldsymbol{Y}; v_{\tilde{N}})
- f_{X}(\boldsymbol{Y}; v_{\tilde{N}})\|_{2}^{2} \pto 0
\end{equation}
in the sublinear sparsity limit. 
\end{lemma}
\begin{IEEEproof}
Let $\tilde{a}_{n}=\sqrt{k}a_{n}$, 
$\tilde{U}_{n} = U_{n} + \nu_{N}\tilde{a}_{n}$, and 
$\tilde{\Omega}_{n}=\sqrt{k}\Omega_{n}\sim\mathcal{N}(0, v_{\tilde{N}})$. 
Repeating the proof of Lemma~\ref{lemma_MSE}, we have 
\begin{align}
&\| f_{X}(\nu_{N}\boldsymbol{a} + \boldsymbol{Y}; v_{\tilde{N}})
- f_{X}(\boldsymbol{Y}; v_{\tilde{N}})\|_{2}^{2} 
\nonumber \\
&= \sum_{n\notin\mathcal{S}}\left\{
 f_{X}(\nu_{N}a_{n} + \Omega_{n}; v_{\tilde{N}})
 - f_{X}(\Omega_{n}; v_{\tilde{N}})
\right\}^{2}
\nonumber \\
&+ \sum_{n\in\mathcal{S}}\left\{
 \Delta f_{X}(\tilde{a}_{n}, \tilde{U}_{n} + \tilde{\Omega}_{n}; v_{\tilde{N}}) 
\right\}^{2}, 
\end{align}
with 
\begin{equation} \label{Delta_f}
\Delta f_{X}(\tilde{a}_{n}, \tilde{Y}; v_{\tilde{N}}) 
=  f_{X}\left(
  \frac{\tilde{Y}}{\sqrt{k}}; v_{\tilde{N}}
 \right)
 - f_{X}\left(
  \frac{\tilde{Y} - \nu_{N}\tilde{a}_{n}}{\sqrt{k}}; v_{\tilde{N}}
 \right).
\end{equation} 
Using (\ref{former_limit_p}) and 
the upper bound $\{f_{X}(\nu_{N}a_{n} + \Omega_{n}; v_{\tilde{N}})
- f_{X}(\Omega_{n}; v_{\tilde{N}})\}^{2}
\leq2\{f_{X}^{2}(\nu_{N}a_{n} + \Omega_{n}; v_{\tilde{N}})
+ f_{X}^{2}(\Omega_{n}; v_{\tilde{N}})\}$, 
we find that the first term converges in probability to zero in the 
sublinear sparsity limit. Thus, Lemma~\ref{lemma_difference} is obtained by 
proving the following convergence: 
\begin{equation} \label{difference}
\sum_{n\in\mathcal{S}}\mathbb{E}\left[
 \left.
  \Delta f_{X}^{2}(\tilde{a}_{n}, \tilde{U}_{n} + \tilde{\Omega}_{n}; v_{\tilde{N}})
 \right| \boldsymbol{a}
\right] \pto 0 
\end{equation}
since the convergence in mean implies the convergence in probability. 
See Appendix~\ref{proof_lemma_difference} for the proof of 
(\ref{difference}). 
\end{IEEEproof}

Lemma~\ref{lemma_difference} implies the third assumption in 
Assumption~\ref{assumption_inner} when the Bayesian estimator is used as 
the inner denoiser.  

Finally, we investigate the last assumption in 
Assumption~\ref{assumption_inner}. 
We first prove the following intermediate result: 

\begin{lemma} \label{lemma_xi}
Suppose that Assumption~\ref{assumption_lemma} holds. Then, we have 
\begin{equation} \label{xi_limit_former}
\boldsymbol{\Omega}^{\mathrm{T}}f_{X}(\boldsymbol{Y}; v_{\tilde{N}})
- \mathbb{E}\left[
 \boldsymbol{\Omega}^{\mathrm{T}}f_{X}(\boldsymbol{Y}; v_{\tilde{N}})
\right]
\pto 0,
\end{equation}
\begin{equation} \label{xi_limit_latter}
(\boldsymbol{\Omega}')^{\mathrm{T}}f_{X}(\boldsymbol{Y}; v_{\tilde{N}})
- \mathbb{E}\left[
 (\boldsymbol{\Omega}')^{\mathrm{T}}f_{X}(\boldsymbol{Y}; v_{\tilde{N}})
\right]
\pto 0
\end{equation}
in the sublinear sparsity limit. 
\end{lemma}
\begin{IEEEproof}
See Appendix~\ref{proof_lemma_xi} for the proof of the former 
convergence~(\ref{xi_limit_former}). 
We prove that the former convergence~(\ref{xi_limit_former}) 
implies the latter convergence~(\ref{xi_limit_latter}). 
By definition, we can represent $\boldsymbol{\Omega}'\sim\mathcal{N}
(\boldsymbol{0}, v_{k,N}'\boldsymbol{I}_{N})$ with 
$\boldsymbol{Z}$ independent of $\boldsymbol{\Omega}$ as 
\begin{equation} \label{Omega'_representation}
\boldsymbol{\Omega}' = \frac{\mathrm{cov}}{v}\boldsymbol{\Omega}
+ \boldsymbol{Z}, \quad 
\boldsymbol{Z}\sim\mathcal{N}(\boldsymbol{0}, \tilde{v}_{k,N}\boldsymbol{I}_{N}),
\end{equation}
with $\tilde{v}_{k,N}=\tilde{v}/\{k\log(N/k)\}$ and 
$\tilde{v}=v' - \mathrm{cov}^{2}/v\geq0$. It is straightforward to confirm 
$\mathbb{E}[\boldsymbol{\Omega}'\boldsymbol{\Omega}^{\mathrm{T}}]
=\mathrm{cov}_{k,N}\boldsymbol{I}_{N}$ and $\mathbb{E}[\boldsymbol{\Omega}'
(\boldsymbol{\Omega}')^{\mathrm{T}}]=v_{k,N}'\boldsymbol{I}_{N}$. 
Using the representation~(\ref{Omega'_representation}) yields 
\begin{equation} \label{xi_limit_latter_tmp}
(\boldsymbol{\Omega}')^{\mathrm{T}}f_{X}(\boldsymbol{Y}; v_{\tilde{N}})
= \frac{\mathrm{cov}}{v}\boldsymbol{\Omega}^{\mathrm{T}}
f_{X}(\boldsymbol{Y}; v_{\tilde{N}})
+ \boldsymbol{Z}^{\mathrm{T}}f_{X}(\boldsymbol{Y}; v_{\tilde{N}}). 
\end{equation}
The former convergence~(\ref{xi_limit_former}) implies that 
the first term in (\ref{xi_limit_latter_tmp}) converges in probability 
to its expectation in the sublinear sparsity limit. 

For the second term in (\ref{xi_limit_latter_tmp}), we repeat the derivation  
of (\ref{unnormalized_square_error_tmp}) to obtain 
\begin{align}
&\boldsymbol{Z}^{\mathrm{T}}f_{X}(\boldsymbol{Y}; v_{\tilde{N}})
= \sum_{n\notin\mathcal{S}}Z_{n}f_{X}(\Omega_{n}; v_{\tilde{N}}) 
\nonumber \\
&+ \frac{1}{k}\sum_{n\in\mathcal{S}}\tilde{Z}_{n}
\sqrt{k}f_{X}(k^{-1/2}U_{n} + \Omega_{n}; v_{\tilde{N}}), 
\label{xi_limit_latter_second} 
\end{align}
with $\tilde{Z}_{n}=\sqrt{k}Z_{n}\sim\mathcal{N}(0, \tilde{v}/\log(N/k))$. 
Since the second term is the sum of zero-mean i.i.d.\ random variables, 
we use the weak law of large numbers to find that the second term in 
(\ref{xi_limit_latter_second}) converges in probability to zero 
in the sublinear sparsity limit. 
For the first term in (\ref{xi_limit_latter_second}), we have 
\begin{align}
\mathbb{E}\left[
 \left(
  \sum_{n\notin\mathcal{S}}Z_{n}f_{X}(\Omega_{n}; v_{\tilde{N}}) 
 \right)^{2}
\right] 
&= \tilde{v}_{k,N}\sum_{n\notin\mathcal{S}}\mathbb{E}\left[
 f_{X}^{2}(\Omega_{n}; v_{\tilde{N}})
\right]
\nonumber \\
&\to0, 
\end{align}
where the last convergence follows from (\ref{former_limit_general}). Thus, 
the first term in (\ref{xi_limit_latter_second}) converges in probability 
to zero in the sublinear sparsity limit. Combining these results, 
we arrive at the latter convergence~(\ref{xi_limit_latter}). 
Thus, Lemma~\ref{lemma_xi} holds.  
\end{IEEEproof}

Lemma~\ref{lemma_xi} is used as part of technical results to justify the 
last assumption in Assumption~\ref{assumption_inner}. 
The boundedness of $\mathbb{E}[ \boldsymbol{\Omega}^{\mathrm{T}}
f_{X}(\boldsymbol{Y}; v_{\tilde{N}})]$ can be confirmed with 
Stein's lemma~\cite{Stein72}. 

\begin{lemma} \label{lemma_Delta_xi}
Suppose that Assumption~\ref{assumption_lemma} holds. 
If $\mathrm{Pr}(|U|=\sqrt{2v})=0$ holds, then we have  
\begin{equation} \label{Delta_xi_limit_former}
\boldsymbol{\Omega}^{\mathrm{T}}\left\{
 f_{X}(\nu_{N}\boldsymbol{a} + \boldsymbol{Y}; v_{\tilde{N}})
 - f_{X}(\boldsymbol{Y}; v_{\tilde{N}})
\right\}\pto 0, 
\end{equation}
\begin{equation} \label{Delta_xi_limit_latter}
(\boldsymbol{\Omega}')^{\mathrm{T}}\left\{
 f_{X}(\nu_{N}\boldsymbol{a} + \boldsymbol{Y}; v_{\tilde{N}})
 - f_{X}(\boldsymbol{Y}; v_{\tilde{N}})
\right\}\pto 0 
\end{equation}
in the sublinear sparsity limit. 
\end{lemma}
\begin{IEEEproof}
We first prove that the former limit~(\ref{Delta_xi_limit_former}) 
implies the latter limit~(\ref{Delta_xi_limit_latter}). 
Let $\tilde{\boldsymbol{a}}=\sqrt{k}\boldsymbol{a}$ and 
$\tilde{\boldsymbol{Y}}=\sqrt{k}\boldsymbol{Y}$. 
Using $\Delta f_{X}$ in (\ref{Delta_f}) and 
the representation~(\ref{Omega'_representation}) yields 
\begin{align}
&(\boldsymbol{\Omega}')^{\mathrm{T}}\left\{
 f_{X}(\nu_{N}\boldsymbol{a} + \boldsymbol{Y}; v_{\tilde{N}})
 - f_{X}(\boldsymbol{Y}; v_{\tilde{N}})
\right\}
\nonumber \\
&= \frac{\mathrm{cov}}{v}\boldsymbol{\Omega}^{\mathrm{T}}
\Delta f_{X}(\tilde{\boldsymbol{a}}, 
\nu_{N}\tilde{\boldsymbol{a}} + \tilde{\boldsymbol{Y}}; v_{\tilde{N}}) 
\nonumber \\
&+ \boldsymbol{Z}^{\mathrm{T}}
\Delta f_{X}(\tilde{\boldsymbol{a}}, \nu_{N}\tilde{\boldsymbol{a}} 
+ \tilde{\boldsymbol{Y}}; v_{\tilde{N}}).  
\label{Delta_xi_limit_latter_tmp}
\end{align} 
The first term in (\ref{Delta_xi_limit_latter_tmp}) converges in probability 
to zero in the sublinear sparsity limit if the former 
limit~(\ref{Delta_xi_limit_former}) is correct. 

For the second term in (\ref{Delta_xi_limit_latter_tmp}), we use 
$\Delta f_{X}$ in (\ref{Delta_f}) to evaluate the conditional variance as 
\begin{align}
&\mathbb{V}\left[
 \left.
  \boldsymbol{Z}^{\mathrm{T}}\Delta f_{X}(\tilde{\boldsymbol{a}}, 
  \nu_{N}\tilde{\boldsymbol{a}} + \tilde{\boldsymbol{Y}}; v_{\tilde{N}})
 \right| \tilde{\boldsymbol{Y}}, \boldsymbol{a}
\right]
\nonumber \\
&= \sum_{n=1}^{N}\mathbb{E}[Z_{n}^{2}]
\{\Delta f_{X}(\tilde{a}_{n}, \nu_{N}\tilde{a}_{n} + \tilde{Y}_{n}; 
v_{\tilde{N}})\}^{2}
\nonumber \\
&= \tilde{v}/\{k\log(N/k)\}\left\|
 f_{X}(\nu_{N}\boldsymbol{a} + \boldsymbol{Y}; v_{\tilde{N}}) 
 - f_{X}(\boldsymbol{Y}; v_{\tilde{N}})
\right\|_{2}^{2}
\nonumber \\
&\pto 0
\end{align}
in the sublinear sparsity limit, where the last convergence follows 
from Lemma~\ref{lemma_difference} under the assumption 
$\mathrm{Pr}(|U|=\sqrt{2v})=0$.  
Thus, the second term in (\ref{Delta_xi_limit_latter_tmp}) converges 
in probability 
to the conditional expectation $\mathbb{E}[\boldsymbol{Z}^{\mathrm{T}}
\Delta f_{X}(\boldsymbol{a}, \boldsymbol{Y}; v_{\tilde{N}})
| \boldsymbol{a}]=0$ in the sublinear 
sparsity limit. From these results, we find that the former 
limit~(\ref{Delta_xi_limit_former}) implies the latter 
limit~(\ref{Delta_xi_limit_latter}).  

We prove the former limit~(\ref{Delta_xi_limit_former}). 
Repeating the proof of Lemma~\ref{lemma_difference} yields 
\begin{align}
&\boldsymbol{\Omega}^{\mathrm{T}}\left\{
 f_{X}(\nu_{N}\boldsymbol{a} + \boldsymbol{Y}; v_{\tilde{N}})
 - f_{X}(\boldsymbol{Y}; v_{\tilde{N}})
\right\}
\nonumber \\
&= \sum_{n\notin\mathcal{S}}\Omega_{n}\Delta f_{X}
(\tilde{a}_{n}, \nu_{N}\tilde{a}_{n} + \tilde{\Omega}_{n}; v_{\tilde{N}})
\nonumber \\
&+ \sum_{n\in\mathcal{S}}\Omega_{n}
\Delta f_{X}(\tilde{a}_{n}, \tilde{U}_{n} + \tilde{\Omega}_{n}; v_{\tilde{N}}), 
\label{Delta_xi_tmp}
\end{align}
with $\Delta f_{X}$ in (\ref{Delta_f}), 
$\tilde{U}_{n} = U_{n} + \nu_{N}\tilde{a}_{n}$, and 
$\tilde{\Omega}_{n}=\sqrt{k}\Omega_{n}$. 
Thus, the former limit~(\ref{Delta_xi_limit_former})
is obtained by proving the following convergence 
in probability in the sublinear sparsity limit: 
\begin{equation} \label{Delta_xi_former}
\sum_{n\notin\mathcal{S}}\mathbb{E}\left[
 \left. 
  \left|
   \Omega_{n}\Delta f_{X}(\tilde{a}_{n}, 
   \nu_{N}\tilde{a}_{n} + \tilde{\Omega}_{n}; v_{\tilde{N}})
  \right|
 \right| \boldsymbol{a} 
\right]
\pto 0,  
\end{equation}
\begin{equation}
\sum_{n\in\mathcal{S}}\Omega_{n}\Delta f_{X}(\tilde{a}_{n}, 
\tilde{U}_{n} + \tilde{\Omega}_{n}; v_{\tilde{N}})
\pto 0. 
\label{Delta_xi_latter}
\end{equation}
See Appendix~\ref{proof_lemma_Delta_xi} for the proof of 
(\ref{Delta_xi_former}). 

For the latter limit~(\ref{Delta_xi_latter}), 
we use the Cauchy-Schwarz inequality to obtain 
\begin{align}
&\left|
 \sum_{n\in\mathcal{S}}\Omega_{n}\Delta f_{X}(\tilde{a}_{n}, 
 \tilde{U}_{n} + \tilde{\Omega}_{n}; v_{\tilde{N}})
\right|
\nonumber \\
&\leq \sum_{n\in\mathcal{S}}\Omega_{n}^{2}\sum_{n\in\mathcal{S}}
\Delta f_{X}^{2}(\tilde{a}_{n}, \tilde{U}_{n} + \tilde{\Omega}_{n}; v_{\tilde{N}}). 
\end{align}
Since Lemma~\ref{lemma_difference} under the assumption 
$\mathrm{Pr}(|U|=\sqrt{2v})=0$ implies that the second factor 
converges in probability to zero, it is sufficient to prove the boundedness 
in probability of the first factor in the sublinear sparsity limit. 
Evaluating the expectation of the first factor with 
$\Omega_{n}\sim\mathcal{N}(0, v_{k,N})$ as  
\begin{equation}
\sum_{n\in\mathcal{S}}\mathbb{E}\left[
 \Omega_{n}^{2}
\right]
= \frac{v}{\log(N/k)} \to0. 
\end{equation}
This convergence in mean implies that the first factor converges in 
probability to zero in the sublinear sparsity limit. Thus, 
Lemma~\ref{lemma_Delta_xi} holds. 
\end{IEEEproof}

Lemmas~\ref{lemma_xi} and \ref{lemma_Delta_xi} imply the last assumption 
in Assumption~\ref{assumption_inner} when the Bayesian estimator is used as 
the inner denoiser. 

Lemmas~\ref{lemma_MSE}, \ref{lemma_difference}, and~\ref{lemma_Delta_xi} are 
used to ignore the influence of a small error 
$o(1)\boldsymbol{a}$ that appears in Bolthausen's conditioning 
technique~\cite{Bolthausen14}. For linear sparsity~\cite{Bayati11}, we could 
consider quantities divided by $N$, instead of the unnormalized 
quantities in these lemmas. Consequently, the corresponding results 
are trivial for all Lipschitz-continuous functions $f_{X}$. 
For sublinear sparsity, however, we need to consider the unnormalized 
quantities. The lemmas are non-trivial statements that depends heavily 
on additional properties of the inner denoiser, such as 
the Bayesian estimator $f_{X}$.  

The assumption $\mathrm{Pr}(|U|=\sqrt{2v})$ is inevitable 
in Lemmas~\ref{lemma_MSE}, \ref{lemma_difference}, 
and~\ref{lemma_Delta_xi}. To weaken this assumption, we need additional 
assumptions on the convergence speed of $\nu_{N}$ in these lemmas. 
The convergence speed should be investigated by evaluating 
the detailed order of $o(1)$ terms in state evolution while such research 
is outside the scope of this paper.     

\subsection{State Evolution for Bayesian GAMP}
\subsubsection{Bayesian Denoisers} \label{sec4c}
As the outer and inner denoisers in GAMP, 
we consider Bayesian denoisers in terms of the minimum MSE (MMSE). 
For the inner denoiser the Bayesian estimator is used. 
To formulate a Bayesian inner denoiser, from the state evolution 
recursion~(\ref{v_in_bar}), (\ref{mu_tt}), and (\ref{mu_0t}) we consider 
the following Gaussian measurement: 
\begin{equation}
\boldsymbol{y}_{t} 
= \bar{\eta}_{t}\boldsymbol{x} + \boldsymbol{\omega}_{t}, 
\quad \boldsymbol{\omega}_{t}\sim\mathcal{N}\left(
 \boldsymbol{0}, \frac{\bar{v}_{\Out,t}}{\delta k\log (N/k)}\boldsymbol{I}_{N}
\right). 
\end{equation}
Associating the scaled measurement $\boldsymbol{y}_{t}/\bar{\eta}_{t}$ 
with $\boldsymbol{Y}$ in (\ref{AWGN_measurement}), we obtain the Bayesian 
inner denoiser,
\begin{equation} \label{Bayes_inner_denoiser}
f_{\In,t}(\boldsymbol{y}_{t}; \bar{v}_{\Out,t}) = f_{X}\left(
 \frac{\boldsymbol{y}_{t}}{\bar{\eta}_{t}}; 
 \frac{\bar{v}_{\Out,t}}{\delta\bar{\eta}_{t}^{2}\log (N/k)}
\right),
\end{equation}
with the Bayesian estimator $f_{X}$ in (\ref{posterior_mean_estimator_X}).  

For the outer denoiser the Bayes-optimal denoiser is used to maximize the 
effective SNR  
$M\bar{\eta}_{t}^{2}P/\bar{v}_{\Out,t}$~\cite{Takeuchi241}. Consider the 
following measurement model:  
\begin{equation}
Y = g(Z, W), 
\end{equation}
\begin{equation} \label{outer_measurement}
Z  + B_{0} = 0, \quad Z_{t} = Z + B_{t}, 
\end{equation}
where $B_{0}\sim\mathcal{N}(0, P)$ and $B_{t}$ are independent of $W$ in 
Assumption~\ref{assumption_w} and zero-mean Gaussian random variables with 
covariance $\mathbb{E}[B_{0}B_{t}] = P - \mu_{0,t}$ and 
$\mathbb{E}[B_{t}^{2}]=\bar{v}_{\In,t}$. By definition, $Z$ and $Z_{t}$ are 
zero-mean Gaussian random variables with covariance (\ref{Z_tt}) and 
(\ref{Z_0t}): $\mathbb{E}[ZZ_{t}]=-\mathbb{E}[B_{0}(-B_{0}+B_{t})]=\mu_{0,t}$  
and $\mathbb{E}[Z_{t}^{2}]=\mathbb{E}[(-B_{0}+B_{t})^{2}]
= P -2(P - \mu_{0,t}) + \bar{v}_{\In,t}= \mu_{t,t}$ due to the 
definition of $\bar{v}_{\In,t}$ in (\ref{v_in_bar}). 

Let $f_{Z}(z_{t}, y; \bar{v}_{\In,t}) 
= \mathbb{E}[Z | Z_{t}=z_{t}, Y=y]: \mathbb{R}^{2}\to\mathbb{R}$ denote 
the posterior mean estimator of $Z$ given $Z_{t}=z_{t}$ and $Y=y$. 
Then, the Bayes-optimal outer denoiser is formulated as~\cite{Takeuchi241} 
\begin{equation} \label{Bayes_outer_denoiser}
f_{\Out,t}(\boldsymbol{z}_{t}, \boldsymbol{y}; \bar{v}_{\In,t})
= \frac{\boldsymbol{z}_{t} - f_{Z}(\boldsymbol{z}_{t}, \boldsymbol{y}; 
\bar{v}_{\In,t})}{\bar{v}_{\In,t}}. 
\end{equation}
This paper refers to GAMP using the Bayesian inner 
denoiser~(\ref{Bayes_inner_denoiser}) and the Bayes-optimal outer 
denoiser~(\ref{Bayes_outer_denoiser}) as Bayesian GAMP. 

The state evolution recursion is simplified for Bayesian GAMP, because of 
the consistency $\tilde{v}_{\In,t}=\bar{v}_{\In,t}$ and the identity 
$\bar{\eta}_{t}=1$. 
The simplified state evolution recursion with the initial condition 
$\bar{v}_{\In,0}=P$ is given by 
\begin{equation} \label{Bayes_v_out_bar}
\frac{1}{\bar{v}_{\Out,t}} = 
\mathbb{E}\left[
 \frac{\partial f_{\Out,t}}{\partial z_{t}}(Z_{t}, g(Z, W); \bar{v}_{\In,t}) 
\right],
\end{equation}
\begin{equation} \label{Bayes_v_in_bar}
\bar{v}_{\In,t+1} = \mathbb{E}\left[
 U^{2}1\left(
  U^{2}<\frac{2\bar{v}_{\Out,t}}{\delta}
 \right)
\right], 
\end{equation}
where $U$ represents the scaled non-zero signal elements in 
Assumption~\ref{assumption_x}. 

\begin{corollary} \label{corollary_Bayes}
Suppose that Assumptions~\ref{assumption_x}--\ref{assumption_SE}, and
Assumption~\ref{assumption_lemma} hold. 
If $\mathrm{Pr}(|U|=\sqrt{2\bar{v}_{\Out,t}/\delta})=0$ holds for all 
$t\in\{0,\ldots,T\}$, then Bayesian GAMP 
satisfies the consistency $\tilde{v}_{\In,t+1}=\bar{v}_{\In,t+1}$ for all 
$t\in\{-1,\ldots,T\}$ and the identity $\bar{\eta}_{t}=1$ for all 
$t\in\{0,\ldots,T\}$. Furthermore, the unnormalized square error 
$\|\hat{\boldsymbol{x}}_{t+1} - \boldsymbol{x}\|_{2}^{2}$ for Bayesian GAMP 
converges in probability to $\bar{v}_{\In,t+1}$---given via the state evolution 
recursion~(\ref{Bayes_v_out_bar}) and (\ref{Bayes_v_in_bar})---in 
the sublinear sparsity limit for all $t\in\{0,\ldots, T\}$. 
\end{corollary}
\begin{IEEEproof}
See Appendix~\ref{proof_corollary_Bayes}. 
\end{IEEEproof}

The assumption $\mathrm{Pr}(|U|=\sqrt{2\bar{v}_{\Out,t}/\delta})=0$ is of 
course satisfied for absolutely continuous $U$, 
such as $U\sim\mathcal{N}(0, P)$. For $U$ having discrete components, 
the assumption is broken for at most countable points of $\delta>0$ 
when the other parameters $\gamma\in[0, 1)$ and $P/\sigma^{2}>0$ are fixed. 
In this sense, Corollary~\ref{corollary_Bayes} is applicable to almost all 
points on the space of parameters $(\delta, \gamma, P/\sigma^{2})
\in(0, \infty)\times[0, 1)\times(0, \infty)$. 

We know that $\bar{v}_{\Out,t}$ in (\ref{Bayes_v_out_bar}) is monotonically 
increasing with respect to $\bar{v}_{\In,t}$~\cite{Cobo24,Takeuchi241}. 
Furthermore, $\bar{v}_{\In,t+1}$ in (\ref{Bayes_v_in_bar}) is obviously 
non-decreasing with respect to $\bar{v}_{\Out,t}$. From 
$\bar{v}_{\In,1}\leq\mathbb{E}[U^2]=\bar{v}_{\In,0}$ and $\bar{v}_{\In,t+1}\geq0$, 
we follow \cite[Lemma~1]{Takeuchi15} to prove that the state evolution 
recursion~(\ref{Bayes_v_out_bar}) and 
(\ref{Bayes_v_in_bar}) for Bayesian GAMP converges to a fixed point 
as $t\to\infty$. In particular, $\bar{v}_{\In,\infty}$ can be zero when the 
non-zero signal $U$ never occurs in a neighborhood of the origin. Thus, 
this paper defines the threshold for signal reconstruction as follows: 

\begin{definition}[Reconstruction Threshold] 
\label{def_reconstruction_threshold}
The reconstruction threshold $\delta^{*}>0$ for Bayesian GAMP is defined 
as the infimum of $\delta'$ such that $\bar{v}_{\In,\infty}=0$ holds 
for all $\delta>\delta'$. 
\end{definition}

By definition, exact signal reconstruction is asymptotically achieved in 
terms of the $\ell_{2}$-norm for all $\delta>\delta^{*}$ since 
$\bar{v}_{\In,\infty}$ is the unnormalized square error. 

The reconstruction threshold depends on the distribution of the non-zero 
signal element $U$. Let $u_{\mathrm{min}}\geq0$ denote the essential minimum of 
the non-zero signal amplitude $|U|$, i.e.\ the supremum of $u\geq0$ such 
that $\mathrm{Pr}(|U|>u)=1$ holds. Intuitively, increasing the signal power 
$P$ improves the reconstruction performance. Thus, $U$ with the minimum 
power $P=u_{\mathrm{min}}^{2}$ should be the worst non-zero signal to maximize  
the reconstruction threshold. This intuition is proved for the linear 
measurement.

\subsubsection{Linear Measurement}
To compare the reconstruction threshold with existing results, 
consider the linear measurement function $g(z, w) = z + w$ and the 
Gaussian noise vector $\boldsymbol{w}\sim\mathcal{N}(\boldsymbol{0}, 
\sigma^{2}\boldsymbol{I}_{M})$. In this case, Assumption~\ref{assumption_w} 
holds for $W\sim\mathcal{N}(0, \sigma^{2})$. Furthermore, the Bayes-optimal 
outer denoiser~(\ref{Bayes_outer_denoiser}) is given by  
\begin{equation}
f_{\Out,t}(z_{t}, y; \bar{v}_{\In,t}) = \frac{z_{t} - y}
{\sigma^{2} + \bar{v}_{\In,t}}, 
\end{equation}
which implies that the state evolution recursion~(\ref{Bayes_v_out_bar}) 
reduces to 
\begin{equation} \label{linear_v_out_bar}
\bar{v}_{\Out,t} = \sigma^{2} + \bar{v}_{\In,t}. 
\end{equation}

\begin{proposition} \label{proposition_threshold}
Suppose that the essential minimum $u_{\mathrm{min}}$ of the non-zero signal 
amplitude is strictly positive. Consider the linear measurement function 
$g(z, w)=z + w$ and the Gaussian noise vector 
$\boldsymbol{w}\sim\mathcal{N}(\boldsymbol{0}, \sigma^{2}\boldsymbol{I}_{M})$. 
Then, the reconstruction threshold $\delta^{*}$ is maximized for the non-zero 
signal element $U$ satisfying $P=u_{\min}^{2}$. In particular, 
the reconstruction threshold for this worst signal is given by  
$\delta^{*}=2(1 + \sigma^{2}/u_{\mathrm{min}}^{2})$. 
\end{proposition}
\begin{IEEEproof}
Consider the change of variables $y_{t}=\bar{v}_{\In,t}$ and 
$x_{t}=2\bar{v}_{\Out,t}/\delta$. It is straightforward to confirm that 
the state evolution recursion~(\ref{Bayes_v_in_bar}) and 
(\ref{linear_v_out_bar}) reduce to 
\begin{equation} \label{y}
y_{t+1} = \phi(x_{t}), \quad 
\phi(x) = \mathbb{E}\left[
 U^{2}1(U^{2} < x)
\right], 
\end{equation}
\begin{equation} \label{x}
y_{t} = \psi(x_{t}), \quad 
\psi(x) = \frac{\delta}{2}x - \sigma^{2}, 
\end{equation}
with the initial condition $y_{0}=P$. 
The convergence point $(x_{\infty}, y_{\infty})$ is the fixed point of the 
state evolution recursion~(\ref{y}) and (\ref{x}) with the maximum 
$y_{\infty}\in[0, P]$. 

We first evaluate the reconstruction threshold for $U$ with 
$P=u_{\mathrm{min}}^{2}$. Since $\mathrm{Pr}(|U|=u_{\mathrm{min}})=1$ holds, 
the state evolution recursion~(\ref{y}) reduces to 
\begin{equation} \label{y_tmp}
y_{t+1} = u_{\mathrm{min}}^{2}1\left(
 x_{t} > u_{\mathrm{min}}^{2} 
\right). 
\end{equation}
Thus, the convergence point $y_{\infty}$ becomes zero if and only if 
the state evolution recursion~(\ref{x}) satisfies 
$y_{t} = \delta x_{t}/2 - \sigma^{2}>u_{\mathrm{min}}^{2}$ 
at $x_{t}=u_{\mathrm{min}}^{2}$. This observation implies that the reconstruction 
threshold $\delta^{*}$ for $U$ with $P=u_{\mathrm{min}}^{2}$ 
is given by $\delta^{*}=2(1 + \sigma^{2}/u_{\mathrm{min}}^{2})$. 

We next prove that $y_{\infty}=0$ holds for all $U$ and 
$\delta>\delta^{*}$. Since $u_{\mathrm{min}}$ 
is strictly positive, we can use the inequality $1(U^{2}<x)\leq x/U^{2}$ 
to obtain the upper bound $\phi(x)\leq x$ on (\ref{y}) for all $x\geq0$. 
For all $\delta>\delta^{*}$, however, 
the straight line $y=\psi(x)$ in (\ref{x}) has the slope greater than $1$, 
i.e.\ $\delta/2>1 + \sigma^{2}/u_{\mathrm{min}}^{2}>1$. Furthermore, 
$\psi(u_{\mathrm{min}}^{2})> u_{\mathrm{min}}^{2}$ holds at $x=u_{\mathrm{min}}^{2}$. 
These observations imply $\psi(x)>\phi(x)$ for all $x\geq u_{\mathrm{min}}^{2}$. 
Thus, $y_{\infty}=0$ holds for all $\delta>\delta^{*}$, regardless of the 
distribution of $U$. In other words, $U$ with $P=u_{\mathrm{min}}^{2}$ is the 
worst signal to achieve the maximum reconstruction threshold 
$\delta^{*}=2(1 + \sigma^{2}/u_{\mathrm{min}}^{2})$. 
\end{IEEEproof}

To utilize Corollary~\ref{corollary_Bayes}, we need 
$2\bar{v}_{\Out,t}/\delta\neq u_{\mathrm{min}}^{2}$ for all $t\in\{0,\ldots,T\}$. 
This condition implies that Corollary~\ref{corollary_Bayes} cannot be used 
at the reconstruction threshold $\delta=\delta^{*}$. From the definition 
$y_{t}=\bar{v}_{\In,t}$, in fact, (\ref{y_tmp}) and 
 Assumption~\ref{assumption_SE} imply $\bar{v}_{\In,t}=u_{\mathrm{min}}^{2}$ 
for all $t\in\{0,\ldots,t\}$. Substituting this into 
(\ref{linear_v_out_bar}) and using the condition 
$2\bar{v}_{\Out,t}/\delta\neq u_{\mathrm{min}}^{2}$, after some algebra, we 
arrive at $\delta\neq\delta^{*}$. 

The reconstruction threshold in Proposition~\ref{proposition_threshold} 
corresponds to the prefactor in the information-theoretically optimal 
sample complexity scaling~${\cal O}(k\log (N/k))$~\cite{Wainwright092,Fletcher09,Aeron10,Scarlett17,Gamarnik17,Reeves192}. Existing papers~\cite{Aeron10} and 
\cite{Scarlett17} considered different power normalization similar to that 
in \cite{Truong23} and a slightly different sublinear sparsity regime from 
those in this paper, respectively. Existing results in 
\cite{Wainwright092,Fletcher09,Gamarnik17,Reeves192} are relevant to those 
in this paper. 

In terms of achievability, Wainwright~\cite[Theorem 1(b)]{Wainwright092} 
proved that, for any fixed non-zero signal elements, exact support recovery is 
possible if $\delta$ is larger than $(2048 + \epsilon)
\max\{1, \sigma^{2}/u_{\mathrm{min}}^{2}\} + o(1)$ for any $\epsilon>0$. 
An optimum bound was proved in \cite{Reeves192} for constant non-zero signals 
with $P=u_{\mathrm{min}}^{2}$ in the high SNR regime: Exact signal recovery is 
possible as $P/\sigma^{2}\to\infty$ if and only if $\delta$ is larger than 
the inverse $2/\log(1 + P/\sigma^{2})$ of the Gaussian channel capacity. 
Comparing the reconstruction threshold $\delta^{*}=2(1+\sigma^{2}/P)$ 
with this optimum result, we find that Bayesian GAMP is strictly suboptimal 
in the high SNR regime. 

In terms of converse, Wainwright~\cite[Theorem 2]{Wainwright092} proved 
that there is an instance of non-zero signal elements such that the average 
error probability for support recovery is larger than $1/2$ for all support 
recovery algorithms if $\delta$ is smaller than 
$\sigma^{2}/(4u_{\mathrm{min}}^{2})$. Also, 
Fletcher {\em et al}.~\cite{Fletcher09} proved that the average error 
probability tends to $1$ for the maximum likelihood (ML) support recovery 
if $\delta$ is smaller than $2\sigma^{2}/u_{\mathrm{min}}^{2}$, which is smaller 
than the reconstruction threshold 
$\delta^{*}=2(1+ \sigma^{2}/u_{\mathrm{min}}^{2})$ for Bayesian GAMP. 

\subsubsection{1-Bit Compressed Sensing}
Consider 1-bit compressed sensing, i.e.\ the sign measurement function 
$g(z, w)=\mathrm{sgn}(z + w)$ and the Gaussian noise vector 
$\boldsymbol{w}\sim\mathcal{N}(\boldsymbol{0}, \sigma^{2}\boldsymbol{I}_{M})$. 
In this case, the Bayes-optimal outer denoiser~(\ref{Bayes_outer_denoiser}) 
for $y=1$ is given by 
\begin{align}
f_{\Out,t}(z_{t}, -1&;\bar{v}_{\In,t}) 
= p_{\mathrm{G}}(z_{t}; \sigma^{2} + \bar{v}_{\In,t}) 
\nonumber \\
&\cdot\left\{
 \int p_{\mathrm{G}}(z - z_{t}; \bar{v}_{\In,t})Q\left(
  \frac{z}{\sigma}
 \right)dz
\right\}^{-1},
\end{align}
where $Q(x)$ denotes the complementary cumulative distribution 
function of the standard Gaussian distribution. The symmetry 
$f_{\Out,t}(z_{t},1;\bar{v}_{\In,t})=-f_{\Out,t}(-z_{t}, -1; \bar{v}_{\In,t})$ 
is available for $y=-1$. The partial 
derivative of $f_{\Out,t}$ with respect to $z_{t}$ reduces to 
\begin{equation}
\frac{\partial f_{\Out,t}}{\partial z_{t}}
(z_{t}, y;\bar{v}_{\In,t})
= f_{\Out,t}\left(
 f_{\Out,t} - \frac{z_{t}}{\sigma^{2} + \bar{v}_{\In,t}}
\right). 
\end{equation}
Furthermore, $\bar{v}_{\Out,t}$ in (\ref{Bayes_v_out_bar}) is given by 
\begin{equation} \label{1bCS_v_out_bar}
\frac{1}{\bar{v}_{\Out,t}} 
= 2\mathbb{E}\left[
 Q\left(
  \frac{Z_{t}}{\sqrt{\sigma^{2} + \bar{v}_{\In,t}}}
 \right)\frac{\partial f_{\Out,t}}{\partial z_{t}}(Z_{t}, -1;\bar{v}_{\In,t})
\right], 
\end{equation}
with $Z_{t}\sim\mathcal{N}(0, 1 - \bar{v}_{\In,t})$. 

For the noiseless case $\sigma^{2}=0$, the measurement vector 
$\boldsymbol{y}=\mathrm{sgn}(\boldsymbol{A}\boldsymbol{x})$ is invariant 
even if $\boldsymbol{x}$ is replaced with $C\boldsymbol{x}$ for any 
constant $C\neq0$. This invariance implies that the signal norm 
$\|\boldsymbol{x}\|_{2}$ cannot be estimated from the measurement vector 
unless a priori information on $\boldsymbol{x}$ is available. Thus, it is 
fair to compare reconstruction algorithms for 1-bit compressed sensing 
in terms of the squared norm for the normalized error 
\begin{equation} \label{squared_norm}
\left\|
 \frac{\boldsymbol{x}}{\|\boldsymbol{x}\|_{2}} 
 - \frac{\hat{\boldsymbol{x}}}{\|\hat{\boldsymbol{x}}\|_{2}}
\right\|_{2}^{2}, 
\end{equation}
with an estimator $\hat{\boldsymbol{x}}$ of the 
signal vector. 

For Bayesian GAMP, the squared norm~(\ref{squared_norm}) for the 
normalized error is equal to the unnormalized square 
error~(\ref{Bayes_v_in_bar}) in the sublinear sparsity limit. This can be 
understood as follows: From Assumption~\ref{assumption_x} the signal vector 
$\boldsymbol{x}$ has the unit norm with probability $1$ in the sublinear 
sparsity limit. Since the Bayesian inner denoiser utilizes the true prior 
distribution of the non-zero signal elements, Bayesian GAMP should be able 
to estimate the signal norm $\|\boldsymbol{x}\|_{2}$ accurately.

\section{Numerical Results} \label{sec5}
\subsection{Numerical Conditions}
In all numerical results, Assumption~\ref{assumption_x} is postulated 
for the $k$-sparse signal vector $\boldsymbol{x}$. In particular, we use 
a standard Gaussian random variable $U\sim\mathcal{N}(0, 1)$ with $P=1$ 
to represent the non-zero signals. The sensing matrix $\boldsymbol{A}$ is 
a standard Gaussian random matrix, as postulated in 
Assumption~\ref{assumption_A}.  
The Gaussian noise vector $\boldsymbol{w}\sim\mathcal{N}(\boldsymbol{0},
\sigma^{2}\boldsymbol{I}_{M})$ with variance $\sigma^{2}>0$ is used to satisfy 
Assumption~\ref{assumption_w}. 

We consider Bayesian GAMP, which is GAMP using the Bayes-optimal outer 
denoiser and Bayesian inner denoiser defined in Section~\ref{sec2}. 
In simulating Bayesian GAMP, $\bar{\xi}_{\In,t}$ in (\ref{z_t}) was replaced 
with $\xi_{\In,t}$ in (\ref{xi_in}). Owing to this replacement, we do not 
need to solve the state evolution recursion in simulating Bayesian GAMP.    

\begin{figure}[t]
\begin{center}
\includegraphics[width=\hsize]{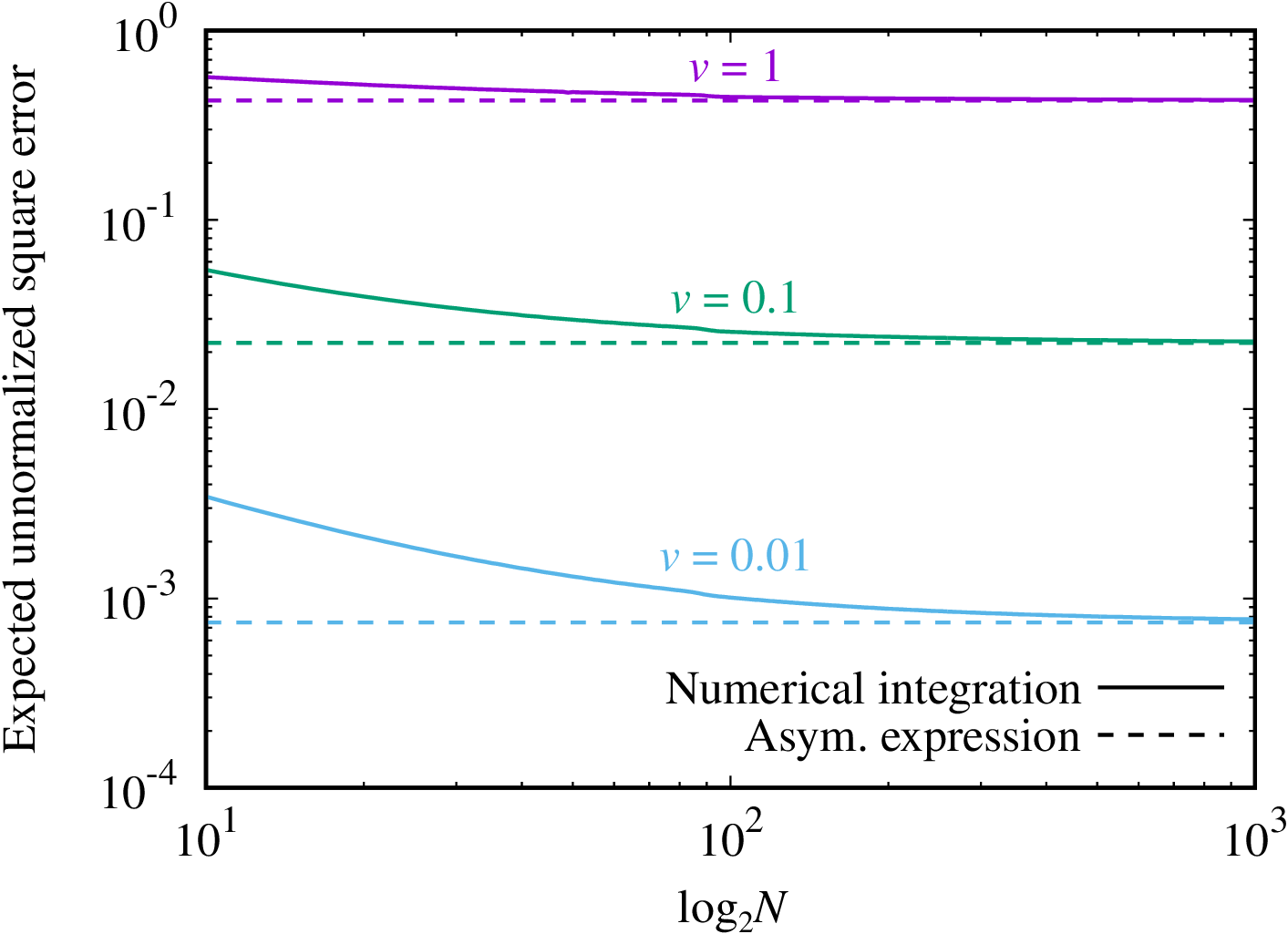}
\caption{
The expected unnormalized square error 
$\mathbb{E}[\|\boldsymbol{X} - f_{X}(\boldsymbol{Y}; v_{\tilde{N}})\|_{2}^{2}]$ 
versus $\log_{2}N$ in Lemma~\ref{lemma_MSE} for $k=N^{0.25}$. The horizontal 
dashed lines show the asymptotic expression~(\ref{unnormalized_square_error}) 
in the sublinear sparsity limit. 
}
\label{fig1} 
\end{center}
\end{figure}

\subsection{Linear Measurement}
\subsubsection{State Evolution}
We first verify the asymptotic expression~(\ref{unnormalized_square_error}) 
for the unnormalized square error in Lemma~\ref{lemma_MSE}. 
Figure~\ref{fig1} shows the expected unnormalized square error 
$\mathbb{E}[\|\boldsymbol{X} - f_{X}(\boldsymbol{Y}; v_{\tilde{N}})\|_{2}^{2}]$ 
with $v_{\tilde{N}}=v/\log(N/k)$. The expected unnormalized square error 
converges to the asymptotic expression~(\ref{unnormalized_square_error}) 
for all $v$ as $N$ increases. However, the convergence speed is extremely 
slow: We need $N=2^{1000}$ for the asymptotic expression to provide 
a good approximation for the expected unnormalized square error. 

\begin{figure}[t]
\begin{center}
\includegraphics[width=\hsize]{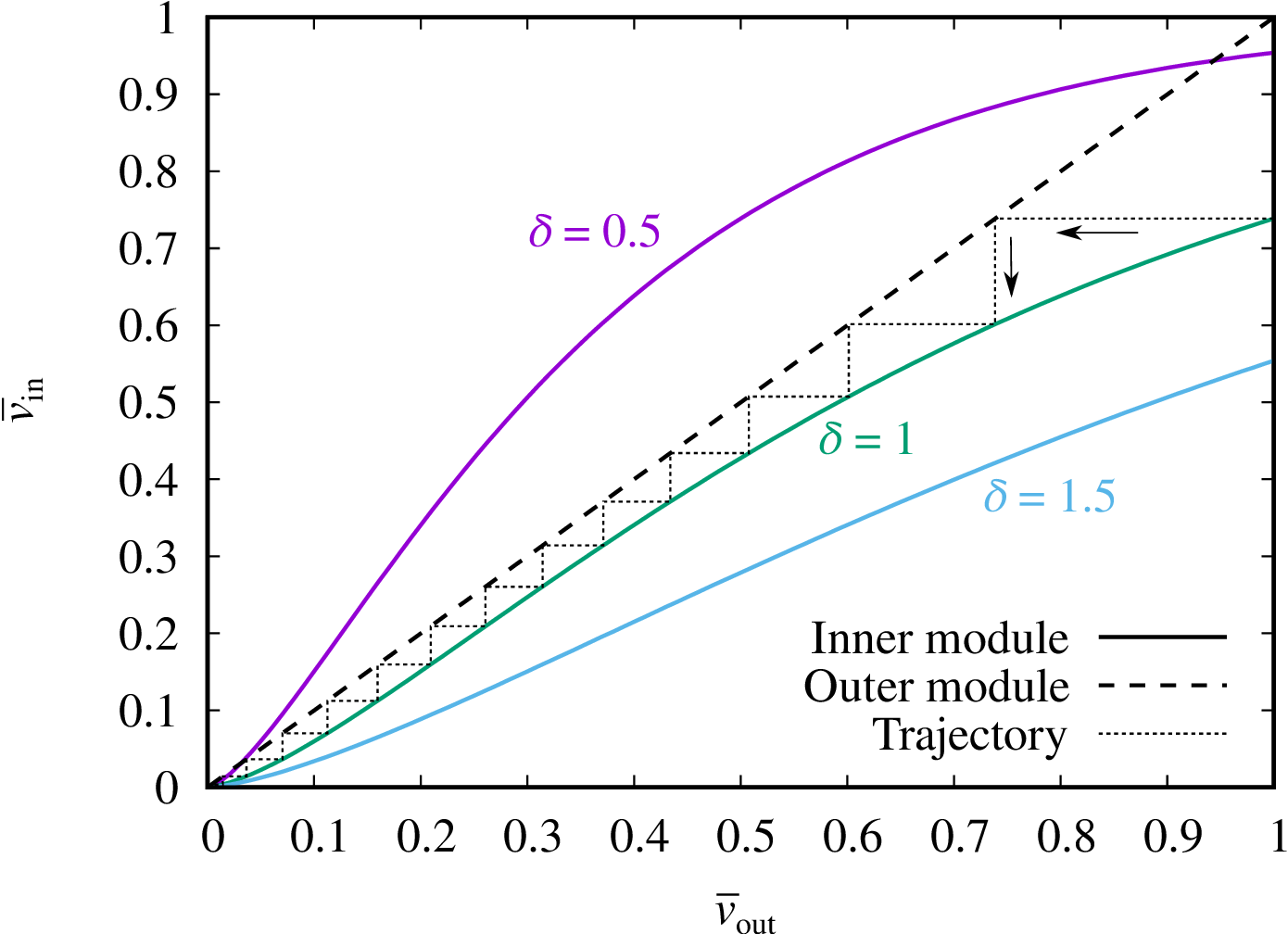}
\caption{
EXIT-like chart of Bayesian AMP for the linear measurement and 
$1/\sigma^{2}=40$~dB. The outer module shows 
the straight line~(\ref{linear_v_out_bar}) while the inner module 
represents (\ref{Bayes_v_in_bar}) for $M/\{k\log (N/k)\}\to\delta$ in the 
sublinear sparsity limit. The chart is independent of the parameter 
$\log k/\log N\to\gamma\in[0, 1)$. 
}
\label{fig2} 
\end{center}
\end{figure}

We next focus on a chart to understand the dynamics of the state evolution 
recursion for the linear measurement, like the extrinsic information 
transfer (EXIT) chart. As shown in Fig.~\ref{fig2}, $\bar{v}_{\In,t}$ in 
(\ref{Bayes_v_in_bar}) converges to the unique fixed 
point $\bar{v}_{\In,\infty}\approx 0$ for $\delta=1$ and $\delta=1.5$ as 
$t\to\infty$. On the other hand, $\bar{v}_{\In,t}$ tends to the fixed point 
in the upper right for $\delta=0.5$. Since the state evolution 
recursion~(\ref{Bayes_v_in_bar}) is a monotonically non-increasing function of 
$\delta$ for fixed $\bar{v}_{\Out,t}$, we arrive at the following definition 
of the weak reconstruction threshold: 

\begin{definition}[Weak Reconstruction Threshold]
\label{def_weak_reconstruction_threshold}
The weak reconstruction threshold $\delta_{\mathrm{w}}^{*}>0$ for Bayesian 
AMP is defined as the infimum of $\delta'$ such that the state evolution 
recursion~(\ref{Bayes_v_out_bar}) and (\ref{Bayes_v_in_bar}) have a unique 
fixed point for all $\delta>\delta'$. 
\end{definition}

The reconstruction threshold in Definition~\ref{def_reconstruction_threshold} 
requires exact signal reconstruction $\bar{v}_{\In,\infty}=0$, as well as 
the uniqueness of the fixed point. The weak reconstruction threshold 
$\delta_{\mathrm{w}}^{*}$ indicates that the performance of Bayesian AMP 
changes discontinuously at $\delta=\delta_{\mathrm{w}}^{*}$. 

\subsubsection{Numerical Simulation} 
Bayesian AMP was simulated for the linear measurement. As baselines, 
orthogonal matching pursuit (OMP)~\cite{Tropp07} and FISTA~\cite{Beck09} 
were also simulated. FISTA is an iterative algorithm 
to solve the Lasso problem
\begin{equation} \label{Lasso}
\min_{\boldsymbol{x}\in\mathbb{R}^{N}}
\frac{1}{2M}\|\boldsymbol{y} - \boldsymbol{A}\boldsymbol{x}\|_{2}^{2} 
+ \lambda\|\boldsymbol{x}\|_{1}
\end{equation} 
for some $\lambda>0$. To improve the convergence property of FISTA, 
gradient-based restart~\cite{O'Donoghue15} was used. Furthermore, the 
parameter $\lambda$ was optimized for each $\delta$ via exhaustive search 
to minimize the unnormalized square error in the last iteration. 

\begin{figure}[t]
\begin{center}
\includegraphics[width=\hsize]{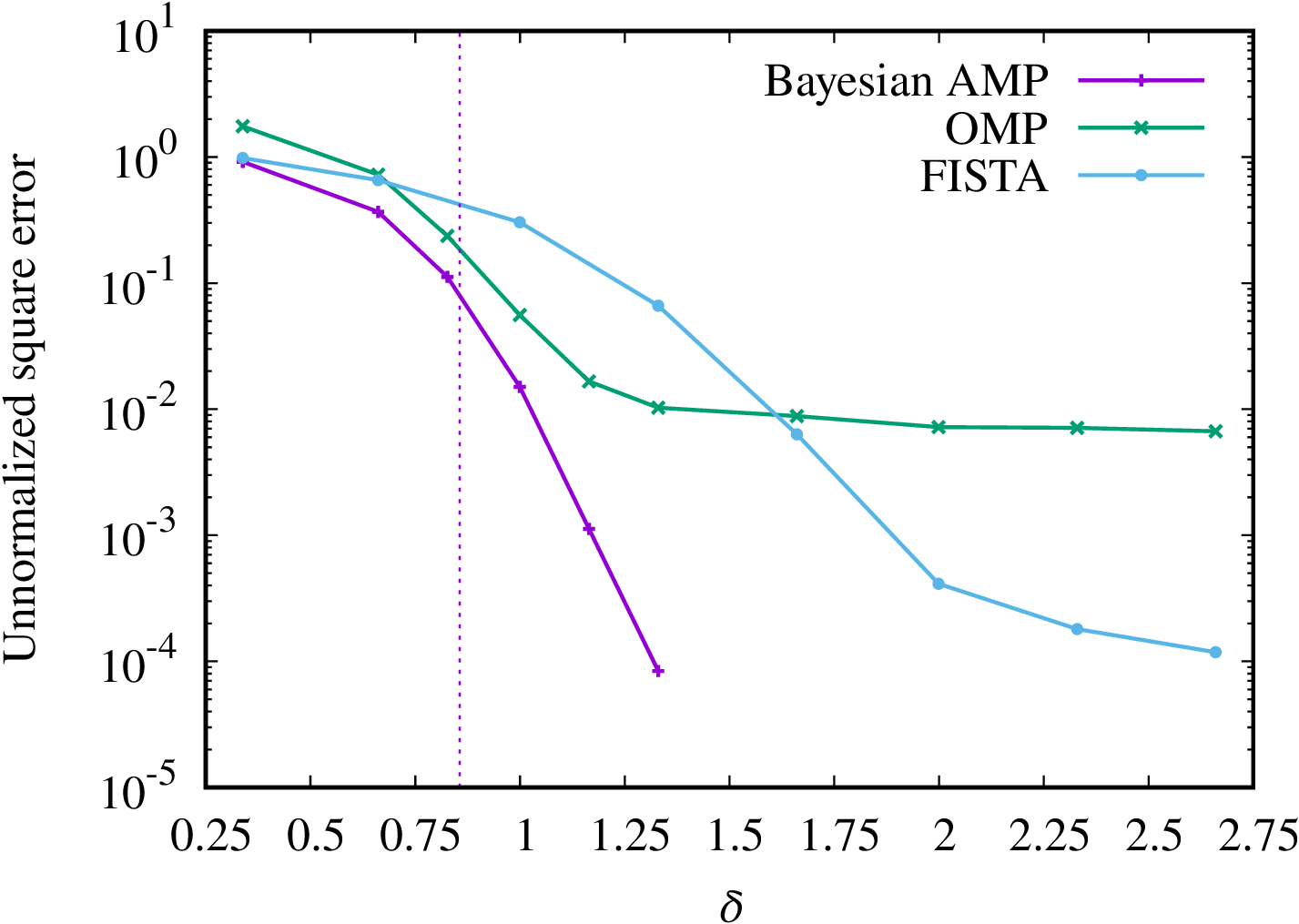}
\caption{
Unnormalized square error versus $\delta=M/\{k\log (N/k)\}$ for the linear 
measurement, $k=16$, $N=2^{16}$, and $1/\sigma^{2}=40$~dB. 
Bayesian AMP with $20$ iterations is compared 
to FISTA with backtracking~\cite{Beck09}, gradient-based 
restart~\cite{O'Donoghue15}, $10^{3}$ iterations, and optimized $\lambda$ 
in (\ref{Lasso}) for each $\delta$, as well as OMP~\cite{Tropp07} with $k$ 
iterations. The vertical dotted line shows the weak reconstruction threshold 
in Definition~\ref{def_weak_reconstruction_threshold} for Bayesian AMP. 
$10^{4}$ independent trials were simulated for all algorithms, with the only 
exception of $10^{5}$ trials for Bayesian AMP with $\delta\approx 1.33$. 
}
\label{fig3} 
\end{center}
\end{figure}

Figure~\ref{fig3} shows the unnormalized square error of the three algorithms 
for the linear measurement. OMP has the error floor for large $\delta$ while 
it is comparable to Bayesian AMP around the weak reconstruction threshold. 
This observation is consistent to the fact that, once the position of 
non-zero signals is detected incorrectly, OMP has no procedure to correct 
the detection error. 

FISTA is inferior to Bayesian AMP while it achieves smaller unnormalized 
square error than OMP for large $\delta$. The poor performance of FISTA 
for large $\delta$ is due to its slow convergence: $10^{3}$ is not many 
iterations enough for FISTA to reach the true Lasso solution.  
Furthermore, FISTA is the worst among the three algorithms 
around the weak reconstruction threshold. This may be because FISTA uses 
no a priori information on the sparse signal vector. 

Bayesian AMP achieves the best performance for all $\delta$. Interestingly, 
the weak reconstruction threshold provides a reasonably good prediction for 
the location of the so-called waterfall region for Bayesian AMP, in which 
the unnormalized square error decreases quickly as $\delta$ increases.  
This observation implies that qualitative changes, such as the number of 
fixed points for the state evolution recursion, are robust to the finite size 
effect, while the state evolution recursion cannot provide quantitatively 
reliable prediction for the unnormalized square error, as indicated from 
Fig.~\ref{fig1}. 

\begin{figure}[t]
\begin{center}
\includegraphics[width=\hsize]{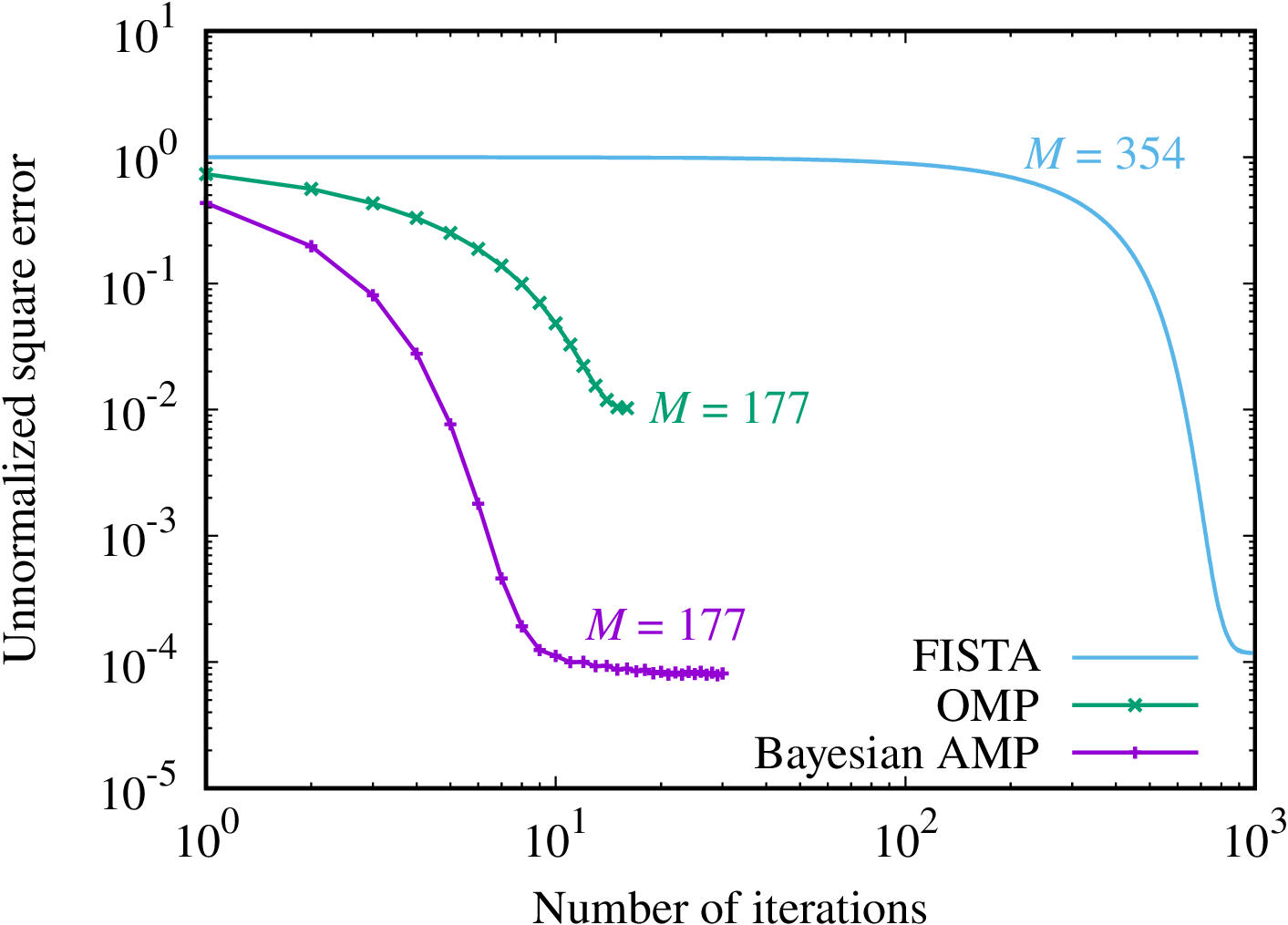}
\caption{
Unnormalized square error versus the number of iterations for the linear 
measurement, $k=16$, $N=2^{16}$, and $1/\sigma^{2}=40$~dB. 
Bayesian AMP is compared to OMP~\cite{Tropp07} and FISTA with 
backtracking~\cite{Beck09}, gradient-based restart~\cite{O'Donoghue15}, and 
optimized $\lambda=\sqrt{0.8\sigma^{2}M^{-1}\log N}$~\cite{Wainwright091}. 
$10^{5}$ independent trials were simulated for Bayesian AMP while 
$10^{4}$ independent trials were for OMP and FISTA.  
}
\label{fig4} 
\end{center}
\end{figure}

Figure~\ref{fig4} shows the convergence properties of the three algorithms. 
Bayesian AMP converges to a fixed point much more quickly than FISTA. 
The convergence speed of FISTA depends strongly on the parameter $\lambda$ 
in (\ref{Lasso}).  
Since $\lambda$ was optimized so as to minimize the unnormalized square error 
at the last iteration, FISTA converges slowly. If we used $\lambda$ smaller 
than the optimized value in Fig.~\ref{fig4}, FISTA could not converge even 
after $10^{3}$ iterations. When we use larger $\lambda$, 
FISTA converges more quickly but cannot achieve the final unnormalized square 
error for the optimized value in Fig.~\ref{fig4}.

\subsection{1-bit Compressed Sensing}
\subsubsection{State Evolution}
\begin{figure}[t]
\begin{center}
\includegraphics[width=\hsize]{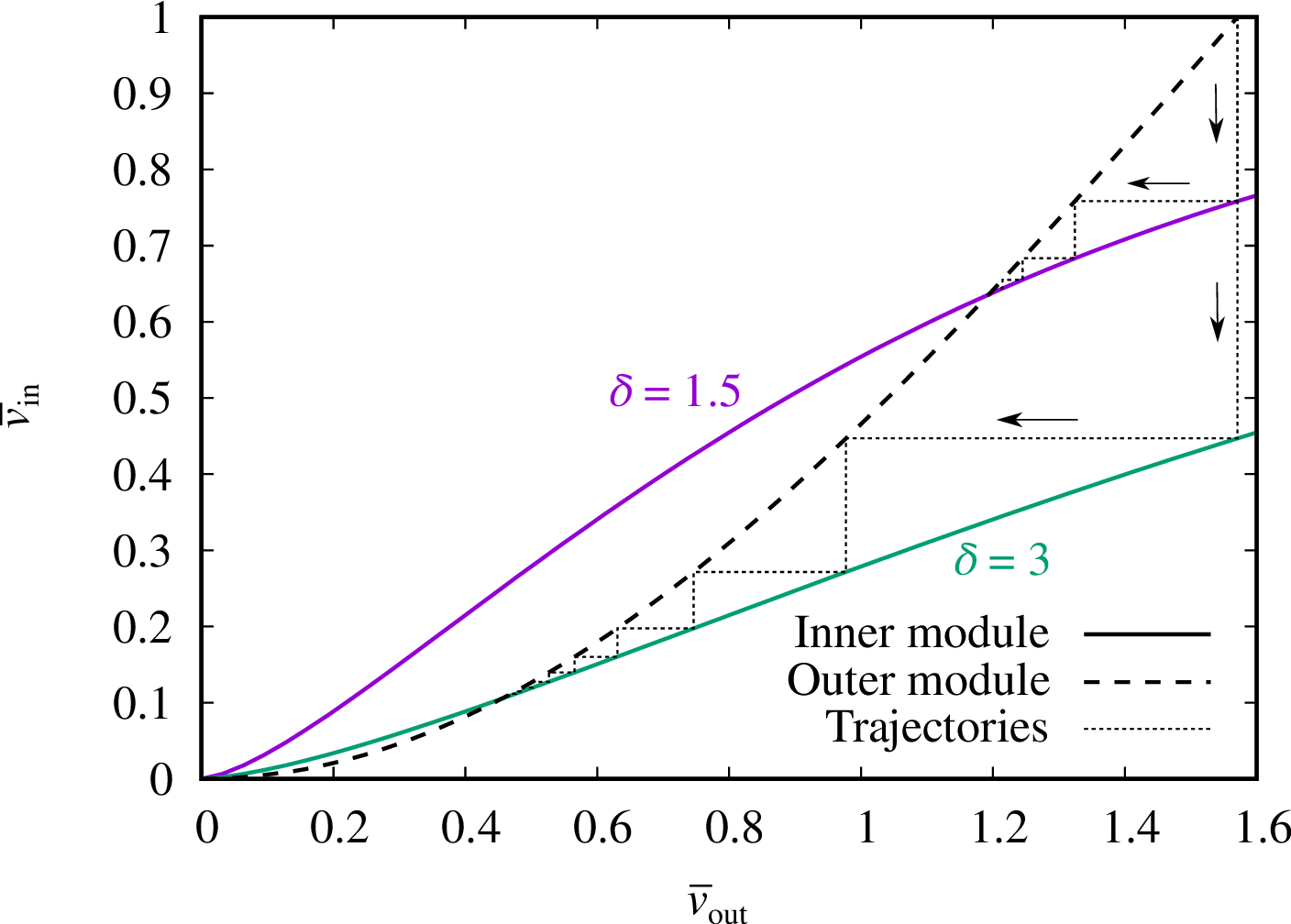}
\caption{
EXIT-like chart of Bayesian GAMP for 1-bit compressed sensing and 
$\sigma^{2}=0$. The outer module shows the inverse function of 
(\ref{1bCS_v_out_bar}) while the inner module represents 
(\ref{Bayes_v_in_bar}) for $M/\{k\log (N/k)\}\to\delta$ in the sublinear 
sparsity limit. The chart is independent of the parameter 
$\log k/\log N\to\gamma\in[0, 1)$. 
}
\label{fig5} 
\end{center}
\end{figure}

We focus on 1-bit compressed sensing for the noiseless case $\sigma^{2}=0$. 
Figure~\ref{fig5} shows an EXIT-like chart for the state evolution recursion 
of Bayesian GAMP. The two curves for the outer and inner modules have two 
intersections $(\bar{v}_{\Out,\infty}, \bar{v}_{\In,\infty})$ at 
$\bar{v}_{\Out,\infty}=0$ and $\bar{v}_{\Out,\infty}>0$, of which the latter is the 
convergence point of the state evolution. 

The EXIT-like chart for 1-bit compressed sensing is qualitatively different 
from that for the linear measurement in Fig.~\ref{fig2}: Numerical 
evaluation implied that the state evolution 
recursion~(\ref{Bayes_v_in_bar}) and (\ref{1bCS_v_out_bar}) have two 
fixed points for all $\delta>0$. In particular, the convergence point of 
Bayesian GAMP moves continuously toward the origin as $\delta$ increases. 
Thus, the reconstruction threshold in 
Definition~\ref{def_reconstruction_threshold} is 
$\delta^{*}=\infty$ for the sparse signal vector with non-zero Gaussian 
elements.

\subsubsection{Numerical Simulation}
Bayesian GAMP was simulated for 1-bit compressed sensing in the noiseless 
case $\sigma^{2}=0$. As baselines, BIHT~\cite{Matsumoto22} and 
GLasso~\cite{Plan13,Thrampoulidis15} were also simulated. 
FISTA~\cite{Beck09} was used to solve the Lasso problem~(\ref{Lasso}) for 
GLasso. All algorithms converged quickly for 1-bit compressed sensing, 
as opposed to the linear measurement, so that $20$ iterations were used for 
all algorithms. 

\begin{figure}[t]
\begin{center}
\includegraphics[width=\hsize]{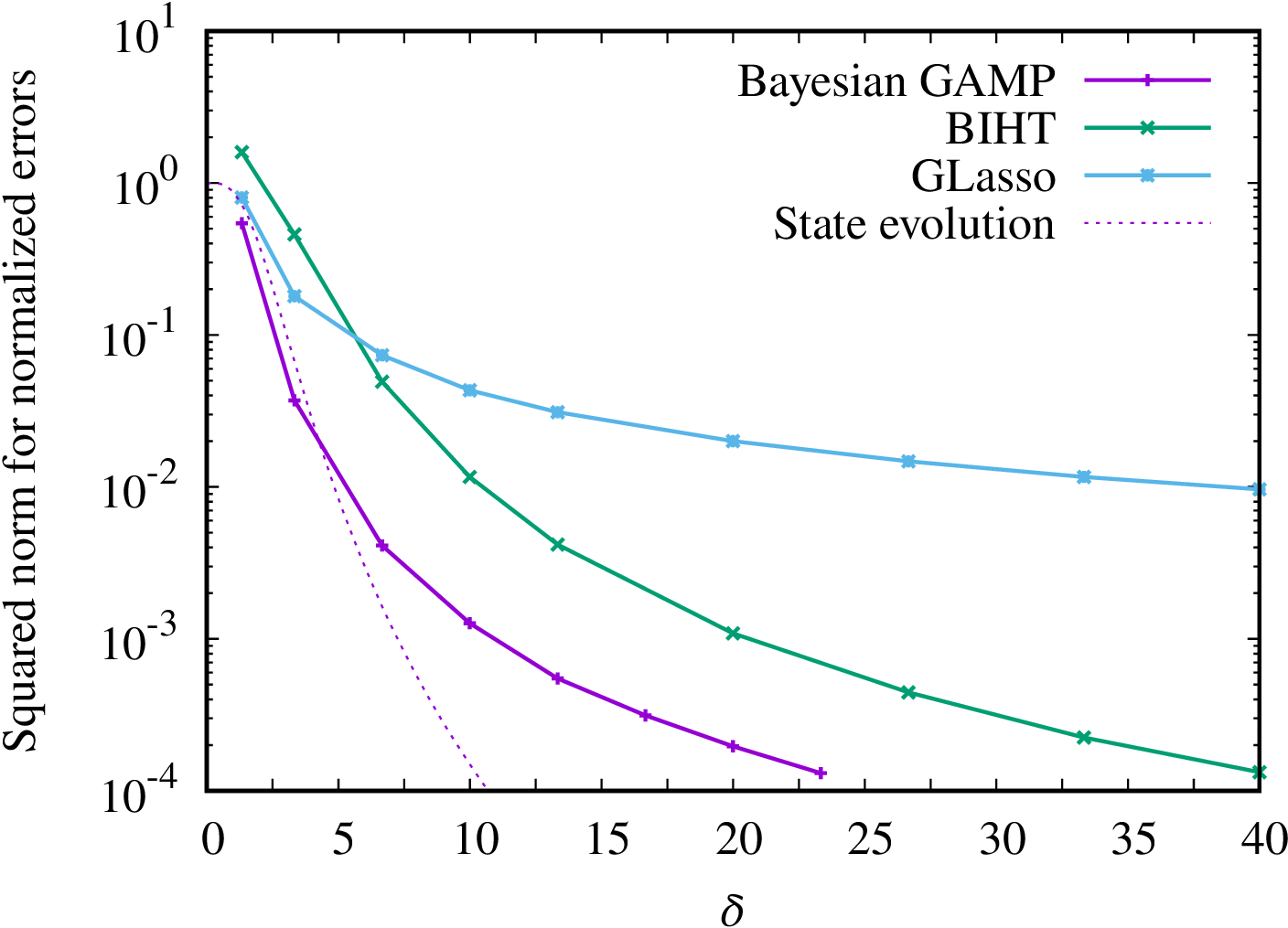}
\caption{
Squared norm~(\ref{squared_norm}) for the normalized error versus 
$\delta=M/\{k\log (N/k)\}$ for 1-bit compressed sensing, $k=16$, 
$N=2^{16}$, and $\sigma^{2}=0$. 
Bayesian GAMP is compared to BIHT~\cite{Matsumoto22} and 
GLasso~\cite{Plan13,Thrampoulidis15}, as well as the state evolution for 
Bayesian GAMP. GLasso was implemented by using FISTA 
with backtracking~\cite{Beck09}, gradient-based restart~\cite{O'Donoghue15}, 
and optimized $\lambda$ in (\ref{Lasso}) for each $\delta$. 
$20$ iterations and $10^{4}$ independent trials were performed 
for all algorithms. 
}
\label{fig6} 
\end{center}
\end{figure}

As shown in Fig.~\ref{fig6}, Bayesian GAMP achieves the smallest squared 
norm~(\ref{squared_norm}) among the three algorithms. Interestingly, the 
state evolution result provides a reasonably accurate prediction for 
finite-sized systems when $\delta$ is small, while it cannot predict the 
occurrence of the error floor in the large $\delta$ regime. As a result, 
state evolution allows us to estimate roughly the location of waterfall 
region in which the squared norm~(\ref{squared_norm}) for the normalized error 
decreases rapidly. 

\section{Conclusions} \label{sec6}
This paper has proposed GAMP for reconstruction of an unknown signal vector 
with sublinear sparsity from the generalized linear measurement. 
State evolution has been used to evaluate the asymptotic dynamics of GAMP 
in terms of the unnormalized square error. The overall flow in state evolution 
is the same as that for the linear sparsity. To justify each proof 
step in state evolution. however, stronger assumptions for the inner denoiser 
are required than those for the linear sparsity. This paper has proved the 
required assumptions for the Bayesian inner denoiser. The state evolution and 
numerical results have implied that Bayesian GAMP is useful not only 
for the linear sparsity but also for the sublinear sparsity. 

A challenge is to specify the class of inner denoisers that satisfy 
Assumption~\ref{assumption_inner}. It implies a smaller class 
than that of Lipschitz-continuous denoisers for the linear sparsity. 
Toward solving this challenging problem, a possible direction in future 
research would be to replace the Bayesian inner denoiser with 
non-separable denoisers. 

Another challenge is finite size analysis of GAMP for the sublinear sparsity.  
As shown in Fig.~\ref{fig1}, the sublinear sparsity limit needs too large 
systems to provide an accurate approximation for finite-sized systems. 
As a result, state evolution only provides a reasonably good prediction for 
the position of the reconstruction threshold. Finite size analysis would be 
able to provide an accurate prediction for the transient dynamics of 
GAMP for the sublinear sparsity. 

%\appendix
\appendices

\section{Proof of Theorem~\ref{theorem_SE}}
\label{proof_theorem_SE}
\subsection{Overview}
The proof of Theorem~\ref{theorem_SE} consists of three steps: A first step 
is to formulate an error model that describes the dynamics of estimation 
errors for GAMP. The error model should be formulated so that the asymptotic 
Gaussianity of the estimation errors can be proved via state evolution in the 
next step. If the asymptotic Gaussianity is not realizable, the Onsager 
correction in GAMP should be re-designed. Fortunately, we can use the 
Onsager correction in the original GAMP for linear sparsity to prove the 
asymptotic Gaussianity for sublinear sparsity. 

A second step is state evolution of the error model formulated in the 
first step. Bolthausen's conditioning technique~\cite{Bolthausen14} is used 
to prove the asymptotic Gaussianity of the estimation errors. 
In his technique, the conditional distribution of the sensing matrix 
$\boldsymbol{A}$ given all previous messages is utilized to evaluate the 
distribution of the current messages. 

The last step is the derivation of state evolution recursion for GAMP. 
The state evolution recursion is obtained as a corollary of the main 
technical result in the second step. 

A challenge is in the proof of properties for the inner denoiser 
in the second step. For the linear sparsity, we can repeat the proof of 
properties for the outer denoiser to prove them for the inner 
denoiser~\cite{Rangan11,Javanmard13,Takeuchi241}. For the sublinear 
sparsity, on the other hand, specific properties of the inner denoiser have 
to be used to evaluate unnormalized quantities for the inner denoiser. 
To solve this issue, we use Assumption~\ref{assumption_inner} and 
Lemma~\ref{lemma_covariance_error}, of which the latter is presented 
in Appendix~\ref{appen_state_evolution} along with the other existing lemmas. 

\subsection{Error Model}
Let $\boldsymbol{h}_{t}= \boldsymbol{x}_{t} - \bar{\eta}_{t}\boldsymbol{x}$ and 
$\boldsymbol{q}_{t+1}=\hat{\boldsymbol{x}}_{t+1} - \bar{\eta}_{t+1}\boldsymbol{x}$ 
denote estimation errors before and after inner denoising~(\ref{x_hat_t}), 
respectively, with $\bar{\eta}_{t}$ in (\ref{eta_bar}). Similarly, 
define $\boldsymbol{b}_{t}=\boldsymbol{z}_{t}-\bar{\eta}_{t}\boldsymbol{z}$ and 
$\boldsymbol{m}_{t}=\xi_{\Out,t}^{-1}\hat{\boldsymbol{z}}_{t}$ before and after 
outer denoising~(\ref{z_hat_t}). In the first step, 
we prove that they satisfy the following error model: 
\begin{equation} \label{b}
\boldsymbol{b}_{t} = \boldsymbol{A}\boldsymbol{q}_{t} 
+ \bar{\xi}_{\In,t-1}\boldsymbol{m}_{t-1}, 
\end{equation}
\begin{equation} \label{m}
\boldsymbol{m}_{t} = \frac{1}{\xi_{\Out,t}}f_{\Out,t}\left(
 \boldsymbol{b}_{t} + \bar{\eta}_{t}\boldsymbol{z}, 
 \boldsymbol{y}; v_{\In,t}
\right), 
\end{equation}
\begin{equation} \label{h}
\boldsymbol{h}_{t} = \boldsymbol{q}_{t} 
- \frac{1}{M}\boldsymbol{A}^{\mathrm{T}}\boldsymbol{m}_{t}, 
\end{equation}
\begin{equation} \label{q}
\boldsymbol{q}_{t+1} = f_{\In,t}(\bar{\eta}_{t}\boldsymbol{x} 
+ \boldsymbol{h}_{t}; v_{\Out,t}) 
- \bar{\eta}_{t+1}\boldsymbol{x},  
\end{equation}
with the initial condition 
$\boldsymbol{b}_{0}=\boldsymbol{A}\boldsymbol{q}_{0}$ and 
$\boldsymbol{q}_{0}= -\bar{\eta}_{0}\boldsymbol{x}$. 
It is convenient to introduce the convention $f_{\In,-1}=0$, 
which allows us to obtain $\boldsymbol{q}_{0}= -\bar{\eta}_{0}\boldsymbol{x}$ 
by letting $t=-1$ in (\ref{q}). 

We obtain the representation of $\boldsymbol{m}_{t}$ in (\ref{m}) 
by substituting $\boldsymbol{z}_{t}
=\boldsymbol{b}_{t}+\bar{\eta}_{t}\boldsymbol{z}$ into 
$\boldsymbol{z}_{t}$ in (\ref{z_hat_t}). 
Similarly, we have $\boldsymbol{q}_{t+1}$ in (\ref{q}) 
by substituting $\boldsymbol{x}_{t}
=\bar{\eta}_{t}\boldsymbol{x} + \boldsymbol{h}_{t}$ into 
$\boldsymbol{q}_{t+1}=f_{\In,t}(\boldsymbol{x}_{t}; v_{\Out,t}) 
- \bar{\eta}_{t+1}\boldsymbol{x}$ obtained from the definition of 
$\hat{\boldsymbol{x}}_{t+1}$ in (\ref{x_hat_t}). 

The representation of $\boldsymbol{b}_{t}$ in (\ref{b}) is obtained by 
applying the definition of $\boldsymbol{z}_{t}$ in (\ref{z_t}) to 
$\boldsymbol{b}_{t}=\boldsymbol{z}_{t}-\bar{\eta}_{t}\boldsymbol{z}$ and 
using $\boldsymbol{z}=\boldsymbol{A}\boldsymbol{x}$, 
$\boldsymbol{q}_{t}=\hat{\boldsymbol{x}}_{t} - \bar{\eta}_{t}\boldsymbol{x}$, 
and $\boldsymbol{m}_{t}=\xi_{\Out,t}^{-1}\hat{\boldsymbol{z}}_{t}$. Finally, 
the representation of $\boldsymbol{h}_{t}$ in (\ref{h}) follows from 
$\boldsymbol{h}_{t}= \boldsymbol{x}_{t} - \bar{\eta}_{t}\boldsymbol{x}$, 
the definition of $\boldsymbol{x}_{t}$ in (\ref{x_t}),  
$\boldsymbol{q}_{t}=\hat{\boldsymbol{x}}_{t} - \bar{\eta}_{t}\boldsymbol{x}$, 
and $\boldsymbol{m}_{t}=\xi_{\Out,t}^{-1}\hat{\boldsymbol{z}}_{t}$. 

\subsection{State Evolution} \label{appen_state_evolution}
In the second step for the proof of Theorem~\ref{theorem_SE}, we use 
Bolthausen's conditioning technique~\cite{Bolthausen14} to evaluate the 
dynamics of the error model~(\ref{b})--(\ref{q}) via state evolution. 
In the technique, the conditional distribution of $\boldsymbol{A}$ given 
all previous messages is utilized. To represent previous messages concisely, 
we define the matrix $\boldsymbol{B}_{t}=[\boldsymbol{b}_{0},\ldots,
\boldsymbol{b}_{t-1}]\in\mathbb{R}^{M\times t}$. Similarly, we define the matrices 
$\boldsymbol{M}_{t}\in\mathbb{R}^{M\times t}$, 
$\boldsymbol{H}_{t}\in\mathbb{R}^{N\times t}$, and 
$\boldsymbol{Q}_{t}\in\mathbb{R}^{N\times t}$. 

The random vectors $\Theta=\{\boldsymbol{x}, \boldsymbol{w}\}$ are fixed  
throughout state evolution analysis. Let $\mathfrak{E}_{t',t}
=\{\boldsymbol{B}_{t'}, \boldsymbol{M}_{t'}, \boldsymbol{H}_{t}, 
\boldsymbol{Q}_{t+1}\}$. The set $\mathfrak{E}_{t,t}$ contains the messages 
computed just before updating the message $\boldsymbol{b}_{t}$ in (\ref{b}). 
On the other hand, $\mathfrak{E}_{t+1,t}$ includes the messages computed just 
before updating $\boldsymbol{h}_{t}$ in (\ref{h}). The conditional distribution 
of $\boldsymbol{A}$ given $\mathfrak{E}_{t,t}$ or $\mathfrak{E}_{t+1,t}$, as 
well as $\Theta$, is evaluated with the following existing lemma: 

\begin{lemma}[\cite{Bayati11}] \label{lemma_conditioning}
Suppose that Assumption~\ref{assumption_A} holds. For some integers 
$t\leq M$ and $t'\leq N$, let $\boldsymbol{X}\in\mathbb{R}^{M\times t}$, 
$\boldsymbol{U}\in\mathbb{R}^{N\times t}$, 
$\boldsymbol{Y}\in\mathbb{R}^{N\times t'}$, and 
$\boldsymbol{V}\in\mathbb{R}^{M\times t'}$ satisfy 
the following constraints:
\begin{equation}
\boldsymbol{X} = \boldsymbol{A}\boldsymbol{U}, \quad 
\boldsymbol{Y} = \boldsymbol{A}^{\mathrm{T}}\boldsymbol{V}. 
\end{equation}
\begin{itemize}
\item If $\boldsymbol{U}$ has full rank, 
then the conditional distribution of $\boldsymbol{A}$ given  
$\boldsymbol{X}$ and $\boldsymbol{U}$ is represented as 
\begin{equation} \label{conditional_A0}
\boldsymbol{A}\sim \boldsymbol{X}\boldsymbol{U}^{\dagger}
+ \tilde{\boldsymbol{A}}\boldsymbol{P}_{\boldsymbol{U}}^{\perp},
\end{equation}
where $\tilde{\boldsymbol{A}}$ is independent of 
$\{\boldsymbol{X}, \boldsymbol{U}\}$ and has independent standard 
Gaussian elements. 

\item If Both $\boldsymbol{U}$ and $\boldsymbol{V}$ have full rank, 
then the conditional distribution of $\boldsymbol{A}$ given  
$\mathfrak{E}=\{\boldsymbol{X}, \boldsymbol{U}, \boldsymbol{Y}, 
\boldsymbol{V}\}$ is represented as 
\begin{align}
\boldsymbol{A} &\sim 
\boldsymbol{X}\boldsymbol{U}^{\dagger}
+ (\boldsymbol{V}^{\dagger})^{\mathrm{T}}
\boldsymbol{Y}^{\mathrm{T}}\boldsymbol{P}_{\boldsymbol{U}}^{\perp} 
+ \boldsymbol{P}_{\boldsymbol{V}}^{\perp}\tilde{\boldsymbol{A}}
\boldsymbol{P}_{\boldsymbol{U}}^{\perp} 
\label{conditional_A1} \\
&= (\boldsymbol{V}^{\dagger})^{\mathrm{T}}
\boldsymbol{Y}^{\mathrm{T}}
+ \boldsymbol{P}_{\boldsymbol{V}}^{\perp}\boldsymbol{X}
\boldsymbol{U}^{\dagger}
+ \boldsymbol{P}_{\boldsymbol{V}}^{\perp}\tilde{\boldsymbol{A}}
\boldsymbol{P}_{\boldsymbol{U}}^{\perp}, 
\label{conditional_A2}
\end{align}
where $\tilde{\boldsymbol{A}}$ is independent of $\mathfrak{E}$ and 
has independent standard Gaussian elements. 
\end{itemize}
\end{lemma}

The Onsager correction in GAMP is designed so as to cancel the influence 
from the first two terms in (\ref{conditional_A1}) or (\ref{conditional_A2}).  
Stein's lemma is useful in designing the Onsager correction. 

\begin{lemma}[Stein's Lemma~\cite{Stein72,Takeuchi241}] \label{lemma_Stein}
Suppose that $\{Z_{\tau}\}_{\tau=1}^{t}$ are zero-mean Gaussian random variables. 
Then, for any piecewise Lipschitz-continuous function 
$f:\mathbb{R}^{t}\to\mathbb{R}$ we have 
\begin{equation} \label{Stein}
\mathbb{E}[Z_{t'}f(Z_{1},\ldots,Z_{t})] = 
\sum_{\tau=1}^{t}\mathbb{E}[Z_{t'}Z_{\tau}]\mathbb{E}\left[
 \frac{\partial f}{\partial Z_{\tau}}(Z_{1},\ldots,Z_{t})
\right]. 
\end{equation}
\end{lemma} 

The Lipschitz-continuity is not assumed for the inner denoiser. Thus, 
Lemma~\ref{lemma_Stein} is replaced with the following Stein lemma 
for the inner denoiser: 
\begin{lemma} \label{lemma_Stein_inner}
Suppose that Assumption~\ref{assumption_inner} is satisfied and follow the 
notation in Assumption~\ref{assumption_inner}. Then, we have 
\begin{align}
&\mathbb{E}\left[
 \omega_{n, \tau}f_{\In,t}(\bar{\eta}_{t}x_{n} + \omega_{n,t}; \bar{v}_{\Out,t})
\right]
\nonumber \\
&= \mathbb{E}\left[
 f_{\In,t}'(\bar{\eta}_{t}x_{n} + \omega_{n,t}; \bar{v}_{\Out,t})
\right]\mathbb{E}[\omega_{n,\tau}\omega_{n,t}] 
\label{Stein_identity}
\end{align}
for all $n\in\{1,\ldots,N\}$ and $\tau\in\{0,\ldots,t\}$. 
\end{lemma}
\begin{IEEEproof}
We follow a proof in \cite[Lemma~2]{Takeuchi21}, in which the 
Lipschitz-continuity assumption for $f_{\In,t}$ was used to prove the 
boundedness of both sides in (\ref{Stein_identity}), as well as the 
almost everywhere differentiability of $f_{\In,t}$. From 
Assumption~\ref{assumption_inner}, these properties 
are justified for the inner denoiser $f_{\In,t}$. Thus, we repeat the 
proof in \cite[Lemma~2]{Takeuchi21} to arrive at 
Lemma~\ref{lemma_Stein_inner}.    
\end{IEEEproof}

The following lemma is utilized to ignore the influence of a small error 
that originates from the first two terms in 
(\ref{conditional_A1}) or (\ref{conditional_A2}). 

\begin{lemma} \label{lemma_covariance_error}
For a random vector $\boldsymbol{\omega}_{i}\in\mathbb{R}^{N}$, with 
$i\in\{0, 1\}$, suppose that 
a separable function 
$\boldsymbol{\psi}_{i}=[\psi_{i,1},\ldots,\psi_{i,N}]^{\mathrm{T}}: 
\mathbb{R}^{N}\to\mathbb{R}^{N}$, i.e.\ 
$[\psi_{i}(\boldsymbol{\omega})]_{n}=\psi_{i,n}(\omega_{n})$ satisfies 
the boundedness in probability 
 $\lim_{N\to\infty}
\mathrm{Pr}(\|\boldsymbol{\psi}_{i}(\boldsymbol{\omega}_{i})\|_{2}<C)=1$ 
for some $C>0$. Let $\Delta\boldsymbol{\psi}_{i}
=\boldsymbol{\psi}_{i}(\boldsymbol{\epsilon}_{i}+\boldsymbol{\omega}_{i})
- \boldsymbol{\psi}_{i}(\boldsymbol{\omega}_{i})$ 
for a random vector $\boldsymbol{\epsilon}_{i}\in\mathbb{R}^{N}$. 
If $\|\Delta\boldsymbol{\psi}_{i}\|_{2}$ converges in probability 
to zero for $i\in\{0, 1\}$ as $N\to\infty$, then we have 
\begin{align}
\boldsymbol{\psi}_{i}^{\mathrm{T}}(\boldsymbol{\epsilon}_{i} 
+ \boldsymbol{\omega}_{i})
\boldsymbol{\psi}_{0}(\boldsymbol{\epsilon}_{0} + 
\boldsymbol{\omega}_{0})
- \boldsymbol{\psi}_{i}^{\mathrm{T}}(\boldsymbol{\omega}_{i})
\boldsymbol{\psi}_{0}(\boldsymbol{\omega}_{0})
\pto 0
\end{align}
in the limit $N\to\infty$ for $i\in\{0, 1\}$. 
\end{lemma}
\begin{IEEEproof}
By definition, we have  
\begin{align}
\boldsymbol{\psi}_{i}^{\mathrm{T}}(\boldsymbol{\epsilon}_{i} 
+ \boldsymbol{\omega}_{i})
\boldsymbol{\psi}_{0}(\boldsymbol{\epsilon}_{0}+\boldsymbol{\omega}_{0})
= \boldsymbol{\psi}_{i}^{\mathrm{T}}(\boldsymbol{\omega}_{i})
\boldsymbol{\psi}_{0}(\boldsymbol{\omega}_{0})
\nonumber \\
+ \boldsymbol{\psi}_{i}^{\mathrm{T}}(\boldsymbol{\omega}_{i})
\Delta\boldsymbol{\psi}_{0}
+ \Delta\boldsymbol{\psi}_{i}^{\mathrm{T}}
\boldsymbol{\psi}_{0}(\boldsymbol{\omega}_{0})
+ \Delta\boldsymbol{\psi}_{i}^{\mathrm{T}}
\Delta\boldsymbol{\psi}_{0}. 
\end{align}
Since $\|\boldsymbol{\psi}_{i}(\boldsymbol{\omega}_{i})\|_{2}$ 
has been assumed to be bounded in probability, we use the Cauchy-Schwarz 
inequality for the last three terms to find that the assumption 
$\|\Delta\boldsymbol{\psi}_{i}\|_{2}\pto0$ implies the convergence of 
these terms to zero. 
Thus, Lemma~\ref{lemma_covariance_error} holds. 
\end{IEEEproof}

Lemma~\ref{lemma_covariance_error} requires two strong assumptions; 
the boundedness in probability of 
$\|\boldsymbol{\psi}_{i}(\boldsymbol{\omega}_{i})\|_{2}$ and the 
convergence in probability $\|\Delta\boldsymbol{\psi}_{i}\|_{2}\pto0$. 
In state evolution 
analysis, we use $\boldsymbol{\psi}_{0}(\boldsymbol{\omega}_{t}) 
= f_{\In,t}(\bar{\eta}_{t}\boldsymbol{x} + \boldsymbol{\omega}_{t}; 
v_{\Out,t}) - \bar{\eta}_{t}\boldsymbol{x}$, which satisfies 
these assumptions from Assumption~\ref{assumption_inner}.  

To present the main technical result in state evolution, we define a recursive 
system of random variables that describes the asymptotic dynamics of the 
error model~(\ref{b})--(\ref{q}). Let $Z\sim\mathcal{N}(0, P)$ denote a 
Gaussian random variable that is independent of $W$ and represents the 
asymptotic behavior of $\boldsymbol{z}$ in (\ref{m}). Furthermore, we define 
$B_{0}=-\bar{\eta}_{0}Z$. Note that $\bar{\eta}_{0}$ in (\ref{eta_bar}) depends 
on $\bar{\xi}_{\Out,0}$ and $\bar{\zeta}_{0}$ in (\ref{xi_out_bar}) and 
(\ref{zeta_bar}), which are uniquely determined from the randomness of 
$Z$ and $W$, because of $B_{0}+\bar{\eta}_{0}Z=0$.  

Let $\{B_{\tau}\in\mathbb{R}\}_{\tau=1}^{t}$ denote zero-mean Gaussian random 
variables with covariance $\mathbb{E}[B_{\tau'}B_{\tau}]=Q_{\tau',\tau}$, in which 
we define $Q_{0,0}=\mathbb{E}[\|\boldsymbol{q}_{0}\|^{2}]
=\bar{\eta}_{0}^{2}P$, with $\boldsymbol{q}_{0}=-\bar{\eta}_{0}\boldsymbol{x}$ 
and Assumption~\ref{assumption_x}, while $Q_{\tau',\tau}$ for $\tau'>0$ or 
$\tau>0$ is defined shortly. The random variables $\{B_{\tau}\}_{\tau=0}^{t}$ 
represent the asymptotic dynamics of 
$\{\boldsymbol{b}_{\tau}\}_{\tau=0}^{t}$ in (\ref{b}). 
We use these random variables to define 
\begin{equation} \label{M}
M_{t} = \frac{1}{\bar{\xi}_{\Out,t}}f_{\Out,t}\left(
 B_{t} + \bar{\eta}_{t}Z, g(Z, W); \tilde{v}_{\In,t}
\right), 
\end{equation}
with $\tilde{v}_{\In,t}$ in (\ref{v_in_tilde}), 
which corresponds to $\boldsymbol{m}_{t}$ in (\ref{m}). 

Let $\{\boldsymbol{\omega}_{\tau}\in\mathbb{R}^{N}\}_{\tau=0}^{t}$ denote 
zero-mean Gaussian random vectors with covariance 
$\mathbb{E}[\boldsymbol{\omega}_{\tau}\boldsymbol{\omega}_{\tau'}^{\mathrm{T}}]
=M^{-1}\mathbb{E}[M_{\tau'}M_{\tau}]\boldsymbol{I}_{N}$. They are independent 
of $\boldsymbol{x}$ and represent the asymptotic dynamics of 
$\{\boldsymbol{h}_{\tau}\}_{\tau=0}^{t}$ in (\ref{h}). 
We use these random vectors to define  
\begin{align} 
Q_{t'+1,t+1} = &\lim_{N\to\infty}\mathbb{E}[
\left\{
 f_{\In,t'}(\bar{\eta}_{t'}\boldsymbol{x} + \boldsymbol{\omega}_{t'}; 
 \bar{v}_{\Out,t'}) - \bar{\eta}_{t'+1}\boldsymbol{x}
\right\}^{\mathrm{T}}
\nonumber \\
&\cdot\left\{
 f_{\In,t}(\bar{\eta}_{t}\boldsymbol{x} + \boldsymbol{\omega}_{t}; 
 \bar{v}_{\Out,t}) - \bar{\eta}_{t+1}\boldsymbol{x}
\right\}], \label{Q_tau_t}
\end{align}
\begin{align}
&Q_{0,t+1} = Q_{t+1,0} 
\nonumber \\
&= - \bar{\eta}_{0}\lim_{N\to\infty}\mathbb{E}[
\boldsymbol{x}^{\mathrm{T}}\left\{
 f_{\In,t}(\bar{\eta}_{t}\boldsymbol{x} + \boldsymbol{\omega}_{t}; 
 \bar{v}_{\Out,t}) - \bar{\eta}_{t+1}\boldsymbol{x}
\right\}], \label{Q_0_t}
\end{align}
with $\bar{v}_{\Out,t}$ in (\ref{v_out_bar}). 
The notational convention $f_{\In,-1}=0$ allows us to obtain (\ref{Q_0_t}) 
by letting $t'=-1$ in (\ref{Q_tau_t}). 
The covariance $Q_{t'+1,t+1}$ corresponds to the unnormalized 
covariance $\mathbb{E}[\boldsymbol{q}_{t'+1}^{\mathrm{T}}\boldsymbol{q}_{t+1}]$ 
in the sublinear sparsity limit. 

The following theorem is the main technical result in state evolution. 

\begin{theorem} \label{theorem_SE_tech}
Define $\bar{\xi}_{\In,t}$ in (\ref{z_t}) and (\ref{v_in}) as (\ref{xi_in_bar}) 
and suppose that Assumptions~\ref{assumption_x}--\ref{assumption_inner} are 
satisfied. Then, the following results hold in the 
sublinear sparsity limit for all $\tau\in\{0,1,\ldots,T\}$:  
\begin{enumerate}[label=(O\alph*)]
\item \label{Oa}
Let $\boldsymbol{\beta}_{\tau}=\boldsymbol{Q}_{\tau}^{\dagger}\boldsymbol{q}_{\tau}$ 
and $\boldsymbol{q}_{\tau}^{\perp} = \boldsymbol{P}_{\boldsymbol{Q}_{\tau}}^{\perp}
\boldsymbol{q}_{\tau}$. 
For $\tau>0$, the conditional distribution of $\boldsymbol{b}_{\tau}$ given 
$\mathfrak{E}_{\tau,\tau}$ and $\Theta$ is represented as 
\begin{equation}
\boldsymbol{b}_{\tau} \sim \boldsymbol{B}_{\tau}\boldsymbol{\beta}_{\tau} 
+ \boldsymbol{M}_{\tau}\boldsymbol{o}(1)
+ \tilde{\boldsymbol{A}}_{\tau}\boldsymbol{q}_{\tau}^{\perp},
\end{equation}
where $\tilde{\boldsymbol{A}}_{\tau}$ is independent of 
$\{\mathfrak{E}_{\tau,\tau}, \Theta\}$ and has independent standard Gaussian 
elements. 

\item \label{Ob}
For all $\tau'\in\{0,\ldots,\tau\}$, 
\begin{equation}
\frac{1}{M}\boldsymbol{b}_{\tau'}^{\mathrm{T}}\boldsymbol{b}_{\tau} 
\pto Q_{\tau',\tau}.
\end{equation}

\item \label{Oc}
\begin{equation}
(\boldsymbol{b}_{0},\ldots,\boldsymbol{b}_{\tau},\boldsymbol{w})
\plto (B_{0}, \ldots, B_{\tau}, W),
\end{equation}
\begin{equation} \label{xi_out_convergence}
\xi_{\Out,\tau}\pto \bar{\xi}_{\Out,\tau}, 
\end{equation}
\begin{equation}
v_{\Out,\tau}\pto \bar{v}_{\Out,\tau}. 
\end{equation}

\item \label{Od}
For all $\tau'\in\{0,\ldots,\tau\}$, 
\begin{equation}
\frac{1}{M}\boldsymbol{b}_{\tau'}^{\mathrm{T}}\boldsymbol{m}_{\tau} 
\pto Q_{\tau',\tau}. 
\end{equation}

\item \label{Oe}
There is some $\epsilon>0$ such that 
\begin{equation}
\mathrm{Pr}\left(
 \lambda_{\mathrm{min}}(M^{-1}\boldsymbol{M}_{\tau+1}^{\mathrm{T}}
\boldsymbol{M}_{\tau+1}) > \epsilon 
\right) \to 1.
\end{equation}
\end{enumerate}
\begin{enumerate}[label=(I\alph*)]
\item \label{Ia}
Let $\boldsymbol{\alpha}_{\tau}=\boldsymbol{M}_{\tau}^{\dagger}
\boldsymbol{m}_{\tau}$ and $\boldsymbol{m}_{\tau}^{\perp}
=\boldsymbol{P}_{\boldsymbol{M}_{\tau}}^{\perp}\boldsymbol{m}_{\tau}$. Then, 
the conditional distribution of $\boldsymbol{h}_{\tau}$ given 
$\mathfrak{E}_{\tau+1,\tau}$ and $\Theta$ is represented as 
\begin{equation}
\boldsymbol{h}_{0} \sim 
o(1)\boldsymbol{q}_{0} + \frac{1}{M}\tilde{\boldsymbol{A}}_{0}^{\mathrm{T}}
\boldsymbol{m}_{0},
\end{equation}
\begin{equation}
\boldsymbol{h}_{\tau} \sim \boldsymbol{H}_{\tau}\boldsymbol{\alpha}_{\tau} 
+ \boldsymbol{Q}_{\tau+1}\boldsymbol{o}(1)
+ \frac{1}{M}\tilde{\boldsymbol{A}}_{\tau}^{\mathrm{T}}\boldsymbol{m}_{\tau}^{\perp} 
\end{equation}
for $\tau>0$, where $\tilde{\boldsymbol{A}}_{\tau}$ is independent of 
$\{\mathfrak{E}_{\tau+1,\tau}, \Theta\}$ and has independent standard Gaussian 
elements. 

\item \label{Ib}
For all $\tau'\in\{0,\ldots,\tau\}$, 
\begin{equation}
\frac{M}{N}\boldsymbol{h}_{\tau'}^{\mathrm{T}}\boldsymbol{h}_{\tau} 
\pto \mathbb{E}[M_{\tau'}M_{\tau}]. 
\end{equation}

\item \label{Ic}
For all $\tau'\in\{-1,\ldots,\tau\}$, 
\begin{equation}
\boldsymbol{q}_{\tau'+1}^{\mathrm{T}}\boldsymbol{q}_{\tau+1}
\pto Q_{\tau'+1,\tau+1}, 
\end{equation}
\begin{equation}
v_{\In,\tau+1}\pto\tilde{v}_{\In,\tau+1}. 
\end{equation}

\item \label{Id} 
For all $\tau'\in\{0,\ldots,\tau\}$, 
\begin{equation} \label{hx}
\boldsymbol{h}_{\tau}^{\mathrm{T}}\boldsymbol{x} \pto 0, 
\end{equation}
\begin{equation} \label{hq}
\boldsymbol{h}_{\tau'}^{\mathrm{T}}\boldsymbol{q}_{\tau+1} 
\pto \bar{\xi}_{\In,\tau}\mathbb{E}[M_{\tau'}M_{\tau}]. 
\end{equation}

\item \label{Ie}
There is some $\epsilon>0$ such that 
\begin{equation}
\mathrm{Pr}\left(
 \lambda_{\mathrm{min}}(\boldsymbol{Q}_{\tau+2}^{\mathrm{T}}
\boldsymbol{Q}_{\tau+2}) > \epsilon 
\right) \to 1. 
\end{equation}
\end{enumerate}
\end{theorem}
\begin{IEEEproof}
See Appendix~\ref{proof_theorem_SE_tech}. 
\end{IEEEproof}

Properties~\ref{Oa}--\ref{Oe} for the outer module are equivalent to 
those for linear sparsity~\cite{Takeuchi241}. The main differences between 
the sublinear and linear sparsity are in Properties~\ref{Ia}--\ref{Ie} for 
the inner module. Properties~\ref{Ia} and \ref{Ib} imply that the MSE 
$N^{-1}\|\boldsymbol{h}_{t}\|_{2}^{2}$ for the outer module 
is ${\cal O}(M^{-1})$. The unnormalized quantities are evaluated in 
Properties~\ref{Ic}--\ref{Ie}. To omit a property in the inner module that 
corresponds to (\ref{xi_out_convergence}) in the outer module, this paper 
uses deterministic $\bar{\xi}_{\In,t}$ in (\ref{xi_in_bar}). Property~\ref{Ic} 
consists of  minimal properties required in the inner module.  

Theorem~\ref{theorem_SE} can be proved as a corollary of 
Theorem~\ref{theorem_SE_tech}. 

\begin{IEEEproof}[Proof of Theorem~\ref{theorem_SE}]
We evaluate the unnormalized square error 
$\|\hat{\boldsymbol{x}}_{t+1} - \boldsymbol{x}\|^{2}$. 
From Assumption~\ref{assumption_x} we use the weak law of large numbers 
for the initial condition $\boldsymbol{q}_{0}= -\bar{\eta}_{0}\boldsymbol{x}$ 
to have 
\begin{equation} \label{q_0_convergence}
\|\boldsymbol{q}_{0}\|_{2}^{2} 
= \frac{\bar{\eta}_{0}^{2}}{k}\sum_{n\in\mathcal{S}}(\sqrt{k}x_{n})^{2}
\pto \bar{\eta}_{0}^{2}P = Q_{0,0}. 
\end{equation}
Using the definition of $\hat{\boldsymbol{x}}_{t+1}$ in (\ref{x_hat_t}), 
$\boldsymbol{x}_{t}=\bar{\eta}_{t}\boldsymbol{x} + \boldsymbol{h}_{t}$, and 
the definition of $\boldsymbol{q}_{t+1}$ in (\ref{q}) yields 
\begin{align}
&\|\hat{\boldsymbol{x}}_{t+1} - \boldsymbol{x}\|^{2}
= \|\boldsymbol{q}_{t+1} + (1 - \bar{\eta}_{t+1})
\bar{\eta}_{0}^{-1}\boldsymbol{q}_{0}\|^{2} 
\nonumber \\
&\pto Q_{t+1,t+1} + 2\frac{1 - \bar{\eta}_{t+1}}{\bar{\eta}_{0}}Q_{0,t+1}
+ (1 - \bar{\eta}_{t+1})^{2}P,
\end{align}
where the last convergence follows from Property~\ref{Ic} in 
Theorem~\ref{theorem_SE_tech} and (\ref{q_0_convergence}). Substituting the 
definitions of $Q_{t+1,t+1}$ and $Q_{0,t+1}$ in (\ref{Q_tau_t}) and (\ref{Q_0_t}), 
after some algebra, 
we find $\|\hat{\boldsymbol{x}}_{t+1} - \boldsymbol{x}\|^{2}\pto 
\bar{v}_{\In,t+1}$ given in (\ref{v_in_bar}). 

To complete the proof of Theorem~\ref{theorem_SE}, we confirm 
$\mathbb{E}[M_{t}^{2}]=\bar{v}_{\Out,t}$. From the definitions of 
$\bar{v}_{\Out,t}$ and $M_{t}$ in (\ref{v_out_bar}) and (\ref{M}), 
it is sufficient to derive the covariance~(\ref{Z_tt}) and 
(\ref{Z_0t}) for $Z_{t} = B_{t} + \bar{\eta}_{t}Z$ and $Z$. 
Using $Z= -\bar{\eta}_{0}^{-1}B_{0}$ yields 
\begin{align}
&\mathbb{E}[Z_{t+1}^{2}]
= \mathbb{E}\left[
 \left(
  B_{t+1} - \frac{\bar{\eta}_{t+1}}{\bar{\eta}_{0}}B_{0}
 \right)^{2}
\right]
\nonumber \\
&= Q_{t+1,t+1} 
- 2\frac{\bar{\eta}_{t+1}}{\bar{\eta}_{0}}Q_{0,t+1} 
+ \frac{\bar{\eta}_{t+1}^{2}}{\bar{\eta}_{0}^{2}}Q_{0,0}
= \mu_{t+1,t+1},
\end{align}
with $\mu_{t+1,t+1}$ in (\ref{mu_tt}). 
In the derivation of the second equality, we have used the definition 
$\mathbb{E}[B_{\tau}B_{t}] = Q_{\tau,t}$. The last equality follows from the 
definition of $Q_{\tau,t}$ in (\ref{Q_tau_t}) and (\ref{Q_0_t}).  
Similarly, we repeat the same proof to obtain $\mathbb{E}[ZZ_{t+1}]
=\mu_{0,t+1}$ with $\mu_{0,t+1}$ in (\ref{mu_0t}). 
Thus, we arrive at Theorem~\ref{theorem_SE}. 
\end{IEEEproof}

\section{Proof of Theorem~\ref{theorem_SE_tech}} 
\label{proof_theorem_SE_tech} 
\subsection{Proof by Induction}
The proof of Theorem~\ref{theorem_SE_tech} is by induction with respect 
to $\tau\in\{0,1,\ldots,T\}$. See Appendix~\ref{proof_outer_0} for the proofs 
of Properties~\ref{Oa}--\ref{Oe} for $\tau=0$. Properties~\ref{Ia}--\ref{Ie} 
for $\tau=0$ are proved in Appendix~\ref{proof_inner_0}. For some 
$t\in\{1,\ldots,T\}$, suppose that Properties~\ref{Oa}-\ref{Oe} and 
\ref{Ia}--\ref{Ie} are correct for all $\tau<t$. 
See Appendix~\ref{proof_outer} for the proofs of Properties~\ref{Oa}--\ref{Oe} 
for $\tau=t$ under this induction hypothesis. Properties~\ref{Ia}--\ref{Ie} 
for $\tau=t$ are proved in Appendix~\ref{proof_inner}. By induction, 
Theorem~\ref{theorem_SE_tech} holds for all $\tau\in\{0, 1,\ldots,T\}$. 

\subsection{Outer Module for $\tau=0$} 
\label{proof_outer_0}
\begin{IEEEproof}[Proof of \ref{Ob}]
Assumption~\ref{assumption_A} implies that 
$\boldsymbol{b}_{0}=\boldsymbol{A}\boldsymbol{q}_{0}$ 
conditioned on $\boldsymbol{q}_{0}$ has zero-mean i.i.d.\ Gaussian elements 
with variance $\|\boldsymbol{q}_{0}\|_{2}^{2}$. Thus, we use the weak law of 
large numbers to obtain 
\begin{equation}
\frac{1}{M}\|\boldsymbol{b}_{0}\|_{2}^{2} 
\peq \|\boldsymbol{q}_{0}\|_{2}^{2} + o(1)
\end{equation}
conditioned on $\boldsymbol{q}_{0}$ in the sublinear sparsity limit. 
Combining this result and (\ref{q_0_convergence}), 
we find that Property~\ref{Ob} holds for $\tau=0$. 
\end{IEEEproof}

\begin{IEEEproof}[Proof of \ref{Oc}]
The first convergence in Property~\ref{Oc} for $\tau=0$ is trivial from 
$\boldsymbol{b}_{0}\sim\mathcal{N}(\boldsymbol{0}, 
\|\boldsymbol{q}_{0}\|_{2}^{2}\boldsymbol{I}_{M})$ 
conditioned on $\boldsymbol{q}_{0}$, Property~\ref{Ob} for $\tau=0$, 
and Assumption~\ref{assumption_w}. 
The second convergence for $\tau=0$ follows from 
$v_{\In,0}=\tilde{v}_{\In,0}=P$, Assumption~\ref{assumption_outer}, 
the first convergence, and 
\cite[Lemma~5]{Bayati11}. Finally, we use the definition of 
$\hat{\boldsymbol{z}}_{0}$ in (\ref{z_hat_t}) and  
the first two properties in Property~\ref{Oc} for $\tau=0$ 
to find that $v_{\Out,0}$ in (\ref{v_out}) converges in probability to 
$\bar{v}_{\Out,0}$ in (\ref{v_out_bar}). 
\end{IEEEproof}

\begin{IEEEproof}[Proof of \ref{Od}]
Assumption~\ref{assumption_outer} implies that $\boldsymbol{m}_{0}$ in 
(\ref{m}) is a piecewise Lipschitz-continuous function. 
Thus, $\boldsymbol{b}_{0}^{\mathrm{T}}\boldsymbol{m}_{0}$ is a piecewise 
second-order pseudo-Lipschitz function. We can use Property~\ref{Oc} for 
$\tau=0$ and the initial conditions $v_{\In,0}=\tilde{v}_{\In,0}=P$ to have 
\begin{align}
\frac{1}{M}\boldsymbol{b}_{0}^{\mathrm{T}}\boldsymbol{m}_{0} 
&\peq \bar{\xi}_{\Out,0}^{-1}\mathbb{E}\left[
 B_{0}f_{\Out,0}(0, g(Z, W); \tilde{v}_{\In,0})
\right] + o(1) \nonumber \\
&= - \bar{\xi}_{\Out,0}^{-1}\bar{\zeta}_{0}\mathbb{E}[B_{0}Z] + o(1) 
= \mathbb{E}[B_{0}^{2}] + o(1)
\nonumber \\
&= Q_{0,0} + o(1), 
\end{align}
with $\bar{\zeta}_{0}$ in (\ref{zeta_bar}), 
where the second and third equalities follow from Lemma~\ref{lemma_Stein} 
and the definition $Z=-\bar{\xi}_{\Out,0}\bar{\zeta}_{0}^{-1}B_{0}$, respectively. 
Thus, Property~\ref{Od} holds for $\tau=0$. 
\end{IEEEproof}

\begin{IEEEproof}[Proof of \ref{Oe}]
From Assumption~\ref{assumption_outer} we can use Property~\ref{Oc} for 
$\boldsymbol{m}_{0}$ in (\ref{m}) to obtain 
\begin{equation} \label{m_0_convergence} 
\frac{1}{M}\|\boldsymbol{m}_{0}\|_{2}^{2} 
\pto \mathbb{E}[M_{0}^{2}]> 0, 
\end{equation}
with $M_{0}$ in (\ref{M}), 
where the strict inequality follows from Assumption~\ref{assumption_SE}. 
Thus, Property~\ref{Oe} holds for $\tau=0$. 
\end{IEEEproof}

\subsection{Inner Module for $\tau=0$} 
\label{proof_inner_0}
\begin{IEEEproof}[Proof of \ref{Ia}]
Using Lemma~\ref{lemma_conditioning} for the constraint 
$\boldsymbol{b}_{0}=\boldsymbol{A}\boldsymbol{q}_{0}$ yields 
\begin{equation}
\boldsymbol{A}
\sim \frac{\boldsymbol{b}_{0}\boldsymbol{q}_{0}^{\mathrm{T}}}
{\|\boldsymbol{q}_{0}\|_{2}^{2}}
- \tilde{\boldsymbol{A}}_{0}\boldsymbol{P}_{\boldsymbol{q}_{0}}^{\perp}
\end{equation}
conditioned on $\mathfrak{E}_{1,0}$ and $\Theta$, in which 
$\tilde{\boldsymbol{A}}_{0}$ is independent of $\{\mathfrak{E}_{1,0}, \Theta\}$ 
and has independent standard Gaussian elements. 
Substituting this representation into the definition of $\boldsymbol{h}_{0}$ 
in (\ref{h}), we have 
\begin{equation} \label{h_0_tmp}
\boldsymbol{h}_{0} \sim \left(
 1 - \frac{\boldsymbol{b}_{0}^{\mathrm{T}}\boldsymbol{m}_{0}}
 {M\|\boldsymbol{q}_{0}\|_{2}^{2}}
\right)\boldsymbol{q}_{0}
+ \frac{1}{M}\boldsymbol{P}_{\boldsymbol{q}_{0}}^{\perp}
\tilde{\boldsymbol{A}}_{0}^{\mathrm{T}}\boldsymbol{m}_{0}
\end{equation}
conditioned on $\mathfrak{E}_{1,0}$ and $\Theta$. 

We first evaluate the first term in (\ref{h_0_tmp}). 
Using Property~\ref{Od} for $\tau=0$ and 
the convergence~(\ref{q_0_convergence}) in probability, we obtain
\begin{equation}
\frac{\boldsymbol{b}_{0}^{\mathrm{T}}\boldsymbol{m}_{0}}
{M\|\boldsymbol{q}_{0}\|_{2}^{2}}
\peq \frac{Q_{0,0} + o(1)}{Q_{0,0} + o(1)} 
\peq 1 + o(1),
\end{equation}
which implies that the first term in (\ref{h_0_tmp}) reduces to 
$o(1)\boldsymbol{q}_{0}$. 

We next evaluate the second term in (\ref{h_0_tmp}). Using 
the definition $\boldsymbol{P}_{\boldsymbol{q}_{0}}^{\perp}=\boldsymbol{I}_{N} 
- \|\boldsymbol{q}_{0}\|_{2}^{-2}\boldsymbol{q}_{0}\boldsymbol{q}_{0}^{\mathrm{T}}$
yields 
\begin{equation} 
\frac{1}{M}\boldsymbol{P}_{\boldsymbol{q}_{0}}^{\perp}
\tilde{\boldsymbol{A}}_{0}^{\mathrm{T}}\boldsymbol{m}_{0}
= \frac{1}{M}\tilde{\boldsymbol{A}}_{0}^{\mathrm{T}}\boldsymbol{m}_{0}
- \frac{1}{\sqrt{M}}\frac{\boldsymbol{q}_{0}^{\mathrm{T}}
\tilde{\boldsymbol{A}}_{0}^{\mathrm{T}}\boldsymbol{m}_{0}}
{\sqrt{M}\|\boldsymbol{q}_{0}\|_{2}^{2}}\boldsymbol{q}_{0}. 
\end{equation}
We evaluate the coefficient of $\boldsymbol{q}_{0}$ in the second term as 
\begin{align}
&\left(
 \frac{\boldsymbol{q}_{0}^{\mathrm{T}}
 \tilde{\boldsymbol{A}}_{0}^{\mathrm{T}}\boldsymbol{m}_{0}}
 {\sqrt{M}\|\boldsymbol{q}_{0}\|_{2}^{2}}
\right)^{2}
= \frac{M^{-1}\boldsymbol{m}_{0}^{\mathrm{T}}\tilde{\boldsymbol{A}}_{0}
\boldsymbol{P}_{\boldsymbol{q}_{0}}^{\parallel}
\tilde{\boldsymbol{A}}_{0}^{\mathrm{T}}\boldsymbol{m}_{0}}
{\|\boldsymbol{q}_{0}\|_{2}^{2}}
\nonumber \\
&\peq \frac{M^{-1}\|\boldsymbol{m}_{0}\|_{2}^{2}
\mathrm{Tr}(\boldsymbol{P}_{\boldsymbol{q}_{0}}^{\parallel}) + o(1)}
{\|\boldsymbol{q}_{0}\|_{2}^{2}} 
\peq \frac{\mathbb{E}[M_{0}^{2}]}{Q_{0,0}} + o(1). 
\end{align}
Here, the second equality follows from the weak law of large numbers, 
because of $\tilde{\boldsymbol{A}}_{0}^{\mathrm{T}}\boldsymbol{m}_{0}\sim
\mathcal{N}(\boldsymbol{0},\|\boldsymbol{m}_{0}\|_{2}^{2}\boldsymbol{I})$ 
conditioned on $\boldsymbol{m}_{0}$. The last equality is obtained from 
(\ref{q_0_convergence}) and (\ref{m_0_convergence}). 
Thus, Property~\ref{Ia} holds for $\tau=0$. 
\end{IEEEproof}

\begin{IEEEproof}[Proof of \ref{Ib}]
Using Property~\ref{Ia} for $\tau=0$ and the weak law of large numbers, 
as well as (\ref{q_0_convergence}), we have
\begin{equation}
\frac{M}{N}\|\boldsymbol{h}_{0}\|_{2}^{2} 
\peq \frac{\boldsymbol{m}_{0}^{\mathrm{T}}\tilde{\boldsymbol{A}}_{0}
\tilde{\boldsymbol{A}}_{0}^{\mathrm{T}}\boldsymbol{m}_{0} }{MN}+ o(1)
\peq \frac{\|\boldsymbol{m}_{0}\|_{2}^{2}}{M} + o(1) 
\end{equation}
conditioned on $\mathfrak{E}_{1,0}$ and $\Theta$ 
in the sublinear sparsity limit. Thus, we use (\ref{m_0_convergence}) to 
arrive at Property~\ref{Ib} for $\tau=0$.  
\end{IEEEproof}

\begin{IEEEproof}[Proof of \ref{Ic}]
We present a general proof for Property~\ref{Ic} with $\tau=0$. 
The presented proof is applicable to that for the general case $\tau>0$.
We first prove the latter convergence. 
Using $v_{\Out,\tau}\pto \bar{v}_{\Out,\tau}$ in Property~\ref{Oc} for $\tau=0$, 
we find that $v_{\In,\tau+1}$ in (\ref{v_in}) converges in probability to 
$\tilde{v}_{\In,\tau+1}$ in (\ref{v_in_tilde}) for $\tau=0$. 
 
We next prove the former convergence in Property~\ref{Ic} for $\tau=0$. 
Using the definition of $\boldsymbol{q}_{t+1}$ in (\ref{q}) for $t=0$ yields 
\begin{align}
\boldsymbol{q}_{\tau'+1}^{\mathrm{T}}\boldsymbol{q}_{t+1} 
= (\bar{\eta}_{\tau'} - \bar{\eta}_{\tau'+1})(\bar{\eta}_{t} - \bar{\eta}_{t+1})
\|\boldsymbol{x}\|^{2}
\nonumber \\
+ \left\{
 f_{\In,\tau'}(\bar{\eta}_{\tau'}\boldsymbol{x} + \boldsymbol{h}_{\tau'}; 
 v_{\Out,\tau'}) - \bar{\eta}_{\tau'}\boldsymbol{x}
\right\}^{\mathrm{T}}
\nonumber \\
\cdot\left\{
 f_{\In,t}(\bar{\eta}_{t}\boldsymbol{x} + \boldsymbol{h}_{t}; 
 v_{\Out,t}) - \bar{\eta}_{t}\boldsymbol{x}
\right\}
\nonumber \\
+ (\bar{\eta}_{t} - \bar{\eta}_{t+1})\left\{
 f_{\In,\tau'}(\bar{\eta}_{\tau'}\boldsymbol{x} + \boldsymbol{h}_{\tau'}; 
 v_{\Out,\tau'}) - \bar{\eta}_{\tau'}\boldsymbol{x}
\right\}^{\mathrm{T}}\boldsymbol{x}
\nonumber \\
+ (\bar{\eta}_{\tau'} - \bar{\eta}_{\tau'+1})
\boldsymbol{x}^{\mathrm{T}}\left\{
 f_{\In,t}(\bar{\eta}_{t}\boldsymbol{x} + \boldsymbol{h}_{t}; 
 v_{\Out,t}) - \bar{\eta}_{t}\boldsymbol{x}
\right\} \label{qq}
\end{align}
for $\tau'\in\{-1, 0\}$, with $f_{\In,-1}=0$ and $\bar{\eta}_{-1}=0$. 
In (\ref{q_0_convergence}), we have proved that $\|\boldsymbol{x}\|^{2}$ in 
the first term converges in probability to $Q_{0,0}$ in the sublinear 
sparsity limit. 

To evaluate the last term in (\ref{qq}), 
we utilize Lemma~\ref{lemma_covariance_error}.  
From Property~\ref{Oc} for $\tau=0$ we have $v_{\Out,0}\pto\bar{v}_{\Out,0}$. 
For $t=0$, let $\boldsymbol{\psi}_{1}(\boldsymbol{\omega}_{1})=\boldsymbol{x}$ 
and $\boldsymbol{\psi}_{0}(\boldsymbol{\omega}_{t}) 
= f_{\In,t}(\bar{\eta}_{t}\boldsymbol{x} + \boldsymbol{\omega}_{t}; v_{\Out,t}) 
- \bar{\eta}_{t}\boldsymbol{x}$ with 
$\boldsymbol{\omega}_{t}=M^{-1}\tilde{\boldsymbol{A}}_{0}^{\mathrm{T}}
\boldsymbol{m}_{0}\sim\mathcal{N}(\boldsymbol{0}, 
M^{-2}\|\boldsymbol{m}_{0}\|^{2}\boldsymbol{I}_{N})$ given 
$\mathfrak{E}_{1,0}$ and $\Theta$, as well as 
$\boldsymbol{\epsilon}_{0}=o(1)\boldsymbol{q}_{0}$. 
Since $\boldsymbol{\psi}_{1}$ is 
independent of $\boldsymbol{\omega}_{1}$, the function $\boldsymbol{\psi}_{1}$ 
satisfies $\|\Delta\boldsymbol{\psi}_{1}\|_{2}=0$ and the boundedness 
assumption for $\|\boldsymbol{\psi}_{1}\|^{2}$ in 
Lemma~\ref{lemma_covariance_error}. Furthermore, 
Assumption~\ref{assumption_inner} implies that 
$\|\boldsymbol{\psi}_{0}(\boldsymbol{\omega}_{t})\|_{2}^{2}$ is bounded 
in probability and that$\|\Delta\boldsymbol{\psi}_{0}\|_{2}\pto0$ holds. 
Since (\ref{q_0_convergence}) implies the 
boundedness in probability of $\|\boldsymbol{q}_{0}\|_{2}$, we use 
Property~\ref{Ia} for $\tau=0$ and Lemma~\ref{lemma_covariance_error} to 
obtain  
\begin{align}
&\boldsymbol{x}^{\mathrm{T}}\left\{
 f_{\In,t}(\bar{\eta}_{t}\boldsymbol{x} + \boldsymbol{h}_{t}; v_{\Out,t}) 
 - \bar{\eta}_{t}\boldsymbol{x}
\right\}
\nonumber \\
&\peq \boldsymbol{x}^{\mathrm{T}}\left\{
 f_{\In,t}(\bar{\eta}_{t}\boldsymbol{x} + \boldsymbol{\omega}_{t}; 
 \bar{v}_{\Out,t}) - \bar{\eta}_{t}\boldsymbol{x}
\right\} + o(1)
\nonumber \\
&\peq \mathbb{E}\left[
 \boldsymbol{x}^{\mathrm{T}}\left\{
  f_{\In,t}(\bar{\eta}_{t}\boldsymbol{x} + \boldsymbol{\omega}_{t};
  \bar{v}_{\Out,t}) - \bar{\eta}_{t}\boldsymbol{x}
 \right\}
\right] + o(1),
\end{align}
where the last equality follows from Assumption~\ref{assumption_inner}. 
Similarly, the third term in (\ref{qq}) can be evaluated in the same 
manner. 

Finally, we evaluate the second term in (\ref{qq}). 
Let $\boldsymbol{\psi}_{1}(\boldsymbol{\omega}_{\tau'}) 
= f_{\In,\tau'}(\bar{\eta}_{\tau'}\boldsymbol{x} + \boldsymbol{\omega}_{\tau'};
v_{\Out,\tau'}) - \bar{\eta}_{\tau'}\boldsymbol{x}$ and 
$\boldsymbol{\psi}_{0}(\boldsymbol{\omega}_{t}) 
= f_{\In,t}(\bar{\eta}_{t}\boldsymbol{x} + \boldsymbol{\omega}_{t}; 
v_{\Out,t}) - \bar{\eta}_{t}\boldsymbol{x}$ in 
Lemma~\ref{lemma_covariance_error}. Assumption~\ref{assumption_inner}
implies that $\|\boldsymbol{\psi}_{1}(\boldsymbol{\omega}_{\tau'})\|_{2}^{2}$ 
and $\|\boldsymbol{\psi}_{0}(\boldsymbol{\omega}_{t})\|_{2}^{2}$  
are bounded in probability and that $\|\Delta\boldsymbol{\psi}_{i}\|_{2}\pto0$ 
holds for $i\in\{0, 1\}$. Thus, we can use 
Lemma~\ref{lemma_covariance_error} to obtain  
\begin{align}
&\left\{
 f_{\In,\tau'}(\bar{\eta}_{\tau'}\boldsymbol{x} + \boldsymbol{h}_{\tau'}; 
 v_{\Out,\tau'}) - \bar{\eta}_{\tau'}\boldsymbol{x}
\right\}^{\mathrm{T}}
\nonumber \\
&\cdot\left\{
 f_{\In,t}(\bar{\eta}_{t}\boldsymbol{x} + \boldsymbol{h}_{t}; v_{\Out,t}) 
 - \bar{\eta}_{t}\boldsymbol{x}
\right\}
\nonumber \\
&\peq \left\{
 f_{\In,\tau'}(\bar{\eta}_{\tau'}\boldsymbol{x} + \boldsymbol{\omega}_{\tau'}; 
 \bar{v}_{\Out,\tau'}) - \bar{\eta}_{\tau'}\boldsymbol{x}
\right\}^{\mathrm{T}}
\nonumber \\
&\cdot\left\{
 f_{\In,t}(\bar{\eta}_{t}\boldsymbol{x} + \boldsymbol{\omega}_{t}; 
 \bar{v}_{\Out,t}) - \bar{\eta}_{t}\boldsymbol{x}
\right\}
+ o(1),
\end{align}
which converges in probability to its expectation in the sublinear sparsity 
limit from Assumption~\ref{assumption_inner}. Applying these results to 
(\ref{qq}) and using the definition of $Q_{\tau'+1,t+1}$ in (\ref{Q_tau_t}), 
after some algebra, we arrive at the former convergence in Property~\ref{Ic} 
for $\tau=0$. 
\end{IEEEproof}

\begin{IEEEproof}[Proof of \ref{Id}]
We first confirm the former convergence in probability 
$\boldsymbol{h}_{0}^{\mathrm{T}}\boldsymbol{x}\pto0$. 
From Assumption~\ref{assumption_x} we represent 
$\boldsymbol{h}_{0}^{\mathrm{T}}\boldsymbol{x}$ as 
\begin{equation}
\boldsymbol{h}_{0}^{\mathrm{T}}\boldsymbol{x}
= \sum_{n\in\mathcal{S}}h_{n,0}x_{n}, 
\end{equation}
where $\mathcal{S}\subset\{1,\ldots,N\}$ denotes the support of 
$\boldsymbol{x}$ with $|\mathcal{S}|=k$. We use Property~\ref{Ia} for 
$\tau=0$ to evaluate the conditional variance of 
$\boldsymbol{h}_{0}^{\mathrm{T}}\boldsymbol{x}$ as 
\begin{align}
&\mathbb{V}\left[
 \left. 
  \boldsymbol{h}_{0}^{\mathrm{T}}\boldsymbol{x}
 \right| \mathfrak{E}_{1,0}, \Theta
\right]
= \mathbb{V}\left[
 \left.
  \sum_{n\in\mathcal{S}}\omega_{n,0}x_{n}
 \right| \mathfrak{E}_{1,0}, \Theta
\right] + o(1)
\nonumber \\
&= \sum_{n\in\mathcal{S}}x_{n}^{2}\mathbb{E}\left[
 \left.
  \omega_{n,0}^{2}
 \right| \mathfrak{E}_{1,0}, \Theta
\right] + o(1)
\nonumber \\
&\peq \frac{\|\boldsymbol{x}\|^{2}}{M}\mathbb{E}[M_{0}^{2}] + o(1) 
\pto 0, 
\end{align}
with $\boldsymbol{\omega}_{0}=M^{-1}\tilde{\boldsymbol{A}}^{\mathrm{T}}
\boldsymbol{m}_{0}\sim\mathcal{N}(\boldsymbol{0}, 
M^{-2}\|\boldsymbol{m}_{0}\|^{2}\boldsymbol{I}_{N})$ given 
$\mathfrak{E}_{1,0}$ and $\Theta$. In the derivation of the third equality, 
we have used (\ref{m_0_convergence}). The last convergence follows from 
(\ref{q_0_convergence}). Thus, we have the convergence in probability 
\begin{equation}
\boldsymbol{h}_{0}^{\mathrm{T}}\boldsymbol{x} 
\peq \mathbb{E}[\boldsymbol{h}_{0}^{\mathrm{T}}\boldsymbol{x}] + o(1) 
= \mathbb{E}[\boldsymbol{\omega}_{0}^{\mathrm{T}}\boldsymbol{x}] + o(1) 
= o(1),
\end{equation}
where the second inequality follows from Property~\ref{Ia} for $\tau=0$. 

We next prove the latter convergence~(\ref{hq}) in probability for $\tau=0$.  
Using the definition of $\boldsymbol{q}_{1}$ in (\ref{q}), 
$\boldsymbol{h}_{0}^{\mathrm{T}}\boldsymbol{x}\pto0$, Property~\ref{Ia} 
for $\tau=0$, and Assumption~\ref{assumption_inner}, as well as 
Property~\ref{Oc} for $\tau=0$, we have 
\begin{align}
&\boldsymbol{h}_{0}^{\mathrm{T}}\boldsymbol{q}_{1} 
\peq \mathbb{E}\left[
 \boldsymbol{\omega}_{0}^{\mathrm{T}}
 f_{\In,0}(\bar{\eta}_{0}\boldsymbol{x} + \boldsymbol{\omega}_{0}; 
 \bar{v}_{\Out,0})
\right] + o(1)
\nonumber \\
&= \sum_{n=1}^{N}\mathbb{E}\left[
 \omega_{n,0}f_{\In,0}(\bar{\eta}_{0}x_{n} + \omega_{n,0}; \bar{v}_{\Out,0})
\right] + o(1).
\end{align}
From Assumption~\ref{assumption_inner} and (\ref{m_0_convergence}), 
we can utilize Lemma~\ref{lemma_Stein_inner} to obtain 
\begin{align}
\boldsymbol{h}_{0}^{\mathrm{T}}\boldsymbol{q}_{1} 
&\peq \sum_{n=1}^{N}\frac{\mathbb{E}[M_{0}^{2}]}{M}
\mathbb{E}\left[
 f_{\In,0}'(\bar{\eta}_{0}x_{n} + \omega_{n,0}; \bar{v}_{\Out,0})
\right] + o(1)
\nonumber \\
&= \frac{\mathbb{E}[M_{0}^{2}]}{M}
\boldsymbol{1}^{\mathrm{T}}\mathbb{E}\left[
 f_{\In,0}'(\bar{\eta}_{0}\boldsymbol{x} + \boldsymbol{\omega}_{0}; 
 \bar{v}_{\Out,0})
\right] + o(1)
\nonumber \\
&\peq \bar{\xi}_{\In,0}\mathbb{E}[M_{0}^{2}] + o(1),
\end{align}
where the last equality follows from the definition of $\bar{\xi}_{\In,t}$ 
in (\ref{xi_in_bar}). Thus, Property~\ref{Id} holds for $\tau=0$. 
\end{IEEEproof}

\begin{IEEEproof}[Proof of \ref{Ie}]
Repeat an existing proof in \cite[Proof of (76) for $\tau=0$]{Takeuchi20} 
for the unnormalized quantity $\boldsymbol{Q}_{2}^{\mathrm{T}}\boldsymbol{Q}_{2}$ 
under Assumption~\ref{assumption_SE}. 
\end{IEEEproof}

\subsection{Outer Module for $\tau=t$}
\label{proof_outer}
\begin{IEEEproof}[Proof of \ref{Oa}]
Let 
\begin{equation}
\boldsymbol{\Lambda}_{t-1} 
= \mathrm{diag}\{\xi_{\In,0},\ldots,\xi_{\In,t-2}\}. 
\end{equation}
From the induction hypotheses~\ref{Oe} and \ref{Ie} for $\tau<t$ we find 
that $\boldsymbol{M}_{t}$ and $\boldsymbol{Q}_{t}$ have full rank. 
Thus, we can use Lemma~\ref{lemma_conditioning} for the constraints 
\begin{equation}
\boldsymbol{B}_{t} 
- (\boldsymbol{0}, \boldsymbol{M}_{t-1}\boldsymbol{\Lambda}_{t-1})
= \boldsymbol{A}\boldsymbol{Q}_{t}, 
\end{equation}
\begin{equation}
M(\boldsymbol{Q}_{t} - \boldsymbol{H}_{t}) 
= \boldsymbol{A}^{\mathrm{T}}\boldsymbol{M}_{t}
\end{equation}
to obtain 
\begin{align}
\boldsymbol{A}
&\sim [\boldsymbol{B}_{t} 
- (\boldsymbol{0}, \boldsymbol{M}_{t-1}\boldsymbol{\Lambda}_{t-1})]
\boldsymbol{Q}_{t}^{\dagger}
- M(\boldsymbol{M}_{t}^{\dagger})^{\mathrm{T}} 
\boldsymbol{H}_{t}^{\mathrm{T}}\boldsymbol{P}_{\boldsymbol{Q}_{t}}^{\perp} 
\nonumber \\
&+ \boldsymbol{P}_{\boldsymbol{M}_{t}}^{\perp}
\tilde{\boldsymbol{A}}_{t}\boldsymbol{P}_{\boldsymbol{Q}_{t}}^{\perp}
\end{align}
conditioned on $\mathfrak{E}_{t,t}$ and $\Theta$, 
where $\tilde{\boldsymbol{A}}_{t}$ is independent of 
$\{\mathfrak{E}_{t,t}, \Theta\}$ and has independent standard Gaussian 
elements. In the derivation of the second term, 
we have used $\boldsymbol{Q}_{t}^{\mathrm{T}}
\boldsymbol{P}_{\boldsymbol{Q}_{t}}^{\perp} =\boldsymbol{O}$. 
Substituting this representation into the definition of $\boldsymbol{b}_{t}$ 
in (\ref{b}) yields 
\begin{align}
\boldsymbol{b}_{t} 
&\sim \boldsymbol{B}_{t}\boldsymbol{\beta}_{t} 
- (\boldsymbol{0}, \boldsymbol{M}_{t-1}\boldsymbol{\Lambda}_{t-1})
\boldsymbol{\beta}_{t}
- M(\boldsymbol{M}_{t}^{\dagger})^{\mathrm{T}} 
\boldsymbol{H}_{t}^{\mathrm{T}}\boldsymbol{q}_{t}^{\perp}
\nonumber \\
&+ \bar{\xi}_{\In,t-1}\boldsymbol{m}_{t-1} 
+ \boldsymbol{P}_{\boldsymbol{M}_{t}}^{\perp}
\tilde{\boldsymbol{A}}_{t}\boldsymbol{q}_{t}^{\perp}, \label{b_t}
\end{align}
with $\boldsymbol{\beta}_{t}=\boldsymbol{Q}_{t}^{\dagger}\boldsymbol{q}_{t}$ 
and $\boldsymbol{q}_{t}^{\perp} = \boldsymbol{P}_{\boldsymbol{Q}_{t}}^{\perp}
\boldsymbol{q}_{t}$. 

We evaluate the third term in (\ref{b_t}). Using the representation 
$\boldsymbol{q}_{t}^{\perp} 
= \boldsymbol{q}_{t} - \boldsymbol{P}_{\boldsymbol{Q}_{t}}^{\parallel}
\boldsymbol{q}_{t} = \boldsymbol{q}_{t} - \boldsymbol{Q}_{t}
\boldsymbol{\beta}_{t}$ yields 
\begin{equation}
M(\boldsymbol{M}_{t}^{\dagger})^{\mathrm{T}}
\boldsymbol{H}_{t}^{\mathrm{T}}\boldsymbol{q}_{t}^{\perp} 
=M(\boldsymbol{M}_{t}^{\dagger})^{\mathrm{T}}\left(
 \boldsymbol{H}_{t}^{\mathrm{T}}\boldsymbol{q}_{t} 
 - \boldsymbol{H}_{t}^{\mathrm{T}}\boldsymbol{Q}_{t}\boldsymbol{\beta}_{t}
\right). 
\end{equation}
For the first term in the parentheses, 
we use the induction hypothesis~\ref{Id} for $\tau=t-1$ to obtain 
\begin{align}
\left[
 \boldsymbol{H}_{t}^{\mathrm{T}}\boldsymbol{q}_{t} 
\right]_{t'}
&\peq \bar{\xi}_{\In, t-1}\mathbb{E}[M_{t'}M_{t-1}] + o(1)
\nonumber \\
&\peq \frac{\bar{\xi}_{\In, t-1}}{M}
\boldsymbol{m}_{t'}^{\mathrm{T}}
\boldsymbol{m}_{t-1} + o(1)
\end{align}
for all $t'\in\{0,\ldots,t-1\}$, where the last equality follows from 
the induction hypotheses~\ref{Oc} for $\tau=t-1$, as well as 
the definitions of $\boldsymbol{m}_{t}$ and $M_{t}$ in (\ref{m}) and (\ref{M}). 
Thus, we have
\begin{align}
M(\boldsymbol{M}_{t}^{\dagger})^{\mathrm{T}}
\boldsymbol{H}_{t}^{\mathrm{T}}\boldsymbol{q}_{t}
&\peq \bar{\xi}_{\In, t-1}\boldsymbol{m}_{t-1}
+ \boldsymbol{M}_{t}\left(
 \frac{\boldsymbol{M}_{t}^{\mathrm{T}}\boldsymbol{M}_{t}}{M}
\right)^{-1}
\boldsymbol{o}(1) 
\nonumber \\
&\peq \bar{\xi}_{\In, t-1}\boldsymbol{m}_{t-1}
+ \boldsymbol{M}_{t}\boldsymbol{o}(1),
\end{align}
where the last equality follows from the induction hypothesis~\ref{Oe} 
for $\tau=t-1$. 
Repeating the same derivation for the second term, we obtain 
\begin{align}
M(\boldsymbol{M}_{t}^{\dagger})^{\mathrm{T}}
\boldsymbol{H}_{t}^{\mathrm{T}}\boldsymbol{Q}_{t}\boldsymbol{\beta}_{t}
= M(\boldsymbol{M}_{t}^{\dagger})^{\mathrm{T}}\sum_{t'=0}^{t-1}
[\boldsymbol{\beta}_{t}]_{t'}\boldsymbol{H}_{t}^{\mathrm{T}}\boldsymbol{q}_{t'}
\nonumber \\
\peq \sum_{t'=1}^{t-1}[\boldsymbol{\beta}_{t}]_{t'}
\bar{\xi}_{\In,t'-1}\boldsymbol{m}_{t'-1}
+ \boldsymbol{M}_{t}\boldsymbol{o}(1),
\end{align}
where we have used the induction hypothesis~(\ref{hx}) for all $\tau<t$. 
Combining these results, we arrive at 
\begin{equation} \label{b_t_tmp}
\boldsymbol{b}_{t} \sim \boldsymbol{B}_{t}\boldsymbol{\beta}_{t} 
+ \boldsymbol{M}_{t}\boldsymbol{o}(1)
+ \tilde{\boldsymbol{A}}_{t}\boldsymbol{q}_{t}^{\perp}
- \boldsymbol{P}_{\boldsymbol{M}_{t}}^{\parallel}
\tilde{\boldsymbol{A}}_{t}\boldsymbol{q}_{t}^{\perp}
\end{equation}
conditioned on $\mathfrak{E}_{t,t}$ and $\Theta$, where we have used 
$\boldsymbol{P}_{\boldsymbol{M}_{t}}^{\perp}=\boldsymbol{I} 
- \boldsymbol{P}_{\boldsymbol{M}_{t}}^{\parallel}$. 

To complete the proof of Property~\ref{Oa} for $\tau=t$, we represent 
the last term in (\ref{b_t_tmp}) as $\boldsymbol{P}_{\boldsymbol{M}_{t}}^{\parallel}
\tilde{\boldsymbol{A}}_{t}\boldsymbol{q}_{t}^{\perp}
=\boldsymbol{M}_{t}\boldsymbol{a}_{t}$
with $\boldsymbol{a}_{t}=
(\boldsymbol{M}_{t}^{\mathrm{T}}\boldsymbol{M}_{t})^{-1}
\boldsymbol{M}_{t}^{\mathrm{T}}\tilde{\boldsymbol{A}}_{t}\boldsymbol{q}_{t}^{\perp}$. 
Evaluating the squared norm of $\boldsymbol{a}_{t}$ yields  
\begin{align}
&\|\boldsymbol{a}_{t}\|_{2}^{2}
= \frac{1}{M^{2}}(\boldsymbol{q}_{t}^{\perp})^{\mathrm{T}}
\tilde{\boldsymbol{A}}_{t}^{\mathrm{T}}\boldsymbol{M}_{t}\left(
 \frac{\boldsymbol{M}_{t}^{\mathrm{T}}\boldsymbol{M}_{t}}{M}
\right)^{-2}
\boldsymbol{M}_{t}^{\mathrm{T}}\tilde{\boldsymbol{A}}_{t}\boldsymbol{q}_{t}^{\perp}
\nonumber \\
&\peq \frac{1}{M}\left\{
 \|\boldsymbol{q}_{t}^{\perp}\|_{2}^{2}\mathrm{Tr}\left[
  \left(
   \frac{1}{M}\boldsymbol{M}_{t}^{\mathrm{T}}\boldsymbol{M}_{t}
  \right)^{-1}
 \right] + o(1)
\right\}\pto0
\end{align}
in the sublinear sparsity limit. In the derivation of the second equality, 
we have used $\tilde{\boldsymbol{A}}_{t}\boldsymbol{q}_{t}^{\perp}
\sim\mathcal{N}(\boldsymbol{0}, 
\|\boldsymbol{q}_{t}^{\perp}\|^{2}\boldsymbol{I})$ and the weak law of 
large numbers. The last convergence follows from the induction 
hypotheses~\ref{Oe} and \ref{Ic} for $\tau=t-1$.  
Thus, Property~\ref{Oa} holds for $\tau=t$. 
\end{IEEEproof}

\begin{IEEEproof}[Proof of \ref{Ob}]
We first consider the case $\tau'\in\{0,\ldots, t-1\}$. 
Using Property~\ref{Oa} for $\tau=t$ yields 
\begin{align}
&\frac{1}{M}\boldsymbol{b}_{\tau'}^{\mathrm{T}}\boldsymbol{b}_{t}
\sim \frac{1}{M}\boldsymbol{b}_{\tau'}^{\mathrm{T}}\boldsymbol{B}_{t}
\boldsymbol{\beta}_{t} 
+ \frac{1}{M}\boldsymbol{b}_{\tau'}^{\mathrm{T}}\boldsymbol{M}_{t}\boldsymbol{o}(1)
+ \frac{1}{M}\boldsymbol{b}_{\tau'}^{\mathrm{T}}\tilde{\boldsymbol{A}}_{t}
\boldsymbol{q}_{t}^{\perp}
\nonumber \\
&\peq \frac{1}{M}\boldsymbol{b}_{\tau'}^{\mathrm{T}}\boldsymbol{B}_{t}
\boldsymbol{\beta}_{t} + o(1)
\peq \boldsymbol{q}_{\tau'}^{\mathrm{T}}\boldsymbol{Q}_{t}
\boldsymbol{\beta}_{t} + o(1)
= \boldsymbol{q}_{\tau'}^{\mathrm{T}}\boldsymbol{q}_{t} + o(1)
\nonumber \\
&\peq Q_{\tau',t} + o(1)
\end{align}
conditioned on $\mathfrak{E}_{t,t}$ and $\Theta$. 
Here, the first equality follows from the induction hypothesis~\ref{Od} 
for all $\tau<t$ and the weak law of large numbers. The second equality 
is obtained from the induction hypotheses~\ref{Ob} and \ref{Ic} 
for all $\tau<t$. The second last equality is due to the definition 
$\boldsymbol{\beta}_{t}=\boldsymbol{Q}_{t}^{\dagger}\boldsymbol{q}_{t}$. 
The last equality follows from the induction hypothesis~\ref{Ic} 
for all $\tau<t$.

We next consider the case $\tau'=t$. Repeating the same proof as that 
for $\tau'<t$ yields 
\begin{align}
&\frac{1}{M}\|\boldsymbol{b}_{t}\|_{2}^{2} 
\sim \frac{1}{M}\boldsymbol{\beta}_{t}^{\mathrm{T}}\boldsymbol{B}_{t}^{\mathrm{T}}
\boldsymbol{B}_{t}\boldsymbol{\beta}_{t} 
+ \frac{1}{M}(\boldsymbol{q}_{t}^{\perp})^{\mathrm{T}}
\tilde{\boldsymbol{A}}_{t}^{\mathrm{T}}
\tilde{\boldsymbol{A}}_{t}\boldsymbol{q}_{t}^{\perp}
+ o(1)
\nonumber \\
&\peq \boldsymbol{\beta}_{t}^{\mathrm{T}}\boldsymbol{Q}_{t}^{\mathrm{T}}
\boldsymbol{Q}_{t}\boldsymbol{\beta}_{t}
+ \|\boldsymbol{q}_{t}^{\perp}\|_{2}^{2} +o(1)
= \|\boldsymbol{q}_{t}\|_{2}^{2} + o(1) 
\nonumber \\
&\peq Q_{t,t} + o(1)
\end{align}
conditioned on  $\mathfrak{E}_{t,t}$ and $\Theta$. In the derivation of 
the first convergence in probability, we have used the induction 
hypotheses~\ref{Ob} and \ref{Ic} for all $\tau<t$, as well as  
the weak law of large numbers. The second last equality follows from 
$\boldsymbol{\beta}_{t}=\boldsymbol{Q}_{t}^{\dagger}\boldsymbol{q}_{t}$ and 
$\boldsymbol{q}_{t}^{\perp} = \boldsymbol{P}_{\boldsymbol{Q}_{t}}^{\perp}
\boldsymbol{q}_{t}$. The last equality is due to the induction 
hypothesis~\ref{Ic} for $\tau=t-1$.  
Thus, Property~\ref{Ob} holds for $\tau=t$. 
\end{IEEEproof}

\begin{IEEEproof}[Proof of \ref{Oc}]
See \cite[p.~779]{Bayati11} for the proof of the first convergence with  
$\tau=t$. The induction hypothesis $v_{\In,t}\pto\tilde{v}_{\In,t}$ in 
Property~\ref{Ic} for $\tau=t-1$ allows us to generalize the proof of 
the last two properties for $\tau=0$ to the case $\tau=t$. 
Thus, Property~\ref{Oc} holds for $\tau=t$.  
\end{IEEEproof}

\begin{IEEEproof}[Proof of \ref{Od}]
Using the definition of $\boldsymbol{m}_{t}$ in (\ref{m}), 
the induction hypothesis~\ref{Ic} for $\tau=t-1$, 
and Property~\ref{Oc} for $\tau=t$ yields 
\begin{align}
\frac{1}{M}\boldsymbol{b}_{\tau'}^{\mathrm{T}}\boldsymbol{m}_{t} 
&\peq \frac{1}{\bar{\xi}_{\Out,t}}\mathbb{E}\left[
 B_{\tau'}f_{\Out,t}\left(
  B_{t} + \bar{\eta}_{t}Z, g(Z, W); \tilde{v}_{\In,t}
 \right)
\right] 
\nonumber \\
&+ o(1). 
\end{align}
From Assumption~\ref{assumption_outer} we can use Lemma~\ref{lemma_Stein}  
to obtain 
\begin{align}
&\frac{1}{\bar{\xi}_{\Out,t}}\mathbb{E}\left[
 B_{\tau'}f_{\Out,t}\left(
  B_{t} + \bar{\eta}_{t}Z, g(Z, W); \tilde{v}_{\In,t}
 \right)
\right]
\nonumber \\
&= \frac{1}{\bar{\xi}_{\Out,t}}\left\{
 \bar{\xi}_{\Out,t}\mathbb{E}[B_{\tau'}B_{t}] 
 + \left(
  \bar{\xi}_{\Out,t}\bar{\eta}_{t} - \bar{\zeta}_{t} 
 \right)\mathbb{E}[B_{\tau'}Z]
\right\}
\nonumber \\
&= \mathbb{E}[B_{\tau'}B_{t}] = Q_{\tau', t}. 
\end{align}
with $\bar{\xi}_{\Out,t}$, $\bar{\eta}_{t}$, and 
$\bar{\zeta}_{t}$ in (\ref{xi_out_bar},) (\ref{eta_bar}), and 
(\ref{zeta_bar}), respectively. 
Combining these results, we arrive at Property~\ref{Od} for $\tau=t$. 
\end{IEEEproof}

\begin{IEEEproof}[Proof of \ref{Oe}]
Repeat an existing proof in \cite[Proof of (65) for $\tau=t$]{Takeuchi20} 
under Assumption~\ref{assumption_SE}. 
\end{IEEEproof}

\subsection{Inner Module for $\tau=t$} 
\label{proof_inner} 
\begin{IEEEproof}[Proof of \ref{Ia}] 
From the induction hypotheses~\ref{Oe} and \ref{Ie} for $\tau=t-1$ we find 
that $\boldsymbol{M}_{t}$ and $\boldsymbol{Q}_{t+1}$ have full rank. 
Thus, we can use Lemma~\ref{lemma_conditioning} for the constraints 
\begin{equation}
\boldsymbol{B}_{t+1} 
- (\boldsymbol{0}, \boldsymbol{M}_{t}\boldsymbol{\Lambda}_{t})
= \boldsymbol{A}\boldsymbol{Q}_{t+1}, 
\end{equation}
\begin{equation}
M(\boldsymbol{Q}_{t} - \boldsymbol{H}_{t}) 
= \boldsymbol{A}^{\mathrm{T}}\boldsymbol{M}_{t}, 
\end{equation}
we have 
\begin{align}
\boldsymbol{A}&\sim M(\boldsymbol{M}_{t}^{\dagger})^{\mathrm{T}}
(\boldsymbol{Q}_{t} - \boldsymbol{H}_{t})^{\mathrm{T}}
+ \boldsymbol{P}_{\boldsymbol{M}_{t}}^{\perp}
\boldsymbol{B}_{t+1}\boldsymbol{Q}_{t+1}^{\dagger}
\nonumber \\
&- \boldsymbol{P}_{\boldsymbol{M}_{t}}^{\perp}\tilde{\boldsymbol{A}}_{t}
\boldsymbol{P}_{\boldsymbol{Q}_{t+1}}^{\perp}
\end{align}
conditioned on $\mathfrak{E}_{t+1,t}$ and $\Theta$, 
where $\tilde{\boldsymbol{A}}_{t}$ is independent of 
$\{\mathfrak{E}_{t+1,t}, \Theta\}$ and has independent standard Gaussian 
elements. In the derivation of the second term, we have used 
$\boldsymbol{P}_{\boldsymbol{M}_{t}}^{\perp}\boldsymbol{M}_{t}=\boldsymbol{O}$.  
Substituting this expression into 
the definition of $\boldsymbol{h}_{t}$ in (\ref{h}) yields 
\begin{align}
\boldsymbol{h}_{t} &\sim \boldsymbol{H}_{t}\boldsymbol{\alpha}_{t}
- \boldsymbol{Q}_{t}\boldsymbol{\alpha}_{t}
- \frac{1}{M}(\boldsymbol{Q}_{t+1}^{\dagger})^{\mathrm{T}}
\boldsymbol{B}_{t+1}^{\mathrm{T}}\boldsymbol{m}_{t}^{\perp} 
+ \boldsymbol{q}_{t} 
\nonumber \\
&+ \frac{1}{M}\boldsymbol{P}_{\boldsymbol{Q}_{t+1}}^{\perp}
\tilde{\boldsymbol{A}}_{t}^{\mathrm{T}}\boldsymbol{m}_{t}^{\perp}, \label{h_t}
\end{align}
with $\boldsymbol{\alpha}_{t}=\boldsymbol{M}_{t}^{\dagger}\boldsymbol{m}_{t}$ and 
$\boldsymbol{m}_{t}^{\perp}=\boldsymbol{P}_{\boldsymbol{M}_{t}}^{\perp}
\boldsymbol{m}_{t}$. 

We evaluate the third term in (\ref{h_t}). 
Using the representation $\boldsymbol{m}_{t}^{\perp} 
= \boldsymbol{m}_{t} - \boldsymbol{P}_{\boldsymbol{M}_{t}}^{\parallel}
\boldsymbol{m}_{t} = \boldsymbol{m}_{t} - \boldsymbol{M}_{t}
\boldsymbol{\alpha}_{t}$ yields 
\begin{equation}
\frac{(\boldsymbol{Q}_{t+1}^{\dagger})^{\mathrm{T}}
\boldsymbol{B}_{t+1}^{\mathrm{T}}\boldsymbol{m}_{t}^{\perp}}{M}
=(\boldsymbol{Q}_{t+1}^{\dagger})^{\mathrm{T}}\left(
 \frac{\boldsymbol{B}_{t+1}^{\mathrm{T}}\boldsymbol{m}_{t}}{M} 
 - \frac{\boldsymbol{B}_{t+1}^{\mathrm{T}}\boldsymbol{M}_{t}
 \boldsymbol{\alpha}_{t}}{M}
\right).
\end{equation}
For the first term in the parentheses, 
we use Property~\ref{Od} for $\tau=t$ to obtain 
\begin{equation}
\frac{1}{M}\left[
 \boldsymbol{B}_{t+1}^{\mathrm{T}}\boldsymbol{m}_{t} 
\right]_{t'}
\peq Q_{t',t} + o(1) 
\peq \boldsymbol{q}_{t'}^{\mathrm{T}}\boldsymbol{q}_{t} + o(1)
\end{equation}
for all $t'\in\{0,\ldots,t\}$, 
where the last equality follows from the induction 
hypothesis~\ref{Ic} for $\tau=t-1$. Thus, we have
\begin{align}
(\boldsymbol{Q}_{t+1}^{\dagger})^{\mathrm{T}}
\frac{\boldsymbol{B}_{t+1}^{\mathrm{T}}\boldsymbol{m}_{t}}{M} 
&\peq \boldsymbol{q}_{t} + \boldsymbol{Q}_{t+1}\left(
 \boldsymbol{Q}_{t+1}^{\mathrm{T}}\boldsymbol{Q}_{t+1}
\right)^{-1}\boldsymbol{o}(1)
\nonumber \\
&\peq \boldsymbol{q}_{t} + \boldsymbol{Q}_{t+1}\boldsymbol{o}(1),  
\end{align}
where the last equality follows from the induction hypothesis~\ref{Ie} 
for $\tau=t-1$. 
Repeating the same derivation for the second term, we obtain 
\begin{align}
&(\boldsymbol{Q}_{t+1}^{\dagger})^{\mathrm{T}}
\frac{\boldsymbol{B}_{t+1}^{\mathrm{T}}\boldsymbol{M}_{t}\boldsymbol{\alpha}_{t}}{M}
= \sum_{t'=0}^{t-1}[\boldsymbol{\alpha}_{t}]_{t'}
(\boldsymbol{Q}_{t+1}^{\dagger})^{\mathrm{T}}
\frac{\boldsymbol{B}_{t+1}^{\mathrm{T}}\boldsymbol{m}_{t'}}{M}
\nonumber \\
&\peq \sum_{t'=0}^{t-1}[\boldsymbol{\alpha}_{t}]_{t'}\boldsymbol{q}_{t'}
+ \boldsymbol{Q}_{t+1}\boldsymbol{o}(1)
= \boldsymbol{Q}_{t}\boldsymbol{\alpha}_{t} 
+ \boldsymbol{Q}_{t+1}\boldsymbol{o}(1). 
\end{align}
Combining these results, we arrive at 
\begin{equation} \label{h_t_tmp}
\boldsymbol{h}_{t} \sim \boldsymbol{H}_{t}\boldsymbol{\alpha}_{t} 
+ \boldsymbol{Q}_{t+1}\boldsymbol{o}(1)
+ \frac{1}{M}\tilde{\boldsymbol{A}}_{t}^{\mathrm{T}}\boldsymbol{m}_{t}^{\perp}
- \frac{1}{M}\boldsymbol{P}_{\boldsymbol{Q}_{t+1}}^{\parallel}
\tilde{\boldsymbol{A}}_{t}^{\mathrm{T}}\boldsymbol{m}_{t}^{\perp}
\end{equation}
conditioned on $\mathfrak{E}_{t+1,t}$ and $\Theta$, where we have used 
$\boldsymbol{P}_{\boldsymbol{Q}_{t+1}}^{\perp}=\boldsymbol{I} 
- \boldsymbol{P}_{\boldsymbol{Q}_{t+1}}^{\parallel}$. 

To complete the proof of Property~\ref{Ia} for $\tau=t$, we represent 
the last term in (\ref{h_t_tmp}) as 
$M^{-1}\boldsymbol{P}_{\boldsymbol{Q}_{t+1}}^{\parallel}
\tilde{\boldsymbol{A}}_{t}^{\mathrm{T}}\boldsymbol{m}_{t}^{\perp}
=\boldsymbol{Q}_{t+1}\tilde{\boldsymbol{a}}_{t}$
with $\tilde{\boldsymbol{a}}_{t}=
M^{-1}(\boldsymbol{Q}_{t+1}^{\mathrm{T}}\boldsymbol{Q}_{t+1})^{-1}
\boldsymbol{Q}_{t+1}^{\mathrm{T}}\tilde{\boldsymbol{A}}_{t}^{\mathrm{T}}
\boldsymbol{m}_{t}^{\perp}$. 
Evaluating the squared norm of $\tilde{\boldsymbol{a}}_{t}$ yields  
\begin{align}
&\|\tilde{\boldsymbol{a}}_{t}\|_{2}^{2}
= \frac{1}{M}\frac{(\boldsymbol{m}_{t}^{\perp})^{\mathrm{T}}
\tilde{\boldsymbol{A}}_{t}}{\sqrt{M}}\boldsymbol{Q}_{t+1}\left(
 \boldsymbol{Q}_{t+1}^{\mathrm{T}}\boldsymbol{Q}_{t+1}
\right)^{-2}\boldsymbol{Q}_{t+1}^{\mathrm{T}}
\frac{\tilde{\boldsymbol{A}}_{t}^{\mathrm{T}}\boldsymbol{m}_{t}^{\perp}}
{\sqrt{M}}
\nonumber \\
&\peq \frac{1}{M}\left\{
 \frac{\|\boldsymbol{m}_{t}^{\perp}\|_{2}^{2}}{M}\mathrm{Tr}\left[
  \left(
   \boldsymbol{Q}_{t+1}^{\mathrm{T}}\boldsymbol{Q}_{t+1}
  \right)^{-1}
 \right] + o(1)
\right\}
\pto0
\end{align}
in the sublinear sparsity limit. In the derivation of the second equality, 
we have used $M^{-1/2}\tilde{\boldsymbol{A}}_{t}^{\mathrm{T}}
\boldsymbol{m}_{t}^{\perp}\sim\mathcal{N}(\boldsymbol{0}, 
M^{-1}\|\boldsymbol{m}_{t}^{\perp}\|^{2}\boldsymbol{I})$ and the weak law of 
large numbers. The last convergence follows from Property~\ref{Oc} for 
$\tau=t$ and the induction hypothesis~\ref{Ie} for $\tau=t-1$.  
Thus, Property~\ref{Ia} holds for $\tau=t$. 
\end{IEEEproof}

\begin{IEEEproof}[Proof of \ref{Ib}]
We first consider the case $\tau'\in\{0,\ldots, t-1\}$. 
Using Property~\ref{Ia} for $\tau=t$ yields 
\begin{align}
&\frac{M}{N}\boldsymbol{h}_{\tau'}^{\mathrm{T}}\boldsymbol{h}_{t}
\sim \frac{M}{N}\boldsymbol{h}_{\tau'}^{\mathrm{T}}\boldsymbol{H}_{t}
\boldsymbol{\alpha}_{t} 
+ \frac{M}{N}\boldsymbol{h}_{\tau'}^{\mathrm{T}}\boldsymbol{Q}_{t+1}\boldsymbol{o}(1)
+ \frac{\boldsymbol{h}_{\tau'}^{\mathrm{T}}\tilde{\boldsymbol{A}}_{t}^{\mathrm{T}}
\boldsymbol{m}_{t}^{\perp}}{N}
\nonumber \\
&\peq \frac{M}{N}\boldsymbol{h}_{\tau'}^{\mathrm{T}}\boldsymbol{H}_{t}
\boldsymbol{\alpha}_{t} + o(1)
\peq \frac{1}{M}\boldsymbol{m}_{\tau'}^{\mathrm{T}}\boldsymbol{M}_{t}
\boldsymbol{\alpha}_{t} + o(1)
\nonumber \\
&= \frac{1}{M}\boldsymbol{m}_{\tau'}^{\mathrm{T}}\boldsymbol{m}_{t} + o(1) 
\peq \mathbb{E}[M_{\tau'}M_{t}] + o(1) 
\end{align}
conditioned on $\mathfrak{E}_{t+1,t}$ and $\Theta$. Here, 
the first equality follows from the induction hypothesis~\ref{Id} 
for all $\tau<t$ and the weak law of large numbers. 
The second equality is obtained from the induction hypotheses~\ref{Oc} 
and \ref{Ib} for all $\tau<t$. The third equality is due to 
$\boldsymbol{\alpha}_{t}=\boldsymbol{M}_{t}^{\dagger}\boldsymbol{m}_{t}$.
The last equality follows from Property~\ref{Oc} for $\tau=t$. 

We next consider the case $\tau'=t$. Repeating the same proof as that 
for $\tau'<t$ yields 
\begin{align}
&\frac{M}{N}\|\boldsymbol{h}_{t}\|_{2}^{2} 
\sim \frac{M}{N}\boldsymbol{\alpha}_{t}^{\mathrm{T}}\boldsymbol{H}_{t}^{\mathrm{T}}
\boldsymbol{H}_{t}\boldsymbol{\alpha}_{t} 
+ \frac{(\boldsymbol{m}_{t}^{\perp})^{\mathrm{T}}\tilde{\boldsymbol{A}}_{t}
\tilde{\boldsymbol{A}}_{t}^{\mathrm{T}}\boldsymbol{m}_{t}^{\perp}}{MN}
+ o(1)
\nonumber \\
&\peq \frac{1}{M}\boldsymbol{\alpha}_{t}^{\mathrm{T}}\boldsymbol{M}_{t}^{\mathrm{T}}
\boldsymbol{M}_{t}\boldsymbol{\alpha}_{t}
+ \frac{\|\boldsymbol{m}_{t}^{\perp}\|_{2}^{2}}{M} + o(1)
= \frac{\|\boldsymbol{m}_{t}\|_{2}^{2}}{M} + o(1) 
\nonumber \\
&\peq \mathbb{E}[M_{t}^{2}] + o(1) 
\end{align}
conditioned on $\mathfrak{E}_{t+1,t}$ and $\Theta$.  
In the derivation of the first convergence in probability, we have used 
the induction hypotheses~\ref{Oc} and \ref{Ib} for all $\tau<t$, as well as 
the weak law of large numbers. The second equality follows 
from $\boldsymbol{\alpha}_{t}=\boldsymbol{M}_{t}^{\dagger}\boldsymbol{m}_{t}$ and 
$\boldsymbol{m}_{t}^{\perp} = \boldsymbol{P}_{\boldsymbol{M}_{t}}^{\perp}
\boldsymbol{m}_{t}$. The last equality is obtained from 
Property~\ref{Oc} for $\tau=t$. Thus, Property~\ref{Ib} holds for $\tau=t$. 
\end{IEEEproof}

\begin{IEEEproof}[Proof of \ref{Ic}]
Define $\{\boldsymbol{\omega}_{\tau}\in\mathbb{R}^{N}\}_{\tau=0}^{t}$ 
conditioned on $\boldsymbol{M}_{t+1}$ as 
\begin{equation}
\boldsymbol{\omega}_{0} = \frac{1}{M}\tilde{\boldsymbol{A}}_{0}^{\mathrm{T}}
\boldsymbol{m}_{0}, 
\end{equation}
\begin{equation} \label{omega_t} 
\boldsymbol{\omega}_{t} = \sum_{\tau=0}^{t-1}[\boldsymbol{\alpha}_{t}]_{\tau}
\boldsymbol{\omega}_{\tau} + \frac{1}{M}\tilde{\boldsymbol{A}}_{t}^{\mathrm{T}}
\boldsymbol{m}_{t}^{\perp}
\end{equation}
for $t>0$, where 
$\{\tilde{\boldsymbol{A}}_{\tau}\}_{\tau=0}^{t}$ are independent standard 
Gaussian matrices. By definition, $\{\boldsymbol{\omega}_{\tau}\}_{\tau=0}^{t}$ 
are zero-mean correlated Gaussian random vectors. We repeat 
the proof of Property~\ref{Ib} to find $\mathbb{E}[\boldsymbol{\omega}_{\tau}
\boldsymbol{\omega}_{\tau'}^{\mathrm{T}} | \boldsymbol{M}_{t+1}]
= M^{-2}\boldsymbol{m}_{\tau'}^{\mathrm{T}}\boldsymbol{m}_{\tau}\boldsymbol{I}_{N}$.  

Property~\ref{Ic} for $\tau=t$ is obtained by repeating the proof of 
Property~\ref{Ic} for $\tau=0$. 
To utilize Lemma~\ref{lemma_covariance_error} in the proof of 
the former convergence, we need to prove 
$\|\boldsymbol{h}_{\tau} - \boldsymbol{\omega}_{\tau}\|_{2}\pto 0$ for all  
$\tau\in\{0,\ldots, t\}$. 

The proof is by induction. For $\tau=0$, we use 
Property~\ref{Ia} for $\tau=0$ and (\ref{q_0_convergence}) to obtain 
$\|\boldsymbol{h}_{0} - \boldsymbol{\omega}_{0}\|_{2} 
= o(\|\boldsymbol{q}_{0}\|_{2})\peq o(1)$. Suppose that 
$\|\boldsymbol{h}_{\tau} - \boldsymbol{\omega}_{\tau}\|_{2}\pto 0$ holds 
for all  $\tau\in\{0,\ldots, t-1\}$. Using Property~\ref{Ia} for $\tau=t$ 
and the definition of $\boldsymbol{\omega}_{t}$ in (\ref{omega_t}) yields 
\begin{align}
&\|\boldsymbol{h}_{t} - \boldsymbol{\omega}_{t}\|_{2}
= \left\|
 \sum_{\tau=0}^{t-1}[\boldsymbol{\alpha}_{t}]_{\tau}(\boldsymbol{h}_{\tau} 
 - \boldsymbol{\omega}_{\tau})
 + \boldsymbol{Q}_{t+1}\boldsymbol{o}(1) 
\right\|_{2} \nonumber \\
&\leq \sum_{\tau=0}^{t-1}[\boldsymbol{\alpha}_{t}]_{\tau}\|\boldsymbol{h}_{\tau} 
- \boldsymbol{\omega}_{\tau}\|_{2}
+ \| \boldsymbol{Q}_{t+1}\boldsymbol{o}(1) \|_{2} 
\pto 0 
\end{align}
conditioned on $\mathfrak{E}_{t+1,t}$ and $\Theta$, 
where the last convergence follows from the induction hypotheses 
$\|\boldsymbol{h}_{\tau} - \boldsymbol{\omega}_{\tau}\|_{2}\pto 0$ and 
\ref{Ic} for all $\tau<t$. By induction, 
$\|\boldsymbol{h}_{\tau} - \boldsymbol{\omega}_{\tau}\|_{2}\pto 0$ holds 
for all $\tau\in\{0,\ldots, t\}$. 
Thus, Property~\ref{Ic} holds for $\tau=t$. 
\end{IEEEproof}

\begin{IEEEproof}[Proof of \ref{Id}]
We first confirm the former convergence~(\ref{hx}) for $\tau=t$. 
Using Property~\ref{Ia} for $\tau=0$ yields 
\begin{equation}
\boldsymbol{x}^{\mathrm{T}}\boldsymbol{h}_{t}
\sim \boldsymbol{x}^{\mathrm{T}}\boldsymbol{H}_{\tau}\boldsymbol{\alpha}_{\tau} 
+ \boldsymbol{x}^{\mathrm{T}}\boldsymbol{Q}_{\tau+1}\boldsymbol{o}(1)
+ \frac{1}{M}\boldsymbol{x}^{\mathrm{T}}\tilde{\boldsymbol{A}}_{\tau}^{\mathrm{T}}
\boldsymbol{m}_{\tau}^{\perp}
\to 0
\end{equation}
conditioned on $\mathfrak{E}_{t+1,t}$ and $\Theta$, where the first, second, 
and last terms converge in probability to zero in the sublinear sparsity 
limit, because of the induction hypothesis~\ref{Id} for all $\tau<t$, 
the induction hypothesis~\ref{Ic} for all $\tau<t$ and 
(\ref{q_0_convergence}), and of the weak law of large numbers, respectively. 

We next prove the latter convergence~(\ref{hq}) for $\tau=t$. 
From Property~\ref{Ia} for $\tau=t$, 
$\|\boldsymbol{h}_{\tau} - \boldsymbol{\omega}_{\tau}\|_{2}\pto 0$ for all  
$\tau\in\{0,\ldots, t\}$, and $v_{\Out,t}\pto\bar{v}_{\Out,t}$ in 
Property~\ref{Oc} for $\tau=t$, we can utilize 
Assumption~\ref{assumption_inner} for $\boldsymbol{q}_{t+1}$ in (\ref{q}) and 
the induction hypothesis~(\ref{hx}) for $\tau=\tau'$ to obtain 
\begin{align}
&\boldsymbol{h}_{\tau'}^{\mathrm{T}}\boldsymbol{q}_{t+1} 
\peq \sum_{n=1}^{N}\mathbb{E}\left[
 \omega_{n, \tau'}f_{\In,t}(\bar{\eta}_{t}x_{n} + \omega_{n,t}; \bar{v}_{\Out,t})
\right] + o(1)
\nonumber \\
&= \sum_{n=1}^{N}\mathbb{E}[f_{\In,t}'(\bar{\eta}_{t}x_{n} + \omega_{n,t}; 
\bar{v}_{\Out,t})]
\mathbb{E}[\omega_{n,\tau'}\omega_{n,t}] + o(1)
\nonumber \\
&= \bar{\xi}_{\In,t}\mathbb{E}[M_{\tau'}M_{t}] + o(1).
\end{align}
In the derivation of the second equality, we have used 
Lemma~\ref{lemma_Stein_inner} under Assumption~\ref{assumption_inner}. 
The last equality follows from the definition of 
$\bar{\xi}_{\In,t}$ in (\ref{xi_in_bar}) and  
$\mathbb{E}[\omega_{n,\tau'}\omega_{n,t} | \boldsymbol{M}_{t+1}]
= M^{-2}\boldsymbol{m}_{\tau'}^{\mathrm{T}}\boldsymbol{m}_{t}$, and 
Property~\ref{Oc} for $\tau=t$. 
Thus, Property~\ref{Id} holds for $\tau=t$. 
\end{IEEEproof}

\begin{IEEEproof}[Proof of \ref{Ie}]
Repeat an existing proof in \cite[Proof of (76) for $\tau=t$]{Takeuchi20} 
for the unnormalized quantity $\boldsymbol{Q}_{t+1}^{\mathrm{T}}
\boldsymbol{Q}_{t+1}$ under Assumption~\ref{assumption_SE}. 
\end{IEEEproof}

\section{Proof of (\ref{former_limit_general})}
\label{proof_former_limit} 
\subsection{Formulation}
We represent $\mathbb{E}[|\Omega_{n}|^{j} 
f_{X}^{2}(\nu_{N}a_{n} + \Omega_{n}; v_{\tilde{N}}) | \boldsymbol{a}]$ with 
$\Omega_{n}\sim\mathcal{N}(0, v_{k,N})$ and $v_{k,N}=v_{\tilde{N}}/k$ as 
\begin{align}
&\mathbb{E}[|\Omega_{n}|^{j}f_{X}^{2}(\nu_{N}a_{n} + \Omega_{n}; v_{\tilde{N}}) 
| \boldsymbol{a}]
\nonumber \\
&= \int |w|^{j}f_{X}^{2}(\nu_{N}a_{n} + w; v_{\tilde{N}})p_{\mathrm{G}}(w; v_{k,N})dw
\nonumber \\
&= k^{-j/2}\int |\tilde{w}|^{j}
f_{X}^{2}(\nu_{N}a_{n} + k^{-1/2}\tilde{w}; v_{\tilde{N}})
p_{\mathrm{G}}(\tilde{w}; v_{\tilde{N}})d\tilde{w}.
\label{former_tmp}
\end{align}
We decompose the interval of integration into four disjoint sets, 
\begin{equation} \label{W1}
\mathcal{W}_{n,1} =\{ w\in\mathbb{R}: |w|<\sqrt{2\alpha_{N}v} 
- 2|\nu_{N}\tilde{a}_{n}|\}, 
\end{equation}
\begin{equation} \label{W2}
\mathcal{W}_{n,2} =\{ w\in\mathbb{R}: |w|\in[\sqrt{2\alpha_{N}v}
- 2|\nu_{N}\tilde{a}_{n}|, \sqrt{2v} - 2|\nu_{N}\tilde{a}_{n}|]\}, 
\end{equation}
\begin{equation} \label{W3}
\mathcal{W}_{n,3} 
= \{ w\in\mathbb{R}: |w|\in(\sqrt{2v} - 2|\nu_{N}\tilde{a}_{n}|, \sqrt{2v}]\},
\end{equation}
\begin{equation} \label{W4}
\mathcal{W}_{n,4} =\{ w\in\mathbb{R}: |w|>\sqrt{2v}\}, 
\end{equation}
with $\alpha_{N}=1-\epsilon_{N}$  
and $\epsilon_{N}=\log\log (N/k)/\log(N/k)$.\footnote{
In the proof of Lemma~\ref{lemma_MSE} we always consider sufficiently 
large $N$ such that $\log\log (N/k)/\log (N/k)\in(0, 1)$ is satisfied. 
} 
In the following appendices we evaluate the integral in (\ref{former_tmp}) 
over each set in the sublinear sparsity limit.

In the region $\mathcal{W}_{n,1}$ of integration, the posterior 
mean estimator $f_{A}(w; v_{\tilde{N}})$ is small enough to prove 
(\ref{former_limit_general}).  
In the last region $\mathcal{W}_{n,4}$ of integration, the probability 
$\mathrm{Pr}(|\tilde{\Omega}|>\sqrt{2v})$ for 
$\tilde{\Omega}\sim\mathcal{N}(0, v_{\tilde{N}})$ is small enough to prove 
(\ref{former_limit_general}). 
The remaining regions correspond to thin transition 
regions between the two regions. 

\subsection{First Set $\mathcal{W}_{n,1}$}
We prove that the integral in (\ref{former_tmp}) over $\mathcal{W}_{n,1}$ is 
$o((N-k)^{-1})$ in the sublinear sparsity limit. 
Using the definition of $f_{X}$ in (\ref{posterior_mean_estimator_X}) yields 
\begin{align}
&k^{-j/2}\int_{\mathcal{W}_{n,1}}|w|^{j}f_{X}^{2}(\nu_{N}a_{n} + k^{-1/2}w; v_{\tilde{N}})
p_{\mathrm{G}}(w; v_{\tilde{N}})dw
\nonumber \\
&\leq \frac{\bar{f}_{U,n,1}^{2}(v_{\tilde{N}})}{k^{1 + j/2}}
\int_{\mathcal{W}_{n,1}}|w|^{j}f_{A}^{2}(\nu_{N}\tilde{a}_{n} + w; v_{\tilde{N}})
p_{\mathrm{G}}(w; v_{\tilde{N}})dw,  
\label{former_first_tmp}
\end{align}
with 
\begin{equation}
\bar{f}_{U,n,1}(v_{\tilde{N}})
=\sup_{w\in\mathcal{W}_{n,1}}|f_{U}(\nu_{N}\tilde{a}_{n} + w; v_{\tilde{N}})|,
\end{equation} 
which is bounded from above by some $\bar{f}_{U,1}>0$ for all $N$ and $n$, 
because of the uniform Lipschitz-continuity assumption for $f_{U}$ and 
$|\nu_{N}\tilde{a}_{N}+w|<\sqrt{2\alpha_{N}v}$ for all $w\in\mathcal{W}_{n,1}$.  

To evaluate the upper bound~(\ref{former_first_tmp}), we utilize an intuition 
in which $f_{A}(w; v_{\tilde{N}})$ should tend to zero in the sublinear sparsity 
limit for all $|w|<\sqrt{2\alpha_{N}v}$. In exploiting this intuition 
rigorously, we use the following upper bound for $f_{A}$ in 
(\ref{posterior_mean_estimator_A}):
\begin{align}
f_{A}(w; v_{\tilde{N}}) 
&< \frac{k\mathbb{E}_{U}[p_{\mathrm{G}}(w - U; v_{\tilde{N}})]}
{(N - k)p_{\mathrm{G}}(w; v_{\tilde{N}})}
\nonumber \\ 
&\leq \frac{k}{(N - k)\sqrt{2\pi v_{\tilde{N}}}p_{\mathrm{G}}(w; v_{\tilde{N}})}. 
\label{f_A_upper_bound}
\end{align}
The last upper bound is slightly loose: For 
$U\sim\mathcal{N}(0, \sigma_{U}^{2})$ we have 
$\mathbb{E}_{U}[p_{\mathrm{G}}(w - U; v_{\tilde{N}})]
=p_{\mathrm{G}}(y; \sigma_{U}^{2}+v_{\tilde{N}})
\leq(2\pi\sigma_{U}^{2})^{-1/2}$ while the upper bound 
$\mathbb{E}_{U}[p_{\mathrm{G}}(w - U; v_{\tilde{N}})]\leq(2\pi v_{\tilde{N}})^{-1/2}$ 
has been used in (\ref{f_A_upper_bound}). 
However, this loose bound is sufficient to prove Lemma~\ref{lemma_MSE}. 

We evaluate the upper bound~(\ref{former_first_tmp}). 
Applying the upper bound~(\ref{f_A_upper_bound}) to the integral 
in (\ref{former_first_tmp}) yields 
\begin{align}
&\frac{1}{k^{1+j/2}}
\int_{\mathcal{W}_{n,1}}|w|^{j}f_{A}^{2}(\nu_{N}\tilde{a}_{n} + w; 
v_{\tilde{N}})p_{\mathrm{G}}(w; v_{\tilde{N}})dw
\nonumber \\
%&< \frac{k^{1-j/2}}{(N-k)^{2}\sqrt{2\pi v_{\tilde{N}}}}\int_{\mathcal{W}_{n,1}}|w|^{j}e^{\{2(\nu_{N}\tilde{a}_{n} + w)^{2}-w^{2}\}/(2v_{\tilde{N}})}dw \nonumber \\
&< \frac{k^{1-j/2}e^{-\nu_{N}^{2}\tilde{a}_{n}^{2}/v_{\tilde{N}}}}
{(N-k)^{2}\sqrt{2\pi v_{\tilde{N}}}}
\int_{\mathcal{W}_{n,1}}|w|^{j}\exp\left\{
 \frac{(w + 2\nu_{N}\tilde{a}_{n})^{2}}{2v_{\tilde{N}}}
\right\}dw
\label{former_first_tmp2} \\
&< \frac{2k^{1-j/2}(2\alpha_{N}v)^{(j+1)/2}}
{(N-k)^{2}\sqrt{2\pi v_{\tilde{N}}}}
\exp\left(
 \frac{\alpha_{N}v}{v_{\tilde{N}}}
\right)
\nonumber \\
&= \frac{2\sqrt{\alpha_{N}}(2\alpha_{N}v)^{j/2}}
{\sqrt{\pi}}
\frac{k^{1-j/2}(N/k)^{\alpha_{N}}
\sqrt{\log(N/k)}}{(N-k)^{2}},
\end{align}
with $v_{\tilde{N}}=v/\log(N/k)$, which is a uniform upper bound for all 
$n$. Combining these results, we arrive at 
\begin{align}
&\frac{1}{k^{j/2}}\sum_{n\notin\mathcal{S}}\int_{\mathcal{W}_{n,1}}|w|^{j}
f_{X}^{2}(\nu_{N}a_{n} + k^{-1/2}w; v_{\tilde{N}})p_{\mathrm{G}}(w; v_{\tilde{N}})dw 
\nonumber \\
&< \frac{2\bar{f}_{U,1}^{2}\sqrt{\alpha_{N}}(2\alpha_{N}v)^{j/2}}
{\sqrt{\pi}}
\frac{k^{1-j/2}(N/k)^{\alpha_{N}}\sqrt{\log(N/k)}}{N-k}
\to 0  
\end{align}
in the sublinear sparsity limit for all $j\geq0$, 
because of $\alpha_{N}=1-\log\log (N/k)/\log(N/k)$.

\subsection{Second Set $\mathcal{W}_{n,2}$}
We prove evaluate the integral in (\ref{former_tmp}) over $\mathcal{W}_{n,2}$ 
in the sublinear sparsity limit. Since 
$\mathcal{W}_{n,2}=\emptyset$ holds for $2|\nu_{N}\tilde{a}_{n}|\geq\sqrt{2v}$, 
without loss of generality, we can assume 
$2|\nu_{N}\tilde{a}_{n}|<\sqrt{2v}$.
Repeating the derivation of (\ref{former_first_tmp2}) yields 
\begin{align}
&k^{-j/2}\int_{\mathcal{W}_{n,2}}
|w|^{j}f_{X}^{2}(\nu_{N}a_{n} + k^{-1/2}w; v_{\tilde{N}})p_{\mathrm{G}}(w; v_{\tilde{N}})dw
\nonumber \\
&\leq \frac{\bar{f}_{U,n,2}^{2}(v_{\tilde{N}})k^{1-j/2}
e^{-\nu_{N}^{2}\tilde{a}_{n}^{2}/v_{\tilde{N}}}}
{(N-k)^{2}\sqrt{2\pi v_{\tilde{N}}}}
\nonumber \\
&\cdot\int_{\mathcal{W}_{n,2}}
|w|^{j}\exp\left\{
 \frac{(w + 2\nu_{N}\tilde{a}_{n})^{2}}{2v_{\tilde{N}}}
\right\}dw
\end{align}
with 
\begin{equation}
\bar{f}_{U,n,2}(v_{\tilde{N}})
=\sup_{w\in\mathcal{W}_{n,2}}|f_{U}(\nu_{N}\tilde{a}_{n} + w; v_{\tilde{N}})|,
\end{equation}
which is bounded from above by some $\bar{f}_{U,2}>0$ for all $N$ and $n$, 
because of the uniform Lipschitz-continuity for $f_{U}$ and 
$|\nu_{N}\tilde{a}_{n} + w|\leq \sqrt{2v} - |\nu_{N}\tilde{a}_{n}|<\sqrt{2v}$ 
for all $w\in\mathcal{W}_{n,2}$. 

Let $\mathcal{N}_{\mathrm{e}}=\{n\in\{1,\ldots,N\}: 
2|\nu_{N}\tilde{a}_{n}| > \sqrt{2\alpha_{N}v}\}$ denote the set of 
exceptional indices for which $2|\nu_{N}\tilde{a}_{n}|$ is larger than 
$\sqrt{2\alpha_{N}v}$. Repeating the proof of (\ref{latter_limit_last_term}) 
yields  
\begin{equation} \label{latter_limit_last_term_all}
\frac{1}{k}\sum_{n=1}^{N}\nu_{N}^{2}\tilde{a}_{n}^{2}
= \nu_{N}^{2}\|\boldsymbol{a}\|_{2}^{2}\pto0, 
\end{equation}
which implies 
$|\mathcal{N}_{\mathrm{e}}|\peq o(k)$. For $n\in\mathcal{N}_{\mathrm{e}}$ we obtain 
\begin{align}
&\int_{\mathcal{W}_{n,2}}
|w|^{j}\exp\left\{
 \frac{(w + 2\nu_{N}\tilde{a}_{n})^{2}}{2v_{\tilde{N}}}
\right\}dw
\nonumber \\
&< 2\sqrt{2 v}(2v)^{j/2}\exp\left(
 \frac{v}{v_{\tilde{N}}}
\right)
= 2(2v)^{(j+1)/2}\frac{N}{k}. 
\end{align}
For $n\notin\mathcal{N}_{\mathrm{e}}$, on the other hand,  
we have the following upper bound :
\begin{align}
&\int_{\mathcal{W}_{n,2}}
|w|^{j}\exp\left\{
 \frac{(w + 2\nu_{N}\tilde{a}_{n})^{2}}{2v_{\tilde{N}}}
\right\}dw
\nonumber \\
&< 2(\sqrt{2 v} - \sqrt{2\alpha_{N}v})
(2v)^{j/2}\exp\left(
 \frac{v}{v_{\tilde{N}}}
\right)
\nonumber \\
&= \frac{2(2v)^{(j+1)/2}}{1 + \sqrt{\alpha_{N}}}
\frac{N\log\log (N/k)}{k\log(N/k)}
\end{align}
for $\alpha_{N}=1 - \log\log (N/k)/\log(N/k)$. 
Combining these upper bounds, as well as $\mathcal{N}_{\mathrm{e}}\peq o(k)$, 
we arrive at  
\begin{align}
&\frac{1}{k^{j/2}}\sum_{n\not\in\mathcal{S}}\int_{\mathcal{W}_{n,2}}
|w|^{j}f_{X}^{2}(\nu_{N}a_{n} + k^{-1/2}w; v_{\tilde{N}})p_{\mathrm{G}}(w; v_{\tilde{N}})dw
\nonumber \\
&< |\mathcal{N}_{\mathrm{e}}|
\frac{2(2v)^{j/2}\bar{f}_{U,2}^{2}N\sqrt{\log(N/k)}}{\sqrt{\pi}k^{j/2}(N-k)^{2}}
\nonumber \\
&+ \frac{2(2v)^{j/2}\bar{f}_{U,2}^{2}}
{\sqrt{\pi}(1 + \sqrt{\alpha_{N}})}
\frac{N\log\log (N/k)}{k^{j/2}(N-k)\sqrt{\log (N/k)}}
\pto 0
\end{align}
in the sublinear sparsity limit for all $j\geq0$.

\subsection{Third Set $\mathcal{W}_{n,3}$}
We evaluate the integral in (\ref{former_tmp}) over $\mathcal{W}_{n,3}$ 
in the sublinear sparsity limit. 
Using the definition of $f_{X}(\nu_{N}a_{n} + k^{-1/2}w;v_{\tilde{N}})$ 
in (\ref{posterior_mean_estimator_X}) and 
the trivial upper bound $f_{A}(w;v_{\tilde{N}})\leq 1$ yields 
\begin{align}
&k^{-j/2}\int_{\mathcal{W}_{n,3}}|w|^{j}f_{X}^{2}(\nu_{N}a_{n} + k^{-1/2}w; v_{\tilde{N}})
p_{\mathrm{G}}(w; v_{\tilde{N}})dw
\nonumber \\
&\leq \frac{1}{k^{1+j/2}}\int_{\mathcal{W}_{n,3}}
|w|^{j}f_{U}^{2}(\nu_{N}\tilde{a}_{n} + w; v_{\tilde{N}})
p_{\mathrm{G}}(w; v_{\tilde{N}})dw. 
\end{align}

We utilize the fact that $\mathcal{W}_{n,3}$ in (\ref{W3}) is a thin set. 
Let $\tilde{\mathcal{N}}_{\mathrm{e}}=\{n\in\{1,\ldots,N\}: 
\sqrt{2v} - 2|\nu_{N}\tilde{a}_{n}| \leq \sqrt{3v/2}\}$. 
From (\ref{latter_limit_last_term_all}), we find 
$|\tilde{\mathcal{N}}_{\mathrm{e}}|\peq o(k)$. 
We have the following upper bound for $n\notin\mathcal{N}_{\mathrm{e}}$:
\begin{align}
&\int_{\mathcal{W}_{n,3}}
|w|^{j}f_{U}^{2}(\nu_{N}\tilde{a}_{n} + w; v_{\tilde{N}})
p_{\mathrm{G}}(w; v_{\tilde{N}})dw
\nonumber \\
&\leq \bar{f}_{U,n,3}^{2}(v_{\tilde{N}})\int_{\mathcal{W}_{n,3}}
|w|^{j}p_{\mathrm{G}}(w; v_{\tilde{N}})dw
\nonumber \\
&< \frac{4\bar{f}_{U,n,3}^{2}(v_{\tilde{N}})|\nu_{N}\tilde{a}_{n}|(2v)^{(j-1)/2}}
{\sqrt{\pi}}(k/N)^{3/4}\sqrt{\log(N/k)},  
\end{align}
with 
\begin{equation}
\bar{f}_{U,n,3}(v_{\tilde{N}})
=\sup_{w\in\mathcal{W}_{n,3}}|f_{U}(\nu_{N}\tilde{a}_{n} + w; v_{\tilde{N}})|,
\end{equation}
which is bounded from above by some $\bar{f}_{U,3}>0$ for all $N$ and 
$n\notin\tilde{\mathcal{N}}_{\mathrm{e}}$, because of the uniform 
Lipschitz-continuity for $f_{U}$ and 
$|\nu_{N}\tilde{a}_{n} + w|<|\nu_{N}\tilde{a}_{n}| + \sqrt{2v}
< (\sqrt{2v} - \sqrt{3v/2})/2 + \sqrt{2v}$ for all 
$w\in\mathcal{W}_{n,3}$ and $n\notin\tilde{\mathcal{N}}_{\mathrm{e}}$. 

For $n\in\mathcal{N}_{\mathrm{e}}$, on the other hand, we use the uniform 
Lipschitz-continuity for $f_{U}$: There are some $N$-independent constants 
$L>0$ and $C>0$ such that $|f_{U}(y; v_{\tilde{N}})|\leq L|y| + C$ holds 
for all $y\in\mathbb{R}$ and $N\in\mathbb{N}$. Thus, we have 
\begin{align}
&\int_{\mathcal{W}_{n,3}}|w|^{j}f_{U}^{2}(\nu_{N}\tilde{a}_{n} + w; v_{\tilde{N}})
p_{\mathrm{G}}(w; v_{\tilde{N}})dw 
\nonumber \\
&<\int_{\mathcal{W}_{n,3}}|w|^{j}\{L|\nu_{N}\tilde{a}_{n} + w| + C\}^{2}
p_{\mathrm{G}}(w; v_{\tilde{N}})dw 
\nonumber \\
&<2(2v)^{j/2}\{2L^{2}(\nu_{N}^{2}\tilde{a}_{n}^{2} + 2v) + C^{2}\}
\int_{\mathcal{W}_{n,3}}p_{\mathrm{G}}(w; v_{\tilde{N}})dw 
\nonumber \\
&< 2(2v)^{j/2}\{2L^{2}(\nu_{N}^{2}\tilde{a}_{n}^{2} + 2v) + C^{2}\}.  
\end{align}
where we have used the inequality $(x+y)^{2}\leq2(x^{2}+y^{2})$ 
for all $x, y\in\mathbb{R}$ repeatedly.
Combining these results, we arrive at 
\begin{align}
&k^{-j/2}\sum_{n\notin\mathcal{S}}
\int_{\mathcal{W}_{n,3}}|w|^{j}f_{X}^{2}(\nu_{N}a_{n} + k^{-1/2}w; v_{\tilde{N}})
p_{\mathrm{G}}(w; v_{\tilde{N}})dw
\nonumber \\
&< \frac{4(2v)^{(j-1)/2}\bar{f}_{U,3}^{2}(k/N)^{3/4}
\sqrt{\log(N/k)}}{\sqrt{\pi}k^{1+j/2}}\sum_{n\notin\mathcal{S}}
|\nu_{N}\tilde{a}_{n}| 
\nonumber \\
&+ \frac{2(2v)^{j/2}}{k^{1+j/2}}\left\{
 2L^{2}\sum_{n\in\tilde{\mathcal{N}}_{\mathrm{e}}}\nu_{N}^{2}\tilde{a}_{n}^{2}
 + (4L^{2}v + C^{2})|\tilde{\mathcal{N}}_{\mathrm{e}}| 
\right\}
\pto 0.
\end{align}
In the derivation of the last convergence, we have used 
(\ref{latter_limit_last_term_all}), $|\tilde{\mathcal{N}}_{\mathrm{e}}|=o(k)$, 
and the following upper bound obtained from the Cauchy-Schwarz inequality and 
$\tilde{a}_{n}=\sqrt{k}a_{n}$: 
\begin{equation} \label{Cauchy_Schwarz}
\sum_{n\notin\mathcal{S}}|\nu_{N}\tilde{a}_{n}|
\leq \left\{
 k(N-k)\nu_{N}^{2}\sum_{n\notin\mathcal{S}}a_{n}^{2}
\right\}^{1/2} \peq o(\sqrt{kN}),
\end{equation}
because of the boundedness in probability of $\|\boldsymbol{a}\|_{2}$. 

\subsection{Last Set $\mathcal{W}_{n,4}$}
We prove that the integral in (\ref{former_tmp}) over $\mathcal{W}_{n,4}$ is 
$o((N-k)^{-1})$ in the sublinear sparsity limit. 
We utilize an intuition in which the probability 
$\mathrm{Pr}(|\tilde{\Omega}|\geq\sqrt{2v})$ should 
tend to zero in the sublinear sparsity limit. This intuition allows us to 
bound $f_{X}^{2}$ roughly. Using the definition of 
$f_{X}(\nu_{N}a_{n} + k^{-1/2}w;v_{\tilde{N}})$ 
in (\ref{posterior_mean_estimator_X}) and 
the trivial upper bound $f_{A}(w;v_{\tilde{N}})\leq 1$ yields 
\begin{align}
&k^{-j/2}\int_{\mathcal{W}_{n,4}}|w|^{j}f_{X}^{2}(\nu_{N}a_{n} + k^{-1/2}w; v_{\tilde{N}})
p_{\mathrm{G}}(w; v_{\tilde{N}})dw
\nonumber \\
&\leq \frac{1}{k^{1+j/2}}\int_{\mathcal{W}_{n,4}}
|w|^{j}f_{U}^{2}(\nu_{N}\tilde{a}_{n} + w; v_{\tilde{N}})
p_{\mathrm{G}}(w; v_{\tilde{N}})dw.
\end{align}
The uniform Lipschitz-continuity assumption for $f_{U}$ implies 
that there are some $N$-independent constants $L>0$ and $C>0$ such that 
$|f_{U}(w; v_{\tilde{N}})|\leq L|w| + C$ holds 
for all $w\in\mathbb{R}$ and $N\in\mathbb{N}$. Thus, we have 
\begin{align}
&\int_{\mathcal{W}_{n,4}}
|w|^{j}f_{U}^{2}(\nu_{N}\tilde{a}_{n} + w; v_{\tilde{N}})
p_{\mathrm{G}}(w; v_{\tilde{N}})dw
\nonumber \\
&\leq 4\int_{\sqrt{2v}}^{\infty}w^{j}
\{2L^{2}(\nu_{N}^{2}\tilde{a}_{n}^{2} + w^{2}) + C^{2}\}
p_{\mathrm{G}}(w; v_{\tilde{N}})dw.
\end{align}
where we have used the inequality $(x+y)^{2}\leq2(x^{2}+y^{2})$ 
for all $x, y\in\mathbb{R}$ repeatedly.

Contributions from $p_{\mathrm{G}}(w; v_{\tilde{N}})$ need to be evaluated 
carefully. Let 
\begin{equation}
J_{n,j} = \int_{\sqrt{2v}}^{\infty}
w^{j}p_{\mathrm{G}}(w; v_{\tilde{N}})dw. 
\end{equation}
We prove the following upper bound: 
\begin{equation} \label{J_bound}
J_{j} < \frac{C_{j}k}{N\sqrt{\log(N/k)}} 
\end{equation}
for some constant $C_{j}>0$, which implies 
\begin{align}
&k^{-j/2}\sum_{n\notin\mathcal{S}}
\int_{\mathcal{W}_{n,4}}|w|^{j}f_{X}^{2}(\nu_{N}a_{n} + k^{-1/2}w; v_{\tilde{N}})
p_{\mathrm{G}}(w; v_{\tilde{N}})dw
\nonumber \\
&<4\sum_{n\notin\mathcal{S}}\frac{2L^{2}C_{j+2} 
+ (2L^{2}\nu_{N}^{2}\tilde{a}_{n}^{2} + C^{2})C_{j}}
{k^{j/2}N\sqrt{\log (N/k)}} \pto 0 
\label{former_limit_general3}
\end{align}
in the sublinear sparsity limit for all $j\geq 0$. 
In the derivation of the last convergence, we have used the boundedness of 
$\|\boldsymbol{a}\|_{2}$ in probability. Thus, (\ref{former_limit_general}) 
holds for all non-negative integers~$j$. 

To complete the proof of (\ref{former_limit_general3}), 
we prove the upper bound~(\ref{J_bound}) by induction. 
For $j=0$ we use the well-known upper bound 
$Q(x)<p_{\mathrm{G}}(x; 1)/x$ on the Q-function for all $x>0$ to obtain  
\begin{equation}
J_{n,0} = Q\left(
 \sqrt{2\log(N/k)}
\right)
< \frac{k}{2N\sqrt{\pi\log(N/k)}}. 
\end{equation}
For $j=1$ we have 
\begin{equation}
J_{1} = [-v_{\tilde{N}}p_{\mathrm{G}}(w; v_{\tilde{N}})]_{\sqrt{2v}}^{\infty}
= \frac{k}{N}\sqrt{\frac{v}{2\pi\log (N/k)}}.
\end{equation}

Suppose that (\ref{J_bound}) is correct for some $j\geq0$. 
Using the integration by parts for the indefinite integral 
$\int wp_{\mathrm{G}}(w; v_{\tilde{N}})dw 
= - v_{\tilde{N}}p_{\mathrm{G}}(w; v_{\tilde{N}}) + {\rm Const.}$ 
yields 
\begin{align}
&J_{j+2} = \left[
- w^{j+1}v_{\tilde{N}}p_{\mathrm{G}}(w; v_{\tilde{N}})
\right]_{\sqrt{2v}}^{\infty}
+ (j + 1)v_{\tilde{N}}J_{j}
\nonumber \\
&= \frac{(2v)^{(j+2)/2}k}{2N\sqrt{\pi\log(N/k)}}
+ \frac{(j + 1)v}{\log(N/k)}J_{j}
\nonumber \\
&< \left\{
 \frac{(2v)^{(j+2)/2}}{2\sqrt{\pi}}
 + \frac{(j + 1)vC_{j}}{\log(N/k)}
\right\}\frac{k}{N\sqrt{\log(N/k)}}, 
\end{align}
where the last inequality follows from the induction 
hypothesis~(\ref{J_bound}). Thus, (\ref{J_bound}) holds for all 
non-negative integers $j$.

\section{Proof of (\ref{latter_limit})}
\label{proof_latter_limit} 
\subsection{Influence of $f_{U}$} 
For notational simplicity, we omit conditioning on $\boldsymbol{a}$. 
In other words, $\boldsymbol{a}$ is regarded as a deterministic vector 
satisfying the boundedness $\|\boldsymbol{a}\|_{2}<\infty$.  

We evaluate the influence of $f_{U}$. Using the definition of 
$\Delta \tilde{U}_{n}(v_{\tilde{N}})$ in (\ref{Delta_U_n_tilde}), 
the definition of $f_{X}$ in (\ref{posterior_mean_estimator_X}), and  
$\Omega_{n}\sim\mathcal{N}(0,v_{k,N})$ yields 
\begin{align}
&\mathbb{E}\left[
 \Delta \tilde{U}_{n}^{2}(v_{\tilde{N}})
\right]
\nonumber \\
&= \mathbb{E}\left[
 \int \left\{
  \tilde{U}_{n} - f_{A}(y; v_{\tilde{N}})f_{U}(y; v_{\tilde{N}})
 \right\}^{2}p_{\mathrm{G}}(y - \tilde{U}_{n}; v_{\tilde{N}})dy
\right], 
\end{align}
with $v_{k,N}=v_{\tilde{N}}/k$.  
Utilizing the following identity:  
\begin{align}
&(\tilde{U}_{n} - f_{A}f_{U})^{2}
= \{(1 - f_{A})\tilde{U}_{n} + f_{A}(\tilde{U}_{n} - f_{U})\}^{2} 
\nonumber \\
&= (1 - f_{A})^{2}\tilde{U}_{n}^{2} 
+ 2(1 - f_{A})\tilde{U}_{n}f_{A}(\tilde{U}_{n} - f_{U}) 
\nonumber \\
&+ f_{A}^{2}(\tilde{U}_{n} - f_{U})^{2}, 
\end{align}
we have  
\begin{align}
&\mathbb{E}\left[
 \Delta \tilde{U}_{n}^{2}(v_{\tilde{N}})
\right]
= \mathbb{E}\left[
 \int (1 - f_{A})^{2}\tilde{U}_{n}^{2}p_{\mathrm{G}}(y - \tilde{U}_{n}; v_{\tilde{N}})dy
\right]
\nonumber \\
&+ 2\mathbb{E}\left[
 \int (1 - f_{A})\tilde{U}_{n}f_{A}(\tilde{U}_{n} - f_{U})
p_{\mathrm{G}}(y  - \tilde{U}_{n}; v_{\tilde{N}})dy
\right]
\nonumber \\
&+ \mathbb{E}\left[
 \int f_{A}^{2}(\tilde{U}_{n} - f_{U})^{2}
p_{\mathrm{G}}(y  - \tilde{U}_{n}; v_{\tilde{N}})dy
\right]. \label{latter_tmp} 
\end{align}

We confirm that the second and last terms in (\ref{latter_tmp}) do not 
contribute to the left-hand side (LHS) of (\ref{latter_limit}) 
in the sublinear sparsity limit. For the last term, 
using the trivial upper bound $f_{A}^{2}\leq 1$ and the change of variables 
$\tilde{y}=(y-\tilde{U}_{n})/\sqrt{v_{\tilde{N}}}$ yields 
\begin{align}
&\frac{1}{k}\sum_{n\in\mathcal{S}}\mathbb{E}\left[
 \int f_{A}^{2}\left\{
  \tilde{U}_{n} - f_{U}(y; v_{\tilde{N}})
 \right\}^{2}p_{\mathrm{G}}(y - \tilde{U}_{n}; v_{\tilde{N}})dy
\right]
\nonumber \\
&\leq \mathbb{E}\left[
 \int F_{0}(\{\tilde{U}_{n}\}, \tilde{y}; v_{\tilde{N}})
 p_{\mathrm{G}}(\tilde{y}; 1)d\tilde{y}
\right], 
\end{align}
with 
\begin{equation}
F_{0}(\{\tilde{U}_{n}\}, \tilde{y}; v_{\tilde{N}}) 
= \frac{1}{k}\sum_{n\in\mathcal{S}}\left\{
 \tilde{U}_{n} - f_{U}(\sqrt{v_{\tilde{N}}}\tilde{y} + \tilde{U}_{n}; v_{\tilde{N}})
\right\}^{2}. 
\end{equation}

To use the reverse Fatou lemma, we evaluate the function 
$F_{0}(\{\tilde{U}_{n}\}, \tilde{y}; v_{\tilde{N}})$. 
The uniform Lipschitz-continuity for $f_{U}$ implies that there are some 
$N$-independent constants $L>0$ and $C>0$ such that 
$|f_{U}(y; v_{\tilde{N}})|\leq L|y| + C$ for all $y\in\mathbb{R}$.  
Using the inequality $(a+b)^{2}\leq 2(a^{2} + b^{2})$ repeatedly, we have 
\begin{align}
&F_{0}(\{\tilde{U}_{n}\}, \tilde{y}; v_{\tilde{N}}) 
\leq\frac{2}{k}\sum_{n\in\mathcal{S}}\left\{
 \tilde{U}_{n}^{2} + f_{U}^{2}(\sqrt{v_{\tilde{N}}}\tilde{y} + \tilde{U}_{n}; 
 v_{\tilde{N}})
\right\}
\nonumber \\
&\leq \frac{2}{k}\sum_{n\in\mathcal{S}}\left\{
 \tilde{U}_{n}^{2} + 2[2L^{2}(v_{\tilde{N}}\tilde{y}^{2} + \tilde{U}_{n}^{2}) 
+ C^{2}]
\right\}
\nonumber \\ 
&\leq \frac{2(1 + 4L^{2})}{k}\sum_{n\in\mathcal{S}}\tilde{U}_{n}^{2}
+ 8L^{2}v_{\tilde{N}}\tilde{y}^{2} + 4C^{2}. 
\end{align}
From $\tilde{U}_{n}=\nu_{N}\tilde{a}_{n} + U_{n}$ and 
(\ref{latter_limit_last_term}) we have 
\begin{equation} \label{U_tilde_2_bound}
\frac{1}{k}\sum_{n\in\mathcal{S}}\tilde{U}_{n}^{2}
\leq\frac{2}{k}\sum_{n\in\mathcal{S}}(\nu_{N}^{2}\tilde{a}_{n}^{2} + U_{n}^{2})
\peq \frac{2}{k}\sum_{n\in\mathcal{S}}U_{n}^{2} + o(1)
\end{equation}
for given $\{U_{n}\}$, so that the integrand 
$F_{0}(\{\tilde{U}_{n}\}, \tilde{y}; v_{\tilde{N}})p_{\mathrm{G}}(\tilde{y}; 1)$ 
is bounded from above by an $N$-independent and integrable function of 
$\{U_{n}\}$ and $\tilde{y}$, because of $\mathbb{E}[U_{n}^{2}]<\infty$.  
Thus, we can use the reverse Fatou lemma to obtain 
\begin{align}
&\limsup_{N\to\infty}\frac{1}{k}\sum_{n\in\mathcal{S}}\mathbb{E}\left[
 \int f_{A}^{2}\left(
  \tilde{U}_{n} - f_{U}
 \right)^{2}p_{\mathrm{G}}(y - \tilde{U}_{n}; v_{\tilde{N}})dy
\right]
\nonumber \\
&\leq \mathbb{E}\left[
 \int\limsup_{N\to\infty}F_{0}(\{\tilde{U}_{n}\}, \tilde{y}; v_{\tilde{N}}) 
 p_{\mathrm{G}}(\tilde{y}; 1)d\tilde{y}
\right]
\nonumber \\
&= \mathbb{E}\left[
 \limsup_{N\to\infty}\frac{1}{k}\sum_{n\in\mathcal{S}}\left\{
  \tilde{U}_{n} - f_{U}(\tilde{U}_{n}; v_{\tilde{N}})
 \right\}^{2}
\right]\to0 \label{reverse_Fatou_bound}
\end{align}
in the sublinear sparsity limit.

In the derivation of the last convergence, 
we have used the consistency for $f_{U}$ in Assumption~\ref{assumption_lemma} 
to evaluate the upper bound~(\ref{reverse_Fatou_bound}) as 
\begin{align}
&\mathbb{E}\left[
 \limsup_{N\to\infty}\frac{1}{k}\sum_{n\in\mathcal{S}}\left\{
  \tilde{U}_{n} - f_{U}(\tilde{U}_{n}; v_{\tilde{N}})
 \right\}^{2}
\right]
\nonumber \\
&\leq \mathbb{E}\left[
 \limsup_{N\to\infty}\frac{o(1)}{k}\sum_{n\in\mathcal{S}}\tilde{U}_{n}^{2}
\right]\to 0, 
\end{align}
where the last convergence follows from the upper bound~(\ref{U_tilde_2_bound}) 
and $\mathbb{E}[U_{n}^{2}]<\infty$. 

We next focus on the second term in (\ref{latter_tmp}). 
Using the upper bound 
$|(1 - f_{A})\tilde{U}_{n}f_{A}(\tilde{U}_{n} - f_{U})|
<|\tilde{U}_{n}(\tilde{U}_{n} - f_{U})|$ and repeating the same proof, we find 
\begin{align}
&\frac{1}{k}\sum_{n\in\mathcal{S}}\left|
 \mathbb{E}\left[
  \int (1 - f_{A})\tilde{U}_{n}f_{A}(\tilde{U}_{n} - f_{U})
 p_{\mathrm{G}}(y  - \tilde{U}_{n}; v_{\tilde{N}})dy
 \right]
\right|
\nonumber \\
&\to 0. 
\end{align}

\subsection{Formulation}
It is sufficient to evaluate contributions from the first term 
in (\ref{latter_tmp}). 
We decompose the interval of integration into three disjoint sets, 
\begin{equation} \label{Y1}
\mathcal{Y}_{1}
= \left\{
 y\in\mathbb{R}: |y| < \sqrt{2\alpha_{N}v}  
\right\}, 
\end{equation}
\begin{equation} \label{Y2}
\mathcal{Y}_{2} 
= \left\{
 y\in\mathbb{R}: |y|\in[\sqrt{2\alpha_{N}v}, \sqrt{2\beta_{N}v}]
\right\}, 
\end{equation}
\begin{equation} \label{Y3}
\mathcal{Y}_{3} 
= \left\{
 y\in\mathbb{R}: |y|>\sqrt{2\beta_{N}v}
\right\}
\end{equation}
for $\epsilon_{N}=\log\log (N/k)/\log (N/k)$, $\alpha_{N}=1-\epsilon_{N}$, and 
$\beta_{N}=1+4\epsilon_{N}$. 
In the region $\mathcal{Y}_{1}$ of integration, 
$f_{A}(y; v_{\tilde{N}})$ is negligibly small. This false negative event 
contributes to the unnormalized square error. In the region $\mathcal{Y}_{3}$ 
of integration, $f_{A}(y; v_{\tilde{N}})$ tends to $1$ with the exception of 
negligibly rare events. The remaining region $\mathcal{Y}_{2}$ 
corresponds to a thin transition region between the two regions. 

In the following appendices, we first prove that the integral  
over $\mathcal{Y}_{1}$ converges to the right-hand side (RHS) of 
(\ref{latter_limit}) if $\nu_{N}=0$ or $\mathrm{Pr}(|U|=\sqrt{2v})=0$ holds. 
Next, without additional assumptions, the integral over $\mathcal{Y}_{2}$ is 
proved to be zero in the sublinear sparsity limit. Finally, we prove that 
the integral over $\mathcal{Y}_{3}$ converges to zero 
if $\nu_{N}=0$ or $\mathrm{Pr}(|U|=\sqrt{2v})=0$ holds. 

\subsection{First Set $\mathcal{Y}_{1}$}
We evaluate the integral for the first term in (\ref{latter_tmp}) over 
$\mathcal{Y}_{1}$. 
Using the change of variables $\tilde{y}=(y-\tilde{U}_{n})/\sqrt{v_{\tilde{N}}}$ 
yields 
\begin{align}
&\frac{1}{k}\sum_{n\in\mathcal{S}}\mathbb{E}\left[
 \int_{\mathcal{Y}_{1}}\{1 - f_{A}(y; v_{\tilde{N}})\}^{2}
 \tilde{U}_{n}^{2}p_{\mathrm{G}}(y-\tilde{U}_{n}; v_{\tilde{N}})dy
\right]
\nonumber \\
&=\mathbb{E}\left[
 \int F_{1}(\{\tilde{U}_{n}\}, \tilde{y}; v_{\tilde{N}})
 p_{\mathrm{G}}(\tilde{y}; 1)d\tilde{y}
\right], \label{latter_first}
\end{align}
with 
\begin{align} 
F_{1}(\{\tilde{U}_{n}\}, \tilde{y}; v_{\tilde{N}}) 
= \frac{1}{k}\sum_{n\in\mathcal{S}}\tilde{U}_{n}^{2}
1(\sqrt{v_{\tilde{N}}}\tilde{y}+\tilde{U}_{n}\in\mathcal{Y}_{1})
\nonumber \\
\cdot\{1 - f_{A}(\sqrt{v_{\tilde{N}}}\tilde{y}+\tilde{U}_{n}; v_{\tilde{N}})\}^{2}. 
\label{function1}
\end{align}

To use the dominated convergence theorem, we first observe that the integrand 
in (\ref{latter_first}) is bounded from above by an $N$-independent and 
integrable function of $\{U_{n}\}$ and $\tilde{y}$, because of 
$F_{1}(\{\tilde{U}_{n}\}, \tilde{y}; v_{\tilde{N}})\leq 
k^{-1}\sum_{n\in\mathcal{S}}\tilde{U}_{n}^{2}$, the upper 
bound~(\ref{U_tilde_2_bound}), and $\mathbb{E}[U_{n}^{2}]<\infty$. 

We next prove that $f_{A}(y; v_{\tilde{N}})$ 
converges uniformly to zero for all 
$|y|<\sqrt{2\alpha_{N}v}$ in the sublinear sparsity limit. 
Using the definition of $f_{A}$ in (\ref{posterior_mean_estimator_A}) yields 
\begin{equation} \label{fA_bound_Y1}
f_{A}(y; v_{\tilde{N}}) 
< \frac{k \mathbb{E}_{U}[e^{-(y-U)^{2}/(2v_{\tilde{N}})}]}
{(N - k)e^{-y^{2}/(2v_{\tilde{N}})}}
< \frac{k(N/k)^{\alpha_{N}}}{N - k}
\end{equation}
for all $|y|<\sqrt{2\alpha_{N}v}$. 
Thus, we find that, for $\alpha_{N}=1 - \log\log (N/k)/\log (N/k)$,   
$f_{A}(y; v_{\tilde{N}})$ converges uniformly to zero for all 
$|y|<\sqrt{2\alpha_{N}v}$ in the sublinear sparsity limit. 

This uniform convergence implies the following almost sure convergence: 
\begin{equation}
F_{1}(\{\tilde{U}_{n}\}, \tilde{y}; v_{\tilde{N}})
- \frac{1}{k}\sum_{n\in\mathcal{S}}\tilde{U}_{n}^{2}
1(|\tilde{U}_{n}|<\sqrt{2v})\ato0
\end{equation} 
for all $|\sqrt{v}_{N}\tilde{y}+\tilde{U}_{n}|<\sqrt{2\alpha_{N}v}$, because of 
$\alpha_{N}=1 - \log\log (N/k)/\log (N/k)\to1$. These observations 
allow us to use the dominated convergence theorem to obtain 
\begin{align}
&\lim_{N\to\infty}\frac{1}{k}\sum_{n\in\mathcal{S}}\mathbb{E}\left[
 \int_{\mathcal{Y}_{1}}\{1 - f_{A}(y; v_{\tilde{N}})\}^{2}
 \tilde{U}_{n}^{2}p_{\mathrm{G}}(y-\tilde{U}_{n}; v_{\tilde{N}})dy
\right]
\nonumber \\
&=\mathbb{E}\left[
 \int\lim_{N\to\infty}F_{1}(\{\tilde{U}_{n}\}, \tilde{y}; 
 v_{\tilde{N}})p_{\mathrm{G}}(\tilde{y}; 1)d\tilde{y} 
\right]
\nonumber \\
&= \lim_{N\to\infty}\frac{1}{k}\sum_{n\in\mathcal{S}}\mathbb{E}\left[
 \tilde{U}_{n}^{2}1(|\tilde{U}_{n}|<\sqrt{2v})
\right] + o(1)
\end{align}
in the sublinear sparsity limit. 

We evaluate the first term under the assumption $\nu_{N}=0$ or 
$\mathrm{Pr}(|U|=\sqrt{2v})=0$. For $\nu_{N}=0$, the first term reduces 
to the RHS of (\ref{latter_limit}), because of $\tilde{U}_{n}
=U_{n}+\nu_{N}\tilde{a}_{n}$.  

For the case $\nu_{N}\neq0$, we have 
\begin{align}
&\frac{1}{k}\sum_{n\in\mathcal{S}}\mathbb{E}\left[
 \tilde{U}_{n}^{2}1(|\tilde{U}_{n}|<\sqrt{2v})
\right]
= \frac{1}{k}\sum_{n\in\mathcal{S}}\mathbb{E}\left[
 U_{n}^{2}1(|\tilde{U}_{n}|<\sqrt{2v})
\right]
\nonumber \\
&+ \frac{2}{k}\sum_{n\in\mathcal{S}}\nu_{N}\tilde{a}_{n}
\mathbb{E}[U_{n}1(|\tilde{U}_{n}|<\sqrt{2v})]
\nonumber \\
&+ \frac{1}{k}\sum_{n\in\mathcal{S}}\nu_{N}^{2}\tilde{a}_{n}^{2}
\mathbb{E}[1(|\tilde{U}_{n}|<\sqrt{2v})]. 
\end{align}
The second and last terms tend to zero in the sublinear sparsity limit, 
because of $1(|\tilde{U}_{n}|<\sqrt{2v})\leq 1$, 
$|\mathbb{E}[U_{n}]|<\infty$, 
(\ref{latter_limit_last_term}), and the Cauchy-Schwarz inequality  
\begin{equation} \label{first_order_summation}
\frac{1}{k}\sum_{n\in\mathcal{S}}|\nu_{N}\tilde{a}_{n}|
\leq\left(
 \frac{1}{k}\sum_{n\in\mathcal{S}}\nu_{N}^{2}\tilde{a}_{n}^{2}
\right)^{1/2}\pto 0. 
\end{equation}

To prove that the first term is asymptotically equivalent to the RHS of 
(\ref{latter_limit}), we evaluate the difference  
\begin{align}
&\left|
 \mathbb{E}\left[
  U_{n}^{2}1(|\tilde{U}_{n}|<\sqrt{2v})
 \right]
 - \mathbb{E}\left[
  U^{2}1(|U|<\sqrt{2v})
 \right] 
\right|
\nonumber \\
&< \mathbb{E}\left[
 U^{2}1(\sqrt{2v} - |\nu_{N}\tilde{a}_{n}| \leq |U| 
 <\sqrt{2v} + |\nu_{N}\tilde{a}_{n}|)
\right]. 
\end{align}
Truncating $|\nu_{N}\tilde{a}_{n}|$ for any $\epsilon>0$ yields 
\begin{align}
&\frac{1}{k}\sum_{n\in\mathcal{S}}\left|
 \mathbb{E}\left[
  U_{n}^{2}1(|\tilde{U}_{n}|<\sqrt{2v})
 \right]
 - \mathbb{E}\left[
  U^{2}1(|U|<\sqrt{2v})
 \right] 
\right|
\nonumber \\
&< \frac{1}{k}\sum_{n\in\mathcal{S}}\mathbb{E}\left[
 U^{2}1\left(
  \left|
   |U| - \sqrt{2v}
  \right| <|\nu_{N}\tilde{a}_{n}|
 \right)1(|\nu_{N}\tilde{a}_{n}|\leq \epsilon)
\right]
\nonumber \\
&+ \frac{\mathbb{E}[U^{2}]}{k}\sum_{n\in\mathcal{S}}
1(|\nu_{N}\tilde{a}_{n}|> \epsilon), 
\end{align}
where the second term follows from 
$1(\sqrt{2v}-  |\nu_{N}\tilde{a}_{n}| \leq |U| 
<\sqrt{2v} + |\nu_{N}\tilde{a}_{n}|)\leq1$. 
Using the upper bound $1(|\nu_{N}\tilde{a}_{n}|>\epsilon)\leq 
\nu_{N}^{2}\tilde{a}_{n}^{2}/\epsilon^{2}$ and (\ref{latter_limit_last_term}), 
we find that the second term tends to zero in the sublinear sparsity limit. 

We prove that for any $\tilde{\epsilon}>0$ there is some $\epsilon>0$ such 
that the first term is bounded from above by $\tilde{\epsilon}$.  
Using the upper bound $|U|\leq\sqrt{2v}+\epsilon$ for 
$|\nu_{N}\tilde{a}_{n}|\leq \epsilon$ yields 
\begin{align}
&\frac{1}{k}\sum_{n\in\mathcal{S}}\mathbb{E}\left[
 U^{2}1\left(
  \left|
   |U| - \sqrt{2v}
  \right| <|\nu_{N}\tilde{a}_{n}|
 \right)
 1(|\nu_{N}\tilde{a}_{n}|\leq \epsilon)
\right]
\nonumber \\
&< (\sqrt{2v} + \epsilon)^{2}\mathrm{Pr}\left(
 \left|
  |U| - \sqrt{2v}
 \right| < \epsilon
\right) < \tilde{\epsilon}. 
\label{first_interval_first_term}
\end{align}
In the derivation of the last upper bound, we have used 
the assumption $\mathrm{Pr}(|U|=\sqrt{2v})=0$, which implies that 
$\mathrm{Pr}(| |U| - \sqrt{2v}| < \epsilon)$ tends to zero as $\epsilon\to0$.   
Thus, we obtain the last upper bound for sufficiently small $\epsilon>0$.

\subsection{Second Set $\mathcal{Y}_{2}$} \label{appen_Y2}
We evaluate the integral for the first term in (\ref{latter_tmp}) 
over $\mathcal{Y}_{2}$. 
Applying the change of variables $y'=-y$ for the negative interval 
$y\in[-\sqrt{2\beta_{N}v}, -\sqrt{2\alpha_{N}v}]$ yields 
\begin{align}
&\mathbb{E}\left[
 \int_{\mathcal{Y}_{2}}
 (1 - f_{A})^{2}\tilde{U}_{n}^{2}p_{\mathrm{G}}(y-\tilde{U}_{n}; v_{\tilde{N}})dy
\right]
\nonumber \\
&= \mathbb{E}\left[
 \int_{\sqrt{2\alpha_{N}v}}^{\sqrt{2\beta_{N}v}}
 \{1 - f_{A}(y; v_{\tilde{N}})\}^{2}\tilde{U}_{n}^{2}
 p_{\mathrm{G}}(y - \tilde{U}_{n}; v_{\tilde{N}})dy
\right]
\nonumber \\
&+ \mathbb{E}\left[
 \int_{\sqrt{2\alpha_{N}v}}^{\sqrt{2\beta_{N}v}}\{1 - f_{A}(-y'; v_{\tilde{N}})\}^{2}
 \tilde{U}_{n}^{2}p_{\mathrm{G}}(y' + \tilde{U}_{n}; v_{\tilde{N}})dy'
\right]. \label{latter_second}
\end{align}
We only evaluate the first term in (\ref{latter_second}) 
since the second term can be evaluated in the same manner. 

We use the upper bound $(1-f_{A})^{2}\leq 1$ and the change of variables 
$y=(2v + 2v_{\tilde{N}}z)^{1/2}\equiv y(z)$ to evaluate the first term 
in (\ref{latter_second}) as 
\begin{align}
&\mathbb{E}\left[
 \int_{\sqrt{2\alpha_{N}v}}^{\sqrt{2\beta_{N}v}}\{1 - f_{A}(y; v_{\tilde{N}})\}^{2}
 \tilde{U}_{n}^{2}p_{\mathrm{G}}(y - \tilde{U}_{n}; v_{\tilde{N}})dy
\right]
\nonumber \\
&\leq\mathbb{E}\left[
 \int_{-\epsilon_{N}\log(N/k)}^{4\epsilon_{N}\log (N/k)} 
 \frac{\sqrt{v_{\tilde{N}}}\tilde{U}_{n}^{2} 
 p_{\mathrm{G}}(y(z) - \tilde{U}_{n}; v_{\tilde{N}})}
 {\sqrt{2\{\log (N/k) + z\}}}dz
\right] \label{latter_second_tmp}
\end{align}
for $\alpha_{N}=1-\epsilon_{N}$ and $\beta_{N}=1+4\epsilon_{N}$. 
From the definition of $y(z)$, for $|z|\leq \epsilon_{N}\log (N/k)$ 
we have 
\begin{align}
&\log\left\{
 \sqrt{2\pi v_{\tilde{N}}}p_{\mathrm{G}}(y(z) - \tilde{U}_{n}; v_{\tilde{N}})
\right\}
\nonumber \\
&= \tilde{U}_{n}\left\{
 \frac{2}{v}\left[
  1 + \frac{z}{\log (N/k)}
 \right]
\right\}^{1/2}\log (N/k) -z 
\nonumber \\
&- \left(
 1 + \frac{\tilde{U}_{n}^{2}}{2v}
\right)\log (N/k)
\leq \frac{\hat{U}_{n}}{\sqrt{2v}}z 
- \frac{\hat{U}_{n}^{2}}{2v}\log (N/k), 
\label{pdf_bound}
\end{align}
with $\hat{U}_{n}=\tilde{U}_{n} - \sqrt{2v}$, 
where the inequality follows from the upper bound $\sqrt{1+x}\leq 1 +x/2$ 
for all $x\in[-1, \infty)$. Applying the upper bound~(\ref{pdf_bound}) 
into (\ref{latter_second_tmp}) yields  
\begin{align}
&\mathbb{E}\left[
 \int_{\sqrt{2\alpha_{N}v}}^{\sqrt{2\beta_{N}v}}\{1 - f_{A}(y; v_{\tilde{N}})\}^{2}
 \tilde{U}_{n}^{2}p_{\mathrm{G}}(y - \tilde{U}_{n}; v_{\tilde{N}})dy
\right]
\nonumber \\
&<\mathbb{E}\left[
 \frac{\tilde{U}_{n}^{2}(N/k)^{-\hat{U}_{n}^{2}/(2v)}}
 {2\sqrt{\pi(1 - \epsilon_{N})\log (N/k)}}
\int_{-\epsilon_{N}\log (N/k)}^{4\epsilon_{N}\log (N/k)}e^{\frac{\hat{U}_{n}z}{\sqrt{2v}}}dz
\right]. \label{latter_second_tmp2}
\end{align}

To upper-bound (\ref{latter_second_tmp2}), 
we use the following upper bound: 
\begin{equation}
\int_{-\epsilon_{N}\log (N/k)}^{4\epsilon_{N}\log (N/k)}e^{\frac{\hat{U}_{n}z}{\sqrt{2v}}}dz
\leq 5\left(
 \frac{N}{k}
\right)^{\frac{4|\hat{U}_{n}|\epsilon_{N}}{\sqrt{2v}}}\epsilon_{N}\log (N/k),
\end{equation}
which is tight for $\hat{U}_{n}=0$. Applying this upper bound to 
(\ref{latter_second_tmp2}), we arrive at
\begin{align}
&\frac{1}{k}\sum_{n\in\mathcal{S}}\mathbb{E}\left[
 \int_{\sqrt{2\alpha_{N}v}}^{\sqrt{2\beta_{N}v}}\{1 - f_{A}(y; v_{\tilde{N}})\}^{2}
 \tilde{U}_{n}^{2}p_{\mathrm{G}}(y - \tilde{U}_{n}; v_{\tilde{N}})dy
\right]
\nonumber \\
&< \frac{5}{2k}\sum_{n\in\mathcal{S}}\mathbb{E}\left[
 \tilde{U}_{n}^{2}\epsilon_{N}\sqrt{\frac{\log (N/k)}{\pi(1 - \epsilon_{N})}}
 \left(
  \frac{N}{k}
 \right)^{\frac{4|\hat{U}_{n}|\epsilon_{N}}{\sqrt{2v}} 
 - \frac{\hat{U}_{n}^{2}}{2v}}
\right]
\nonumber \\
&\leq \frac{5\epsilon_{N}}{2}\left(
 \frac{N}{k}
\right)^{4\epsilon_{N}^{2}}\sqrt{\frac{\log (N/k)}
{\pi(1 - \epsilon_{N})}}\mathbb{E}\left[
 \frac{1}{k}\sum_{n\in\mathcal{S}}\tilde{U}_{n}^{2}
\right] \to 0
\label{latter_second_tmp3}
\end{align}
for $\epsilon_{N}=\log\log (N/k)/\log (N/k)$ in the sublinear sparsity limit. 
In the derivation of the second inequality, 
we have used the following elementary upper bound: 
\begin{equation}
\max_{\hat{u}\in\mathbb{R}}\left(
 \frac{4|\hat{u}|\epsilon_{N}}{\sqrt{2v}} 
 - \frac{\hat{u}^{2}}{2v}
\right)
\leq 4\epsilon_{N}^{2}, 
\end{equation}
where the equality is attained at 
$|\hat{u}| = 2\sqrt{2v}\epsilon_{N}$. The last convergence to zero 
follows from 
the upper bound~(\ref{U_tilde_2_bound}) and $\mathbb{E}[U_{n}^{2}]<\infty$. 

\subsection{Last Set $\mathcal{Y}_{3}$}
\label{proof_last_term}
\subsubsection{Noise Truncation}
We evaluate the integral for the first term in (\ref{latter_tmp}) over 
$\mathcal{Y}_{3}$. 
Using the change of variables $w=y-\tilde{U}_{n}$ yields 
\begin{align}
&\frac{1}{k}\sum_{n\in\mathcal{S}}\mathbb{E}\left[
 \int_{\mathcal{Y}_{3}}\{1 - f_{A}(y; v_{\tilde{N}})\}^{2}\tilde{U}_{n}^{2}
 p_{\mathrm{G}}(y-\tilde{U}_{n}; v_{\tilde{N}})dy
\right]
\nonumber \\
&= \frac{1}{k}\sum_{n\in\mathcal{S}}\mathbb{E}\left[
 \int\{1 - f_{A}(\tilde{U}_{n}+w; v_{\tilde{N}})\}^{2}
 \tilde{U}_{n}^{2}p_{\mathrm{G}}(w; v_{\tilde{N}})dw
\right], \label{latter_last}
\end{align}
where the integral is over 
$\{w\in\mathbb{R}: |\tilde{U}_{n}+w|>\sqrt{2\beta_{N}v}\}$. 

We separate the region of integration into two disjoint sets, 
\begin{equation} \label{W1_u}
\mathcal{W}_{1}(u)=\{w\in\mathbb{R}: w^{2}\leq2\epsilon_{N}v, 
|u+w|>\sqrt{2\beta_{N}v}\},
\end{equation}
\begin{equation}
\mathcal{W}_{2}(u)=\{w\in\mathbb{R}: w^{2}>2\epsilon_{N}v, 
|u+w|>\sqrt{2\beta_{N}v}\}.
\end{equation}
The set $\mathcal{W}_{2}(\tilde{U}_{n})$ contains too rare noise samples to 
contribute to the unnormalized square error. 
Thus, the main challenge is evaluation of the integral in (\ref{latter_last}) 
over the set of the remaining noise samples $\mathcal{W}_{1}(\tilde{U}_{n})$. 

We evaluate the integral in (\ref{latter_last}) over 
$\mathcal{W}_{2}(\tilde{U}_{n})$ in the sublinear sparsity limit. 
Using the trivial upper bound $(1-f_{A})^{2}\leq 1$ and 
$\mathcal{W}_{2}(u)\subset\{w\in\mathbb{R}: w^{2}>2\epsilon_{N}v\}$ yields 
\begin{align}
&\mathbb{E}\left[
 \int_{\mathcal{W}_{2}(\tilde{U}_{n})}\{1 - f_{A}(\tilde{U}_{n}+w; v_{\tilde{N}})\}^{2}
 \tilde{U}_{n}^{2}p_{\mathrm{G}}(w; v_{\tilde{N}})dw
\right]
\nonumber \\
&< \mathbb{E}[\tilde{U}_{n}^{2}]
\int_{w^{2}>2\epsilon_{N} v}p_{\mathrm{G}}(w; v_{\tilde{N}})dw
\nonumber \\
&= 2\mathbb{E}[\tilde{U}_{n}^{2}]Q\left(
 \sqrt{2\epsilon_{N}\log (N/k)}
\right). 
\end{align}
We utilize the upper bound $Q(x)<p_{\mathrm{G}}(x; 1)/x$ 
on the Q-function for all $x>0$ to obtain  
\begin{align}
&\frac{1}{k}\sum_{n\in\mathcal{S}}\mathbb{E}\left[
 \int_{\mathcal{W}_{2}(\tilde{U}_{n})}\{1 - f_{A}(\tilde{U}_{n}+w; v_{\tilde{N}})\}^{2}
 \tilde{U}_{n}^{2}p_{\mathrm{G}}(w; v_{\tilde{N}})dw
\right]
\nonumber \\
&< \frac{2(N/k)^{- \epsilon_{N}}}{\sqrt{4\pi\epsilon_{N}\log (N/k)}}
\mathbb{E}\left[
 \frac{1}{k}\sum_{n\in\mathcal{S}}\tilde{U}_{n}^{2}
\right]
\to 0
\end{align}
for $\epsilon_{N}=\log\log (N/k)/\log (N/k)$ in the sublinear sparsity limit. 
In the proof of the last convergence, we have used the upper 
bound~(\ref{U_tilde_2_bound}) and $\mathbb{E}[U_{n}^{2}]<\infty$.  

To evaluate the integral in (\ref{latter_last}) over 
$\mathcal{W}_{1}(\tilde{U}_{n})$, we use the assumption 
for the cumulative distribution of $U$, which implies that 
the pdf of $U$ is given by 
\begin{equation} \label{pdf_U}
p_{U}(u) = \sum_{i=1}^{\infty}p_{U, i}^{\mathrm{D}}\delta(u-u_{i}) 
+ (1 - p_{U}^{\mathrm{D}})p_{U}^{\mathrm{C}}(u), 
\end{equation}
with 
\begin{equation}
p_{U}^{\mathrm{D}}=\sum_{i=1}^{\infty}p_{U,i}^{\mathrm{D}}\in[0, 1], \quad 
p_{U,i}^{\mathrm{D}}\geq0. 
\end{equation}
In the discrete components of (\ref{pdf_U}), $p_{U,i}^{\mathrm{D}}\in[0, 1]$ 
denotes the probability with which $U$ takes $u_{i}$. 

In the continuous component, the pdf $p_{U}^{\mathrm{C}}(u)$ with a support 
$\mathcal{S}_{U}^{\mathrm{C}}=\{u\in\mathbb{R}: p_{U}^{\mathrm{C}}(u)>0\}$  
is everywhere bounded and almost everywhere continuous. 
Since $\mathcal{S}_{U}^{\mathrm{C}}$ is a Borel set on $\mathbb{R}$, 
without loss of generality, the support $\mathcal{S}_{U}^{\mathrm{C}}$ can be 
decomposed into the union of countable and disjoint closed intervals, 
\begin{equation} \label{support}
\mathcal{S}_{U}^{\mathrm{C}}=\cup_{i\in\mathbb{N}}\mathcal{S}_{U,i}^{\mathrm{C}},
\end{equation}
where disjoint closed intervals 
$\{\mathcal{S}_{U,i}^{\mathrm{C}}\subset\mathbb{R}\}$ 
satisfy $\mathcal{S}_{U,i}^{\mathrm{C}}\cap\mathcal{S}_{U,j}^{\mathrm{C}}
=\emptyset$ for all $i\neq j$. Furthermore, $p_{U}^{\mathrm{C}}$ is continuous 
on each closed interval $\mathcal{S}_{U,i}^{\mathrm{C}}$. 

We focus on the discrete case $p_{U}^{\mathrm{D}}=1$ and the case 
$p_{U}^{\mathrm{D}}<1$, of which the latter is separated into two cases: 
In one case---called the discrete-continuous mixture case---part of 
discrete components $\{u_{i}\}$ are outside the support 
$\mathcal{S}_{U}^{\mathrm{C}}$ of the continuous component. 
In the other case---called the continuous case---there are no discrete 
components outside the support $\mathcal{S}_{U}^{\mathrm{C}}$. We consider 
these three cases separately. In particular, the assumption 
$\nu_{N}=0$ or $\mathrm{Pr}(|U|=\sqrt{2v})=0$ is used in the proof of 
the discrete case. 

\subsubsection{Discrete Case} 
Assume $p_{U}^{\mathrm{D}}=1$. We derive the set of $y\in\mathbb{R}$ such that 
$f_{A}(y; v_{\tilde{N}})$ tends to $1$ in the sublinear sparsity limit. 
Using the definition of $f_{A}$ in (\ref{posterior_mean_estimator_A}) yields 
\begin{equation} \label{posterior_mean_estimator_A_tmp}
f_{A}(y; v_{\tilde{N}}) 
= \left\{
 1 + \frac{(N - k)(N/k)^{-y^{2}/(2v)}}{k\mathbb{E}_{U}[(N/k)^{-(y-U)^{2}/(2v)}]}
\right\}^{-1}. 
\end{equation}
Applying the pdf~(\ref{pdf_U}) of $U$ with $p_{\mathrm{D}}=1$, we have 
\begin{align}
&\mathbb{E}_{U}[(N/k)^{-(y-U)^{2}/(2v)}]
= \sum_{i=1}^{\infty}p_{U,i}^{\mathrm{D}}\left(
 \frac{N}{k}
\right)^{-\frac{(y-u_{i})^{2}}{2v}}
\nonumber \\
&> p_{U,i(y)}^{\mathrm{D}}\left(
 \frac{N}{k}
\right)^{-\frac{(y-u_{i(y)})^{2}}{2v}}
> \left(
 \frac{N}{k}
\right)^{-\epsilon_{N} - \frac{(y-u_{i(y)})^{2}}{2v}} 
\label{E_U_lower_bound_dis}
\end{align}
for $\epsilon_{N} = \log\log (N/k)/\log (N/k)$, with
\begin{equation} \label{i_y}
i(y) = \argmin_{i\in\mathbb{N}: p_{U,i}^{\mathrm{D}}>(N/k)^{-\epsilon_{N}}}(y - u_{i})^{2}. 
\end{equation}
Note that this minimization problem is feasible for the discrete case 
$p_{U}^{\mathrm{D}}=1$. 

We prove $f_{A}(y; v_{\tilde{N}})\to1$ when $y^{2} - (y-u_{i(y)})^{2}
>2(1 + 2\epsilon_{N})v$ holds. Using the lower 
bound~(\ref{E_U_lower_bound_dis}) yields 
\begin{align}
\frac{(N - k)(N/k)^{-y^{2}/(2v)}}{k\mathbb{E}_{U}[(N/k)^{-(y-U)^{2}/(2v)}]}
&< \frac{(N/k)^{1-y^{2}/(2v)}}{(N/k)^{-\epsilon_{N} -(y-u_{i(y)})^{2}/(2v)}}
\nonumber \\
&< (N/k)^{- \epsilon_{N}} 
\end{align}
for $y^{2} - (y-u_{i(y)})^{2}>2(1 + 2\epsilon_{N})v$. 
Applying this upper bound to (\ref{posterior_mean_estimator_A_tmp}), we 
arrive at 
\begin{equation} \label{posterior_mean_estimator_A_lower_bound}
f_{A}(y; v_{\tilde{N}}) > \frac{1}{1 + (N/k)^{-\epsilon_{N}}}
\end{equation}
for all $\{y\in\mathbb{R}: (y-u_{i(y)})^{2} 
< y^{2} - 2(1 + 2\epsilon_{N})v\}$, which tends to $1$ 
for $\epsilon_{N}=\log\log (N/k)/\log (N/k)$ in the sublinear sparsity limit.  

We evaluate the integral in (\ref{latter_last}) over 
$\mathcal{W}_{1}(\tilde{U}_{n})$. 
Utilizing the lower bound on $f_{A}$ in 
(\ref{posterior_mean_estimator_A_lower_bound}) yields 
\begin{align}
&\limsup_{N\to\infty}\frac{1}{k}\sum_{n\in\mathcal{S}}\mathbb{E}\left[
 \int_{\mathcal{W}_{1}(\tilde{U}_{n})}\left(
  1 - f_{A}
 \right)^{2}\tilde{U}_{n}^{2}p_{\mathrm{G}}(w; v_{\tilde{N}})dw
\right]
\nonumber \\
&= \limsup_{N\to\infty}\frac{1}{k}\sum_{n\in\mathcal{S}}\mathbb{E}\left[
 \int_{\mathcal{W}_{1}(\tilde{U}_{n})}dw \left(
  1 - f_{A}
 \right)^{2}\tilde{U}_{n}^{2}p_{\mathrm{G}}(w; v_{\tilde{N}})
\right. \nonumber \\
&\cdot 1\left(
 (\tilde{U}_{n}+w-u_{i(\tilde{U}_{n}+w)})^{2} \geq (\tilde{U}_{n}+w)^{2} 
 - 2(1 + 2\epsilon_{N})v
\right)
\bigg] \label{latter_last_first}
\end{align}
in the sublinear sparsity limit, 
with $f_{A}=f_{A}(\tilde{U}_{n}+w; v_{\tilde{N}})$. 
Applying the upper bound $(1 - f_{A})^{2}\leq 1$ and 
the pdf of $U$ in (\ref{pdf_U}) with $p_{U}^{\mathrm{D}}=1$, we have
\begin{align}
&\limsup_{N\to\infty}\frac{1}{k}\sum_{n\in\mathcal{S}}\mathbb{E}\left[
 \int_{\mathcal{W}_{1}(\tilde{U}_{n})}\left(
  1 - f_{A}
 \right)^{2}\tilde{U}_{n}^{2}p_{\mathrm{G}}(w; v_{\tilde{N}})dw
\right]
\nonumber \\
&\leq \sum_{j=1}^{\infty}\limsup_{N\to\infty}\frac{p_{U,j}^{\mathrm{D}}}{k}
\sum_{n\in\mathcal{S}}\tilde{u}_{n,j}^{2}
\int_{\mathcal{W}_{1}(\tilde{u}_{n,j})} dwp_{\mathrm{G}}(w; v_{\tilde{N}})
\nonumber \\
&\cdot 1\left(
 (\tilde{u}_{n,j}+w-u_{i(\tilde{u}_{n,j}+w)})^{2} 
\geq (\tilde{u}_{n,j} + w)^{2} - (2 + 4\epsilon_{N})v
\right), \label{latter_last_first_tmp}
\end{align}
with $\tilde{u}_{n,j}=u_{j} + \nu_{N}\tilde{a}_{n}$, 
where we have used the reverse Fatou lemma since the terms in the 
summation are bounded from above by summable 
$\{p_{U,j}^{\mathrm{D}}k^{-1}\sum_{n\in\mathcal{S}}\tilde{u}_{n,j}^{2}\}_{j=1}^{\infty}$. 

We prove that the upper bound~(\ref{latter_last_first_tmp}) is equal to 
zero if $\nu_{N}=0$ or $\mathrm{Pr}(|U|=\sqrt{2v})=0$ holds. 
Let $\mathcal{I}=\{i\in\mathbb{N}: 
p_{U,i}^{\mathrm{D}}>(N/k)^{-\epsilon_{N}}\}$. 
Since $p_{U,j}^{\mathrm{D}}$ tends to zero in the sublinear sparsity limit 
for all $j\notin\mathcal{I}$, (\ref{latter_last_first_tmp}) reduces to 
\begin{align}
& \limsup_{N\to\infty}\frac{1}{k}\sum_{n\in\mathcal{S}}\mathbb{E}\left[
 \int_{\mathcal{W}_{1}(\tilde{U}_{n})}\left(
  1 - f_{A}
 \right)^{2}\tilde{U}_{n}^{2}p_{\mathrm{G}}(w; v_{\tilde{N}})dw
\right]
\nonumber \\
&\leq \sum_{j\in\mathcal{I}}\limsup_{N\to\infty}\frac{p_{U,j}^{\mathrm{D}}}{k}
\sum_{n\in\mathcal{S}}\tilde{u}_{n,j}^{2}
\int p_{\mathrm{G}}(w; v_{\tilde{N}})dw, \label{latter_last_first_tmp2}
\end{align} 
where the integral is over 
$\mathcal{W}_{1}(\tilde{u}_{n,j})\cap\mathcal{W}_{\mathrm{d}}(\tilde{u}_{n,j})$, 
with 
\begin{equation} \label{Wd}
\mathcal{W}_{\mathrm{d}}(u) = \left\{
 w\in\mathbb{R}: \frac{(u + w)^{2}}{2} \leq \nu_{N}^{2}\tilde{a}_{n}^{2} + w^{2} 
 + (1 + 2\epsilon_{N})v 
\right\},
\end{equation}
where we have used the following upper bound obtained from the definition 
of $i(y)$ in (\ref{i_y}): 
\begin{align}
&\left(
 \tilde{u}_{n,j}+w-u_{i(u_{j}+w)}
\right)^{2} \leq (\tilde{u}_{n,j} + w - u_{j})^{2} 
\nonumber \\
&= (\nu_{N}\tilde{a}_{n} + w)^{2}  
\leq 2(\nu_{N}^{2}\tilde{a}_{n}^{2} + w^{2})
\end{align}
for all $j\in\mathcal{I}$. 

Define a subset of $\mathcal{W}_{1}(u)$ in (\ref{W1_u}) as 
\begin{align}
\partial\mathcal{W}_{1}(u)=\{w\in\mathbb{R}: w^{2}
\in(2\epsilon_{N}v-\nu_{N}^{2}\tilde{a}_{n}^{2}, 2\epsilon_{N}v], 
\nonumber \\
|u+w|>\sqrt{2\beta_{N}v}\}. 
\end{align}
The two sets $\mathcal{W}_{1}(\tilde{u}_{n,j})\setminus
\partial\mathcal{W}_{1}(\tilde{u}_{n,j})$ 
and $\mathcal{W}_{\mathrm{d}}(\tilde{u}_{n,j})$ in (\ref{Wd}) have  
no intersection since $w^{2}\leq2\epsilon_{N}v-\nu_{N}^{2}\tilde{a}_{n}^{2}$ 
in the set $\mathcal{W}_{1}(\tilde{u}_{n,j})\setminus
\partial\mathcal{W}_{1}(\tilde{u}_{n,j})$ yields 
\begin{equation}
\nu_{N}^{2}\tilde{a}_{n}^{2} + w^{2} + (1 + 2\epsilon_{N})v 
\leq (1 + 4\epsilon_{N})v = \beta_{N}v. 
\end{equation}
This observation implies that (\ref{latter_last_first_tmp2}) reduces to 
\begin{align}
& \limsup_{N\to\infty}\frac{1}{k}\sum_{n\in\mathcal{S}}\mathbb{E}\left[
 \int_{\mathcal{W}_{1}(\tilde{U}_{n})}\left(
  1 - f_{A}
 \right)^{2}\tilde{U}_{n}^{2}p_{\mathrm{G}}(w; v_{\tilde{N}})dw
\right]
\nonumber \\
&\leq \sum_{j\in\mathcal{I}}\limsup_{N\to\infty}\frac{p_{U,j}^{\mathrm{D}}}{k}
\sum_{n\in\mathcal{S}}\tilde{u}_{n,j}^{2}
\int p_{\mathrm{G}}(w; v_{\tilde{N}})dw, \label{latter_last_first_tmp3}
\end{align} 
where the integral is over $\partial\mathcal{W}_{1}(\tilde{u}_{n,j})
\cap\mathcal{W}_{\mathrm{d}}(\tilde{u}_{n,j})$, given by
\begin{align}
&\partial\mathcal{W}_{1}(u)\cap\mathcal{W}_{\mathrm{d}}(u)
\nonumber \\
&= \{
 w\in\mathbb{R}: w^{2}\in(2\epsilon_{N}v - \nu_{N}^{2}\tilde{a}_{n}^{2}, 
 2\epsilon_{N}v], 
\nonumber \\
& 0 < (u + w)^{2} - 2\beta_{N}v 
 < 2\nu_{N}^{2}\tilde{a}_{n}^{2} - (4\epsilon_{N}v - 2w^{2}) 
\}. \label{set_for_discrete}
\end{align}  

To prove that the upper bound~(\ref{latter_last_first_tmp3}) is equal to 
zero, it is sufficient to confirm that the set~(\ref{set_for_discrete}) is 
empty for sufficiently large $N$ when $\nu_{N}=0$ or 
$\mathrm{Pr}(|U|=\sqrt{2v})=0$ holds. The set is obviously 
empty for $\nu_{N}=0$. Thus, we focus on the case $\nu_{N}\neq0$. 
From (\ref{latter_limit_last_term}), we find 
$k^{-1}|\{n\in\mathcal{S}: \nu_{N}^{2}\tilde{a}_{n}^{2}>\epsilon\}|
\pto0$ for all $\epsilon>0$. Thus, without loss of generality, we can assume 
$\nu_{N}^{2}\tilde{a}_{n}^{2}=o(1)$ 
as long as (\ref{latter_last_first_tmp3}) is considered. 
   
We only consider the case $w\geq0$ in (\ref{set_for_discrete}) 
because the case $w<0$ can be treated in the same manner. Combining 
$w\leq\sqrt{2\epsilon_{N}v}$ and the second inequality 
$2\beta_{N}v < (\tilde{u}_{n,j} + w)^{2}$ in 
(\ref{set_for_discrete}) yields
\begin{equation}
\sqrt{2\beta_{N}v} - \tilde{u}_{n,j} < \sqrt{2\epsilon_{N}v}, 
\end{equation}
which implies $\tilde{u}_{n,j}>\sqrt{2\epsilon_{N}v}$ for sufficiently 
large $N$. By completing the square for the second inequality in 
(\ref{set_for_discrete}) with respect to $w$, we find that 
the second inequality in (\ref{set_for_discrete}) is 
satisfied if and only if it holds at $w=\sqrt{2\epsilon_{N}v}$, i.e.\  
\begin{equation}
(\tilde{u}_{n,j} + \sqrt{2\epsilon_{N}v})^{2} 
< 2(\beta_{N}v + \nu_{N}^{2}\tilde{a}_{n}^{2}), 
\end{equation}
with $\tilde{u}_{n,j}=u_{j} + \nu_{N}\tilde{a}_{n}$. 
The obtained two inequalities can be satisfied for sufficiently large $N$ 
only for $u_{j}=\sqrt{2v}$. Since $\mathrm{Pr}(|U|=\sqrt{2v})=0$ is assumed, 
the upper bound~(\ref{latter_last_first_tmp3}) is zero 
for sufficiently large $N$. Thus, the integral in (\ref{latter_last}) over 
$\mathcal{W}_{1}(\tilde{U}_{n})$  
tends to zero in the sublinear sparsity limit for $p_{U}^{\mathrm{D}}=1$ 
if $\nu_{N}=0$ or $\mathrm{Pr}(|U|=\sqrt{2v})=0$ holds. 

\subsubsection{Discrete-Continuous Mixture Case}
Assume $p_{U}^{\mathrm{D}}<1$. We prove $f_{A}(y; v_{\tilde{N}})\to1$ almost 
everywhere for $\{y\in\mathcal{S}_{U}^{\mathrm{C}}: |y|>\sqrt{2\beta_{N}v}\}$.  
Using definition of $f_{A}(y; v_{\tilde{N}})$ in 
(\ref{posterior_mean_estimator_A}) yields 
\begin{equation}
f_{A}(y; v_{\tilde{N}}) 
= \left(
 1 + \frac{k^{-1}(N - k)p_{\mathrm{G}}(y; v_{\tilde{N}})}
 {\mathbb{E}_{U}[p_{\mathrm{G}}(y - U; v_{\tilde{N}})]}
\right)^{-1}. 
\end{equation}
To lower-bound $f_{A}(y; v_{\tilde{N}})$, for the numerator we have 
\begin{equation}
k^{-1}(N - k)p_{\mathrm{G}}(y; v_{\tilde{N}})
< \frac{1}{\sqrt{2\pi v_{\tilde{N}}}}\left(
 \frac{N}{k}
\right)^{1-\beta_{N}} 
\end{equation}
for all $|y|>\sqrt{2\beta_{N}v}$, which tends to zero in the sublinear 
sparsity limit for $\beta_{N}=1 + 4\log\log (N/k)/\log (N/k)$. 
On the other hand, using the pdf~(\ref{pdf_U}) for the denominator yields 
\begin{equation}
\mathbb{E}_{U}[p_{\mathrm{G}}(y-U; v_{\tilde{N}})]
\geq (1 - p_{U}^{\mathrm{D}})
\int p_{\mathrm{G}}(y-u; v_{\tilde{N}})p_{U}^{\mathrm{C}}(u)du. 
\end{equation} 
We use the change of variables $\tilde{u}=(u-y)/\sqrt{v_{\tilde{N}}}$ and 
Fatou's lemma to obtain 
\begin{align}
&\liminf_{N\to\infty}\mathbb{E}_{U}[p_{\mathrm{G}}(y-U; v_{\tilde{N}})]
\nonumber \\
&\geq (1 - p_{U}^{\mathrm{D}})\int p_{\mathrm{G}}(\tilde{u}; 1)
\liminf_{N\to\infty}p_{U}^{\mathrm{C}}(\sqrt{v_{\tilde{N}}}\tilde{u} + y)d\tilde{u} 
\nonumber \\
&=(1 - p_{U}^{\mathrm{D}})p_{U}^{\mathrm{C}}(y)~\hbox{almost everywhere}
\label{E_U_lower_bound_con} 
\end{align}
in the sublinear sparsity limit, where the last equality follows from the 
almost everywhere continuity of $p_{U}^{\mathrm{C}}(u)$. In particular, 
the lower bound~(\ref{E_U_lower_bound_con}) is strictly positive 
for all $y\in\mathcal{S}_{U}^{\mathrm{C}}$.
Combining these results, we find that $f_{A}(y; v_{\tilde{N}})$ converges almost 
everywhere to $1$ for 
$\{y\in\mathcal{S}_{U}^{\mathrm{C}}: |y|>\sqrt{2\beta_{N}v}\}$ 
in the sublinear sparsity limit. 

We evaluate the integral in (\ref{latter_last}) over 
$\mathcal{W}_{1}(\tilde{U}_{n})$. Since 
we have proved $f_{A}(y; v_{\tilde{N}})\to1$ almost everywhere for the continuous 
component, we obtain 
\begin{align}
&\lim_{N\to\infty}\frac{1}{k}\sum_{n\in\mathcal{S}}\mathbb{E}\left[
 \int_{\mathcal{W}_{1}(\tilde{U}_{n})}(1 - f_{A})^{2}\tilde{U}_{n}^{2}
 p_{\mathrm{G}}(w; v_{\tilde{N}})dw
\right]
\nonumber \\
&=\lim_{N\to\infty}\frac{1}{k}\sum_{n\in\mathcal{S}}\mathbb{E}\left[
 \int_{\tilde{\mathcal{W}}_{1}(\tilde{U}_{n})}
 (1 - f_{A})^{2}\tilde{U}_{n}^{2}p_{\mathrm{G}}(w; v_{\tilde{N}})dw
\right] \label{latter_last_first_con}
\end{align}
in the sublinear sparsity limit, 
with $f_{A}=f_{A}(\tilde{U}_{n}+w; v_{\tilde{N}})$ and 
\begin{equation} \label{W_tilde}
\tilde{\mathcal{W}}_{1}(u) = \{w\in\mathcal{W}_{1}(u): 
u+w\notin\mathcal{S}_{U}^{\mathrm{C}}\}, 
\end{equation}
where $\mathcal{W}_{1}(u)$ is defined in (\ref{W1_u}). 

When there is some discrete component $u_{i}$ in (\ref{pdf_U}) such that 
$\tilde{\mathcal{W}}_{1}(u_{i}+\nu_{N}\tilde{a}_{n})\neq\emptyset$ holds, 
we repeat the proof for the discrete case $p_{U}^{\mathrm{D}}=1$ to find that 
the integral in (\ref{latter_last}) over $\mathcal{W}_{1}(\tilde{U}_{n})$ 
tends to zero in the sublinear sparsity limit. 

\subsubsection{Continuous Case}
Assume $p_{U}^{\mathrm{D}}=0$ or 
$\tilde{\mathcal{W}}_{1}(\tilde{u}_{n,i})=\emptyset$ for all 
$i\in\mathbb{N}$ and $n\in\mathcal{S}$. 
We use the pdf of $U$ in (\ref{pdf_U}) and the upper 
bound $(1 - f_{A})^{2}\leq 1$ to evaluate (\ref{latter_last_first_con}) as   
\begin{align}
&\frac{1}{k}\sum_{n\in\mathcal{S}}\mathbb{E}\left[
 \int_{\tilde{\mathcal{W}}_{1}(\tilde{U}_{n})}
 \{1 - f_{A}(\tilde{U}_{n}+w; v_{\tilde{N}})\}^{2}\tilde{U}_{n}^{2}
 p_{\mathrm{G}}(w; v_{\tilde{N}})dw
\right]
\nonumber \\
&\leq\frac{1 - p_{U}^{\mathrm{D}}}{k}\sum_{n\in\mathcal{S}}
\int_{\mathcal{S}_{U}^{\mathrm{C}}}\tilde{u}_{n}^{2}p_{U}^{\mathrm{C}}(u)du
\int_{\tilde{\mathcal{W}}_{1}(\tilde{u}_{n})}p_{\mathrm{G}}(w; v_{\tilde{N}})dw
\nonumber \\
&< \frac{2(1 - p_{U}^{\mathrm{D}})}{k}\sum_{n\in\mathcal{S}}\sum_{i\in\mathbb{N}}
\int_{\mathcal{S}_{U,i}^{\mathrm{C}}}u^{2}p_{U}^{\mathrm{C}}(u)du
\int_{\tilde{\mathcal{W}}_{1}(\tilde{u}_{n})}p_{\mathrm{G}}(w; v_{\tilde{N}})dw
\nonumber \\
&+ \frac{2(1 - p_{U}^{\mathrm{D}})}{k}
\sum_{n\in\mathcal{S}}\nu_{N}^{2}\tilde{a}_{n}^{2}, \label{continuous_case}
\end{align}
with $\tilde{u}_{n}=u + \nu_{N}\tilde{a}_{n}$, 
where the last inequality follows from $\tilde{u}_{n}^{2}
\leq2(u^{2}+\nu_{N}^{2}\tilde{a}_{n}^{2})$ and 
the support representation~(\ref{support}), as well as the normalization of 
$p_{U}^{\mathrm{C}}$ and $p_{\mathrm{G}}$. From (\ref{latter_limit_last_term}),  
the second term on the upper bound~(\ref{continuous_case}) converges 
in probability to zero in the sublinear sparsity limit. 

To evaluate the first term, define a subset of the interior of 
$\mathcal{S}_{U,i}^{\mathrm{C}}$ as 
\begin{align}
\mathcal{S}_{\tilde{U}_{n},i}^{\mathrm{C}} 
= \{u\in\mathcal{S}_{U,i}^{\mathrm{C}}: 
u + \nu_{N}\tilde{a}_{n} + w\in\mathcal{S}_{U}^{\mathrm{C}} 
\nonumber \\
\hbox{for all $w^{2}\leq2\epsilon_{N}v$}\}.
\end{align}
We find $\tilde{\mathcal{W}}_{1}(\tilde{u}_{n})=\emptyset$ in (\ref{W_tilde}) 
for all $u\in\mathcal{S}_{\tilde{U}_{n},i}^{\mathrm{C}}$ to obtain  
\begin{align}
&\frac{1}{k}\sum_{n\in\mathcal{S}}\sum_{i\in\mathbb{N}}
\int_{\mathcal{S}_{U,i}^{\mathrm{C}}}u^{2}p_{U}^{\mathrm{C}}(u)du
\int_{\tilde{\mathcal{W}}_{1}(\tilde{u}_{n})}p_{\mathrm{G}}(w; v_{\tilde{N}})dw
\nonumber \\
&= \frac{1}{k}\sum_{n\in\mathcal{S}}\sum_{i\in\mathbb{N}}
\int_{\partial\mathcal{S}_{\tilde{U}_{n},i}^{\mathrm{C}}}u^{2}p_{U}^{\mathrm{C}}(u)du
\int_{\tilde{\mathcal{W}}_{1}(\tilde{u}_{n})}p_{\mathrm{G}}(w; v_{\tilde{N}})dw
\nonumber \\
&< \frac{1}{k}\sum_{n\in\mathcal{S}}\sum_{i\in\mathbb{N}}
\int_{\partial\mathcal{S}_{\tilde{U}_{n},i}^{\mathrm{C}}}
u^{2}p_{U}^{\mathrm{C}}(u)du, 
\label{latter_last_first_con_tmp}
\end{align}
with $\partial\mathcal{S}_{\tilde{U}_{n},i}^{\mathrm{C}}
=\mathcal{S}_{U,i}^{\mathrm{C}}\setminus
\mathcal{S}_{\tilde{U}_{n},i}^{\mathrm{C}}$. 
To use the reverse Fatou lemma, we confirm the following upper bound: 
\begin{equation}
\frac{1}{k}\sum_{n\in\mathcal{S}}
\int_{\partial\mathcal{S}_{\tilde{U}_{n},i}^{\mathrm{C}}}
u^{2}p_{U}^{\mathrm{C}}(u)du
< \int_{\mathcal{S}_{U,i}^{\mathrm{C}}}
u^{2}p_{U}^{\mathrm{C}}(u)du 
\equiv b_{i}. 
\end{equation}
From $\mathbb{E}[U^{2}]<\infty$, the boundedness 
$\sum_{i\in\mathbb{N}}b_{i}<\infty$ holds. Thus, 
we can use the reverse Fatou lemma to obtain 
\begin{align}
&\limsup_{N\to\infty}\frac{1}{k}\sum_{n\in\mathcal{S}}\sum_{i\in\mathbb{N}}
\int_{\partial\mathcal{S}_{\tilde{U}_{n},i}^{\mathrm{C}}}u^{2}p_{U}^{\mathrm{C}}(u)du
\nonumber \\
&\leq \sum_{i\in\mathbb{N}}\limsup_{N\to\infty}\frac{1}{k}\sum_{n\in\mathcal{S}}
\int_{\partial\mathcal{S}_{\tilde{U}_{n},i}^{\mathrm{C}}}
u^{2}p_{U}^{\mathrm{C}}(u)du
\nonumber \\
&< \sum_{i\in\mathbb{N}}A_{i}\limsup_{N\to\infty}\frac{1}{k}\sum_{n\in\mathcal{S}}
|\partial\mathcal{S}_{\tilde{U}_{n},i}^{\mathrm{C}}|,
\end{align}
with $A_{i}=\sup_{u\in\mathcal{S}_{U,i}^{\mathrm{C}}}u^{2}p_{U}^{\mathrm{C}}(u)$, 
which is bounded from the continuity of $p_{U}^{\mathrm{C}}$ on each 
closed interval $\mathcal{S}_{U,i}^{\mathrm{C}}$.  

To prove the convergence to zero, we define 
$\partial\mathcal{S}_{\tilde{U}_{n},i,i}^{\mathrm{C}}
=\mathcal{S}_{U,i}^{\mathrm{C}}\setminus
\mathcal{S}_{\tilde{U}_{n},i,i}^{\mathrm{C}}$, with 
\begin{equation}
\mathcal{S}_{\tilde{U}_{n},i,i}^{\mathrm{C}}
=\{u\in\mathcal{S}_{\tilde{U}_{n},i}^{\mathrm{C}}: 
u + \nu_{N}\tilde{a}_{n} + w\in\mathcal{S}_{U,i}^{\mathrm{C}}\}. 
\end{equation} 
By definition we have $\mathcal{S}_{\tilde{U}_{n},i,i}^{\mathrm{C}}\subset
\mathcal{S}_{\tilde{U}_{n},i}^{\mathrm{C}}$, so that 
$|\partial\mathcal{S}_{\tilde{U}_{n},i}^{\mathrm{C}}|\leq
|\partial\mathcal{S}_{\tilde{U}_{n},i,i}^{\mathrm{C}}|$ holds. 
Using the upper bound $|\nu_{N}\tilde{a}_{n} + w|^{2}\leq 
2(\nu_{N}^{2}\tilde{a}_{n}^{2} + w^{2})\leq 4\epsilon_{N}v 
+ 2\nu_{N}^{2}\tilde{a}_{n}^{2}$ for all 
$w^{2}\leq2\epsilon_{N}v$ in $\mathcal{S}_{\tilde{U}_{n},i}^{\mathrm{C}}$, 
for the closed interval $\mathcal{S}_{U,i}^{\mathrm{C}}$ we find 
$(\inf\mathcal{S}_{U,i}^{\mathrm{C}} 
+ (4\epsilon_{N}v + 2\nu_{N}^{2}\tilde{a}_{n}^{2})^{1/2}, 
\sup\mathcal{S}_{U,i}^{\mathrm{C}} 
- (4\epsilon_{N}v + 2\nu_{N}^{2}\tilde{a}_{n}^{2})^{1/2})
\subset\mathcal{S}_{\tilde{U}_{n},i,i}^{\mathrm{C}}$, which implies 
$|\partial\mathcal{S}_{\tilde{U}_{n},i,i}^{\mathrm{C}}|\leq 
2\sqrt{4\epsilon_{N}v + 2\nu_{N}^{2}\tilde{a}_{n}^{2}}$. Thus, we use the 
Cauchy-Schwarz inequality to arrive at   
\begin{align}
&\limsup_{N\to\infty}\frac{1}{k}\sum_{n\in\mathcal{S}}
|\partial\mathcal{S}_{\tilde{U}_{n},i}^{\mathrm{C}}|
\leq \limsup_{N\to\infty}\left(
 \frac{1}{k}\sum_{n\in\mathcal{S}}
 |\partial\mathcal{S}_{\tilde{U}_{n},i}^{\mathrm{C}}|^{2}
\right)^{1/2}
\nonumber \\
&\leq 2\limsup_{N\to\infty}\left(
 4\epsilon_{N}v + \frac{2}{k}\sum_{n\in\mathcal{S}}\nu_{N}^{2}\tilde{a}_{n}^{2}
\right)^{1/2}\pto 0,
\end{align}
where the last convergence follows from the definition 
$\epsilon_{N}=\log\log (N/k)/\log(N/k)$ and (\ref{latter_limit_last_term}). 
Combining these results, we find that the integral in (\ref{latter_last}) 
over $\mathcal{W}_{1}(\tilde{U}_{n})$ 
tends to zero in the sublinear sparsity limit.

\section{Proof of (\ref{difference})}
\label{proof_lemma_difference}
\subsection{Formulation}
For notational simplicity, we omit conditioning on $\boldsymbol{a}$. 
In other words, $\boldsymbol{a}$ is regarded as a deterministic vector 
satisfying the boundedness $\|\boldsymbol{a}\|_{2}<\infty$.  
We represent (\ref{difference}) with 
$\tilde{\Omega}_{n}\sim\mathcal{N}(0, v_{\tilde{N}})$ as 
\begin{align}
&\sum_{n\in\mathcal{S}}\mathbb{E}\left[
 \Delta f_{X}^{2}(\tilde{a}_{n}, \tilde{U}_{n} + \tilde{\Omega}_{n}; v_{\tilde{N}})
\right]
\nonumber \\
&= \sum_{n\in\mathcal{S}}\mathbb{E}\left[
 \int \Delta f_{X}^{2}(\tilde{a}_{n}, y; v_{\tilde{N}})
 p_{\mathrm{G}}(y - \tilde{U}_{n}; v_{\tilde{N}})dy
\right]. \label{difference_tmp}
\end{align}

We separate the interval of integration into three disjoint sets,  
\begin{equation} \label{Y1_tilde}
\tilde{\mathcal{Y}}_{n,1}
= \left\{
 y\in\mathbb{R}: |y| < \sqrt{2\alpha_{N}v} - |\nu_{N}\tilde{a}_{n}| 
\right\}, 
\end{equation}
\begin{equation} \label{Y2_tilde}
\tilde{\mathcal{Y}}_{n,2} 
= \left\{
 y\in\mathbb{R}: \left|
  |y| - \sqrt{2\alpha_{N}v}
 \right|\leq|\nu_{N}\tilde{a}_{n}|
\right\}, 
\end{equation}
\begin{equation} \label{Y3_tilde}
\tilde{\mathcal{Y}}_{n,3} 
= \left\{
 y\in\mathbb{R}: |y|> \sqrt{2\alpha_{N}v} + |\nu_{N}\tilde{a}_{n}|
\right\}.  
\end{equation}
The set $\tilde{\mathcal{Y}}_{n,1}$ is a subset of $\mathcal{Y}_{1}$ in 
(\ref{Y1}) while $\tilde{\mathcal{Y}}_{n,3}$ is a subset of 
$\mathcal{Y}_{2}\cup\mathcal{Y}_{3}$ in (\ref{Y2}) and (\ref{Y3}). The set 
$\tilde{\mathcal{Y}}_{n,2}$ is a thin set between $\tilde{\mathcal{Y}}_{n,1}$ 
and $\tilde{\mathcal{Y}}_{n,3}$. The assumption 
$\mathrm{Pr}(|U|=\sqrt{2v})=0$ is used to evaluate the integral in 
(\ref{difference_tmp}) over $\tilde{\mathcal{Y}}_{n,2}$ and 
$\tilde{\mathcal{Y}}_{n,3}$ while it is not used for 
$\tilde{\mathcal{Y}}_{n,1}$.

\subsection{First Set $\tilde{\mathcal{Y}}_{n,1}$}
We prove the following uniform Lipschitz-continuity of 
$f_{X}(\tilde{y}; v_{\tilde{N}})$ for all $\sqrt{k}\tilde{y}\in\mathcal{Y}_{1}$ 
in (\ref{Y1}) with respect to $N$ and $k$: There is some constant $L>0$ 
such that the following inequality holds 
\begin{equation} \label{uniform_Lipschitz_Y1}
|f_{X}(\tilde{x}; v_{\tilde{N}}) - f_{X}(\tilde{y}; v_{\tilde{N}})| 
< L|\tilde{x} - \tilde{y}| 
\end{equation}  
for all $\sqrt{k}\tilde{x}\in\mathcal{Y}_{1}$, 
$\sqrt{k}\tilde{y}\in\mathcal{Y}_{1}$, $N$, and $k$ with $k/N<1$. 
Before proving the uniform Lipschitz-continuity, 
we use it to evaluate the integral in (\ref{difference_tmp}) over 
$\tilde{\mathcal{Y}}_{n,1}$. 
Since $y-\nu_{N}\tilde{a}_{n}$ is in $\mathcal{Y}_{1}$ for 
all $y\in\tilde{\mathcal{Y}}_{n,1}$, we can use the uniform 
Lipschitz-continuity~(\ref{uniform_Lipschitz_Y1}) for 
$\Delta f_{X}(\tilde{a}_{n}, y; v_{\tilde{N}})$ 
in (\ref{Delta_f}) to obtain
\begin{align}
&\sum_{n\in\mathcal{S}}\mathbb{E}\left[
 \int_{\tilde{\mathcal{Y}}_{n,1}} \Delta f_{X}^{2}(\tilde{a}_{n}, y; v_{\tilde{N}})
 p_{\mathrm{G}}(y - \tilde{U}_{n}; v_{\tilde{N}})dy
\right]
\nonumber \\
&< \frac{L^{2}}{k}\sum_{n\in\mathcal{S}}\nu_{N}^{2}\tilde{a}_{n}^{2}
\mathbb{E}\left[
 \int_{\tilde{\mathcal{Y}}_{n,1}}p_{\mathrm{G}}(y - \tilde{U}_{n}; v_{\tilde{N}})dy
\right]
\nonumber \\
&< \frac{L^{2}}{k}\sum_{n\in\mathcal{S}}\nu_{N}^{2}\tilde{a}_{n}^{2}
\pto 0, 
\end{align}
where the last convergence follows from (\ref{latter_limit_last_term}). 

We prove the uniform Lipschitz-continuity~(\ref{uniform_Lipschitz_Y1}). 
Using the mean value theorem yields 
\begin{equation} \label{mean_value_theorem}
f_{X}(\tilde{x}; v_{\tilde{N}}) - f_{X}(\tilde{y}; v_{\tilde{N}}) 
= f_{X}'(\tilde{y}_{*}; v_{\tilde{N}})(\tilde{x} - \tilde{y}) 
\end{equation}
for some $\tilde{y}_{*}$ between $\tilde{x}$ and $\tilde{y}$. 
Thus, it is sufficient to prove the uniform boundedness of 
$|f_{X}'(\tilde{y}; v_{\tilde{N}})|$ for all 
$y=\sqrt{k}\tilde{y}\in\mathcal{Y}_{1}$. 
Using $f_{X}'(\tilde{y}; v_{\tilde{N}})
=v_{\mathrm{post}}(\tilde{y}; v_{\tilde{N}})/v_{k,N}$ 
with (\ref{posterior_variance_X}) yields 
\begin{align}
&f_{X}'(\tilde{y}; v_{\tilde{N}}) 
= \frac{f_{X}^{(2)}(\tilde{y}; v_{\tilde{N}}) 
- f_{X}^{2}(\tilde{y}; v_{\tilde{N}})}{v_{k,N}}
\nonumber \\
&= \frac{f_{A}(y; v_{\tilde{N}})\{f_{U}^{(2)}(y; v_{\tilde{N}}) 
- f_{A}(y; v_{\tilde{N}})f_{U}^{2}(y; v_{\tilde{N}})\}}
{v_{\tilde{N}}}, \label{posterior_mean_estimator_x_deriv}
\end{align}
where we have used the definition of $f_{X,i}$ in 
(\ref{posterior_mean_estimator_X}). 

For all $y\in\mathcal{Y}_{1}$, we use $f_{A}\geq0$ and 
the upper bound~(\ref{fA_bound_Y1}) on $f_{A}$ to obtain 
\begin{equation} \label{derivative_Y1}
|f_{X}'(\tilde{y}; v_{\tilde{N}})| 
< \frac{\bar{f}_{U}^{(2)}k(N/k)^{\alpha_{N}}\log(N/k)}{v(N - k)}
= \frac{\bar{f}_{U}^{(2)}}{v(1 - k/N)} 
\end{equation}
for $\alpha_{N}=1-\epsilon_{N}$ and $\epsilon_{N}=\log\log(N/k)/\log(N/k)$, 
with $\bar{f}_{U}^{(2)}=\sup_{y\in\mathcal{Y}_{1}}|f_{U}^{(2)}(y; v_{\tilde{N}})|<\infty$, 
which is uniformly bounded from Assumption~\ref{assumption_lemma}. 
Since (\ref{derivative_Y1}) is uniformly bounded for $k/N<1$, we arrive at the 
uniform Lipschitz-continuity~(\ref{uniform_Lipschitz_Y1}).

\subsection{Second Set $\tilde{\mathcal{Y}}_{n,2}$}
We prove that the integral in (\ref{difference_tmp}) over 
$\tilde{\mathcal{Y}}_{n,2}$ tends to zero in the sublinear sparsity limit. 
Using the inequality $(x+y)^{2}\leq2(x^{2}+y^{2})$ for 
$\Delta f_{X}^{2}$ in (\ref{Delta_f}), 
$f_{X}$ in (\ref{posterior_mean_estimator_X}), and 
the trivial upper bound $f_{A}\leq1$ to obtain 
\begin{align}
&\mathbb{E}\left[
 \int_{\tilde{\mathcal{Y}}_{n,2}}
 \Delta f_{X}^{2}(\tilde{a}_{n}, y; v_{\tilde{N}})
 p_{\mathrm{G}}(y - \tilde{U}_{n}; v_{\tilde{N}})dy
\right]
\nonumber \\
&< \frac{2}{k}\mathbb{E}\left[
 \int_{\tilde{\mathcal{Y}}_{n,2}}
 f_{U}^{2}(y - \nu_{N}\tilde{a}_{n}; v_{\tilde{N}}) 
 p_{\mathrm{G}}(y - \tilde{U}_{n}; v_{\tilde{N}})dy
\right]
\nonumber \\
&+ \frac{2}{k}\mathbb{E}\left[
 \int_{\tilde{\mathcal{Y}}_{n,2}}
 f_{U}^{2}(y; v_{\tilde{N}}) 
 p_{\mathrm{G}}(y - \tilde{U}_{n}; v_{\tilde{N}})dy
\right]. \label{difference2}
\end{align}

The assumption $\mathrm{Pr}(|U_{n}|=\sqrt{2v})=0$ implies that $|U_{n}|$ has no 
discrete components on the interval $(\sqrt{2v}-\epsilon, 
\sqrt{2v} + \epsilon)$ for some $\epsilon\in(0, \sqrt{v})$. 
Let $\mathcal{N}_{\mathrm{e}}(\epsilon)
=\{n\in\mathcal{S}: |\nu_{N}\tilde{a}_{n}| \geq \epsilon/4\}$. 
From (\ref{latter_limit_last_term}) we find 
$|\mathcal{N}_{\mathrm{e}}(\epsilon)|=o(k)$. 
For $n\in\mathcal{N}_{\mathrm{e}}(\epsilon)$ we use the uniform 
Lipschitz-continuity of $f_{U}$ for (\ref{difference2}) to obtain 
\begin{align}
&\sum_{n\in\mathcal{N}_{\mathrm{e}}(\epsilon)}\mathbb{E}\left[
 \int_{\tilde{\mathcal{Y}}_{n,2}} 
 \Delta f_{X}^{2}(\tilde{a}_{n}, y; v_{\tilde{N}})
 p_{\mathrm{G}}(y - \tilde{U}_{n}; v_{\tilde{N}})dy
\right]
\nonumber \\
&< \frac{2}{k}\sum_{n\in\mathcal{N}_{\mathrm{e}}(\epsilon)}\mathbb{E}\left[
 \int_{\mathbb{R}}
 (L|y - \nu_{N}\tilde{a}_{n}| + C)^{2}
 p_{\mathrm{G}}(y - \tilde{U}_{n}; v_{\tilde{N}})dy
\right]
\nonumber \\
&+ \frac{2}{k}\sum_{n\in\mathcal{N}_{\mathrm{e}}(\epsilon)}\mathbb{E}\left[
 \int_{\mathbb{R}}(L|y| + C)^{2}p_{\mathrm{G}}(y - \tilde{U}_{n}; v_{\tilde{N}})dy
\right] \label{difference_exception}
\end{align}
for some $L>0$ and $C>0$. 

We only evaluate the first term on the upper 
bound~(\ref{difference_exception}) since the second term can be evaluated 
in the same manner. Using the inequality $(x + y)^{2}\leq2(x^{2} + y^{2})$ 
repeatedly yields 
\begin{align}
&\frac{1}{k}\sum_{n\in\mathcal{N}_{\mathrm{e}}(\epsilon)}\mathbb{E}\left[
 \int_{\mathbb{R}}
 (L|y - \nu_{N}\tilde{a}_{n}| + C)^{2}
 p_{\mathrm{G}}(y - \tilde{U}_{n}; v_{\tilde{N}})dy
\right]
\nonumber \\
&< \frac{2}{k}\sum_{n\in\mathcal{N}_{\mathrm{e}}(\epsilon)}\left\{
 2L^{2}[v_{\tilde{N}} + (\mathbb{E}[\tilde{U}_{n}])^{2} 
 + \nu_{N}^{2}\tilde{a}_{n}^{2}] + C^{2}
\right\}\pto 0,
\end{align}
where the last convergence follows from 
$|\mathcal{N}_{\mathrm{e}}(\epsilon)|=o(k)$, 
$(\mathbb{E}[\tilde{U}_{n}])^{2}<\mathbb{E}[\tilde{U}_{n}^{2}]$, 
(\ref{U_tilde_2_bound}), $\mathbb{E}[U_{n}^{2}]<\infty$, 
and (\ref{latter_limit_last_term}). 

For $n\in\mathcal{S}\setminus\mathcal{N}_{\mathrm{e}}(\epsilon)$, 
we consider sufficiently large $N$ such that 
$\sqrt{2\alpha_{N}v} - |\nu_{N}\tilde{a}_{n}| 
> \sqrt{2\alpha_{N}v} - \epsilon/4>0$ holds for all  
$n\in\mathcal{S}\setminus\mathcal{N}_{\mathrm{e}}(\epsilon)$, because of 
$\epsilon<\sqrt{v}$. For such $N$, we evaluate (\ref{difference2}) as 
\begin{align}
&\sum_{n\in\mathcal{S}\setminus\mathcal{N}_{\mathrm{e}}(\epsilon)}\mathbb{E}\left[
 \int_{\tilde{\mathcal{Y}}_{n,2}} 
 \Delta f_{X}^{2}(\tilde{a}_{n}, y; v_{\tilde{N}})
 p_{\mathrm{G}}(y - \tilde{U}_{n}; v_{\tilde{N}})dy
\right]
\nonumber \\
&< \frac{4\bar{f}_{U,n}^{2}(v_{\tilde{N}})}{k}
\sum_{n\in\mathcal{S}\setminus\mathcal{N}_{\mathrm{e}}(\epsilon)}\mathbb{E}\left[
 \int_{\tilde{\mathcal{Y}}_{n,2}}
 p_{\mathrm{G}}(y - \tilde{U}_{n}; v_{\tilde{N}})dy
\right],
\end{align}
with
\begin{equation}
\bar{f}_{U,n}(v_{\tilde{N}})
=\sup_{y\in\tilde{\mathcal{Y}}_{n,2}}\max\left\{
 |f_{U}(y; v_{\tilde{N}})|, 
 |f_{U}(y - \nu_{N}\tilde{a}_{n}; v_{\tilde{N}})|
\right\}, 
\end{equation}
which is bounded from above by some $\bar{f}_{U}>0$ for all $N$ and 
$n\in\mathcal{S}\setminus\mathcal{N}_{\mathrm{e}}(\epsilon)$, because of 
the uniform Lipschitz-continuity for $f_{U}$, 
$|y| < \sqrt{2\alpha_{N}v} + |\nu_{N}\tilde{a}_{n}|
< \sqrt{2v} + \epsilon/4$, and  
$|y - \nu_{N}\tilde{a}_{n}|<\sqrt{2\alpha_{N}v} + 2|\nu_{N}\tilde{a}_{n}|
\leq \sqrt{2v} + \epsilon/2$ for all $y\in\tilde{\mathcal{Y}}_{n,2}$ 
in (\ref{Y2_tilde}) 
and $n\in\mathcal{S}\setminus\mathcal{N}_{\mathrm{e}}(\epsilon)$. 

We separate the region of integration $\tilde{\mathcal{Y}}_{n,2}$ in 
(\ref{Y2_tilde}) into positive and negative intervals, and using 
the change of variables $y'= y - \nu_{N}\tilde{a}_{n}$ in each integral, we have 
\begin{align}
&\frac{1}{k}\sum_{n\in\mathcal{S}\setminus\mathcal{N}_{\mathrm{e}}(\epsilon)}\mathbb{E}\left[
 \int_{\tilde{\mathcal{Y}}_{n,2}}
 p_{\mathrm{G}}(y - \tilde{U}_{n}; v_{\tilde{N}})dy
\right]
\nonumber \\
&< \frac{1}{k}\sum_{n\in\mathcal{S}\setminus\mathcal{N}_{\mathrm{e}}(\epsilon)}
\mathbb{E}\left[
 \int_{\tilde{\mathcal{Y}}_{n,2}^{+}}
 p_{\mathrm{G}}(y' - U_{n}; v_{\tilde{N}})dy'
\right]
\nonumber \\
&+ \frac{1}{k}\sum_{n\in\mathcal{S}\setminus\mathcal{N}_{\mathrm{e}}(\epsilon)}
\mathbb{E}\left[
 \int_{\tilde{\mathcal{Y}}_{n,2}^{-}}
 p_{\mathrm{G}}(y' - U_{n}; v_{\tilde{N}})dy'
\right], \label{difference_major}
\end{align}
with 
\begin{equation} \label{Y2_tilde_+}
\tilde{\mathcal{Y}}_{n,2}^{+}
=\{y'\geq0: \left|
 y'  + \nu_{N}\tilde{a}_{n} - \sqrt{2\alpha_{N}v}
\right| \leq |\nu_{N}\tilde{a}_{n}| \},
\end{equation} 
\begin{equation}
\tilde{\mathcal{Y}}_{n,2}^{-}
=\{y'<0: \left|
 y' + \nu_{N}\tilde{a}_{n} + \sqrt{2\alpha_{N}v}
\right| 
\leq |\nu_{N}\tilde{a}_{n}|\},
\end{equation} 
where we have used the fact that $\tilde{\mathcal{Y}}_{n,2}$ and the interval 
$\{y\in\mathbb{R}: |y|\leq|\nu_{N}\tilde{a}_{n}|\}$ have no intersection.

We only evaluate the first term on the upper bound~(\ref{difference_major}) 
since the second term can be evaluated in the same manner. 
Using the pdf~(\ref{pdf_U}) of $U$ yields 
\begin{align}
&\frac{1}{k}\sum_{n\in\mathcal{S}\setminus\mathcal{N}_{\mathrm{e}}(\epsilon)}\mathbb{E}\left[
 \int_{\tilde{\mathcal{Y}}_{n,2}^{+}}
 p_{\mathrm{G}}(y - U_{n}; v_{\tilde{N}})dy
\right]
\nonumber \\
&= \frac{1 - p_{U}^{\mathrm{D}}}{k}
\sum_{n\in\mathcal{S}\setminus\mathcal{N}_{\mathrm{e}}(\epsilon)}
\int du\int_{\tilde{\mathcal{Y}}_{n,2}^{+}}dy
p_{U}^{\mathrm{C}}(u) p_{\mathrm{G}}(y - u; v_{\tilde{N}})
\nonumber \\
&+ \frac{1}{k}\sum_{n\in\mathcal{S}\setminus\mathcal{N}_{\mathrm{e}}(\epsilon)}
\sum_{i=1}^{\infty}p_{U, i}^{\mathrm{D}}\int_{\tilde{\mathcal{Y}}_{n,2}^{+}}
p_{\mathrm{G}}(y - u_{i}; v_{\tilde{N}})dy. 
\end{align}
For the continuous component, using the change of variables 
$\tilde{u} = (u - y)/\sqrt{v_{\tilde{N}}}$ for fixed $y$ yields 
\begin{align}
&\frac{1}{k}\sum_{n\in\mathcal{S}\setminus\mathcal{N}_{\mathrm{e}}(\epsilon)}
\int du\int_{\tilde{\mathcal{Y}}_{n,2}^{+}}dy
p_{U}^{\mathrm{C}}(u) p_{\mathrm{G}}(y - u; v_{\tilde{N}})
\nonumber \\
&= \frac{1}{k}\sum_{n\in\mathcal{S}\setminus\mathcal{N}_{\mathrm{e}}(\epsilon)}
\int_{\tilde{\mathcal{Y}}_{n,2}^{+}}dy\int d\tilde{u}
p_{U}^{\mathrm{C}}(\sqrt{v_{\tilde{N}}}\tilde{u} + y) p_{\mathrm{G}}(\tilde{u}; 1)
\nonumber \\
&< \frac{2\bar{p}_{U}^{\mathrm{C}}}{k}\sum_{n\in\mathcal{S}}|\nu_{N}\tilde{a}_{n}|
\pto 0,  
\end{align}
with $\bar{p}_{U}^{\mathrm{C}}=\sup_{u\in\mathbb{R}}p_{U}^{\mathrm{C}}(u)$. 
In the proof of the last convergence, we have used 
(\ref{first_order_summation}) and the boundedness assumption 
of $\bar{p}_{U}^{\mathrm{C}}$. 

For the discrete component, we find the uniform upper bound 
$p_{U, i}^{\mathrm{D}}\int_{\tilde{\mathcal{Y}}_{n,2}^{+}}
p_{\mathrm{G}}(y - u_{i}; v_{\tilde{N}})dy\leq p_{U, i}^{\mathrm{D}}$ with respect 
to $n\in\mathcal{S}$, $N$, and $k$, which is summable over $i\in\mathbb{N}$. 
Thus, we can use the reverse Fatou lemma to obtain 
\begin{align}
&\limsup_{N\to\infty}\frac{1}{k}
\sum_{n\in\mathcal{S}\setminus\mathcal{N}_{\mathrm{e}}(\epsilon)}\sum_{i=1}^{\infty}
p_{U, i}^{\mathrm{D}}\int_{\tilde{\mathcal{Y}}_{n,2}^{+}}
p_{\mathrm{G}}(y - u_{i}; v_{\tilde{N}})dy
\nonumber \\
&\leq \sum_{i=1}^{\infty}\limsup_{N\to\infty}\frac{p_{U, i}^{\mathrm{D}}}{k}
\sum_{n\in\mathcal{S}\setminus\mathcal{N}_{\mathrm{e}}(\epsilon)}
\int_{\tilde{\mathcal{Y}}_{n,2}^{+}}p_{\mathrm{G}}(y - u_{i}; v_{\tilde{N}})dy
\end{align}
in the sublinear sparsity limit. 

For any $i$ satisfying 
$u_{i}> \sqrt{\epsilon_{N}v} + \sup\tilde{\mathcal{Y}}_{n,2}^{+}$, we have 
\begin{align}
&\frac{1}{k}\sum_{n\in\mathcal{S}\setminus\mathcal{N}_{\mathrm{e}}(\epsilon)}
\int_{\tilde{\mathcal{Y}}_{n,2}^{+}}p_{\mathrm{G}}(y - u_{i}; v_{\tilde{N}})dy
\nonumber \\
&< \frac{1}{k}\sum_{n\in\mathcal{S}}\frac{2|\nu_{N}\tilde{a}_{n}|}
{\sqrt{2\pi v_{\tilde{N}}}}\exp\left\{
 - \frac{(\sup\tilde{\mathcal{Y}}_{n,2}^{+} - u_{i})^{2}}
 {2v_{\tilde{N}}}
\right\} 
\nonumber \\
&< \frac{2e^{- (\epsilon_{N}/2)\log(N/k)}\sqrt{\log(N/k)}}{k\sqrt{2\pi v}}
\sum_{n\in\mathcal{S}}|\nu_{N}\tilde{a}_{n}|\pto 0,
\end{align}
where the last convergence follows from (\ref{first_order_summation}) and 
$\epsilon_{N}=\log\log(N/k)/\log(N/k)$. 
For any $i$ satisfying $u_{i}< - \sqrt{\epsilon_{N}v} 
+ \inf\tilde{\mathcal{Y}}_{n,2}^{+}$, 
similarly, we find 
\begin{align}
&\frac{1}{k}\sum_{n\in\mathcal{S}\setminus\mathcal{N}_{\mathrm{e}}(\epsilon)}
\int_{\tilde{\mathcal{Y}}_{n,2}^{+}}p_{\mathrm{G}}(y - u_{i}; v_{\tilde{N}})dy
\nonumber \\
&< \frac{1}{k}\sum_{n\in\mathcal{S}}\frac{2|\nu_{N}\tilde{a}_{n}|}
{\sqrt{2\pi v_{\tilde{N}}}}\exp\left\{
 - \frac{(\inf\tilde{\mathcal{Y}}_{n,2}^{+} - u_{i})^{2}}
 {2v_{\tilde{N}}}
\right\} \pto 0.
\end{align}

For the remaining index $i^{*}$ satisfying 
$u_{i^{*}}\in[\inf\tilde{\mathcal{Y}}_{n,2}^{+} - \sqrt{\epsilon_{N}v}, 
\sup\tilde{\mathcal{Y}}_{n,2}^{+} + \sqrt{\epsilon_{N}v}]$, we confirm 
$p_{U,i^{*}}^{\mathrm{D}}=0$. When $\nu_{N}\tilde{a}_{n}\geq0$ holds, we use 
$\tilde{\mathcal{Y}}_{n,2}^{+}$ in (\ref{Y2_tilde_+}) to find that the condition 
reduces to $u_{i^{*}} - \sqrt{2\alpha_{N}v}
\in[ - 2\nu_{N}\tilde{a}_{n}  - \sqrt{\epsilon_{N}v}, \sqrt{\epsilon_{N}v}]
\subset(- \epsilon, \epsilon)$ 
for sufficiently large $N$, because of $|\nu_{N}\tilde{a}_{n}|<\epsilon/4$ 
for $n\in\mathcal{S}\setminus\mathcal{N}_{\mathrm{e}}(\epsilon)$. 
Thus, we have $p_{U,i^{*}}^{\mathrm{D}}=0$. When $\nu_{N}\tilde{a}_{n}<0$ holds, 
similarly, the condition reduces to $u_{i^{*}}-\sqrt{2\alpha_{N}v}
\in[ - \sqrt{\epsilon_{N}v}, \sqrt{\epsilon_{N}v} + 2|\nu_{N}\tilde{a}_{n}|]
\subset(- \epsilon, \epsilon)$ 
for sufficiently large $N$, so that we find $p_{U,i^{*}}^{\mathrm{D}}=0$. 
Combining these results, we arrive at 
\begin{equation}
\frac{1}{k}\sum_{n\in\mathcal{S}\setminus\mathcal{N}_{\mathrm{e}}(\epsilon)}
\sum_{i=1}^{\infty}p_{U, i}^{\mathrm{D}}\int_{\tilde{\mathcal{Y}}_{n,2}^{+}}
p_{\mathrm{G}}(y - u_{i}; v_{\tilde{N}})dy
\pto 0. 
\end{equation}
Thus, the integral in (\ref{difference_tmp}) over 
$\tilde{\mathcal{Y}}_{n,2}$ tends to zero in the sublinear sparsity limit. 

\subsection{Last Set $\tilde{\mathcal{Y}}_{n,3}$}
For any measurable set $\mathcal{Y}\subset\mathbb{R}$,  
we first prove the following identity: 
\begin{align} \label{Y2_identity}
&\sum_{n\in\mathcal{S}}\mathbb{E}\left[
 \int_{\mathcal{Y}} \Delta f_{X}^{2}(\tilde{a}_{n}, y; v_{\tilde{N}})
 p_{\mathrm{G}}(y - \tilde{U}_{n}; v_{\tilde{N}})dy
\right]
\nonumber \\
&\peq \frac{1}{k}\sum_{n\in\mathcal{S}}\mathbb{E}\left[
 \int_{\mathcal{Y}}f_{U,n}^{2}(f_{A,n} - \tilde{f}_{A,n})^{2}
 p_{\mathrm{G}}(y - \tilde{U}_{n}; v_{\tilde{N}})dy
\right] 
\nonumber \\
&+ o(1) 
\end{align}
in the sublinear sparsity limit, where $\tilde{f}_{A,n}$ denotes  
$f_{A}(y - \nu_{N}\tilde{a}_{n}; v_{\tilde{N}})$ while 
$f_{A}(y; v_{\tilde{N}})$ and $f_{U}(y; v_{\tilde{N}})$ are abbreviated as 
$f_{A,n}$ and $f_{U,n}$, respectively. 

For notational simplicity, let 
$\tilde{Y}_{n}=\tilde{U}_{n} + \tilde{\Omega}_{n}$. 
We use $\Delta f_{X}$ in (\ref{Delta_f}) and $f_{X}$ in 
(\ref{posterior_mean_estimator_X}) to obtain 
\begin{align}
&\sum_{n\in\mathcal{S}}\mathbb{E}\left[
 \int_{\mathcal{Y}} \Delta f_{X}^{2}(\tilde{a}_{n}, y; v_{\tilde{N}})
 p_{\mathrm{G}}(y - \tilde{U}_{n}; v_{\tilde{N}})dy
\right]
\nonumber \\
&= \sum_{n\in\mathcal{S}}\mathbb{E}\left[
 \Delta f_{X}^{2}(\tilde{a}_{n}, \tilde{Y}_{n}; v_{\tilde{N}})
 1(\tilde{Y}_{n}\in\mathcal{Y})
\right]
\nonumber \\
&= \frac{1}{k}\sum_{n\in\mathcal{S}}\mathbb{E}\left[
 f_{U,n}^{2}(f_{A,n} - \tilde{f}_{A,n})^{2}1(\tilde{Y}_{n}\in\mathcal{Y})
\right]
\nonumber \\
&+ \frac{2}{k}\sum_{n\in\mathcal{S}}\mathbb{E}\left[
 f_{U,n}(f_{A,n} - \tilde{f}_{A,n})(f_{U,n} - \tilde{f}_{U,n})\tilde{f}_{A,n}
 1(\tilde{Y}_{n}\in\mathcal{Y})
\right]
\nonumber \\
&+ \frac{1}{k}\sum_{n\in\mathcal{S}}\mathbb{E}\left[
 (f_{U,n} - \tilde{f}_{U,n})^{2}\tilde{f}_{A,n}^{2}1(\tilde{Y}_{n}\in\mathcal{Y}) 
\right],  
\end{align}
where $\tilde{f}_{A,n}$ and $\tilde{f}_{U,n}$ denote 
$f_{A}(\tilde{Y}_{n} - \nu_{N}\tilde{a}_{n}; v_{\tilde{N}})$ and 
$f_{U}(\tilde{Y}_{n} - \nu_{N}\tilde{a}_{n}; v_{\tilde{N}})$, respectively,  
while $f_{A}(\tilde{Y}_{n}; v_{\tilde{N}})$ and 
$f_{U}(\tilde{Y}_{n}; v_{\tilde{N}})$ are abbreviated as $f_{A,n}$ and $f_{U,n}$, 
respectively. By definition, the first term is equivalent to the RHS of 
(\ref{Y2_identity}). 

For the last term, we use the uniform Lipschitz-continuity of $f_{U}$ and 
$\tilde{f}_{A,n}^{2}\leq1$ to obtain 
\begin{equation}
\frac{1}{k}\sum_{n\in\mathcal{S}}\mathbb{E}\left[
 (f_{U,n} - \tilde{f}_{U,n})^{2}\tilde{f}_{A,n}^{2}1(\tilde{Y}_{n}\in\mathcal{Y})
\right] 
\leq \frac{L^{2}}{k}\sum_{n\in\mathcal{S}}\nu_{N}^{2}\tilde{a}_{n}^{2}
\end{equation}
for some $L>0$, which converges in probability to zero in the sublinear 
sparsity limit, because of (\ref{latter_limit_last_term}). 
For the second term, similarly, we have 
\begin{align}
&\left|
 \frac{1}{k}\sum_{n\in\mathcal{S}}\mathbb{E}\left[
  f_{U,n}(f_{A,n} - \tilde{f}_{A,n})(f_{U,n} - \tilde{f}_{U,n})\tilde{f}_{A,n}
  1(\tilde{Y}_{n}\in\mathcal{Y})
 \right]
\right|
\nonumber \\
&< \frac{L}{k}\sum_{n\in\mathcal{S}}|\nu_{N}\tilde{a}_{n}|\mathbb{E}\left[
 |f_{U,n}|
\right]
\nonumber \\
&\leq L\left\{
 \frac{1}{k}\sum_{n\in\mathcal{S}}\nu_{N}^{2}\tilde{a}_{n}^{2}
 \frac{1}{k}\sum_{n\in\mathcal{S}}\left(
  \mathbb{E}\left[
   |f_{U,n}|
  \right]
 \right)^{2}
\right\}^{1/2}\pto 0
\end{align}
for some $L>0$. In the derivation of the second inequality, we have used 
the Cauchy-Schwarz inequality. The last convergence follows from 
(\ref{latter_limit_last_term}), the uniform Lipschitz-continuity of $f_{U}$, 
(\ref{U_tilde_2_bound}), 
and $\mathbb{E}[U_{n}^{2}]<\infty$. Thus, we arrive at (\ref{Y2_identity}). 

We next evaluate the integral in (\ref{difference_tmp}) over 
$\tilde{\mathcal{Y}}_{n,3}$. Using (\ref{Y2_identity}) for 
$\mathcal{Y}=\tilde{\mathcal{Y}}_{n,3}$, 
the uniform Lipschitz-continuity of $f_{U}$, and the inequality 
$(f_{A,n} - \tilde{f}_{A,n})^{2}\leq2\{(f_{A,n} - 1)^{2} 
+ (1 - \tilde{f}_{A,n})^{2}\}$, we obtain 
\begin{align} 
&\sum_{n\in\mathcal{S}}\mathbb{E}\left[
 \int_{\tilde{\mathcal{Y}}_{n,3}} \Delta f_{X}^{2}(\tilde{a}_{n}, y; v_{\tilde{N}})
 p_{\mathrm{G}}(y - \tilde{U}_{n}; v_{\tilde{N}})dy
\right]
\nonumber \\
&\peq \frac{2}{k}\sum_{n\in\mathcal{S}}
\mathbb{E}\left[
 \int_{\tilde{\mathcal{Y}}_{n,3}}(L|y| + C)^{2}\left\{
  [f_{A,n}(y; v_{\tilde{N}}) - 1]^{2} 
 \right.
\right.
\nonumber \\
&\left.
 + [1 - \tilde{f}_{A,n}(y - \nu_{N}\tilde{a}_{n}; v_{\tilde{N}})]^{2}
\right\}p_{\mathrm{G}}(y - \tilde{U}_{n}; v_{\tilde{N}})dy
\Biggr] + o(1) 
\nonumber \\
&\pto 0
\end{align}
for some $L>0$ and $C>0$. 

To understand the last convergence, 
we observe $y-\nu_{N}\tilde{a}_{n}\in\mathcal{Y}_{2}\cup\mathcal{Y}_{3}$ 
in (\ref{Y2}) and (\ref{Y3}) for all 
$y\in\tilde{\mathcal{Y}}_{n,3}\subset\mathcal{Y}_{2}\cup\mathcal{Y}_{3}$ 
in (\ref{Y3_tilde}). 
Thus, we can repeat the proofs in Appendix~\ref{appen_Y2} and 
Appendix~\ref{proof_last_term} under the assumption 
$\mathrm{Pr}(|U_{n}|=\sqrt{2v})=0$ to prove the convergence to zero. 

\section{Proof of (\ref{xi_limit_former})}
\label{proof_lemma_xi}
We prove the former convergence~(\ref{xi_limit_former}). 
Repeating the derivation of (\ref{unnormalized_square_error_tmp}), we have 
\begin{align}
\boldsymbol{\Omega}^{\mathrm{T}}f_{X}(\boldsymbol{Y}; v_{\tilde{N}})
&= \sum_{n\notin\mathcal{S}}\Omega_{n}f_{X}(\Omega_{n}; v_{\tilde{N}})
\nonumber \\
&+ \sum_{n\in\mathcal{S}}\Omega_{n}f_{X}(k^{-1/2}U_{n} + \Omega_{n}; v_{\tilde{N}}). 
\label{xi_tmp}
\end{align}
Since the first term is the sum of i.i.d.\ random variables, we evaluate 
its variance as 
\begin{align}
&\mathbb{V}\left[
 \sum_{n\notin\mathcal{S}}\Omega_{n}f_{X}(\Omega_{n}; v_{\tilde{N}})
\right]
= (N - k)\mathbb{V}\left[
 \Omega_{1} f_{X}(\Omega_{1}; v_{\tilde{N}})
\right]
\nonumber \\
&< (N - k)\mathbb{E}\left[
 \Omega_{1}^{2}f_{X}^{2}(\Omega_{1}; v_{\tilde{N}})
\right]\to 0
\end{align}
in the sublinear sparsity limit. In the derivation of the inequality, 
we have used the upper bound $\mathbb{V}[Z]\leq\mathbb{E}[Z^{2}]$ 
for any random variable $Z\in\mathbb{R}$. The last convergence follows 
from (\ref{former_limit_general}) for $j=2$. Thus, the weak law of large 
numbers hold for the first term in (\ref{xi_tmp}). 

We prove the weak law of large numbers for the second term in (\ref{xi_tmp}). 
Since the second term is the sum of i.i.d.\ random variables, 
we evaluate its variance as 
\begin{align}
&\mathbb{V}\left[
 \sum_{n\in\mathcal{S}}\Omega_{n}f_{X}(k^{-1/2}U_{n} + \Omega_{n}; v_{\tilde{N}})
\right]
\nonumber \\
&= k\mathbb{V}\left[
 \Omega_{1} f_{X}(k^{-1/2}U + \Omega_{1}; v_{\tilde{N}})
\right] \nonumber \\
&= \frac{1}{k}\mathbb{V}\left[
 \tilde{\Omega}f_{A}(U + \tilde{\Omega}; v_{\tilde{N}})
 f_{U}(U + \tilde{\Omega}; v_{\tilde{N}})
\right]
\nonumber \\
&\leq \frac{1}{k}\mathbb{E}\left[
 \tilde{\Omega}^{2}f_{U}^{2}(U + \tilde{\Omega}; v_{\tilde{N}})
\right]. \label{second_term_variance}
\end{align}
In the derivation of the second equality, we have used the change of variables 
$\tilde{\Omega}=\sqrt{k}\Omega_{1}\sim\mathcal{N}(0, v_{\tilde{N}})$ and the 
definition of $f_{X}$ in (\ref{posterior_mean_estimator_X}). The last 
inequality follows from $f_{A}\leq 1$. 

We prove that the upper bound~(\ref{second_term_variance}) tends to zero 
in the sublinear sparsity limit. 
Since the uniform Lipschitz-continuity for $f_{U}$ implies 
$|f_{U}(y; v_{\tilde{N}})|
\leq L|y| + C$ for some constants $L>0$ and $C>0$, we have  
\begin{align}
&\mathbb{E}\left[
 \tilde{\Omega}^{2}f_{U}^{2}(U + \tilde{\Omega}; v_{\tilde{N}})
\right]
\leq \mathbb{E}\left[
 \tilde{\Omega}^{2}\{L|U + \tilde{\Omega}| + C\}^{2}
\right]
\nonumber \\
&\leq 2\mathbb{E}\left[
 \tilde{\Omega}^{2}\{2L^{2}(U^{2} + \tilde{\Omega}^{2}) + C^{2}\}
\right] < \infty 
\end{align}
in the sublinear sparsity limit. In the derivation of the second inequality, 
we have used the upper bound $(x+y)^{2}\leq2(x^{2} + y^{2})$ repeatedly. 
The last boundedness follows from 
$\mathbb{E}[U^{2}]<\infty$ and $v_{\tilde{N}}\to0$. 
Thus, the variance~(\ref{second_term_variance}) tends to zero 
in the sublinear sparsity limit. This observation implies the weak law of 
large numbers for the second term in (\ref{xi_tmp}).    

\section{Proof of (\ref{Delta_xi_former})}
\label{proof_lemma_Delta_xi}
\subsection{Truncation}
For notational simplicity, we omit conditioning on $\boldsymbol{a}$. 
In other words, $\boldsymbol{a}$ is regarded as a deterministic vector 
satisfying the boundedness $\|\boldsymbol{a}\|_{2}<\infty$.  
We first separate the indices 
$\bar{\mathcal{S}}=\{1,\ldots,N\}\setminus\mathcal{S}$ into two 
groups. One group consists of indices $\mathcal{N}(\boldsymbol{a})
=\{n\in\bar{\mathcal{S}}: |\nu_{N}\tilde{a}_{n}|>\epsilon\}$ for any 
$\epsilon>0$. 
From (\ref{latter_limit_last_term_all}), we have 
$|\mathcal{N}(\boldsymbol{a})|=o(k)$. The other group 
contains the remaining indices $\bar{\mathcal{S}}\setminus
\mathcal{N}(\boldsymbol{a})$. 

We separate the sum~(\ref{Delta_xi_former}) into two terms: 
\begin{align}
&\sum_{n\notin\mathcal{S}}\mathbb{E}\left[
 \left|
  \Omega_{n}\Delta f_{X}(\tilde{a}_{n}, 
  \nu_{N}\tilde{a}_{n} + \tilde{\Omega}_{n}; v_{\tilde{N}})
 \right|
\right]
\nonumber \\
&= \sum_{n\in\mathcal{N}(\boldsymbol{a})}\mathbb{E}\left[
 \left|
  \Omega_{n}\Delta f_{X}
 \right|
\right]
+ \sum_{n\in\bar{\mathcal{S}}\setminus\mathcal{N}(\boldsymbol{a})}\mathbb{E}\left[
 \left|
  \Omega_{n}\Delta f_{X}
 \right|
\right]. \label{Delta_xi_former_tmp}
\end{align}
Repeating the proof of (\ref{Delta_xi_latter}), we find that the first term 
converges in probability to zero in the sublinear sparsity limit. 

The second term in (\ref{Delta_xi_former_tmp}) is evaluated in a similar 
manner to that for the proof of (\ref{former_limit_general}) in 
Lemma~\ref{lemma_MSE}. We represent $\mathbb{E}[|\Omega_{n} \Delta f_{X}|]$ 
with $\tilde{\Omega}_{n}=\sqrt{k}\Omega_{n}\sim\mathcal{N}(0, v_{\tilde{N}})$ as 
\begin{align}
&\mathbb{E}[|\Omega_{n} \Delta f_{X}(\tilde{a}_{n}, 
\nu_{N}\tilde{a}_{n} + \tilde{\Omega}_{n}; v_{\tilde{N}})| | 
\boldsymbol{a}]
\nonumber \\
&= \frac{1}{\sqrt{k}}\int |w
\Delta f_{X}(\tilde{a}_{n}, \nu_{N}\tilde{a}_{n} + w; 
v_{\tilde{N}})|
p_{\mathrm{G}}(w; v_{\tilde{N}})dw. \label{xi_appen_tmp}
\end{align}
We decompose the interval of integration into the two disjoint sets 
\begin{equation}
\mathcal{W}_{1} = \left\{
 w\in\mathbb{R}: |w|\leq\sqrt{2v}
\right\},
\end{equation} 
\begin{equation} 
\mathcal{W}_{2} = \left\{
 w\in\mathbb{R}: |w|>\sqrt{2v}
\right\},
\end{equation}
of which the later is equivalent to (\ref{W4}). 
In evaluating (\ref{xi_appen_tmp}), the threshold $\sqrt{2v}$ can be 
replaced with any real number larger than $\sqrt{2v}$. 
Thus, this separation for the interval of integration is just the standard 
truncation technique in probability theory. 

\subsection{First Set $\mathcal{W}_{1}$}
We evaluate the integral in (\ref{xi_appen_tmp}) over $\mathcal{W}_{1}$ 
in the sublinear sparsity limit. Let 
\begin{equation}
\Delta f_{A}(\tilde{a}_{n}, w; v_{\tilde{N}})
=f_{A}(\nu_{N}\tilde{a}_{n} + w; v_{\tilde{N}}) - f_{A}(w; v_{\tilde{N}}),
\end{equation}
\begin{equation} 
\Delta f_{U}(\tilde{a}_{n}, w; v_{\tilde{N}})=f_{U}(\nu_{N}\tilde{a}_{n} + w; 
v_{\tilde{N}}) 
- f_{U}(w; v_{\tilde{N}}).
\end{equation} 
Using the following upper bound for $n\in\bar{\mathcal{S}}\setminus
\mathcal{N}(\boldsymbol{a})$: 
\begin{align}
\sqrt{k}|\Delta f_{X}(\tilde{a}_{n}, \nu_{N}\tilde{a}_{n} + w; v_{\tilde{N}})|
&\leq |\Delta f_{A}f_{U}(\nu_{N}\tilde{a}_{n} + w; v_{\tilde{N}})| 
\nonumber \\
&+ |f_{A}(w; v_{\tilde{N}})\Delta f_{U}|
\end{align}
on (\ref{Delta_f})---obtained from the definition of $f_{X}$ in 
(\ref{posterior_mean_estimator_X})---yields 
\begin{align}
&\frac{1}{\sqrt{k}}
\int_{\mathcal{W}_{1}}|w\Delta f_{X}(\tilde{a}_{n}, 
\nu_{N}\tilde{a}_{n} + w; v_{\tilde{N}})|
p_{\mathrm{G}}(w; v_{\tilde{N}})dw
\nonumber \\
&\leq \frac{1}{k}\int_{\mathcal{W}_{1}}|w\Delta f_{A}
f_{U}(\nu_{N}\tilde{a}_{n} + w; v_{\tilde{N}})|p_{\mathrm{G}}(w; v_{\tilde{N}})dw
\nonumber \\
&+ \frac{L|\nu_{N}\tilde{a}_{n}|}{k}\int_{\mathcal{W}_{1}}|w|
f_{A}(w; v_{\tilde{N}})
p_{\mathrm{G}}(w; v_{\tilde{N}})dw \label{Delta_xi_former_first}
\end{align}
for some $L>0$, where we have used the uniform Lipschitz-continuity 
for $f_{U}$. 

To evaluate the first term in (\ref{Delta_xi_former_first}), 
we use the mean value theorem to obtain 
\begin{equation}
f_{A}(\nu_{N}\tilde{a}_{n} + w; v_{\tilde{N}}) = f_{A}(w; v_{\tilde{N}}) 
+ f_{A}'(w + x; v_{\tilde{N}})\nu_{N}\tilde{a}_{n}, 
\end{equation}
for some $|x|\in[0, |\nu_{N}\tilde{a}_{n}|]$. 
Computing the derivative of $f_{A}(y; v_{\tilde{N}})$ in 
(\ref{posterior_mean_estimator_A}) yields 
\begin{equation}
f_{A}'(w; v_{\tilde{N}}) 
= \frac{f_{U}(w; v_{\tilde{N}})}{v_{\tilde{N}}}
\{f_{A}(w; v_{\tilde{N}}) - f_{A}^{2}(w; v_{\tilde{N}})\}. 
\end{equation}
Combining these observations, we have 
\begin{align}
&|\Delta f_{A}(\tilde{a}_{n}, w; v_{\tilde{N}})|
\leq \frac{|\nu_{N}\tilde{a}_{n}|}{v_{\tilde{N}}}
\nonumber \\
&\cdot \max_{|x|\in[0, |\nu_{N}\tilde{a}_{n}|]}|f_{U}(w + x; v_{\tilde{N}})|
f_{A}(w + x; v_{\tilde{N}}). \label{Delta_f_A}
\end{align}
Since the uniform Lipschitz-continuity assumption for $f_{U}$ implies 
the uniform boundedness of 
\begin{align}
&\sup_{w\in\mathcal{W}_{1}}
\sup_{|x|\in[0, |\nu_{N}\tilde{a}_{n}|]}|f_{U}(w + x; v_{\tilde{N}})|
\nonumber \\
&< \sup_{w\in\mathcal{W}_{1}}
\sup_{|x|\in[0, \epsilon]}|f_{U}(w + x; v_{\tilde{N}})|
\equiv\bar{f}_{U,1}(\epsilon)  
\end{align}
for all $n\in\bar{\mathcal{S}}\setminus\mathcal{N}(\boldsymbol{a})$,  
we substitute the upper bound~(\ref{Delta_f_A}) into the first term in 
(\ref{Delta_xi_former_first}) to arrive at  
\begin{align}
&\frac{1}{\sqrt{k}}
\int_{\mathcal{W}_{1}}|w\Delta f_{X}(\tilde{a}_{n}, 
\nu_{N}\tilde{a}_{n} + w; v_{\tilde{N}})|
p_{\mathrm{G}}(w; v_{\tilde{N}})dw
\nonumber \\
&\leq \left(
 \frac{\bar{f}_{U,1}^{2}(\epsilon)}{v_{\tilde{N}}} + L
\right)
\nonumber \\
&\cdot \frac{|\nu_{N}\tilde{a}_{n}|}{k}
\int_{\mathcal{W}_{1}}|w|f_{A}(w + x_{\mathrm{s}}(w); v_{\tilde{N}})
p_{\mathrm{G}}(w; v_{\tilde{N}})dw, 
\label{Delta_xi_former_first_tmp}
\end{align}
with $x_{\mathrm{s}}(w)=\argmax_{|x|\in[0, |\nu_{N}\tilde{a}_{n}|]}
f_{A}(w + x; v_{\tilde{N}})$. 

To evaluate the upper bound~(\ref{Delta_xi_former_first_tmp}), 
we use the upper bound~(\ref{f_A_upper_bound}) on $f_{A}$ for 
the integral in (\ref{Delta_xi_former_first_tmp}),
\begin{align}
&\frac{|\nu_{N}\tilde{a}_{n}|}{k}
\int_{\mathcal{W}_{1}}|w|f_{A}(w + x_{\mathrm{s}}(w); v_{\tilde{N}})
p_{\mathrm{G}}(w; v_{\tilde{N}})dw
\nonumber \\
&< \frac{|\nu_{N}\tilde{a}_{n}|}{(N-k)\sqrt{2\pi v_{\tilde{N}}}}
\int_{\mathcal{W}_{1}}|w|e^{
 \frac{x_{\mathrm{s}}^{2}(w) + 2wx_{\mathrm{s}}(w)}{2v_{\tilde{N}}}
}dw
\label{xi_first_tmp2} \\
&< \frac{4v|\nu_{N}\tilde{a}_{n}|}{(N-k)\sqrt{2\pi v_{\tilde{N}}}}
\exp\left(
 \frac{\nu_{N}^{2}\tilde{a}_{n}^{2} + 2\sqrt{2v}
 |\nu_{N}\tilde{a}_{n}|}{2v_{\tilde{N}}}
\right)
\nonumber \\
&< \frac{4\sqrt{v\log (N/k)}}{(N-k)\sqrt{2\pi}}
|\nu_{N}\tilde{a}_{n}|\left(
 \frac{N}{k}
\right)^{\frac{\nu_{N}^{2}\tilde{a}_{n}^{2}}{2v} 
+ \sqrt{\frac{2}{v}}|\nu_{N}\tilde{a}_{n}|}.
\end{align}
From these results we use $|\nu_{N}\tilde{a}_{n}|\leq\epsilon$ for all 
$n\in\bar{\mathcal{S}}\setminus\mathcal{N}(\boldsymbol{a})$ to arrive at 
\begin{align}
&\frac{1}{\sqrt{k}}\sum_{n\in\bar{\mathcal{S}}\setminus\mathcal{N}(\boldsymbol{a})}
\int_{\mathcal{W}_{1}}\left|
 w\Delta f_{X}\left(
  \tilde{a}_{n}, \nu_{N}\tilde{a}_{n} + w; v_{\tilde{N}}
 \right)
\right|
p_{\mathrm{G}}(w; v_{\tilde{N}})dw
\nonumber \\
&< \frac{4\sqrt{v}}{\sqrt{2\pi}}\left\{
 \frac{\bar{f}_{U,1}^{2}(\epsilon)}{v} 
 + \frac{L}{\log (N/k)}
\right\}
\sum_{n\in\bar{\mathcal{S}}\setminus\mathcal{N}(\boldsymbol{a})}|\nu_{N}\tilde{a}_{n}|
\nonumber \\
&\cdot \frac{\{\log (N/k)\}^{3/2}}{N-k}
\left(
 \frac{N}{k}
\right)^{\frac{\epsilon^{2}}{2v} 
+ \sqrt{\frac{2}{v}}\epsilon}
\pto0
\end{align}
in the sublinear sparsity limit for sufficiently small $\epsilon>0$, 
where the last convergence follows from (\ref{Cauchy_Schwarz}). 

\subsection{Second Set $\mathcal{W}_{2}$}
We evaluate the integral in (\ref{xi_appen_tmp}) over $\mathcal{W}_{2}$ 
in the sublinear sparsity limit. Using the following upper bound:
\begin{align} 
\sqrt{k}|\Delta f_{X}(\tilde{a}_{n}, \nu_{N}\tilde{a}_{n} + w; v_{\tilde{N}})|
&\leq |f_{U}(\nu_{N}\tilde{a}_{n} + w; v_{\tilde{N}})| 
\nonumber \\
&+ |f_{U}(w; v_{\tilde{N}})|
\end{align}
on (\ref{Delta_f})---obtained from the definition of $f_{X}$ 
in (\ref{posterior_mean_estimator_X})---yields 
\begin{align}
&\frac{1}{\sqrt{k}}\int_{\mathcal{W}_{2}}
|w\Delta f_{X}(\tilde{a}_{n}, \nu_{N}\tilde{a}_{n} + w; v_{\tilde{N}})|
p_{\mathrm{G}}(w; v_{\tilde{N}})dw
\nonumber \\
&\leq \frac{1}{k}\int_{\mathcal{W}_{2}}
|wf_{U}(\nu_{N}\tilde{a}_{n} + w; v_{\tilde{N}})|p_{\mathrm{G}}(w; v_{\tilde{N}})dw
\nonumber \\
&+ \frac{1}{k}\int_{\mathcal{W}_{2}}
|wf_{U}(w; v_{\tilde{N}})|p_{\mathrm{G}}(w; v_{\tilde{N}})dw. 
\label{Delta_xi_former_last}
\end{align}

We first evaluate the first term on the upper 
bound~(\ref{Delta_xi_former_last}). 
The uniform Lipschitz-continuity assumption for $f_{U}$ implies that 
there are some $N$-independent constants $L>0$ and $C>0$ such that 
$|f_{U}(\nu_{N}\tilde{a}_{n} + w; v_{\tilde{N}})|
\leq L|\nu_{N}\tilde{a}_{n}| + L|w| + C$ holds 
for all $w\in\mathbb{R}$ and $N\in\mathbb{N}$. Thus, we have 
\begin{align}
&\frac{1}{k}\int_{\mathcal{W}_{2}}
|wf_{U}(\nu_{N}\tilde{a}_{n} + w; v_{\tilde{N}})|p_{\mathrm{G}}(w; v_{\tilde{N}})dw
\nonumber \\
&\leq \frac{2}{k}
\int_{\sqrt{2 v}}^{\infty}
(L|\nu_{N}\tilde{a}_{n}| + Lw + C)wp_{\mathrm{G}}(w; v_{\tilde{N}})dw. 
\end{align}

Contributions from $p_{\mathrm{G}}(w; v_{\tilde{N}})$ need to be evaluated 
carefully. Using the integration by parts for the indefinite integral 
$\int wp_{\mathrm{G}}(w; v_{\tilde{N}})dw 
= - v_{\tilde{N}}p_{\mathrm{G}}(w; v_{\tilde{N}}) + {\rm Const.}$ yields 
\begin{align}
&\int_{\sqrt{2 v}}^{\infty}
(L|\nu_{N}\tilde{a}_{n}| + Lw + C)wp_{\mathrm{G}}(w; v_{\tilde{N}})dw
\nonumber \\
&= \left[
- (L|\nu_{N}\tilde{a}_{n}| + Lw + C)v_{\tilde{N}}p_{\mathrm{G}}(w; v_{\tilde{N}})
\right]_{\sqrt{2 v}}^{\infty}
\nonumber \\
&+ Lv_{\tilde{N}}\int_{\sqrt{2 v}}^{\infty}p_{\mathrm{G}}(w; v_{\tilde{N}})dw
\nonumber \\
&= \frac{(L|\nu_{N}\tilde{a}_{n}| + L\sqrt{2 v} + C)\sqrt{v_{\tilde{N}}}}
{\sqrt{2\pi}}\frac{k}{N}
+ Lv_{\tilde{N}}Q\left(
 \sqrt{\frac{2v}{v_{\tilde{N}}}}
\right). 
\end{align}
We use the well-known upper bound $Q(x)<p_{\mathrm{G}}(x; 1)/x$ 
on the Q-function for all $x>0$ to obtain  
\begin{align}
&\frac{1}{k}\int_{\mathcal{W}_{2}}
|wf_{U}(\nu_{N}\tilde{a}_{n} + w; v_{\tilde{N}})|p_{\mathrm{G}}(w; v_{\tilde{N}})dw
\nonumber \\
&<\frac{2(L|\nu_{N}\tilde{a}_{n}| + L\sqrt{2 v} + C)\sqrt{v}
+ Lv_{\tilde{N}}\sqrt{2}}
{N\sqrt{2\pi\log (N/k)}}. \label{Delta_xi_former_last1}
\end{align}

We next evaluate the second term in (\ref{Delta_xi_former_last}). 
By letting $\nu_{N}\tilde{a}_{n}=0$ in (\ref{Delta_xi_former_last1}), 
we have an upper bound on the second 
term in (\ref{Delta_xi_former_last}), 
\begin{align}
&\frac{1}{k}\int_{\mathcal{W}_{2}}
|wf_{U}(w; v_{\tilde{N}})|p_{\mathrm{G}}(w; v_{\tilde{N}})dw
\nonumber \\
&<\frac{2(L\sqrt{2 v} + C)\sqrt{v} + Lv_{\tilde{N}}\sqrt{2}}
{N\sqrt{2\pi\log (N/k)}}. 
\end{align}
Combining these results, we obtain 
\begin{align} 
&\frac{1}{\sqrt{k}}\sum_{n\in\bar{\mathcal{S}}\setminus\mathcal{N}(\boldsymbol{a})}
\int_{\mathcal{W}_{2}}|w\Delta f_{X}(\tilde{a}_{n}, 
\nu_{N}\tilde{a}_{n}+w; v_{\tilde{N}})|
p_{\mathrm{G}}(w; v_{\tilde{N}})dw 
\nonumber \\
&< \frac{2L\sqrt{v}}{N\sqrt{2\pi\log (N/k)}}
\sum_{n\in\bar{\mathcal{S}}\setminus\mathcal{N}(\boldsymbol{a})}|\nu_{N}\tilde{a}_{n}|
\nonumber \\
&+ \frac{2\{2(L\sqrt{2v} + C)\sqrt{v}
+ Lv_{\tilde{N}}\sqrt{2}\}}{\sqrt{2\pi\log (N/k)}}
\pto 0
\end{align}
in the sublinear sparsity limit, where the last convergence follows from 
(\ref{Cauchy_Schwarz}). 
Thus, we have arrived at the convergence~(\ref{Delta_xi_former}). 

\section{Proof of Corollary~\ref{corollary_Bayes}}
\label{proof_corollary_Bayes}
From Lemmas~\ref{lemma_MSE}--\ref{lemma_Delta_xi} and 
$\mathrm{Pr}(|U|=\sqrt{2\bar{v}_{\Out,t}/\delta})=0$ for all 
$t\in\{0,\ldots,T\}$, Assumption~\ref{assumption_inner} holds. Thus, we can use 
Theorem~\ref{theorem_SE} to prove Corollary~\ref{corollary_Bayes}. 
It is sufficient to prove that the state evolution 
recursion~(\ref{xi_out_bar})--(\ref{Z_0t}) reduces to (\ref{Bayes_v_out_bar}) 
and (\ref{Bayes_v_in_bar}) for Bayesian GAMP, as well as 
the consistency $\tilde{v}_{\In,t+1}=\bar{v}_{\In,t+1}$ and 
the identity $\bar{\eta}_{t}=1$. 

We first note that $\bar{\eta}_{t}=1$ in (\ref{eta_bar}) follows from 
$\tilde{v}_{\In,t}=\bar{v}_{\In,t}$ and \cite{Rangan11}. 
See also \cite[Lemma 10]{Takeuchi241}. Thus, it is sufficient to prove 
$\tilde{v}_{\In,t+1}=\bar{v}_{\In,t+1}$ for $t\in\{-1,\ldots,T\}$. 
The proof is by induction. For $t=-1$ the consistency 
$\tilde{v}_{\In,0}=\bar{v}_{\In,0}=P$ is the initial condition itself. 

Suppose that the consistency $\tilde{v}_{\In,t}=\bar{v}_{\In,t}$ is correct 
for some $t\in\{0,\ldots,T-1\}$. We prove $\tilde{v}_{\In,t+1}=\bar{v}_{\In,t+1}$. 
The induction hypothesis $\tilde{v}_{\In,t}=\bar{v}_{\In,t}$ implies 
the identity $\bar{\eta}_{t}=1$. Substituting $\bar{\xi}_{\In,t}$ 
in (\ref{xi_in_bar}) with $\bar{\eta}_{t}=1$ into $\tilde{v}_{\In,t+1}$ in 
(\ref{v_in_tilde}) and using the definition of $\xi(\cdot; v_{\tilde{N}})$ in 
(\ref{xi_def}) for the Bayesian estimator, we have 
\begin{equation}
\tilde{v}_{\In,t+1} 
= \bar{v}_{\Out,t}
\mathbb{E}[\xi(\boldsymbol{x} + \boldsymbol{\omega}_{t}; v_{\tilde{N}})],
\end{equation}
with $v_{\tilde{N}}=v/\log (N/k)$ and $v=\delta^{-1}\bar{v}_{\Out,t}$, which 
follows from $M=\delta k\log (N/k)$ and 
$\boldsymbol{\omega}_{t}\sim\mathcal{N}(\boldsymbol{0}, 
M^{-1}\bar{v}_{\Out,t}\boldsymbol{I}_{N})$. 
Using the identity~(\ref{xi}) and 
the Bayesian inner denoiser~(\ref{Bayes_inner_denoiser}) with 
$\bar{\eta}_{t}=1$, we arrive at 
\begin{equation}
\tilde{v}_{\In,t+1} 
= \mathbb{E}\left[
 \left\|
  \boldsymbol{x} - f_{\In,t}(\boldsymbol{x} + \boldsymbol{\omega}_{t}; 
  \bar{v}_{\Out,t})
 \right\|^{2}
\right]
= \bar{v}_{\In,t+1}, 
\end{equation}
where the last equality follows from the definition of $\bar{v}_{\In,t+1}$ 
in (\ref{v_in_bar}) and $\bar{\eta}_{t}=1$. Thus, the consistency 
$\tilde{v}_{\In,t+1}=\bar{v}_{\In,t+1}$ holds for all $t\in\{-1,\ldots,T\}$. 

To derive the state evolution recursion~(\ref{Bayes_v_out_bar}), we reproduce 
\cite[Proof of Lemma~3]{Takeuchi241}. Let $f_{\Out,t}= 
\{Z_{t} - f_{Z}(Z_{t}, Y; \bar{v}_{\In,t})\}/\bar{v}_{\In,t}$. From 
the definitions of $\bar{v}_{\Out,t}$ in (\ref{v_out_bar}) with 
$\tilde{v}_{\In,t}=\bar{v}_{\In,t}$ and 
the Bayes-optimal outer denoiser~(\ref{Bayes_outer_denoiser}) we have 
\begin{align}
&\bar{v}_{\In,t}\bar{\xi}_{\Out,t}^{2}\bar{v}_{\Out,t} 
= \mathbb{E}\left[
 (Z_{t} - \mathbb{E}[Z | Z_{t}, Y])f_{\Out,t} 
\right]
\nonumber \\
&= \mathbb{E}\left[
 Z_{t}f_{\Out,t} 
\right]
-\mathbb{E}\left[
 \mathbb{E}[Zf_{\Out,t} | Z_{t}, Y]
\right]
\nonumber \\
&= \mathbb{E}[(Z_{t} - Z)f_{\Out,t}],
\end{align}
where the second equality follows from the fact that 
$f_{\Out,t}$ is the deterministic function of $Z_{t}$ and $Y$. Since 
$Z_{t}$ and $Z$ are zero-mean Gaussian random variables, we use 
Lemma~\ref{lemma_Stein} and the definitions of $\bar{\xi}_{\Out,t}$ 
and $\bar{\zeta}_{t}$ in (\ref{xi_out_bar}) and (\ref{zeta_bar}) to obtain 
\begin{equation}
\bar{v}_{\In,t}\bar{\xi}_{\Out,t}^{2}\bar{v}_{\Out,t} 
= \mathbb{E}[(Z_{t} - Z)Z_{t}]\bar{\xi}_{\Out,t}
- \mathbb{E}[(Z_{t} - Z)Z]\bar{\zeta}_{t}. 
\end{equation} 
Using the identity $\bar{\eta}_{t}=1$ in (\ref{eta_bar}) or equivalently 
$\bar{\zeta}_{t}=\bar{\xi}_{\Out,t}$ yields 
\begin{equation}
\bar{v}_{\Out,t} 
= \frac{\mathbb{E}[(Z_{t} - Z)^{2}]}{\bar{v}_{\In,t}\bar{\xi}_{\Out,t}}
= \frac{\mathbb{E}[B_{t}^{2}]}{\bar{v}_{\In,t}\bar{\xi}_{\Out,t}}
= \frac{1}{\bar{\xi}_{\Out,t}},
\end{equation} 
where the second and last equalities follow from the definition of $B_{t}$ 
in (\ref{outer_measurement}) and $B_{t}\sim\mathcal{N}(0, \bar{v}_{\In,t})$, 
respectively. Thus, we use $\bar{\xi}_{\Out,t}$ in (\ref{xi_out_bar}) 
to arrive at 
the state evolution recursion~(\ref{Bayes_v_out_bar}). 

Finally, we prove the state evolution recursion~(\ref{Bayes_v_in_bar}). 
Since $\boldsymbol{\omega}_{t}\sim\mathcal{N}(\boldsymbol{0}, 
M^{-1}\bar{v}_{\Out,t}\boldsymbol{I}_{N})$ with $M=\delta k\log (N/k)$ holds  
in (\ref{mu_tt}) and (\ref{mu_0t}), 
from Assumption~\ref{assumption_lemma} we can apply Lemma~\ref{lemma_MSE} 
with $v=\delta^{-1}\bar{v}_{\Out,t}$ to $\bar{v}_{\In,t+1}$ in (\ref{v_in_bar}) 
with $\bar{\eta}_{t}=1$ to obtain the state evolution 
recursion~(\ref{Bayes_v_in_bar}). 
Thus, Corollary~\ref{corollary_Bayes} holds. 

\section*{Acknowledgment}
The author thanks the anonymous reviewers for their suggestions that have 
improved the quality of the manuscript greatly. 

\bibliographystyle{IEEEtran}
\bibliography{IEEEabrv,kt-it2024_1}

%\bf{If you include a photo:}\vspace{-33pt}
%\begin{IEEEbiography}[{\includegraphics[width=1in,height=1.25in,clip,keepaspectratio]{fig1}}]{Michael Shell}
%Use $\backslash${\tt{begin\{IEEEbiography\}}} and then for the 1st argument use $\backslash${\tt{includegraphics}} to declare and link the author photo. Use the author name as the 3rd argument followed by the biography text.
%\end{IEEEbiography}



%\vfill

\end{document}